\documentclass{aa}  

\usepackage{graphicx}
\usepackage{hyperref}
\usepackage{natbib} 
\usepackage{xcolor}
\bibpunct{(}{)}{;}{a}{}{,} 

\usepackage[varg]{txfonts}

\newcommand{\kms}{km\,s$^{-1}$}

\newcommand{\cms}{cm$^{-2}$}
\newcommand{\cmc}{cm$^{-3}$}
\newcommand{\rsol}{R$_{\odot}$}
\newcommand{\lsol}{L$_{\odot}$}
\newcommand{\msol}{M$_{\odot}$}
\newcommand{\msolyr}{M$_{\odot}$\,yr$^{-1}$}

\newcommand{\lbol}{$L_{\rm bol}$}
\newcommand{\lint}{$L_{\rm int}$}

\newcommand{\macc}{$\dot{M}_{\rm acc}$}
\newcommand{\Vjet}{$V_{\rm jet}$}
\newcommand{\Rjet}{$R_{\rm jet}$}
\newcommand{\ljet}{$L_{\rm jet}$}
\newcommand{\mjet}{$\dot{M}_{\rm jet}$}
\newcommand{\pjet}{$\dot{P}_{\rm jet}$}
\newcommand{\Fco}{$F_{\rm CO}$}
\newcommand{\Fso}{$F_{\rm SO}$}
\newcommand{\Fsio}{$F_{\rm SiO}$}
\newcommand{\Nco}{$N_{\rm CO}$}
\newcommand{\Nso}{$N_{\rm SO}$}
\newcommand{\Nsio}{$N_{\rm SiO}$}
\newcommand{\Xco}{$X_{\rm CO}$}
\newcommand{\Xso}{$X_{\rm SO}$}
\newcommand{\Xsio}{$X_{\rm SiO}$}

\begin{document}

   \title{The CALYPSO IRAM-PdBI survey of jets from Class 0 protostars}
\subtitle{Exploring whether jets are ubiquitous in young stars\thanks{the calibrated data and continuum and line maps analysed in the paper are part of the IRAM PdBI Large Program CALYPSO and are available on the IRAM webpage at the following link:
\url{https://www.iram.fr/ILPA/LP010/}
} }

   \author{L. Podio
          \inst{1}
          \and
          B. Tabone
          \inst{2}
          \and
          C. Codella
          \inst{1, 3}
          \and
          F. Gueth
          \inst{4}
          \and
          A. Maury
          \inst{5, 6}
          \and
          S. Cabrit
          \inst{7, 3}
          \and
          B. Lefloch
          \inst{3}
          \and
          S. Maret
          \inst{3}
          \and
          A. Belloche
          \inst{8}
          \and
          P. Andr{\'e}
          \inst{5}
          \and
          S. Anderl
          \inst{3}
          \and    
          M. Gaudel
          \inst{5}
          \and
          L. Testi
          \inst{9}
          }

   \institute{
     INAF - Osservatorio Astrofisico di Arcetri, Largo E. Fermi 5, 50125 Firenze, Italy \\ 
     \email{lpodio@arcetri.astro.it}
\and 
Leiden Observatory, Leiden University, PO Box 9513, 2300 RA Leiden, The Netherlands
\and
Univ. Grenoble Alpes, CNRS, IPAG, 38000 Grenoble, France
\and
Institut de Radioastronomie Millim\'etrique (IRAM),38406 Saint-Martin-d’H{\`e}res, France
\and
Laboratoire d’Astrophysique (AIM), CEA, CNRS, Universit{\'e} Paris-Saclay, Universit{'e} Paris Diderot, Sorbonne Paris Cit{\'e}, 91191 Gif-sur-Yvette, France
\and
Harvard-Smithsonian Center for Astrophysics, Cambridge, MA02138, USA
\and
PSL Research University, Sorbonne Universit{\'e}s, Observatoire de Paris, LERMA, CNRS, Paris France
\and
Max-Planck-Institut f{\"u}r Radioastronomie, Auf dem H{\"u}gel 69, 53121 Bonn, Germany
\and
ESO, Karl Schwarzschild Stra{\ss}e 2, 85748 Garching bei M{\"u}nchen, Germany
    }

   \date{Received ; accepted }

 
  \abstract
{}
{As a part of the CALYPSO Large Program, we aim to constrain the properties of protostellar jets and outflows by searching for corresponding emission in a sample of 21 Class 0 protostars covering a broad range of internal luminosities (\lint\, from 0.035 \lsol\, to 47 \lsol).
}
{We analyzed high angular ($\sim 0\farcs5-1\farcs0$) resolution IRAM Plateau de Bure Interferometer (PdBI) observations in three typical outflow and jet tracers, namely: CO ($2-1$), SO ($5_6-4_5$), and SiO ($5-4$). We obtained the detection rate, spatial distribution, kinematics, and collimation of the outflow and jet emission in the three lines. Molecular column densities and abundances, the jet mass-loss and momentum rates, and mechanical luminosity are estimated from the integrated line intensities.}
{
Blue- and red-shifted emission in CO ($2-1$), which probes outflowing gas, was detected in all the sources in the sample and detected for the first time in the cases of SerpS-MM22 and SerpS-MM18b. Collimated high-velocity jets in SiO ($5-4$) were detected in 67\% of the sources (for the first time in IRAS4B2, IRAS4B1, L1448-NB,  SerpS-MM18a) and 77\% of these also showed jet or outflow emission in SO ($5_6-4_5$). The detection rate of jets in SiO and SO increases with internal luminosity. 
In five sources (that is, 24\% of the sample) SO ($5_6-4_5$) is elongated and reveals a velocity gradient perpendicular to the jet direction, hence, it probes the inner envelope or the disk, or both. The detected SiO jets are collimated (typical opening angle, $\alpha \sim 10\degr$) and surrounded by wider angle SO ($\alpha \sim 15\degr$) and CO ($\alpha \sim 25\degr$) emission. The abundance of SO relative to H$_2$ ranges from $<10^{-7}$ up to $10^{-6}$; whereas for SiO, the lower limits were found to range  from $> 2.4 \times 10^{-7}$ to $> 5 \times 10^{-6}$ , with the exception of the jets from IRAS4A1 and IRAS4A2, which show low SiO abundance ($\le2-6 \times 10^{-8}$). The mass-loss rates range from $\sim 7 \times 10^{-8}$ \msolyr\, up to $\sim 3 \times 10^{-6}$ \msolyr\, for $L_{\rm int} \sim 1-50$ \lsol.}
{The CALYPSO IRAM-PdBI survey shows that the outflow phenomenon is ubiquitous in our sample of protostars and that the detection rate of high-velocity collimated jets increases for increasing protostellar accretion, with at least 80\% of the sources with $L_{\rm int} > 1$ \lsol\, driving a jet. The detected protostellar flows exhibit an onion-like structure, where the SiO jet is nested into a wider angle SO and CO outflow. On scales $> 300$~au the SiO jets are less collimated ($4\degr-12\degr$) than atomic jets from Class II sources ($\sim 3\degr$), possibly due to projection effects and contamination by SiO emission from the bow-shocks. On the other hand, velocity asymmetry between the two jet lobes are detected in one third of the sources,
similarly to Class II atomic jets, suggesting that the same launching mechanism is at work.  Most of the jets are SiO rich, which indicates very efficient release of $>1\%-10\%$ of elemental silicon in gas phase likely in dust-free winds, launched from inside the dust sublimation radius. 
The estimated mass-loss rates are larger by up to five orders of magnitude than what was measured for Class II jets, 
indicating that the ejection decreases as the source evolves and accretion fades. Similarly to Class II sources, the mass-loss rates are $\sim 1\%-50\%$ of the mass accretion rate, \macc, suggesting that the correlation between mass ejection and mass accretion holds along the star-formation process from $10^4$ yr up to a few Myr.}

               \maketitle
%

\section{Introduction}
\label{sect:intro}
The earliest stages of the process leading to the formation of a low-mass, Sun-like star are expected to be associated with supersonic (around 100 \kms) collimated jets. This ejection phenomenon is invoked to remove angular momentum from the protostar-disk system and allow the material to  accrete from the disk onto the central object \citep[see, e.g., ][ and references therein]{frank14}. 
Jets, in turn, accelerate the dense material of the cloud surrounding the protostar creating slower ($\sim 10$ \kms) molecular outflows, which are observable on a large scale (typically 0.1 pc) mainly through CO low-J rotational lines \citep[see, e.g., ][]{lada85}.

Although these jets are thought
to be driven by a magneto-centrifugal process which extracts and accelerates the material from the rotating star-disk system, the launching mechanism is still far from being clear.
As a consequence, what is also unclear is the region the jet is launched from: whether it comes from inside the dust sublimation radius at fractions of au from the star or from a larger region of the accretion disk.
In the first case, the molecular jet may be driven by the stellar surface \citep["stellar wind", ][]{glassgold91} or the inner disk region, either by an X-wind \citep{shang07} or by a dust-free magneto-hydrodynamical (MHD) disk wind as was shown in the recent modeling carried out by \citet{tabone20}. In the second case, rather, the molecular jet originates from an extended disk region as in the models of dusty magnetized disk winds \citep{pudritz07,panoglou12}.


In the last 15 years there have been a number of surveys in the (sub-)millimeter range aimed to address different aspects of the star-formation process at its earliest stages, that is, the protostellar multiplicity, the magnetic field topology in star-forming regions, and the presence of disks and outflows. These surveys  mainly targeted continuum emission and CO low-J transitions at intermediate resolution ($3-4\arcsec$) \citep[e.g., ][]{jorgensen07,hull14,lee-k16,tobin16a,tobin18} and have shown that large-scale outflow emission is commonly associated with protostellar objects. However, the lack of angular resolution had not always allowed us to investigate whether all protostellar sources in multiple systems are associated with outflows and to reveal the outflow-accelerating engine, that is, the high-velocity collimated jet, which is directly ejected from the star-disk system and, therefore, is key in regulating the accretion and ejection process and the angular momentum removal.

To date, there have only been a few detailed studies at sub-arcsecond resolution that have targeted the primary molecular jet in a few prototypical protostellar objects, such as in HH 211, HH 212, L1448-C, L1157, or B335, by using selective jet tracers, such as SiO \citep{cabrit07b,cabrit12,codella07,codella14b,hirano06,hirano10,podio15,podio16,lee07a,lee07b,lee09b,lee17b,bjerkeli19}.
These studies revealed that protostellar jets may be as collimated as atomic jets from Class II sources observed at optical/NIR wavelengths \citep{dougados00,bacciotti00,woitas02,cabrit07a,agraamboage11} and they show a similar correlation between the mass ejection and mass accretion rates. In a few cases, they also show signatures of rotation \citep[e.g., ][and references therein]{lee17b,bjerkeli19,lee20}.
Recently, a survey of five molecular outflows in the Serpens star forming region was performed with ALMA \citep{tychoniec19}. However, a statistical study at sub-arcsecond angular resolution of a large sample of protostellar jets is still lacking.



It is now time to enlarge the sample of observed protostellar jets and to perform a statistical study on jets and outflows occurrence and properties using selected molecular line tracers in the mm-spectral window.
To this aim, here we perform a survey of 21 protostars in the most active star-forming regions visible in the northern hemisphere conducted with IRAM PdB array in the context of the CALYPSO large program\footnote{\url{http://irfu.cea.fr/Projects/Calypso}\\ 
\url{https://www.iram-institute.org/EN/content-page-317-7-158-240-317-0.html}}, targeting three lines which are typical outflow and jet tracers, that is, CO ($2-1$), SO ($5_6-4_5$), and SiO ($5-4$).
The first goal of this effort is to answer a simple but crucial question regarding whether jets, and, in general, mass ejection phenomena are commonly observed in Class 0 protostars. The second goal is to derive the jet properties, that is, the jet velocity and width, and the molecular abundances, which are crucial for understanding what region of the disk-protostar system jets are launched from and for constraining models of jet launching, as well as to reconstruct the mass ejection and mass accretion history from the protostellar to the pre-main sequence stage.
Finally, a third goal is to obtain a large database for follow-up observations at the extremely high-spatial resolution as the observations performed for the HH 212 jet \citep[e.g., ][]{codella14b,lee17b}, for instance. 

The paper is structured as follows. The sample of protostars covered by the CALYPSO Large Program and analyzed in the context of this paper is presented in Sect. \ref{sect:sample}. Then in Sect. \ref{sect:obs}, we describe the acquired observations and the data reduction process. The methodology we applied to establish the occurrence of outflows and jets and to derive their properties and the obtained results are presented in Sect. \ref{sect:results}. Then in Sect. \ref{sect:discussion}, we discuss the occurrence and the properties of the detected protostellar jets and we compare them with those of jets from pre-main sequence sources. Finally, we summarize our conclusions in Sect. \ref{sect:conclusions}.

\section{The sample}
\label{sect:sample}

The CALYPSO survey was carried out with the IRAM-PdB interferometer towards 16 fields centered on known Class 0 protostars (i.e., 10$^4$--10$^5$~yr old solar analogue protostars, \citealt{andre00,andre10}), observed at 94 GHz, 219 GHz, and 231 GHz.
The targeted sources are all located in the Gould Belt clouds at $d< 450$ pc.
Seven of the targeted fields are located in the most active sites of star formation in the Perseus cloud: that is,  L1448 (2A, NB, and C objects) and NGC1333 (IRAS2A, SVS13, IRAS4A, and IRAS4B).
In addition, we observed four sources located in different portions of the Serpens Main and South regions: SerpM-S68N, SerpM-SMM4, SerpS-MM18, and SerpS-MM22. 
The selected sample contains also:
(i) three Class 0 sources in Taurus (IRAM04191, L1521-F, and L1527),
(ii) L1157, located in the Cepheus constellation, and driving the prototypical chemically rich outflow, and
(iii) the GF9--2 protostar, located in the east-west filamentary dark cloud GF9.
Several of these fields are associated with more than one protostar (e.g., L1448-NA and L1448-NB are in the same field), or to multiple systems (e.g., L1448-NB1 and L1448-NB2) identified from the analysis of the millimeter continuum emission in the CALYPSO maps by \citet{maury19}.  Therefore the CALYPSO sample consists of 25 Class 0 protostars, four Class I\ and one continuum source whose nature  is still unknown (VLA3), as summarized in Table \ref{tab:sample}. 
In the table, we report for each source the position of the continuum peak at 1.3~mm and the systemic velocity ($V_{\rm sys}$) extracted from the CALYPSO dataset \citep{maury19,maret20}, the distance (d), the internal luminosity ($L_{\rm int}$), the mass of the envelope ($M_{\rm env}$), and the Class (0 or I).
Following \citet{dunham08}, the internal luminosity is derived by Ladjelate et al. (in prep.) using the 70 $\mu$m measurements provided by the {\it Herschel} Gould Belt survey \citep{andre10} and is a more reliable probe of the accretion luminosity than the bolometric luminosity, \lbol, because it is not affected by external heating of the envelope by the interstellar radiation field, $L_{\rm ext}$ ($L_{\rm bol} = L_{\rm int} + L_{\rm ext}$). The latter adds on average a few tenths of a solar luminosity and significantly contributes to \lbol\, in sources with low accretion luminosity.
Protostars belonging to multiple systems are grouped together in Table \ref{tab:sample} and the same systemic velocity and distance of the primary protostar (marked in boldface) is assumed. 
For the binaries with no mass estimate of their individual envelopes, the value corresponds to the fraction of the total envelope mass in proportion of the peak flux densities at 1.3~mm (see \citealt{maury19,belloche20}).
As shown in Table \ref{tab:sample} the CALYPSO sample covers a wide range of internal luminosities (\lint\, from 0.035 to 47 \lsol), and envelope masses ($M_{\rm env}$ from 0.5 to 9.9 \msol), which makes it a unique laboratory for the study of the occurrence and properties of collimated jets as a function of the properties of the driving protostars.

Of the 30 sources identified by \citet{maury19} in the CALYPSO observations, 7 are tentative protostellar candidates, reported in parenthesis in Table \ref{tab:sample}.
Between these, 2 are part of a close binary system, that is, they are located less than one beam away from the primary (L1448-2Ab and L1448-NB2). In these cases, we consider the binary system as a single source for the assessment of the outflow or jet occurrence as it is not possible at the resolution of our dataset to understand whether only one or both components of the close binary systems are driving an outflow or jet.
Moreover, the two Class 0 and the Class I protostars indicated by $^{**}$ in Table \ref{tab:sample} lie outside of the primary beam at 231 GHz (L1448-NW, SVS13C, and SerpM-S68Nc). This causes a strong attenuation in the emission, possibly leading to the non-detection of faint line emission from jets. In fact, previous observations at lower resolutions covering a larger field-of-view report slowly outflowing extended gas in CO ($2-1$) and CO ($1-0$) for two of these sources (the Class 0 protostars L1448-NW and SVS13C, \citealt{lee15,plunkett13}), which is not detected in our CALYPSO CO maps due to attenuation.
As we want to assess the jet occurrence in an homogeneous sample where all sources are observed down to roughly the same sensitivity threshold, we excluded these three sources from our statistics.
Also, VLA3 is not included in the sample for the outflow or jet survey because its protostellar nature and its internal luminosity cannot be derived as it cannot be separated from the much brighter source SVS13A in the Herschel maps. No outflow associated with this continuum source was detected in the CALYPSO maps presented by \citet{lefevre17}.
Based on the above criteria, the sample analyzed to investigate the jet occurrence consists of 21 Class 0 and 3 Class I protostars, which are those reported in Table \ref{tab:jet-occurrence}.
Due to the small number, for the Class I sources the analysis is limited to establish the detection rate of outflows and jets and the derivation of their position angle (PA) (see Sect. \ref{sect:detection-rate}) but they are not further discussed in this paper, which focuses on the properties of jets from Class 0 protostars.

\begin{table*}
\caption{Properties of the protostars identified from the analysis of the CALYPSO continuum maps (see \citealt{maury19}).}
\begin{center}
\begin{tabular}{lccccccc}
\hline \hline
Source  & $\alpha({\rm J2000})$$^a$  & $\delta({\rm J2000})$$^a$ & $V_{\rm sys}$$^b$ & $d^c$  
& $L_{\rm int}$$^d$ & $M_{\rm env}$$^e$ & Class$^f$ \\
        & ($^h$ $^m$ $^s$) & ($\degr$ $\arcmin$ $\arcsec$) & (km s$^{-1}$) & (pc) 
& (L$_{\odot}$) &  (M$_{\odot}$) & \\
\hline
{\bf L1448-2A} & 03 25 22.405 & +30 45 13.26 & +4.2 & 293 & 4.7 (0.5)   & 1.2 & 0 \\
\vspace{0.3cm} 
(L1448-2Ab)    & 03 25 22.355 & +30 45 13.16 &      &     & $<4.7$      & 0.6 & 0 \\ 
{\bf L1448-NB1} & 03 25 36.378 & +30 45 14.77 & +4.9 & 293 & $<3.9$     & 3.3 & 0 \\ 
(L1448-NB2)     & 03 25 36.315 & +30 45 15.15 &      &     & $3.9$      & 1.6 & 0 \\ 
L1448-NA        & 03 25 36.498 & +30 45 21.85 &      &     & 6.4 (0.6)  & 0.8 & I \\ 
\vspace{0.3cm} 
L1448-NW$^{**}$ & 03 25 36.680 & +30 45 33.86 &      &     &  --        & --  & 0 \\ 
{\bf L1448-C}   & 03 25 38.875 & +30 44 05.33 & +5.1 & 293 &  11 (1)    & 1.9 & 0 \\
\vspace{0.3cm} 
L1448-CS        & 03 25 39.132 & +30 43 58.04 &      &     & 3.6        & 0.16& I \\ 
\vspace{0.3cm} 
{\bf IRAS2A1}   & 03 28 55.570 & +31 14 37.07 & +7.5 & 293 & 47 (5)     & 7.9 & 0 \\ 
{\bf SVS13B}    & 03 29 03.078  & +31 15 51.74 & +8.5 & 293 & 3.1 (1.6) & 2.8 & 0 \\
SVS13A          & 03 29 03.756  & +31 16 03.80 &      &     & 44 (5)    & 0.8 & I \\ 
SVS13C$^{**}$   & 03 29 01.980  & +31 15 38.14 &      &     &  --       & --  & 0 \\ 
\vspace{0.3cm} 
(VLA3)          & 03 29 03.378  & +31 16 03.33 &      &     &  --       & --  & unknown\\
{\bf IRAS4A1}   & 03 29 10.537  & +31 13 30.98 & +6.3 & 293 & $<4.7$    & 9.9 & 0 \\ 
\vspace{0.3cm} 
IRAS4A2         & 03 29 10.432  & +31 13 32.12 &      &     & 4.7 (0.5) & 2.3 & 0 \\ 
{\bf IRAS4B1}   & 03 29 12.016  & +31 13 08.02 & +6.8 & 293 & 2.3 (0.3) & 3.3 & 0 \\
\vspace{0.3cm} 
(IRAS4B2)       & 03 29 12.841  & +31 13 06.84 &      &     & $<0.16$   & 1.4 & 0 \\ 
\vspace{0.3cm} 
{\bf IRAM04191} & 04 21 56.899  & +15 29 46.11 & +6.7 & 140 & 0.05 (0.01)& 0.5 & 0 \\ 
\vspace{0.3cm} 
{\bf L1521-F}   & 04 28 38.941  & +26 51 35.14 & +6.6 & 140 & 0.035 (0.01)& 0.7 & 0 \\
\vspace{0.3cm} 
{\bf L1527}     & 04 39 53.875  & +26 03 09.66 & +5.7 & 140 & 0.9 (0.1) &  1.2 & 0 \\
{\bf SerpM-S68N}& 18 29 48.091  & +01 16 43.41 & +9.2 & 436 & 11 (2)    & 11   & 0 \\
$^*$SerpM-S68Nb & 18 29 48.707  & +01 16 55.53 &      &     & 1.8 (0.2) & --   & 0 \\ 
\vspace{0.3cm} 
$^*$(SerpM-S68Nc)$^{**}$& 18 29 48.811& +01 17 04.24& &     & 1.4 (0.2) & --   & I \\ 
{\bf SerpM-SMM4a}& 18 29 56.716 & +01 13 15.65 & +8.8 & 436 &  2.2 (0.2)& 6.7  & 0 \\
\vspace{0.3cm} 
(SerpM-SMM4b)    & 18 29 56.525 & +01 13 11.58 &      &     &  $<2.6$   & 1.0  & 0 \\ 
{\bf SerpS-MM18a}& 18 30 04.118 & --02 03 02.55& +8.1 & 350 &  13 (4)   & 4.5  & 0 \\
\vspace{0.3cm} 
(SerpS-MM18b)    & 18 30 03.541 & --02 03 08.33&      &     &  16 (4)   & 0.9  & 0 \\ 
\vspace{0.3cm} 
{\bf SerpS-MM22} & 18 30 12.310 & --02 06 53.56& +6.2 & 350 & 0.4 (0.2) & 0.9  & 0 \\
\vspace{0.3cm} 
{\bf L1157}      & 20 39 06.269 & +68 02 15.70 & +2.6 & 352 & 4.0 (0.4) & 3.0  & 0 \\
{\bf GF9-2}      & 20 51 29.823 & +60 18 38.44 & -3.0 & 474 &  1.7      & 2.8  & 0 \\
\hline
\end{tabular}
\end{center}
$^a$ Positions of the 1.3 mm continuum peak emission are extracted from the CALYPSO dataset \citep{maury19}. 
$^b$ Systemic velocities, $V_{\rm sys}$, correspond to the mean velocity of C$^{18}$O ($2-1$) emission on the source continuum peak position as fit by \citet{maret20}, except for GF9-2 (from $^{13}$CO ($2-1$)), IRAM04191 \citep{belloche02}, L1521-F (from NH$_3$ (1,1), \citealt{codella97}), and SerpM-SMM4 (from CO ($3-2$), \citealt{dionatos10b}). For multiple systems the systemic velocity of the primary is reported.
$^c$ Distances: \citet{ortiz-leon18a} for the Perseus sources; \citet{zucker19} for the Taurus sources and L1157; \citet{ortiz-leon17,ortiz-leon18b} for SerpM; Palmeirim et al., in prep. for SerpS; C. Zucker, priv. comm. for GF9-2. For multiple systems the distance of the primary is reported.
$^d$ Internal luminosities are derived by Ladjelate et al., in prep., from the 70 $\mu$m flux from the {\it Herschel} Gould Belt survey observations at 8$\arcsec$ spatial resolution \citep{andre10}, except for GF9-2 for which we use the value by \citet{wiesemeyer97} rescaled to the distance given in the fifth column. The uncertainty is in parentheses when available, and is larger for SVS13B because of the proximity to SVS13A. For SerpM-SMM4b the upper limit is given by the bolometric luminosity \citep{aso19}.
$^e$ Envelope mass from \citet{maury19} and references therein. For the binaries with no mass estimate of their individual envelopes, the value corresponds to the fraction of the total envelope mass in proportion of the peak flux densities at 1.3 mm given by \citet{maury19}. The masses have been re-scaled to the distances given in column 5.
$^f$ The classification as Class 0, I, or candidate protostellar object (in parentheses) is based on \citet{maury19}.
$^*$ For SerpM-S68Nb and SerpM-S68Nc we follow the same naming as \citet{maury19}. The names of the two sources are inverted with respect to the classification by \citet{williams00}, also followed by \citet{dionatos10b}.
$^{**}$ protostellar companions which lie outside of the primary beam at 231 GHz ($\sim 21\arcsec$).
\label{tab:sample}
\end{table*}

\section{Observations and data reduction}
\label{sect:obs}

The CALYPSO sources were observed with the IRAM-PdBI during several tracks between September 2010 and March 2013 using the six-antenna array in the most extended (A) and intermediate (C) configurations. The shortest and longest baselines are 16 m and 762 m, respectively, allowing us to recover emission at scales from $\sim$ 8$\arcsec$ down to $\sim$ 0$\farcs$4.
WideX\footnote{\url{https://www.iram-institute.org/EN/content-page-120-4-35-47-118-120.html}}
backends were used to cover the full 3.8 GHz spectral window at the spectral resolution of 2 MHz ($\sim$ 2.6 km s$^{-1}$ at 1.3 mm) for three spectral setups centered at 231, 219, and 94 GHz (corresponding to 1.30, 1.37, and 3.19~mm, respectively).
In addition, higher resolution backends ($\sim$ 0.1 km s$^{-1}$, subsequently smoothed to 1 km s$^{-1}$) were employed to observe the  CO ($2-1$), SO ($5_6-4_5$), and  SiO ($5-4$) lines (see Table \ref{tab:lines}).
The phase root mean square (rms) was typically $\le$ 50$\degr$ and 80$\degr$ for the A and C tracks, respectively, the precipitable water vapor (pwv) was 0.5–-1 mm (A) and $\sim$ 1--2 mm (C), and system temperatures were usually $\sim$ 100--160 K (A) and 150--250 K (C).
Calibration was carried out following standard procedures, using GILDAS-CLIC\footnote{\url{https://www.iram.fr/IRAMFR/GILDAS/}}. Strong quasars such as 3C273 and 3C454.3 were used to calibrate the correlator bandpass, while the absolute flux density scale was mainly estimated by observing MWC349 and 3C84, with a final uncertainty less than 15\%.
The continuum emission was removed from the visibility tables to produce continuum-free line tables.
Self-calibration  was performed on the source continuum emission for all sources with bright dust continuum emission, that is, with continuum peak flux $>80$ mJy beam$^{-1}$ at 231 GHz (see Table 3 of \citealt{maury19}) and the self-calibrated phase solutions were applied to the continuum-subtracted line tables.
We carried out imaging with the GILDAS/MAPPING software, adopting a robust weighting parameter of 1\footnote{For more information on weighting schemes performed by the GILDAS/MAPPING software, see 
\url{https://www.iram.fr/IRAMFR/GILDAS/doc/html/map-html/node32.html}
}
and obtaining typical synthesized beams of $0\farcs4-0\farcs9$, except for the weakest continuum emission sources in the sample (i.e., IRAM04191, L1521F, and GF9-2), for which we used a natural weighting to maximize the sensitivity to point sources obtaining synthesized beams of $0\farcs7-1\farcs0$ \citep{maury19}. These beam sizes correspond to an angular resolution of: $\sim 130-220$ au for the sources in Perseus ($d=293$ pc, \citealt{ortiz-leon18a}) and Taurus (two out of the three sources in Taurus have weak continuum emission, hence the larger beam sizes are compensated by the smaller distance, $d=140$ pc, \citealt{zucker19}, with the exception of L1527, for which an angular resolution of $\sim 60-130$ au is reached); and of $\sim 220-350$ au for the sources located in Serpens ($d=436$ pc for Serpens M, \citealt{ortiz-leon17,ortiz-leon18b}, $d=350$ pc for Serpens S, Palmeirim et al., in prep.), Cepheus ($d=352$ pc, \citealt{zucker19}), and GF9 ($d=474$ pc, C. Zucker, priv. comm.). The synthetized beam and corresponding angular resolution for the 16 targeted fields are summarized in Tabs. \ref{tab:beam_CO}, \ref{tab:beam_SO}, \ref{tab:beam_SiO} and shown in Fig. \ref{fig:jets1}.
The present paper is based on analysis of the 1.3 mm and 1.4 mm observations of CO ($2-1$), SiO ($5-4$), and SO ($5_6-4_5$), which are three standard tracers of molecular jets and outflows, hence they are ideal when tackling the array of open questions on protostellar jets presented in Sect. \ref{sect:intro}. As the jet and outflow emission is faint and spread over a wide range of blue-shifted and red-shifted velocities, we preferentially analyzed the WideX datacubes, which are at lower spectral resolution to increase the signal-to-noise ratio of the line emission. The WideX datacubes were resampled at a resolution of 3.25 \kms (CO ($2-1$)), 3.4 \kms (SO ($5_6-4_5$), and SiO ($5-4$)), reaching a typical rms noise per channel for the final continuum-subtracted datacubes of $\sim$ 2--10 mJy beam$^{-1}$. In some cases, we analyzed the higher resolution datacubes (1 \kms). The spectral resolution and the rms noise per channel of the CO ($2-1$), SO ($5_6-4_5$), and SiO ($5-4$) datacubes for the 16 targeted fields are listed in Tables. \ref{tab:beam_CO}, \ref{tab:beam_SO}, and \ref{tab:beam_SiO}.

\begin{table}
\caption{Properties of the lines targeted by the CALYPSO survey.}
\begin{tabular}[h]{ccccc}
\hline
\hline
Line         & Frequency$^{a}$ & $E_{\rm up}$$^{a}$ & log$_{10}$($A_{\rm ij}$)$^{a}$ & $n_{\rm cr}$$^{b}$ \\
               & (MHz)                 & (K)                       & (s$^{-1}$)                              & (\cmc)    \\    
\hline
CO ($2-1$)       & 230538.000 & 16.6         & $-6.2$                & $7.3 \times 10^3$ \\  
SO ($5_6-4_5$) & 219949.442 & 35.0         & $-3.9$                & $7.7 \times 10^5$ \\  
SiO ($5-4$)      & 217104.980 & 31.3         & $-3.3$                & $1.6 \times 10^6$ \\  
\hline
\end{tabular}\\
\small
$^{a}$ molecular parameters from the CDMS database \citep{muller01}. \\
$^{b}$ the critical densities are computed at T=100 K, using collisional rate coefficients from \citet{yang10} (CO), \citet{lique06} (SO), and \citet{balanca18} (SiO). \\
\label{tab:lines}
\end{table}

\section{Methodology and results}
\label{sect:results}

In this section, we present the methodology we applied in analysing the CALYPSO CO ($2-1$), SO ($5_6-4_5$), and SiO ($5-4$) line cubes and our obtained results (discussed in Sect. \ref{sect:discussion}).
The integrated line emission maps for all the Class 0 sources in our sample are presented in Sect. \ref{sect:detection-rate}, Figure \ref{fig:jets1}. From the maps, we establish the detection rate of outflows and jets (Fig. \ref{fig:jet-occurrence}), we estimate the position angles of the detected flows (Table \ref{tab:jet-occurrence}), and extract position-velocity (PV) diagrams of the emission along them  (Fig. \ref{fig:PV-block1}).
For the sources where an SiO jet is detected at $>10\sigma$ in the integrated maps\footnote{\label{note1} The contours of the integrated maps shown in Fig. \ref{fig:jets1} are from $5\sigma$ with steps of $5\sigma$, therefore a $5\sigma$ detection corresponds to one contour, a $10\sigma$ detection to two contours. The $5\sigma$ level of the integrated maps is different for each source and tracer as it depends on the rms noise per channel of the corresponding datacube (as reported in Tabs. \ref{tab:beam_CO}, \ref{tab:beam_SO}, and \ref{tab:beam_SiO}) and on the velocity interval over which the emission is integrated. The velocity interval of integration and the corresponding $5\sigma$ level are labeled in each map in Fig. \ref{fig:jets1} and spans between $10-50$ mJy\,\kms\,beam$^{-1}$ when line emission is detected only in one channel to $\sim100-1700$ mJy\,\kms\,beam$^{-1}$ when emission is integrated on several velocity channels (up to 21 channels, i.e. $\sim 70$ \kms, for L1448-C). The $5\sigma$ levels reported in Fig.  \ref{fig:jets1} are rounded.}, we estimate the jet properties, that is, their velocity and spatio-kinematical structure (Sect. \ref{sect:kinematics}, Fig. \ref{fig:distri-vrad}), the width and opening angle (Sect. \ref{sect:jet-width}, Figs. \ref{fig:jet-width_all}, \ref{fig:jet-width-main}, \ref{fig:jet-width-hv}), the molecular column densities and abundances (Sect. \ref{sect:jet-abundances}, Table \ref{tab:jets-energetics}, Fig. \ref{fig:jet-abundances}), and the jet energetics, that is, the jet mass-loss and momentum rates and mechanical luminosities (Sect. \ref{sect:jet-energetics}, Table \ref{tab:jets-energetics}, Fig. \ref{fig:jet-energetics}). 



\begin{figure*}
\begin{centering}
\includegraphics[width=13.5cm]{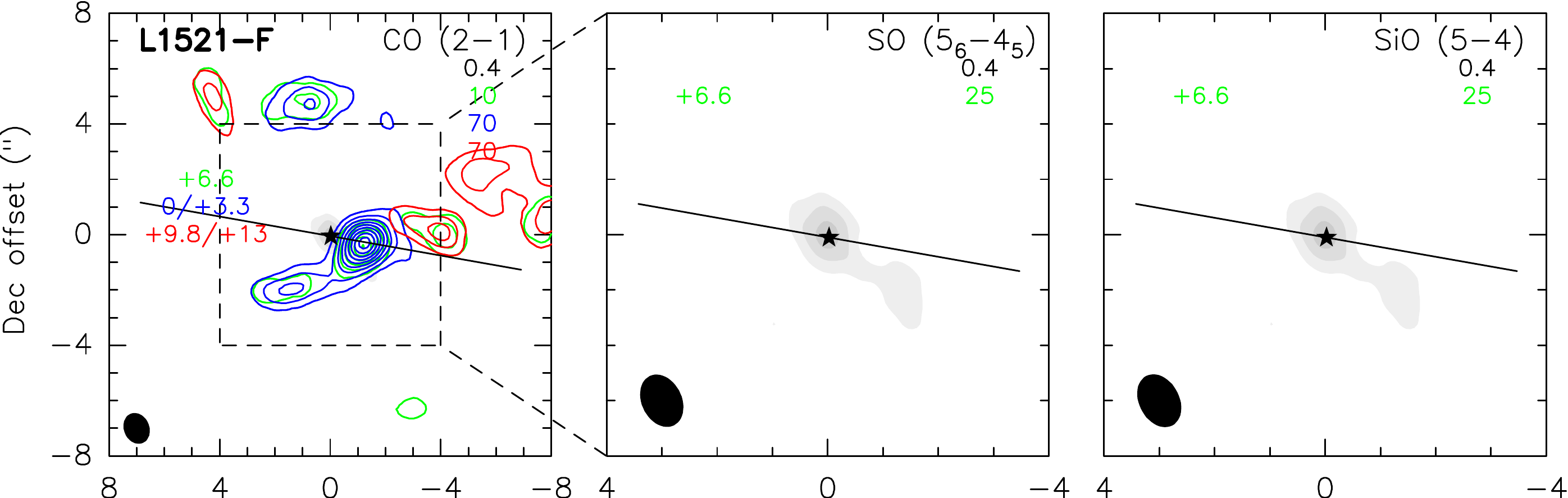}
\vspace{0.2cm}\\
\includegraphics[width=13.5cm]{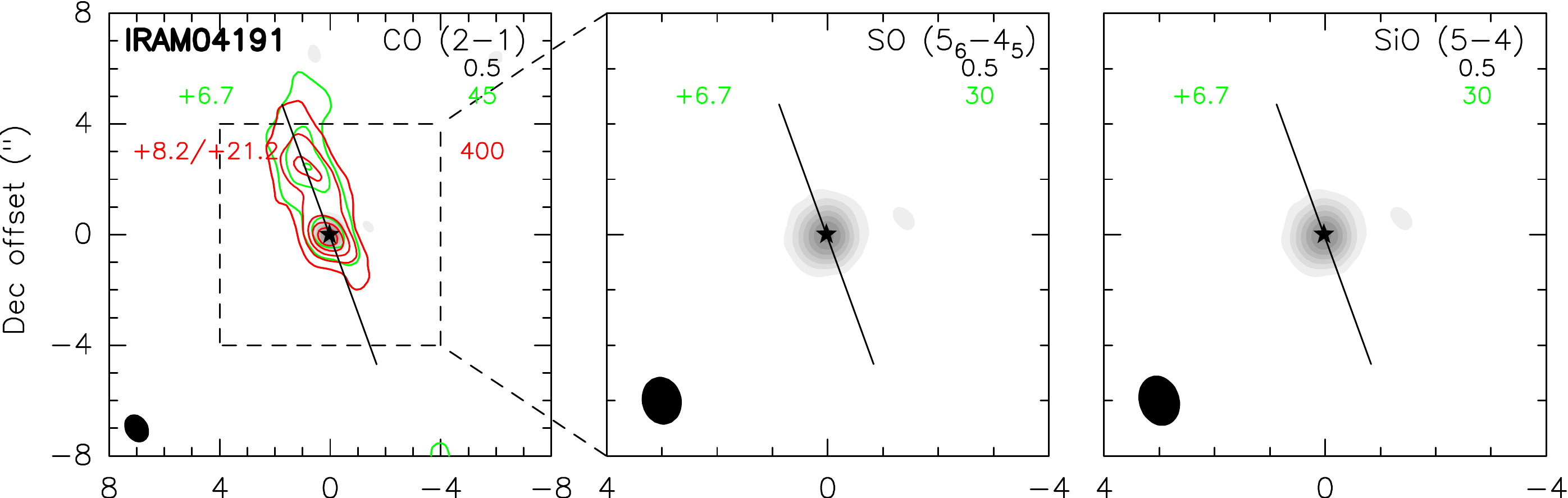}
\vspace{0.2cm}\\
\includegraphics[width=13.5cm]{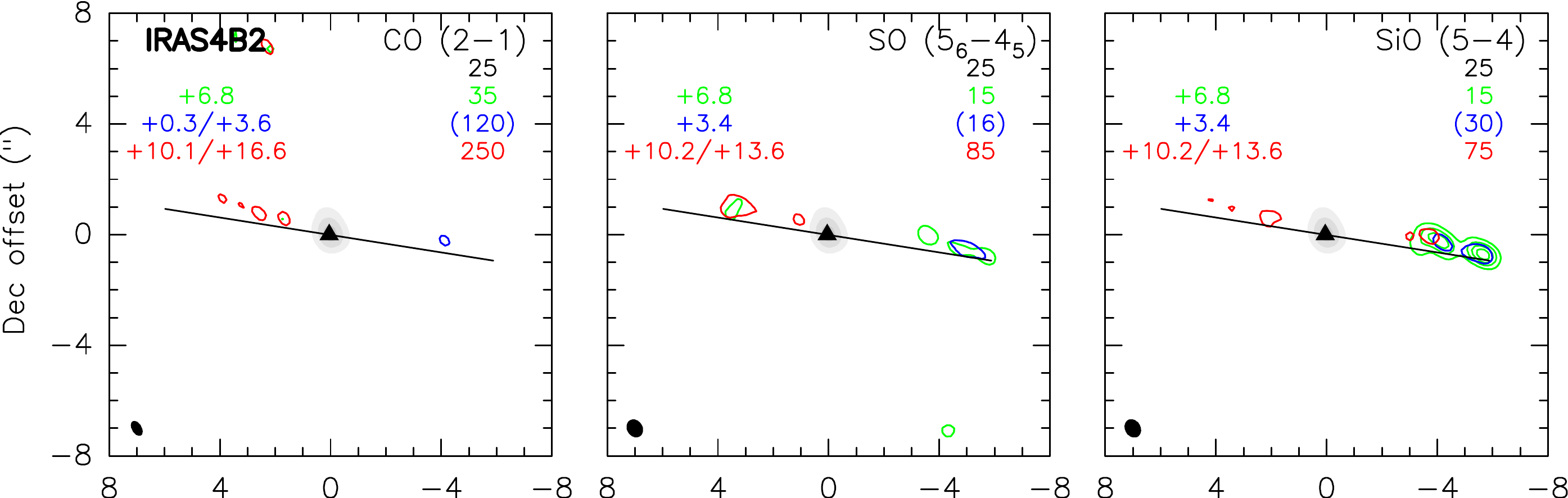}
\vspace{0.2cm}\\
\includegraphics[width=13.5cm]{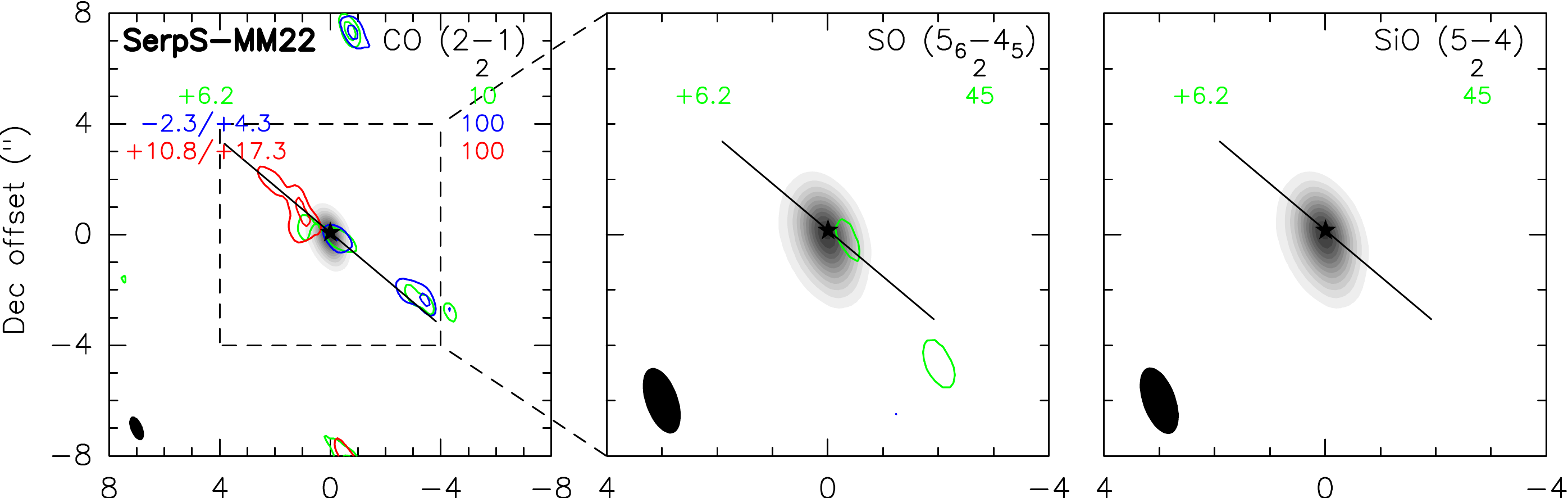}
\vspace{0.2cm}\\
\includegraphics[width=13.5cm]{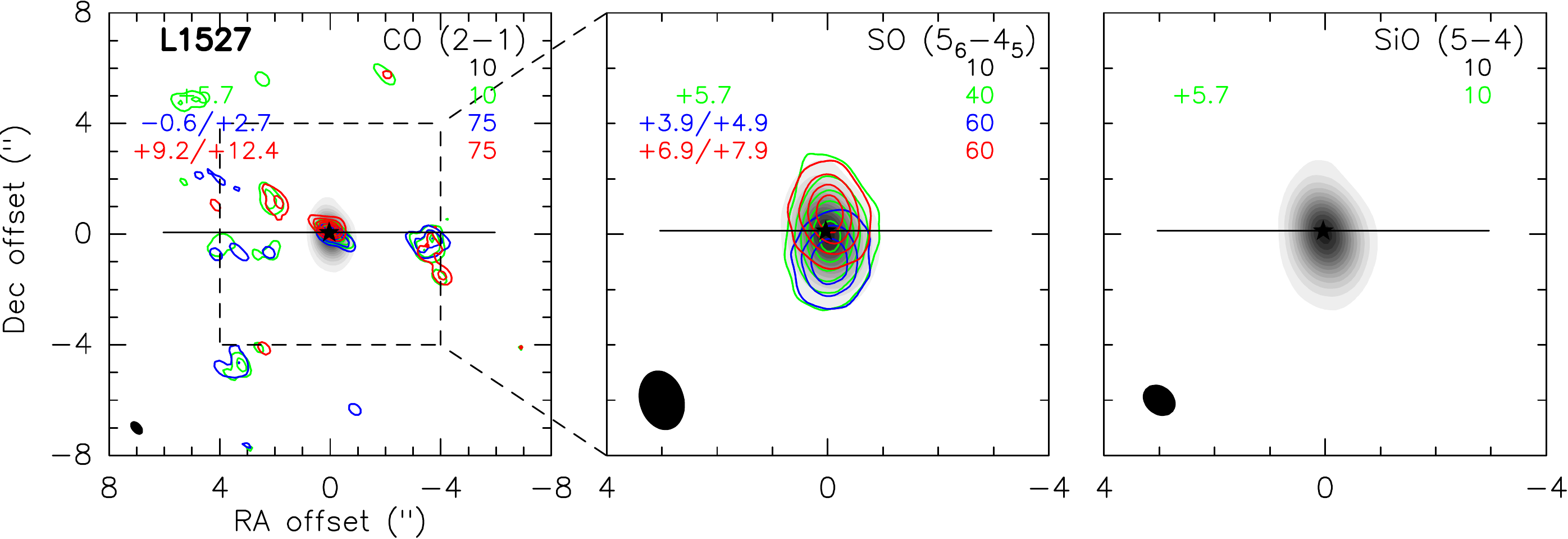}
\caption{Integrated maps of the CO ($2-1$), SO ($5_6-4_5$), and SiO ($5-4$) emission for the Class 0 protostars of the CALYPSO sample ({\it left, center, and right panels}). Continuum at 1.3~mm (grey scale) and integrated line emission at systemic, blue-, and red- shifted velocities (green, blue, and red contours) are shown. The systemic velocity (one channel) and the velocity of the first and last channels over which blue- and red-shifted emission was integrated (in \kms) are labeled in the top-left corner (in green, blue, and red respectively). When the emission is detected on a single channel, its central velocity is labeled. The 5$\sigma$ intensity of the corresponding integrated emission (in mJy \kms\, beam$^{-1}$) is labeled in the top-right corners with the same colour coding. The 5$\sigma$ intensity of the continuum (in mJy beam$^{-1}$) is also labeled in black. The contours are from 5$\sigma$ with steps of 5$\sigma$. When the emission is faint the contours are from 3$\sigma$ with steps of 3$\sigma$ and the corresponding values are indicated in parentheses. The black stars (or triangles) indicate the positions of the protostars (or candidate protostars) identified by \citet{maury19}, the black solid line shows the jet or outflow PA. The beam size is shown in the bottom-left corner.}
\label{fig:jets1}
\end{centering}
\end{figure*}

\begin{figure*}
\setcounter{figure}{0}
\begin{centering}
\includegraphics[width=13.5cm]{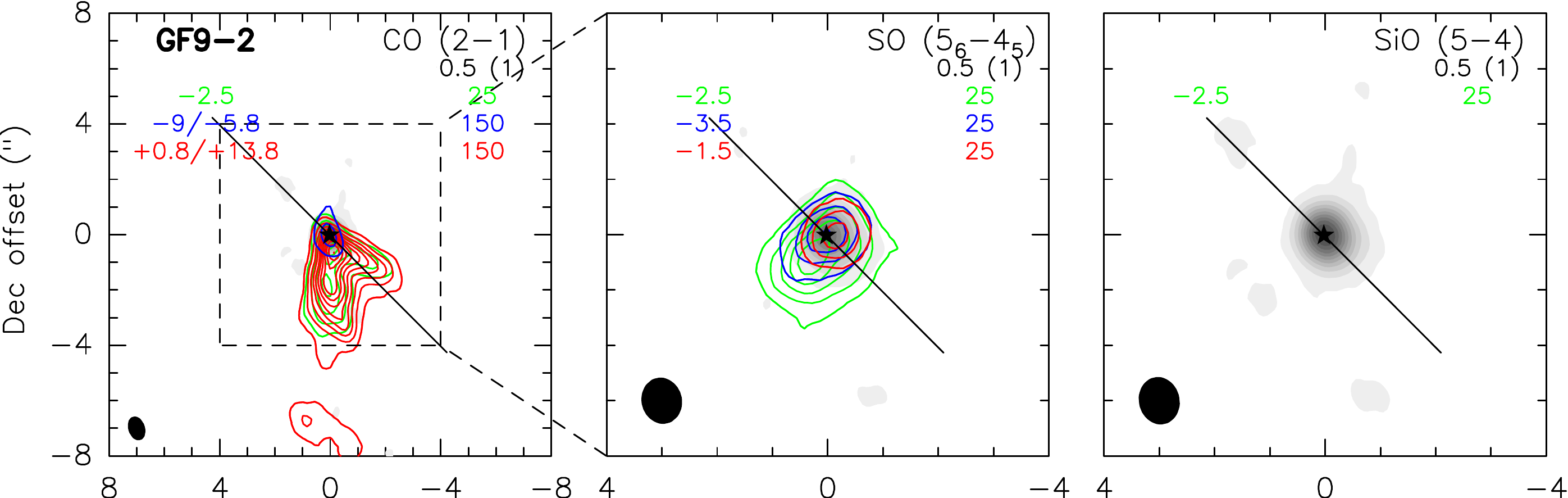}
\vspace{0.2cm}\\
\hspace{0.4cm}
\includegraphics[width=13.1cm]{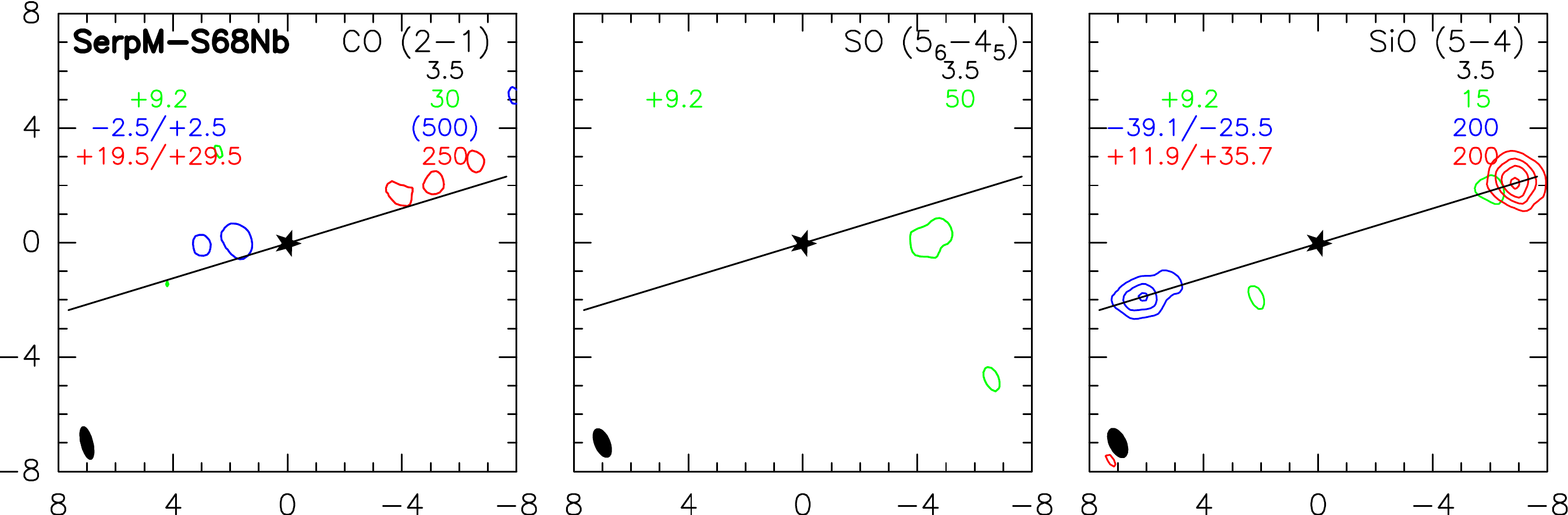}
\vspace{0.2cm}\\
\includegraphics[width=13.3cm]{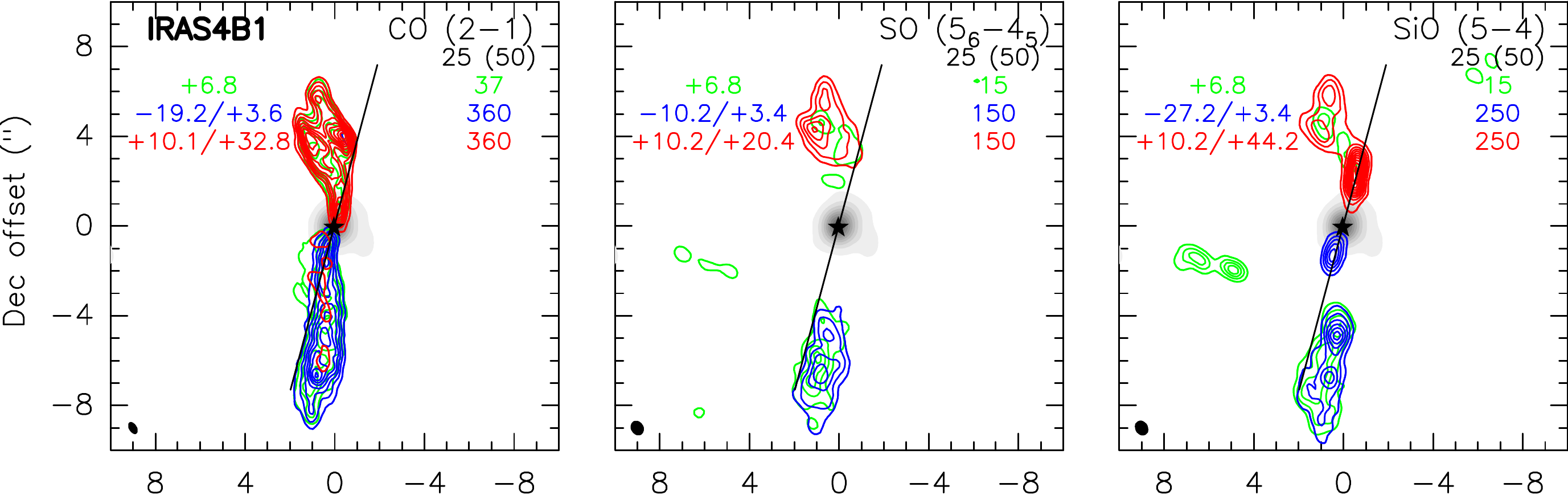}
\vspace{0.2cm}\\
\includegraphics[width=13.5cm]{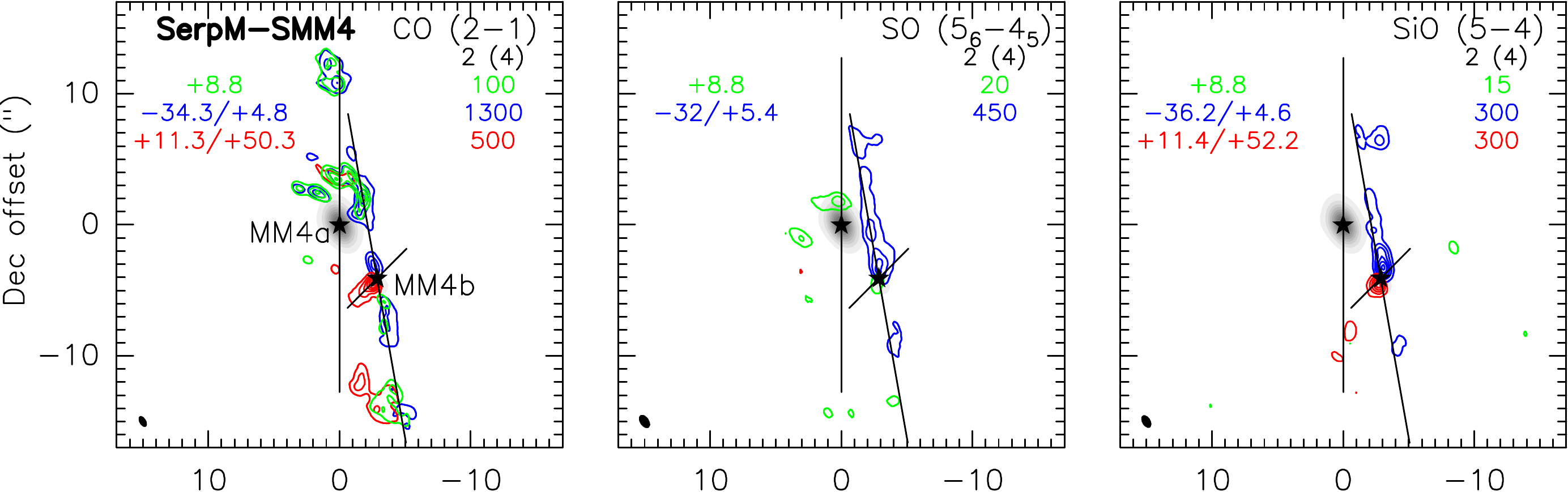}
\vspace{0.2cm}\\
\includegraphics[width=13.5cm]{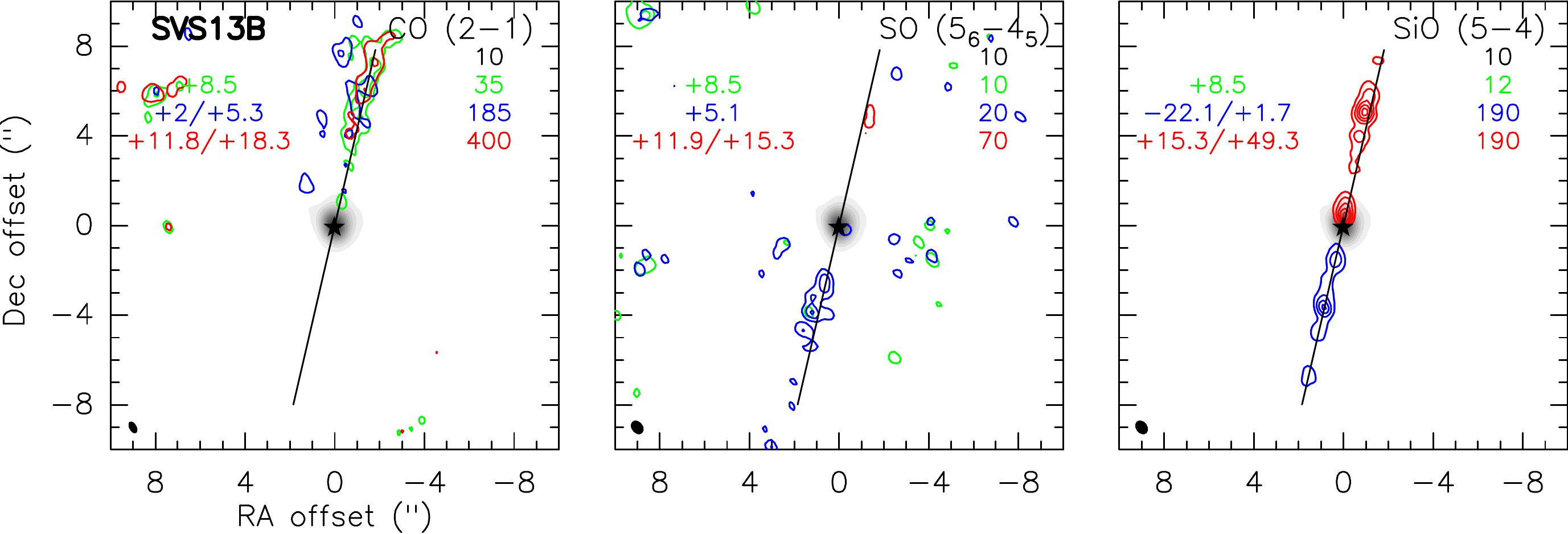}
\caption{{\it Continued}}
\label{fig:jets2}
\end{centering}
\end{figure*}

\begin{figure*}
\setcounter{figure}{0}
\begin{centering}
\includegraphics[width=13.5cm]{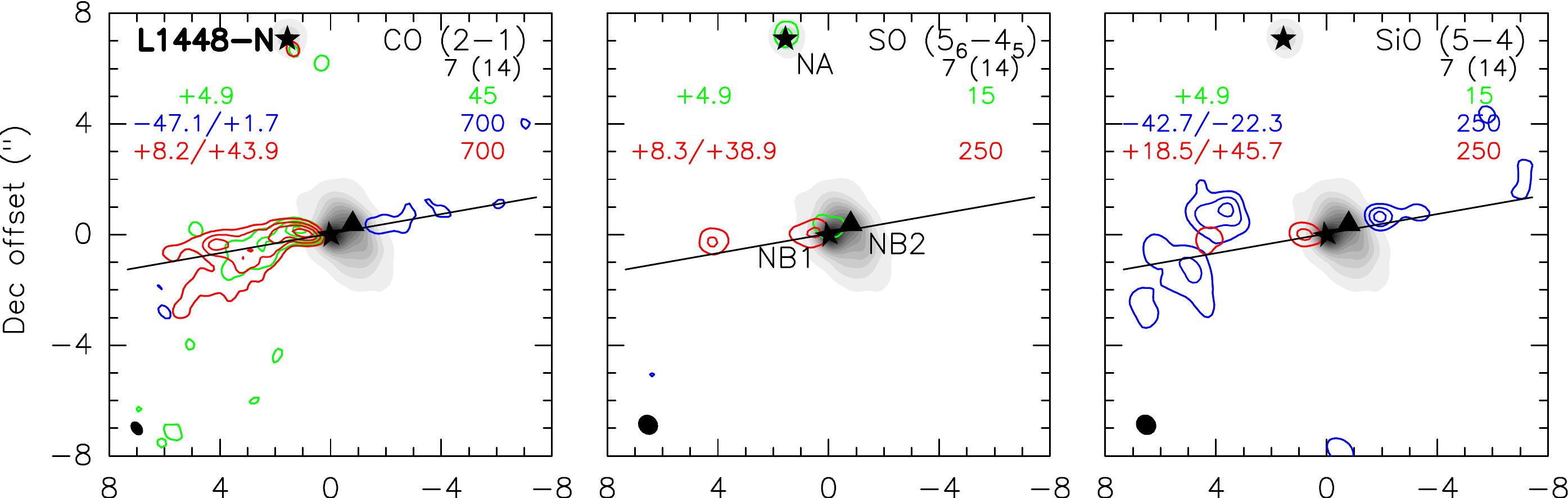}
\vspace{0.2cm}\\
\includegraphics[width=13.5cm]{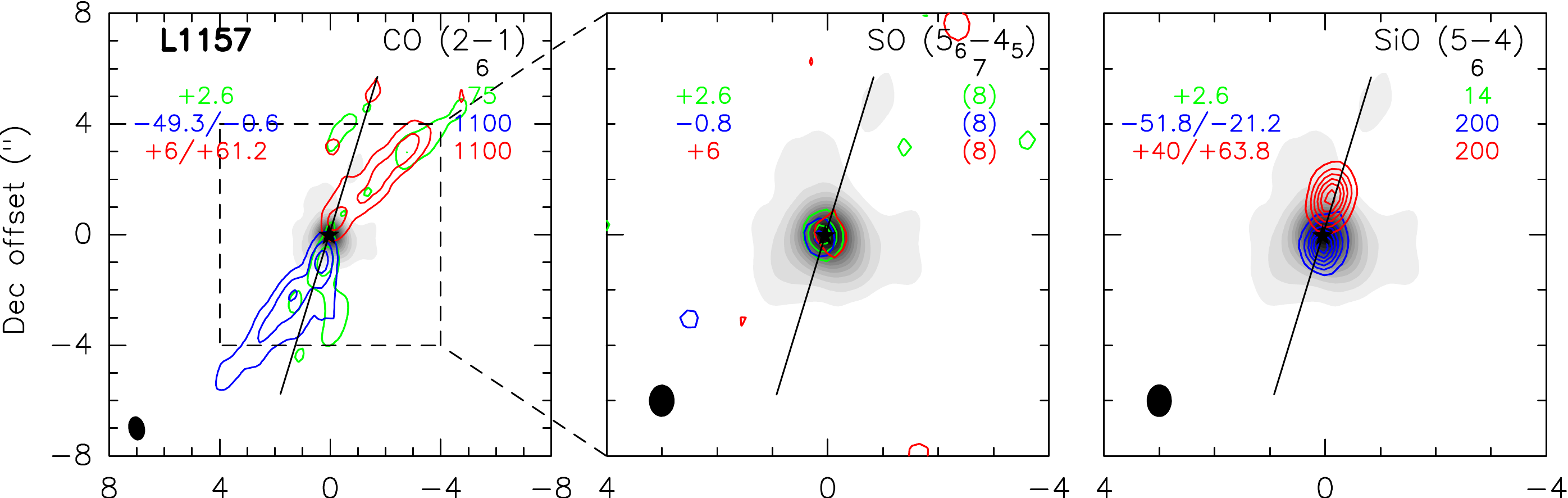}
\vspace{0.2cm}\\
\includegraphics[width=13.5cm]{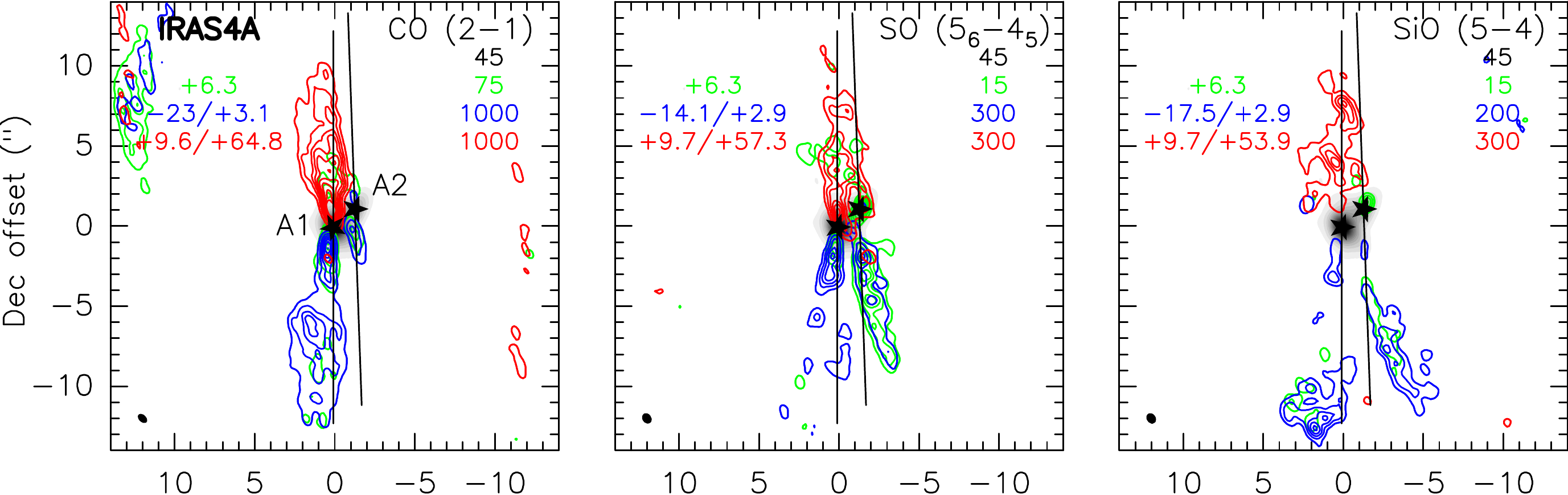}
\vspace{0.2cm}\\
\includegraphics[width=13.5cm]{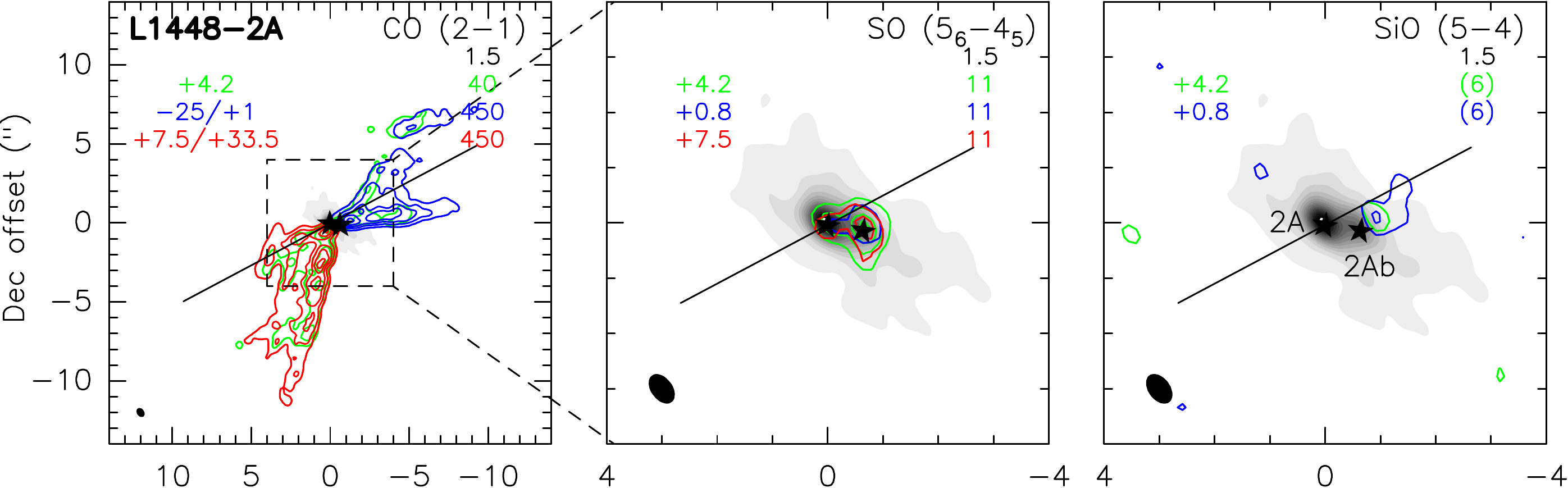}
\vspace{0.2cm}\\
\includegraphics[width=13.5cm]{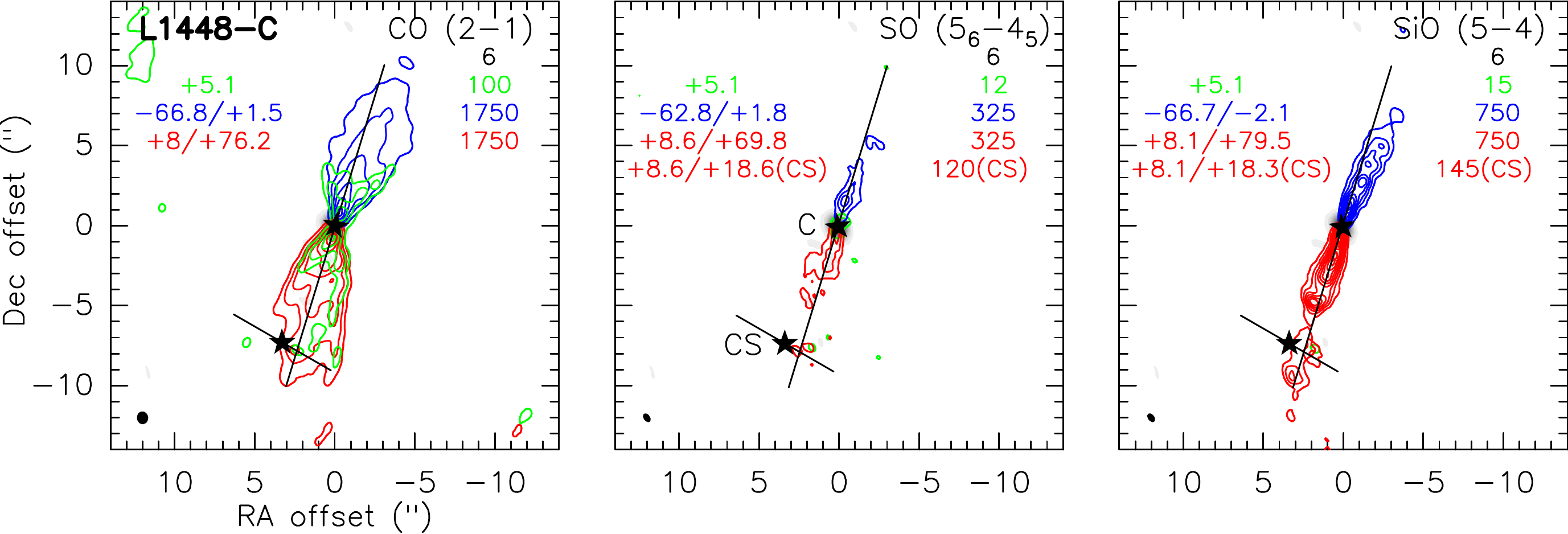}
\caption{{\it Continued}}
\label{fig:jets3}
\end{centering}
\end{figure*}

\begin{figure*}
\setcounter{figure}{0}
\begin{centering}
\includegraphics[width=13.5cm]{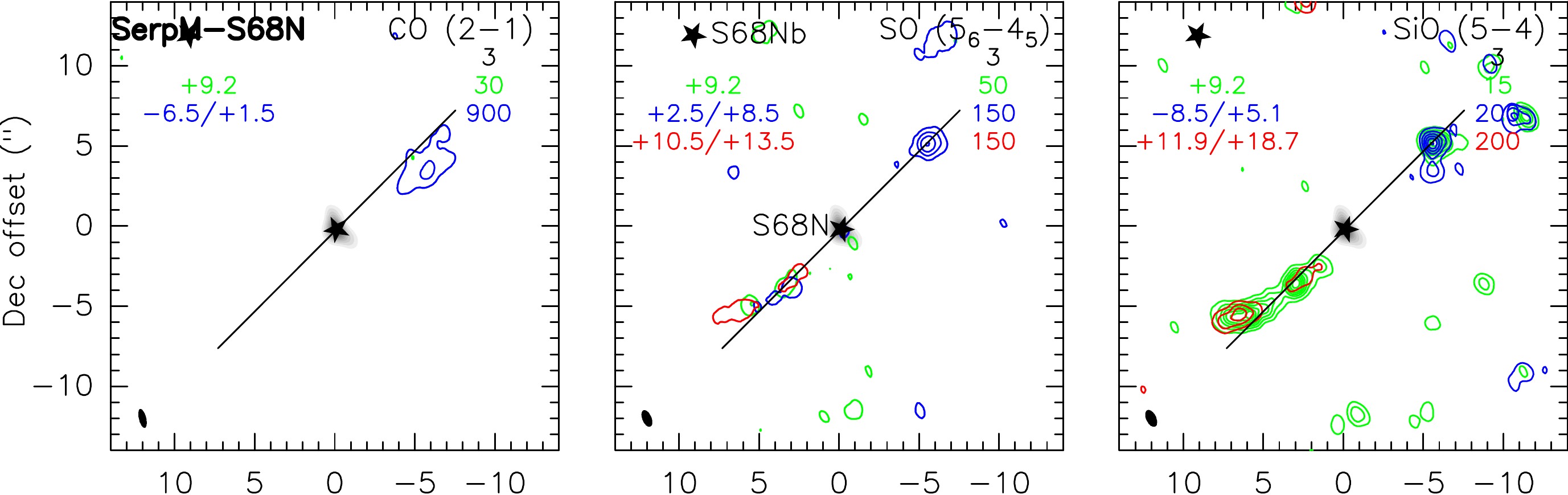}
\vspace{0.2cm}\\
\includegraphics[width=13.5cm]{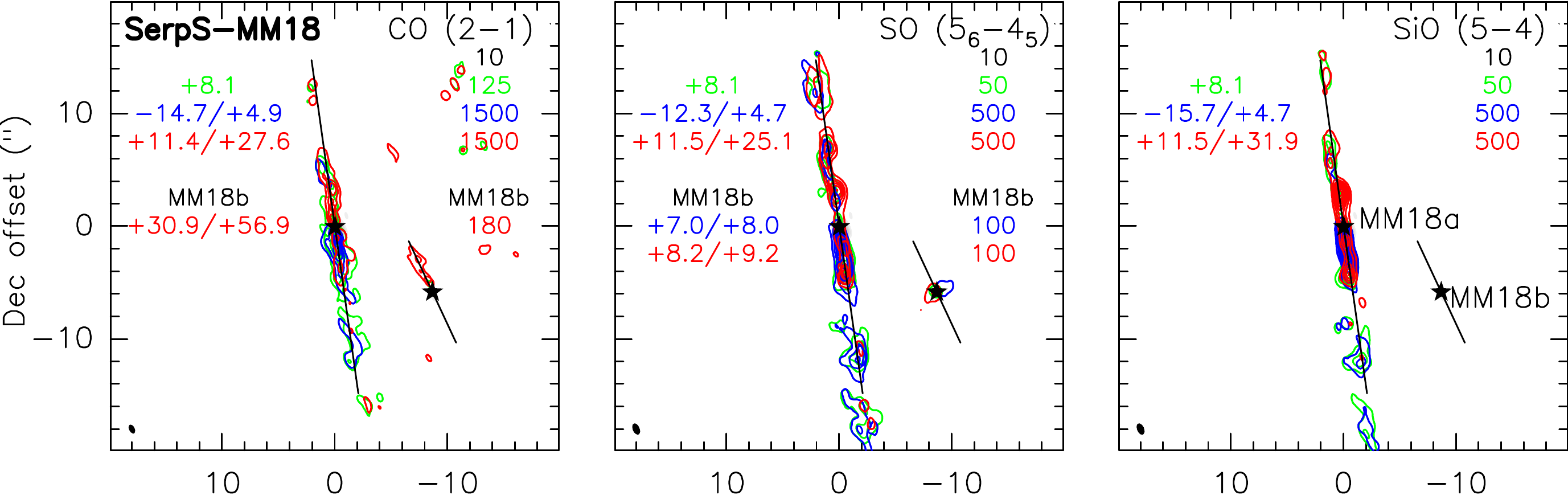}
\vspace{0.2cm}\\
\includegraphics[width=13.5cm]{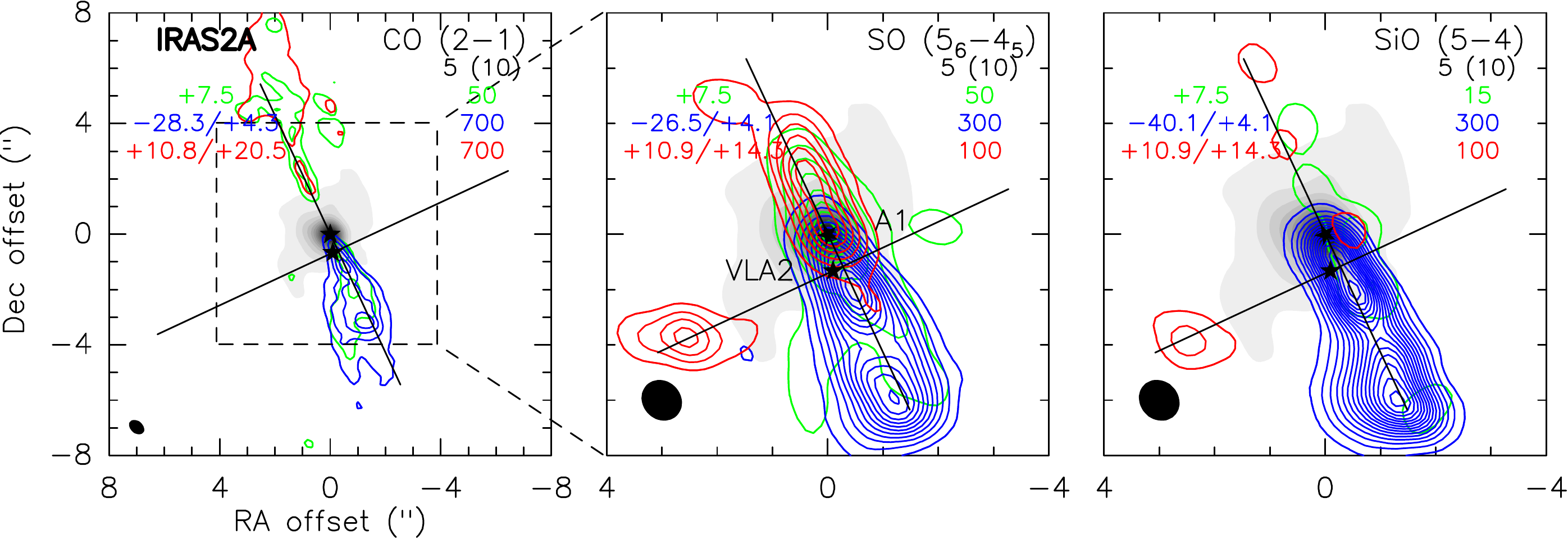}
\caption{{\it Continued}}
\label{fig:jets4}
\end{centering}
\end{figure*}

\subsection{Detection rate of outflows and jets}
\label{sect:detection-rate}

\begin{figure*}
\centering
\includegraphics[width=.95\textwidth]{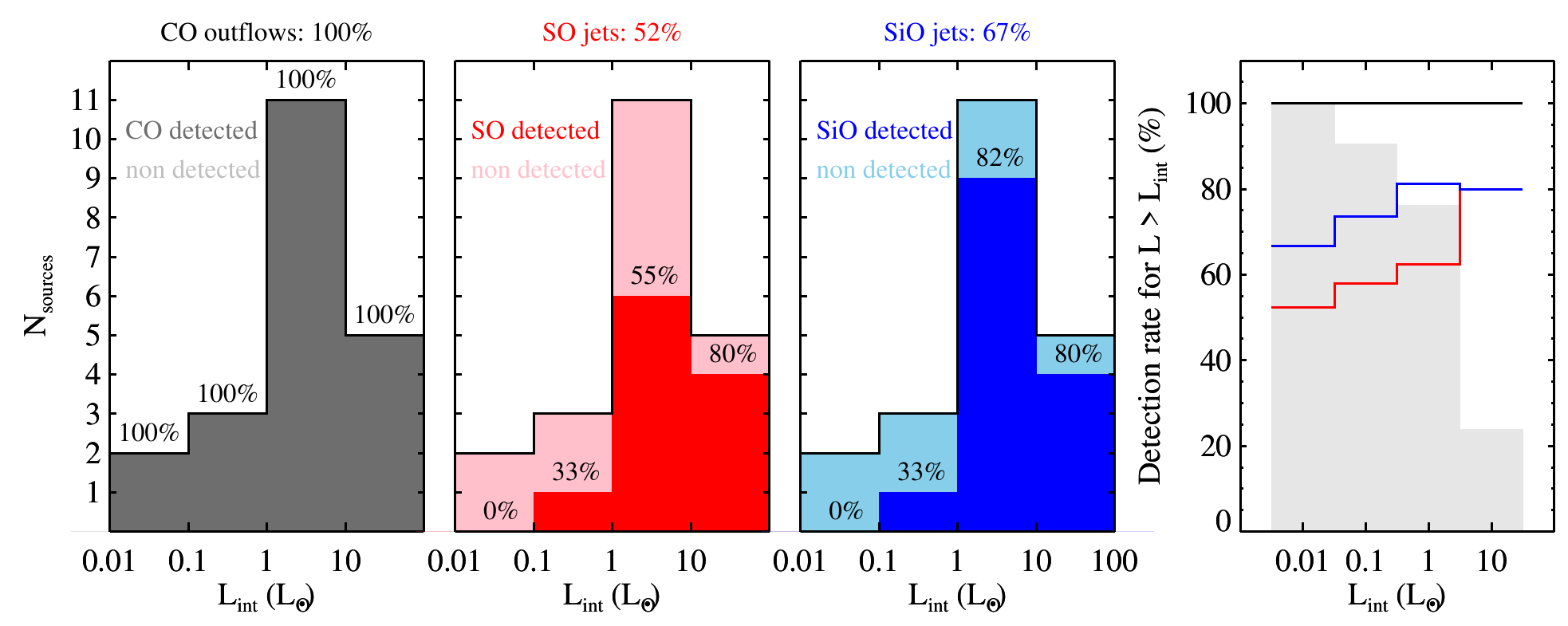}
\caption{Number of sources associated with outflows and jets as traced by CO ($2-1$) (black, {\it left}), SO ($5_6 - 4_5$) (red, {\it middle-left}), and SiO ($5-4$) (blue, {\it middle-right}) as a function of the internal luminosity, $L_{\rm int}$, of the 21 Class 0 protostars in the CALYPSO sample. The detection rate for each $L_{\rm int}$ bin (in logarithmic scale) is reported above the histogram, while the detection rate for the whole sample is labeled on top of each panel. 
The detection rate of jets and outflows for $L > L_{\rm int}$ is shown in the right panel, with the same colour coding. The grey histogram shows the fraction of sources for $L > L_{\rm int}$.}
\label{fig:jet-occurrence}
\end{figure*}

\begin{table*}
  \caption{Detection of outflows and jets as traced by CO ($2-1$), SO ($5_6 - 4_5$), and SiO ($5-4$) lines in the CALYPSO sample of Class 0 and I protostars.}
  \begin{tabular}[h]{cccccccc}
    \hline
    \hline
    Source & $L_{\rm int}$ & COMs$^{a}$ & Disk$^{b}$ &  \multicolumn{3}{c}{Outflow/Jet$^{c}$} & PA$_{\rm jet/outflow}$$^{d}$ \\
           & (\lsol)       &            &            & CO & SO & SiO            & ($\degr$)\\
    \hline
    \multicolumn{8}{c}{{\bf Class 0}}\\
    L1521-F            & $0.035 (0.01)$ &   &   & Y  &    &    & $+260$(B)$^{e}$  \\
    IRAM04191          & $0.05  (0.01)$ &   &   & Y  &    &    & $+20$(R)   \\
    (IRAS4B2)          & $<0.16$        &   &   & Y* & Y* & Y* & $-99$      \\
    SerpS-MM22         & $0.4   (0.2)$  &   &   & Y  &    &    & $+230$     \\
    L1527              & $0.9   (0.1)$  &   & Y & Y  & D  &    & $-90$$^{e}$\\
    GF9-2              & $1.7$          &   &   & Y  &D/E?&    & $50$        \\
    SerpM-S68Nb        & $1.8   (0.2)$  &   &   & Y* &    & Y  & $+108$     \\ 
    SerpM-SMM4a        & $2.2   (0.2)$  &   &   & Y  &    &    & $0$(B)$^{e}$  \\ 
    IRAS4B1            & $2.3   (0.3)$  & Y & y & Y  & Y  & Y  & $+165$     \\
    (SerpM-SMM4b)      & $<2.6$         &   &   & Y  & Y  & Y  & $+10$(B)/$+135$(R)\\ 
    SVS13B             & $3.1   (1.6)$  &   & y & Y* & Y* & Y  & $+167$     \\
    L1448-NB1 (+NB2)   & $3.9$          &   & y & Y  & Y  & Y  & $-80$      \\
    L1157              & $4.0   (0.4)$  & y &   & Y  &D/E?& Y  & $+163$     \\
    IRAS4A1            & $<4.7$         &   &   & Y  & Y  & Y  & $+180$     \\
    IRAS4A2            & $4.7   (0.5)$  & Y &   & Y  & Y  & Y  & $+182$     \\
    L1448-2A (+2Ab)    & $4.7   (0.5)$  &   &   & Y  &D/E?& Y* & $+118^{f}$ \\
    L1448-C            & $11      (1)$  & Y & Y & Y  & Y  & Y  & $-17$      \\
    SerpM-S68N         & $11      (2)$  & y &   & Y  & Y  & Y  & $-45$      \\ 
    SerpS-MM18a        & $13      (4)$  & Y & y & Y  & Y  & Y  & $+188$     \\ 
    (SerpS-MM18b)      & $16      (4)$  &   &   & Y  &D/E?&    & $+205$(R)  \\            
    IRAS2A1            & $47      (5)$  & Y & y & Y  & Y  & Y  & $+205$     \\
    \hline     
    \multicolumn{8}{c}{{\bf Class I}}\\
    L1448-CS            & $3.6$         &    &  & Y  & Y  & Y  & $+60$(R)   \\
    L1448-NA            & $6.4   (0.6)$ &    &  & Y$^{e}$ &  & & $+40$$^{e}$\\
    SVS13A  (+VLA3)     & $44      (5)$ & Y  &  & Y  & Y  & Y  & $+155$     \\
    \hline     
  \end{tabular}\\
\small
$^{a}$ emission from three ("y") or more than three ("Y") complex organic molecules (COMs) based on the CALYPSO survey \citep{belloche20}.\\
$^{b}$ "y" indicates disk candidates, i.e., sources which show a velocity gradient perpendicular to the jet axis (within $\pm 45\degr$) at a few hundred au scales in one of the disk tracers ($^{13}$CO ($2-1$), C$^{18}$O ($2-1$), and SO ($5_6-4_5$)). "Y" indicates the detection of a Keplerian rotating disk \citep[based on the CALYPSO survey, see Table 3 by ][]{maret20}.\\
$^{c}$ "Y" indicates outflow/jet blue- and/or red-shifted emission detected in CO ($2-1$), SO ($5_6-4_5$), and SiO ($5-4$) at $> 10\sigma$ in the integrated maps. The asterix ("Y*") indicates emission at $5-10\sigma$. For SO ($5_6-4_5$): "D" indicates emission from the disk, "D/E?" indicates compact emission at low blue- and red-shifted velocities, and with a velocity gradient perpendicular to the jet (except for L1448-2A), originating from the inner envelope or the disk. \\
$^{d}$ The jet or outflow position angles are given for the blue-shifted lobe from North to East. If the blue- and red-shifted lobes have different PA or if only one of the two lobes is detected, this is indicated by the label (B) or (R).\\ 
$^{e}$ For L1521-F, L1527, and SerpM-SMM4a, we do not detect SiO and SO, and CO emission does not show a clear structure.  The reported PA are taken from \citet{tokuda16,hogerheijde98,tobin12,aso18}. For L1448-NA we do not detect emission in any of the tracer, and the PA is from CO ($2-1$) observations by \citet{lee15}.\\ 
$^{f}$ The jet PA is taken to be in the middle of the CO cavities.\\
\label{tab:jet-occurrence}
\end{table*}

To assess the presence of outflows and jets associated with the protostars identified in the CALYPSO survey (\citealt{maury19}, see Table \ref{tab:sample}), we analyzed the CO ($2-1$), SO ($5_6-4_5$), and SiO ($5-4$) datacubes and searched for emission at blue- and red-shifted velocities with respect to the source systemic velocity, $V_{\rm sys}$, reported in Table \ref{tab:sample}.
Maps of the CO ($2-1$), SO ($5_6-4_5$), and SiO ($5-4$) emission are produced by integrating the data-cubes on the blue- and red-shifted velocity channels where line emission at $\ge5\sigma$ is detected.
Figure \ref{fig:jets1} presents the integrated line emission maps for all the sources in the CALYPSO sample. The $V_{\rm LSR}$ velocity of the first and last channels over which the line emission is integrated is labeled (only one velocity value is given when emission is detected only in one channel), along with the 5$\sigma$ intensity of the integrated emission (on the top-left and top-right corners of each panel, respectively).
We claim that the outflow is detected if the map of CO ($2-1$) shows resolved blue- and/or red-shifted emission at $\ge5\sigma$ associated with the source in the integrated maps\textsuperscript{\ref{note1}}, while blue- or red-shifted SiO ($5-4$) emission is used as a probe of collimated jets. If SO ($5_6-4_5$) is detected at $\ge5\sigma$ along the same direction and on a similar velocity range as SiO, we can claim that the jet is detected also in SO.
If SiO is detected the jet position angle (PA$_{\rm jet}$) is determined from the SiO peaks located closest to the source; when instead SiO is not detected, the outflow PA (PA$_{\rm outflow}$) is inferred from the distribution of CO ($2-1$).
Then, position-velocity diagrams of the CO ($2-1$), SO ($5_6-4_5$), and SiO ($5-4$) emission are extracted along the outflow or jet PA and shown in Fig. \ref{fig:PV-block1}.

The detection of outflows and jets, that is, of blue- and red-shifted emission in the CO ($2-1$), SO ($5_6-4_5$), and SiO ($5-4$) lines, in the CALYPSO sample of Class 0 and I protostars and the estimated PAs are summarized in Table \ref{tab:jet-occurrence}.
The sources are listed by increasing internal luminosity, as $L_{\rm int}$ is a probe of the accretion luminosity, which, in turn, is expected to correlate with the mass ejection rate hence with the jet brightness \citep[see, e.g., ][]{hartigan95}.
In Cols. 3 and 4, we report whether the source is associated with a hot corino or a disk identified by the analysis of the source spectra and of the $^{13}$CO ($2-1$), C$^{18}$O ($2-1$), and SO ($5_6-4_5$) line maps, respectively \citep{belloche20,maret20}. 
Cols. 5, 6, 7, and 8 report the detection of blue- red-shifted emission in the three jet tracers (CO ($2-1$), SO ($5_6-4_5$), SiO ($5-4$)) and the position angle of the detected emission (PA$_{\rm jet/outflow}$). For L1521-F, L1527, and SerpM-SMM4a we do not detect emission in the SiO and SO lines, and the CO emission does not show a clear structure. Therefore, the position angle reported in Table \ref{tab:jet-occurrence} is taken from previous studies \citep{tokuda16,hogerheijde98,tobin12,aso18} (see Sect. \ref{app:notes-on-sources} for the notes on these sources).

Figure \ref{fig:jet-occurrence} shows the number of detections and the detection rate of jets and outflows in each of the three tracers (CO ($2-1$), SO ($5_6-4_5$), SiO ($5-4$)) as a function of the source internal luminosity, \lint, for the Class 0 protostars in our sample. The figure indicates that outflow emission in the CO ($2-1$) line maps is detected in 21 Class 0 protostars out of 21.
Emission in SiO ($5-4$) is detected in 14 sources, which means that at least 67\% of the Class 0 protostars drive an SiO jet. 
Finally, 11 of the 14 protostars with an SiO jet show  emission along the jet PA in SO ($5_6-4_5$), indicating that about 79\% of the SiO jets are also detected in the SO line (with a detection rate of SO jets of 52\% over the whole sample of 21 Class 0 protostars).
Five more sources out of the 21 Class 0 protostars show blue- and red-shifted emission in SO ($5_6-4_5$) (L1527, GF9-2, L1157, L1448-2A, and SerpS-MM18b). In these sources, however, the spatial and velocity distribution of the SO emission is not in agreement with that of CO ($2-1$) (and SiO ($5-4$) in the case of L1157), which probes the outflow (see the integrated maps in Fig. \ref{fig:jets1} and the position-velocity diagrams in Fig. \ref{fig:PV-block1}). The possible origin of the SO ($5_6-4_5$) for these five sources is  discussed in Sect. \ref{sect:discussion}.

Concerning the three Class I protostars in the CALYPSO sample, there is blue- or red-shifted CO ($2-1$) emission (or both)  detected in two sources out of three. The exception is the Class I source L1448-NA for which no line emission is detected in our CALYPSO line maps. However, previous lower resolution observations of CO ($2-1$) show evidence of slow outflowing gas \citep{lee15}. Taking into account this previous detection, CO outflows are detected in three Class I protostars out of three, hence, all Class I sources are associated with outflows. SiO and SO jet emission is detected in two of them (L1448-CS and SVS13A), with the exception of the Class I L1448-NA, hence the detection rate of SiO jets is 67\%, the same as for Class 0 protostars.
We note, however, that the emission from the L1448-CS jet overlaps on the bright emission of the jet driven by L1448-C, impeding us to derive the jet properties (see Appendix \ref{app:notes-on-sources}). On the other hand, the morphology and properties of the jet associated with SVS13A based on the CALYPSO data are already presented in \citet{lefevre17}. 
Therefore, the properties of the jets from the three Class I sources in our sample are not presented in the following sections.


\subsection{Jet spatio-kinematical structure}
\label{sect:kinematics}

Based on the integrated maps in Fig. \ref{fig:jets1} and the PV diagrams in Fig. \ref{fig:PV-block1} we  investigate the spatio-kinematical properties of the flows for the 12 Class 0 sources that exhibit an SiO jet detected at $>10\sigma$ in the integrated maps (see Table \ref{tab:jet-occurrence}). 
For these jets, the CO ($2-1$), SO ($5_6-4_5$), and SiO ($5-4$) line spectra are extracted at the position of the blue- and red-shifted SiO emission peaks (knots) located closest to the driving protostar, denoted as B and R (see Fig. \ref{fig:spec1} in Appendix \ref{app:spectra}). The RA and Dec offsets of the innermost SiO knots, B and R, and their distance from the driving source are given in Table \ref{tab:fluxes}.

The line spectra in Fig. \ref{fig:spec1} show several emission components: SiO ($5-4$), which is commonly used as a selective probe of the jet, emits mainly at high velocity  (HV), that is, at velocities of $15-80$ \kms\, with respect to the source systemic velocity,  which correspond, for a median jet inclination of $60\degr$ to the line of sight, to deprojected velocities of $30-160$ \kms; the bulk of CO ($2-1$) emission, instead, is at low velocity (LV: $<15$ \kms\, with respect to $V_{\rm sys}$). 
This kinematical difference between SiO and CO also reflects in a different spatial distribution of the two molecules. The line maps in Fig. \ref{fig:jets1} and the PV diagrams in Fig. \ref{fig:PV-block1} show that the SiO emission is more collimated and is concentrated on smaller areas of the spatial-velocity space than CO in most of the sources (IRAS4B1, SerpM-SMM4b, L1448NB, L1157, IRAS4A1 and A2, L1448-C, and IRAS2A1). In these sources, SiO probes the collimated jet and, in some cases, it also probes bright terminal shocks with (IRAS4B1, L1448NB) or without (IRAS4A1 and A2) a CO counterpart. 
The CO emission is transversally more open and in some cases, it also extends out to almost twice larger distances than SiO (SerpM-SMM4b, L1157, L1448-2A, IRAS2A1). This suggests that the CO emission is dominated by emission from the outflow, that is, from the surrounding material that is put into motion or entrained by the jet. 
However, in most of the jet sources, CO ($2-1$) has a secondary peak at high velocity and the CO HV component looks co-spatial with SiO in the PV diagrams,  suggesting that in the HV range, CO probes the jet similarly to SiO. There are a few exceptions, for instance, for sources where in the line spectra, there is no clear separation between low and high-velocity components in all three tracers that is likely due to the low inclination of the flow (IRAS4A2, SerpM-S68N, and SerpS-MM18a) and sources where the high-velocity jet emission seen in SiO has no counterpart in CO (SerpM-S68Nb blue lobe,  SVS13B  both  lobes,  L1448-NB blue lobe,  and  SerpM-S68N red lobe). We refer to the latter objects as "CO-poor" jets. 
Furthermore, SO ($5_6-4_5$), when detected in the 12 SiO jet sources, may have different origins: in the jet, it exhibits spatio-kinematical distribution that is very similar to that of SiO in the maps and in the PV diagrams (SerpM-SMM4b, SVS13B, L1448NB, IRAS4A1 and A2, L1448-C, SerpM-S68N, SerpS-MM18a, IRAS2A1), in the terminal shocks (IRAS4B1), or in a compact region around the source, where it likely probes the disk or the inner envelope (L1157, L1448-2A).
A more robust comparison of the spatio-kinematical distribution of the SiO emission with that of the CO and SO LV and HV components requires spatial resolution of $\sim 10$ au, as shown for example for the study of the prototypical protostellar jet from HH 212 \citep{lee18b}.

Based on the spectra we define for each jet lobe the high-velocity ranges where CO ($2-1$) and SO ($5_6-4_5$) are likely to trace the same jet component as SiO ($5-4$). The HV ranges (in $V_{\rm LSR}$) are summarized in Table \ref{tab:fluxes}.
The identification of the high-velocity ranges is further supported by the inspection of the position-velocity diagrams (see Fig. \ref{fig:PV-block1} in Appendix \ref{app:jet-pv}). As explained above, CO emission is mostly dominated by low or intermediate velocity material that corresponds to the wider flow surrounding the jet. However, the high-velocity SiO emission which probes the jet, also has counterparts in CO and SO for many of the sources and they appear co-spatial in the PV diagrams, which suggests that they all trace the same gas component in the jet with little or no contamination by the outflow or by entrained material.
Then we estimate the jet radial velocity in the two innermost knots B and R along the blue- and red-shifted lobes as the velocity of the SiO emission peak in the B and R spectra with respect to the systemic velocity  ($V_{\rm rad}= V_{\rm peak} - V_{\rm sys}$) and we report them in Table \ref{tab:fluxes}. The estimated jet radial velocities are affected by an uncertainty of $\pm 1.7$ \kms\, due to the resolution of our spectra.

\begin{figure}
\centering
\includegraphics[width=.49\textwidth]{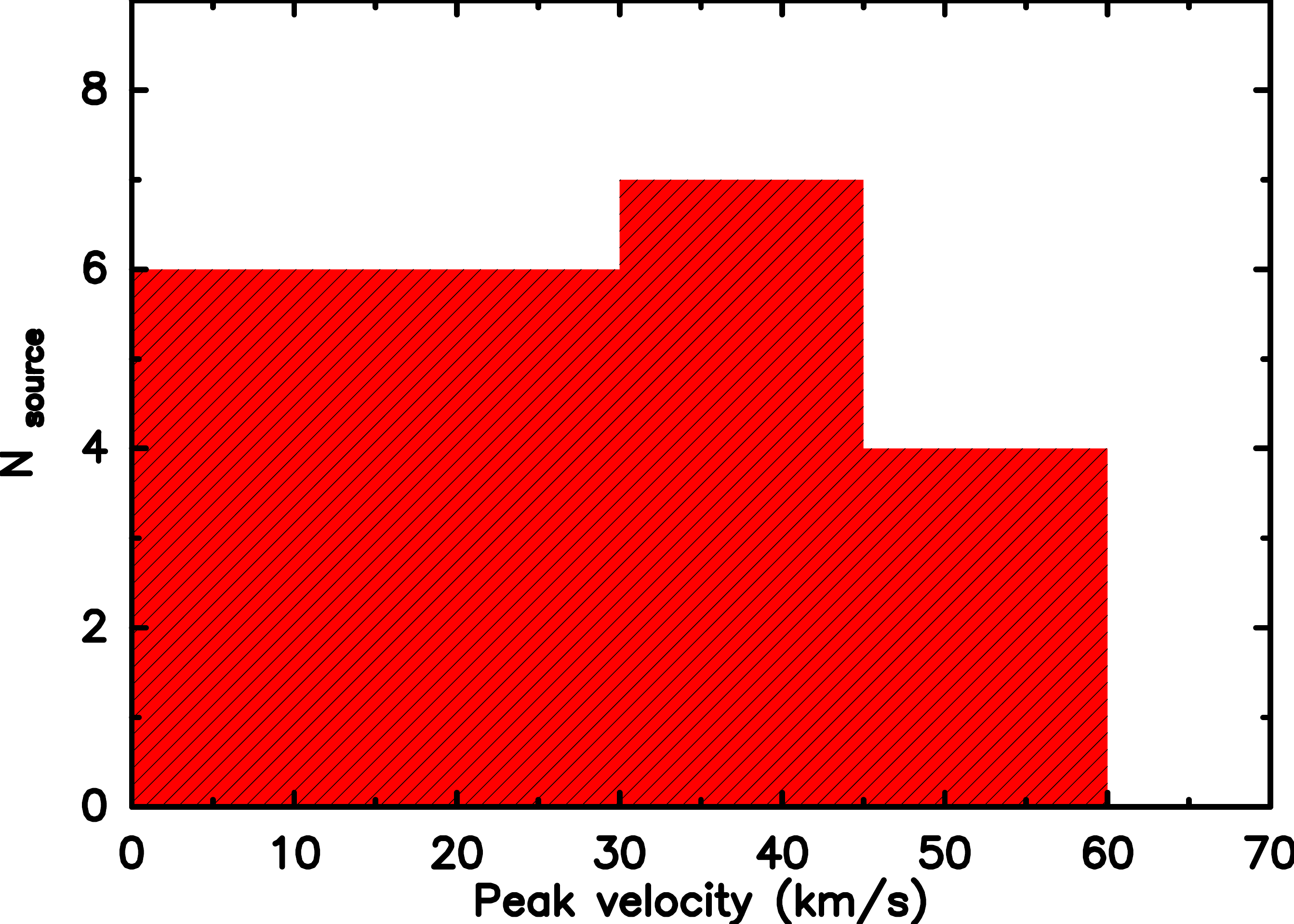} 
\caption{Distribution of jet radial velocities in the CALYPSO sample. The values correspond to the velocity of the SiO emission peak in the spectra extracted at the position of the blue- and red-shifted SiO knots close to the driving protostar, B and R (see Fig. \ref{fig:spec1} in Appendix \ref{app:spectra}).}
\label{fig:distri-vrad}
\end{figure}

The distribution of radial velocities of the jet lobes, $V_{\rm rad}$, inferred from the SiO spectra is shown in Fig. \ref{fig:distri-vrad}. The distribution is flat within the statistical uncertainty, in agreement with a randomly oriented jet distribution. 
This indicates that the jets for which the SiO emission peak is detected at low velocity (IRAS4A2, SerpM-S68N, and SerpS-MM18a) may actually be high-velocity jets seen close to the plane of the sky. The derived median radial velocity of the high-velocity jet component is $30 \pm 10$ \kms.
Estimates of the jet inclination and of the deprojected jet velocity are available only for a few jets in our sample, as detailed in the notes on the individual sources (Appendix \ref{app:notes-on-sources}) and discussed in Sect. \ref{sect:jet-energetics}.
Many of the jets in our sample show asymmetries between the two lobes either
in their morphology or their velocity.
First of all, one out of 12 SiO jets is monopolar, the jet from IRAS2A1 \citep{codella14a}. High-velocity emission in the three tracers is detected only in the blue lobe, while the low-velocity emission associated with the outflow is detected in both lobes.
In the sub-sample of the 11 SiO bipolar jets, morphological asymmetries between the two lobes are also observed (see the integrated maps in Fig. \ref{fig:jets1}).
The jet from SerpM-SMM4b shows different PA (and inclination, \citealt{aso18}) between the blue and red lobes, while 5 out of the 12 jets show an S-shaped morphology, which suggests jet precession around its axis (namely IRAS4B1, L1448-NB, L1157, IRAS4A1, and IRAS4A2). The precession patterns of the jets from L1157 and IRAS4A2 based on the CALYPSO data are discussed in two previous papers \citep{podio16,santangelo15}.
The jet from SerpS-MM18a is wiggling, which is also suggestive of a slight precession, as was previously discussed by \citet{plunkett15b}.
Hence, 50\% of the SiO jets show detectable precession or wiggling at our resolution. This is the first time that the statistical occurrence of this property can be estimated.

Finally, some of the eleven SiO bipolar jets show an   asymmetry in the velocity between the two lobes.
The degree of the velocity asymmetry is quantified by estimating the ratio between the radial velocities of the blue- and red-shifted innermost SiO knots, B and R. In particular, it is computed as the ratio between the velocity of the faster knot, $V_{\rm rad, f}$, over the velocity of the slower knot, $V_{\rm rad, s}$, as reported in Table \ref{tab:fluxes}. The error on this ratio depends on the uncertainty on the determination of the radial velocity of the two jet lobes ($\pm 1.7$ \kms) due to the resolution of our spectra. For the jets of IRAS4A2, SerpM-S68N and SerpS-MM18a, which show very low radial velocities (possibly due to low inclination), the radial velocities ratio, $V_{\rm rad, f}/V_{\rm rad, s}$, is affected by a large uncertainty and this prevents us from drawing a conclusion on whether  the jet is or is not asymmetric in velocity.
Based on the derived $V_{\rm rad, f}/V_{\rm rad, s}$ ratios and their uncertainties, the radial velocity of the blue-shifted and red-shifted lobes differ by factor of $1.4-2.1$ for three jets (namely SVS13B, L1157, and IRAS4A1), and by a factor of $7.8$ for SerpM-S68Nb. As the detection of velocity asymmetries for jets at low inclination is hindered by the low spectral resolution of our data (3.4 \kms), the number of detected velocity asymmetric jets should be considered a lower limit. We conclude that at least 33\% of the observed jets has a velocity asymmetry of $1.3-7.8$.

\subsection{Jet width and opening angle}
\label{sect:jet-width}

\begin{figure*}
\centering
\includegraphics[width=.32\textwidth]{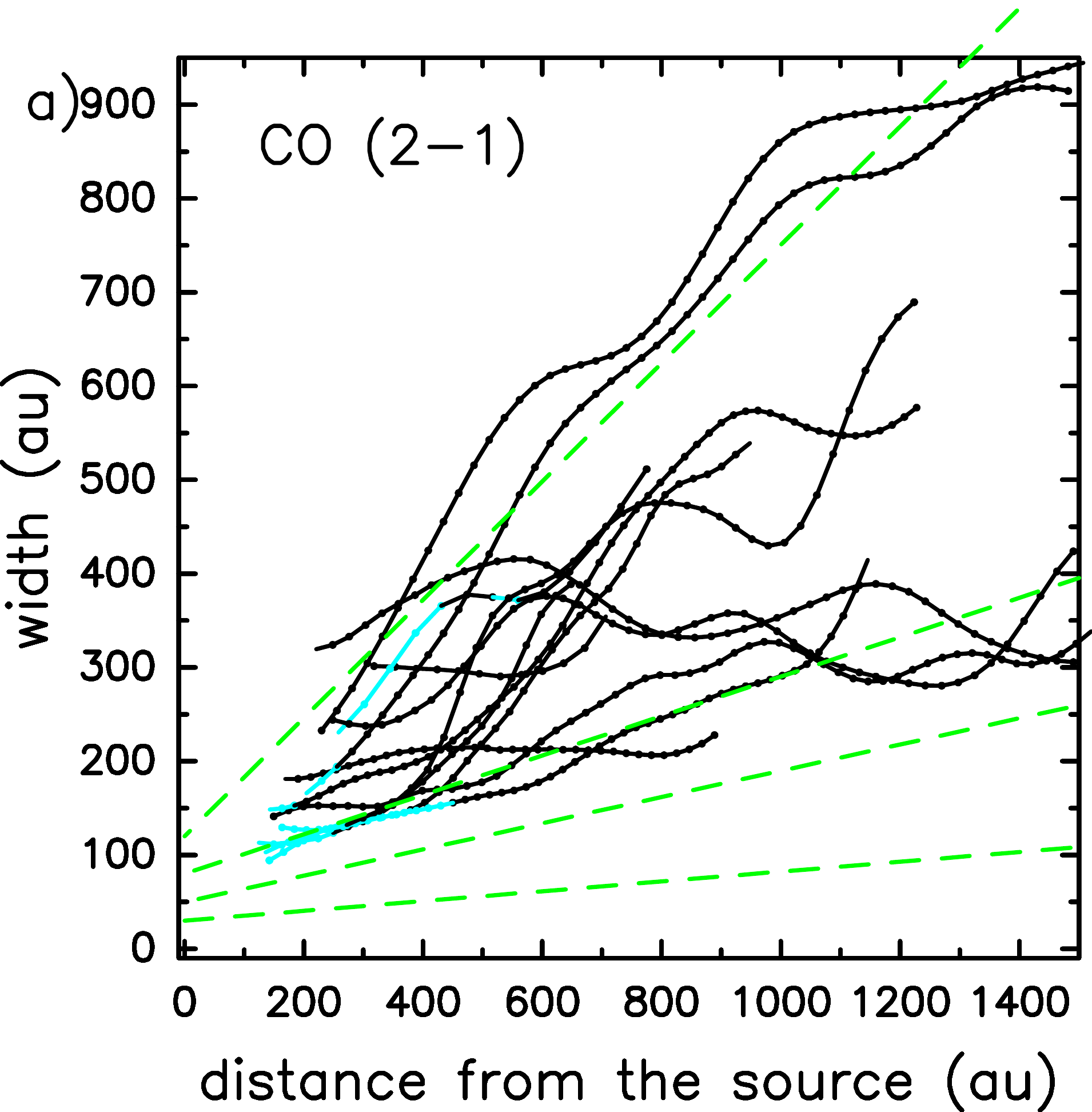}
\includegraphics[width=.32\textwidth]{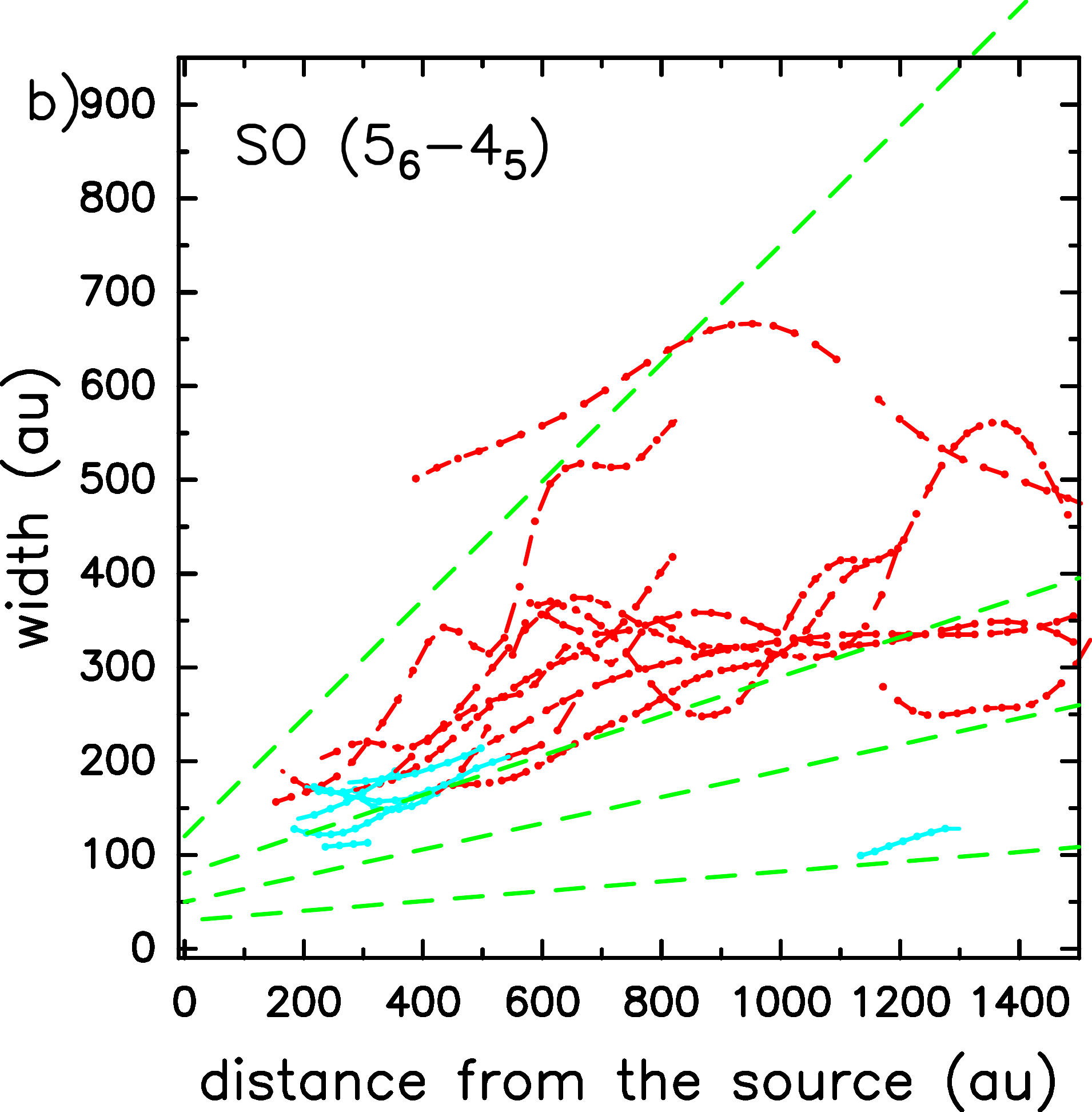}
\includegraphics[width=.32\textwidth]{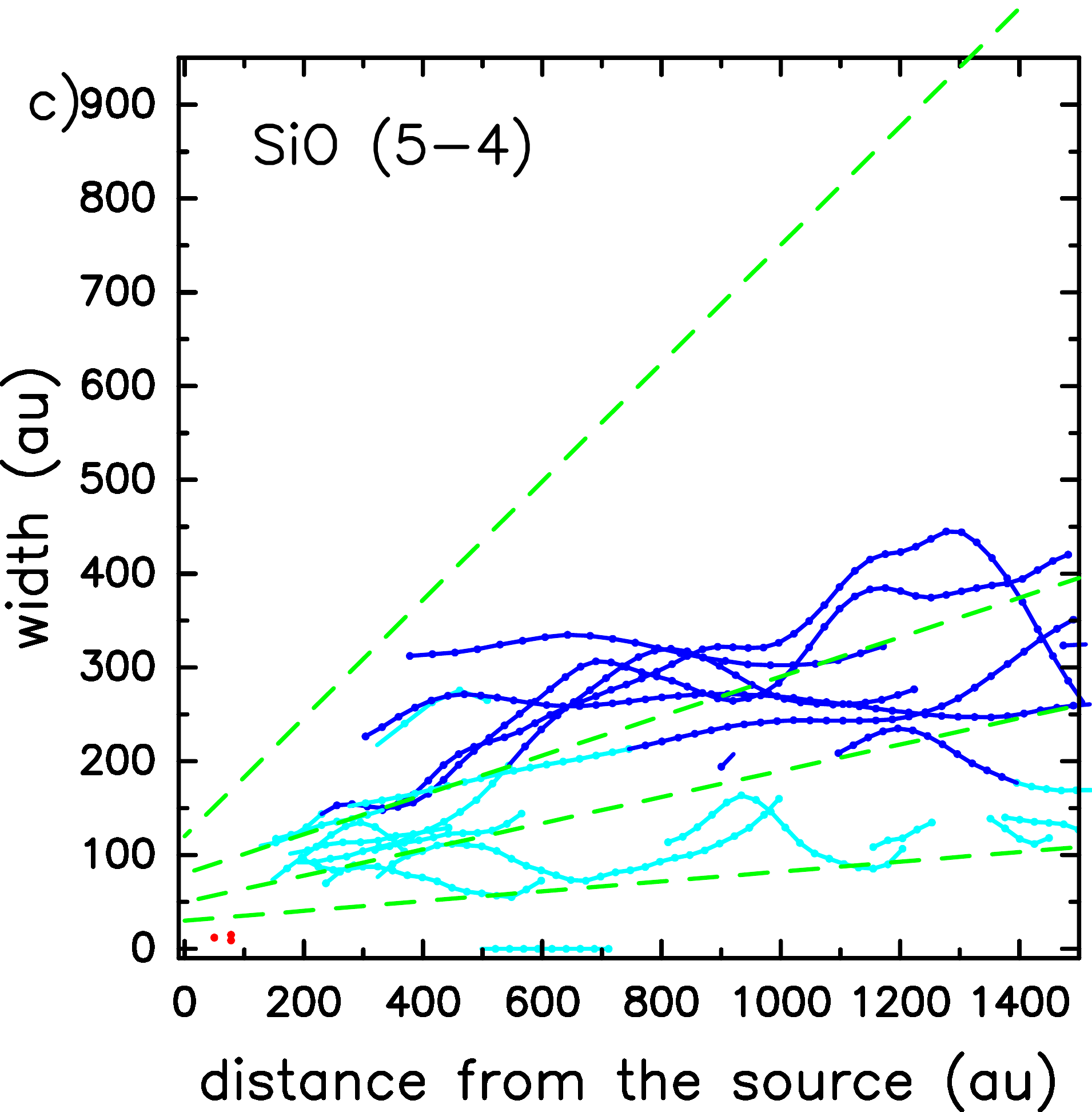}
\caption{Deconvolved widths (2\Rjet) of CO (a), SO (b), and SiO (c) emission along the jet axis within 1500 au from protostars. Only flow detected above 10$\sigma$ are measured. Non-axisymmetric structures that are well spatially resolved are discarded. Cyan lines indicate widths that are smaller that the transverse beam and consequently considered as inclusive upper limits. 
Green dashed lines correspond to straight lines with full opening angles of $\alpha = 3^\circ, 8^\circ, 12^\circ$, and $35^\circ$ and initial width of $30, 50, 80,$ and $120$~au outlining the collimation properties of the flow in different tracers.}
\label{fig:jet-width-main}
\end{figure*}

We investigate the collimation properties of the flows for the 12 Class 0 sources that exhibit an SiO jet detected at $>10\sigma$ in the integrated maps (see Table \ref{tab:jet-occurrence}). The width of the red and blue jet and outflow lobes in each of the three tracers (CO, SO, and SiO) is measured from the emission maps integrated on the red-shifted and blue-shifted velocity ranges, respectively, as indicated in Fig. \ref{fig:jets2}. The widths are measured by fitting the spatial profile of the line emission perpendicular to the jet axis (PA given in Table \ref{tab:jet-occurrence}) with a Gaussian profile at each position along the jet. Deconvolved widths, corresponding to twice the jet radius, 2\Rjet, are then derived by correcting the full width at half maximum (FWHM) of the observed emission, FWHM$_{\rm obs}$, for the size of the beam transverse to the jet axis (see Appendix \ref{app:jet-width}). The deconvolution method allows us to infer jet widths smaller than the beam if the signal-to-noise ratio (S/N) is high enough to measure a FWHM$_{\rm obs}$ larger than the transverse beam size. Therefore, this procedure is applied to each source and each molecular line if detected at $> 10 \sigma$ in the integrated maps. Spatially resolved emission showing asymmetric cavity walls or complex bow-shock structures that strongly deviate from a Gaussian distribution have been discarded (e.g., SiO red-shifted emission in IRAS4A, see Fig. \ref{fig:jets2}). Figure \ref{fig:jet-width-main} collects the jet width, 2\Rjet, in the three tracers (CO ($2-1$), SO ($5_6-4_5$), and SiO ($5-4$)) for the 12 sources driving an SiO jet as a function of the distance from source. The estimates of the flow width for each source separately are presented in Figure \ref{fig:jet-width_all} in Appendix \ref{app:jet-width}.

After deconvolution by the beam, jet widths appear to be always larger than $\simeq 50$~au, even for SiO jets observed very close to the driving source. For example, the beam-corrected width of the SiO jet driven by L1157 is about 80 au at a distance from the source of 100 au. In contrast, ALMA observations of SiO at very high angular resolution \citep[e.g., ][0.02$''$ or 8~au resolution]{lee17b} and VLBI observations \citep[e.g., ][]{claussen98} towards Class 0 jets suggest that jet widths are $<20$~au within 100~au distance. Our beam deconvolution method neglects the effect of residual phase noise and cleaning artifacts which result in an effective resolution that is lower than the resolution implied by the clean beam. Consequently, we choose to consider the widths that are smaller than the size of the transverse beam to be only (inclusive) upper limits on the true width of the jet (cyan points in Fig. \ref{fig:jet-width-main}).
Measurements taken within a beam from the driving source are also discarded since the true flow width may vary a lot within a beam.



Figure \ref{fig:jet-width-main} shows that the width of the flow globally increases with the distance from the source. However, the complex variation of the jet width with distance prevents us from defining an opening angle of the flow for each source. 
For comparison with opening angles measured in atomic Class II and Class I jets on similar scales, following a statistical approach we derive the average collimation properties of the sample by fitting the measured flow widths in Fig. \ref{fig:jet-width-main} with straight lines according to the following equation:
\begin{equation}
    2 R_{\text{jet}} = 2 R_{0} + 2 \tan{(\alpha/2)} z,
    \label{eq:stright-line}
\end{equation}
where $z$ is the distance from source, $\alpha$ is the apparent full opening angle of the flow\footnote{due to projection effects, the true opening angle is smaller: $tan(\alpha_{\rm true}) = tan(\alpha) \times sin(i)$, where $i$ is the inclination to the line of sight.}, and $R_{0}$ is a constant offset. The last quantity is introduced following studies of Class II jets \citep{agraamboage11,dougados00,hartigan04,maurri14}. Since Eq. (\ref{eq:stright-line}) is simply a parametric formula to fit the jet width at distances of $\sim 200-1500$~au from the driving source, $R_{0}$ should not be interpreted as the launching point of the flow.

Figure \ref{fig:jet-width-main} shows that the collimation properties of the velocity-integrated emission depend markedly on the molecular tracer. Opening angles of the CO ($2-1$) emission span a wide range of value between $8^{\circ}$ and $35^{\circ}$ with a median value of $\sim 25^{\circ}$. SO ($5_6-4_5$) emission is more collimated and shows typical opening angles of $\sim 15^{\circ}$. SiO ($5-4$) emission is even more collimated with a typical opening angle of about $10^{\circ}$ and with minimum values as low as $4^{\circ}$.
Thus, the bulk emission of the three lines highlights different regions of the flow, with SiO tracing the most collimated jet, CO the broader component, and SO a flow of intermediate collimation.

\begin{figure}
\centering
\includegraphics[width=.4\textwidth]{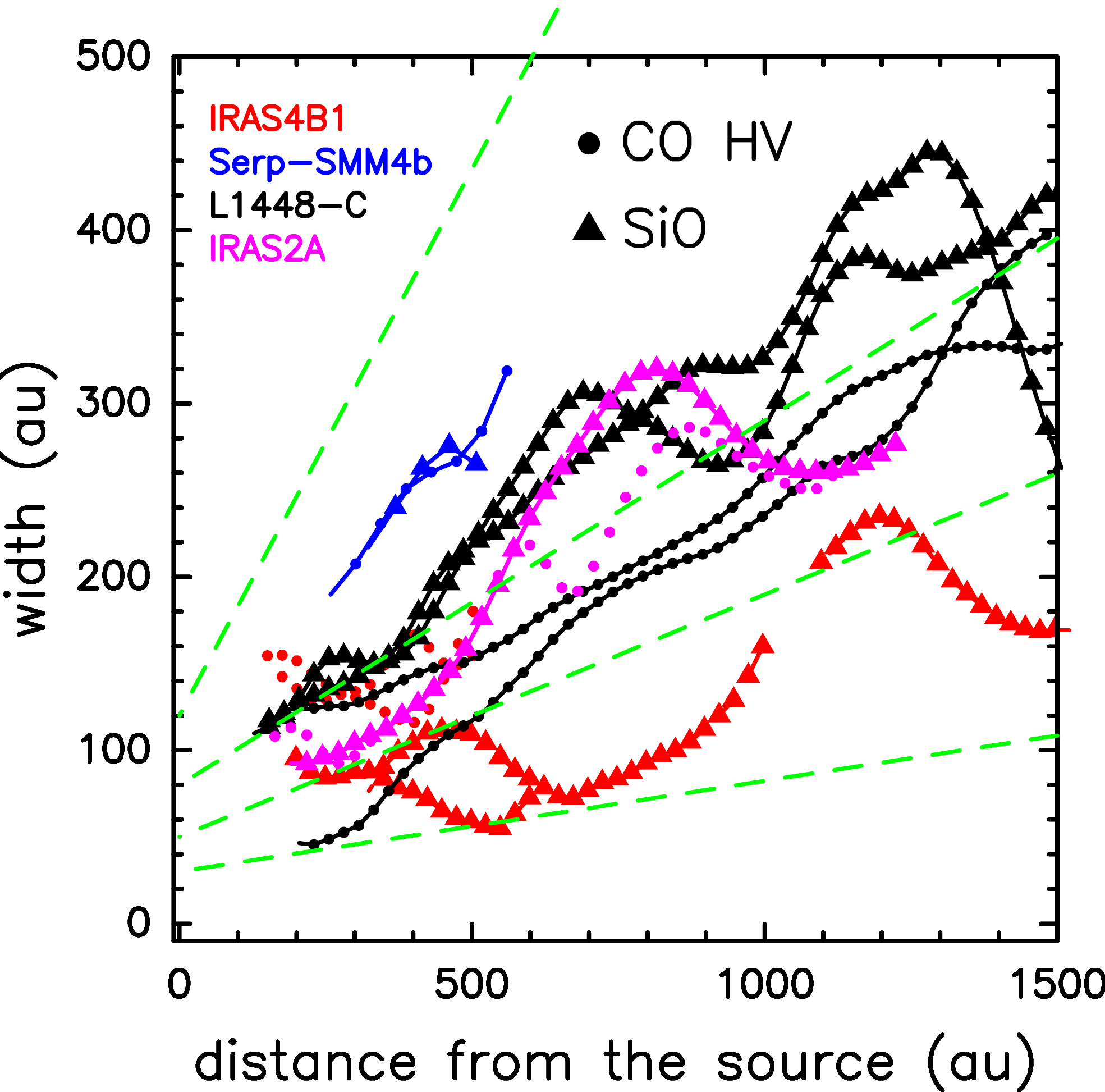}
\caption{Deconvolved widths  (2\Rjet) of SiO (triangles) and high-velocity CO (dots) emission (high-velocity intervals defined in Table \ref{tab:fluxes}). Only the jet lobes where both SiO and high-velocity CO are detected above 10 $\sigma$ at the same positions along the jet are plotted. Colors code the sources. At high velocity, the CO emission appears as collimated as the SiO jet emission. Green dashed lines correspond to straight lines with full opening angles of $\alpha = 3^\circ, 8^\circ, 12^\circ,$ and $35^\circ$ and initial width of $30, 50, 80,$ and $120$~au outlining the collimation properties of the flow.}
\label{fig:jet-width-hv}
\end{figure}

We also measured the width of the high-velocity CO emission and compare the results with the width of the SiO emission in Fig. \ref{fig:jet-width-hv}. 
As explained in Sect. \ref{sect:kinematics} the high-velocity ranges where CO ($2-1$) and SO ($5_6-4_5$) trace the same jet component as SiO ($5-4$) are defined based on the line spectra extracted at the position of the innermost SiO knots (see Table \ref{tab:fluxes} and Fig. \ref{fig:spec1} in Appendix \ref{app:spectra}) and on the inspection of the PV diagrams (see Fig. \ref{fig:PV-block1}).
When selecting only the high-velocity component, the CO emission is as collimated as the SiO emission, which is a further indication that high-velocity CO traces the same component probed by SiO, namely, the jet. The same conclusion applies to SO. 

\begin{table*}
\caption{Properties of the blue- and red-shifted SiO knots closest to the driving sources, B and R, as derived from the spectra in Fig. \ref{fig:spec1}.
}
\begin{tabular}[h]{cccccccccc}
\hline
\hline
Source & & RA$_{\rm off}$, Dec$_{\rm off}$$^{a}$ & Distance$^{b}$ & HV$^{c}$ & $V_{\rm rad}^{d}$ & $V_{\rm rad, f}/V_{\rm rad, s}$$^{e}$ &\Fco$^{f}$ & \Fso$^{f}$ & \Fsio$^{f}$ \\
       & & (\arcsec, \arcsec) & (\arcsec)  & (\kms) & (\kms) &  &(K \kms) & (K \kms) & (K \kms) \\
\hline
SerpM-S68Nb & B & (+5.99, -2.05) & 6.33 & -50/-20  & $-45$ & $7.8\pm2.3$ & $<0.8$ &    --  &   25.0 \\ 
            & R & (-6.91, +2.05) & 7.21 & +17/+40  & $+6$  &             &    6.5 &    --  &   20.3 \\ 
IRAS4B1     & B & (+0.57, -1.27) & 1.39 & -30/-5   & $-17$ & $1.0\pm0.1$ &   27.8 &    --  &   72.1 \\ 
            & R & (-0.53, +2.53) & 2.58 & +16/+50  & $+17$ &             &   55.4 &    5.6 &  169.3 \\ 
SerpM-SMM4b & B & (-0.28, +0.91) & 0.95 & -38/-8   & $-39$ & $1.1\pm0.1$ &   47.0 &   27.8 &   68.3 \\ 
            & R & (+0.23, -0.49) & 0.54 & +30/+60  & $+36$ &             &   74.4 &    8.8 &   69.2 \\ 
SVS13B      & B & (+0.37, -1.61) & 1.65 & -37/+8.5 & $-24$ & $1.4\pm0.2$ & $<3.0$ &    --  &   50.2 \\ 
            & R & (-0.03, +0.39) & 0.39 & +8.5/+58 & $+16$ &             & $<3.0$ &    --  &   77.6 \\ 
L1448NB     & B & (-1.98, +0.53) & 2.05 & -34/-20  & $-34$ & $1.1\pm0.1$ & $<1.7$ &    --  &   41.7 \\ 
            & R & (+0.82, +0.03) & 0.82 & +40/+54  & $+37$ &             &   27.8 &    1.1 &   17.6 \\ 
L1157       & B & (+0.00, -0.22) & 0.22 & -60/-20  & $-34$ & $1.6\pm0.1$ &   22.0 &    --  &  143.9 \\ 
            & R & (-0.10, +0.58) & 0.58 & +30/+70  & $+54$ &             &   45.2 &    --  &  125.4 \\ 
IRAS4A1     & B & (+0.31, -1.47) & 1.50 & -30/-10  & $-16$ & $2.1\pm0.2$ &   77.5 &    9.4 &   16.9 \\  
            & R & (+0.21, +1.33) & 1.35 & +30/+70  & $+34$ &             &  276.6 &   37.6 &   28.5 \\ 
IRAS4A2     & B & (+0.11, -2.31) & 2.31 & -10/+6.3 & $-6$  & $1.1\pm0.4$ &  177.5 &   47.2 &   25.2 \\ 
            & R & (+0.71, +2.29) & 3.40 & +6.3/+40 & $+7$  &             &  145.1 &   58.1 &   55.2 \\ 
L1448-C     & B & (-0.30, +1.10) & 1.14 & -70/-22  & $-51$ & $1.1\pm0.1$ &  160.5 &   64.7 &  501.1 \\ 
            & R & (+0.30, -0.80) & 0.85 & +25/+85  & $+48$ &             &  222.5 &   75.8 &  869.7 \\ 
SerpM-S68N  & B & (-5.46, +5.39) & 7.67 & -7/+5    & $-4$  & $1.5\pm1.1$ &   30.0 &    7.6 &   32.6 \\ 
            & R & (+2.54, -2.91) & 3.86 & +12/+21  & $+3$  &             & $<0.7$ &    --  &    8.5 \\ 
SerpS-MM18a & B & (-0.10, -0.90) & 0.91 & -17/-2   & $-10$ & $1.3\pm0.3$ &  191.1 &   63.7 &  159.2 \\ 
            & R & (+0.30, +2.90) & 2.92 & +21/+32  & $+13$ &             &   98.9 &   34.3 &   71.2 \\ 
IRAS2A1     & B & (-0.58, -1.07) & 1.22 & -32/-9   & $-24$ & --          &  150.3 &  113.2 &  181.6 \\ 
\hline
\end{tabular}\\
\small
$^{a}$ RA and Dec offset of the innermost SiO knots B and R.\\
$^{b}$ Distance of the innermost SiO knots B and R.\\
$^{c}$ High-velocity range (HV) where CO ($2-1$) and SO ($5_6-4_5$) trace jet emission as SiO ($5-4$).\\
$^{d}$ Jet radial velocity in the innermost SiO knots B and R estimated as the velocity of the SiO emission peak in the B and R spectra with respect to the source systemic velocity given in Table \ref{tab:sample}. The estimated jet radial velocities are affected by an uncertainty of $\pm 1.7$ \kms\, due to the resolution of our spectra.\\
$^{e}$ Ratio between the radial velocities, $V_{\rm rad}$, of the blue- and red-shifted innermost knots, B and R. The ratio is computed between the velocity of the faster knot, $V_{\rm rad, f}$, over the velocity of the slower knot, $V_{\rm rad, s}$.\\
$^{f}$ CO ($2-1$), SO ($5_6-4_5$), and SiO ($5-4$) line intensity  integrated on the high-velocity range (HV) (\Fco, \Fso, \Fsio).\\
\label{tab:fluxes}
\end{table*}

\subsection{Molecular column densities and abundances}
\label{sect:jet-abundances}

For the 12 Class 0 protostars driving an SiO jet detected at $>10\sigma$ in the integrated maps (see Table \ref{tab:jet-occurrence}), we estimate the beam-averaged column density and the abundance of CO, SO, and SiO in the high-velocity jet component for the blue- and red-shifted lobes.
To this aim,
the CO ($2-1$), SO ($5_6-4_5$), and SiO ($5-4$) line intensities extracted at the position of the innermost SiO knots, B and R, are integrated on the high-velocity range shown in Fig. \ref{fig:spec1} and summarized in Table \ref{tab:fluxes}.
As shown by the position-velocity diagrams in Fig. \ref{fig:PV-block1} and by the measurements of the flow width in Fig. \ref{fig:jet-width-hv}, in the defined high-velocity ranges, the emission in the CO ($2-1$) and SO ($5_6-4_5$) lines is co-spatial with that of SiO ($5-4$) and shares the same width.
Therefore, we assume that in the HV range the emission in the three tracers originate from the same gas component.  Moreover, based on the high radial velocity of the HV component ($15-80$ \kms) and on its collimation (see Fig. \ref{fig:jet-width-hv}), we assume that the bulk of the HV emission originates primarily from the jet and that the contribution from entrained material along the jet is negligible.
We further assume  local thermodynamic equilibrium (LTE) at a kinetic temperature, $T_{\rm K} = 100$ K, and optically thin emission. Possible deviations from these assumptions are discussed below.
Based on the above assumptions we derive the molecules' beam-averaged column densities in the jet, \Nco, \Nso, and \Nsio, from the line intensities integrated on HV.

The non-LTE analysis of multiple SiO and CO transitions in the large velocity gradient (LVG) approximation in a few prototypical Class 0 jets has shown that the typical gas density and temperature in the high-velocity range are of $10^5-10^7$ cm$^{-3}$ and $30-500$ K \citep[e.g., ][]{gibb04,cabrit07b,podio15,spezzano20}.
Therefore, the assumption of LTE is well justified for the CO ($2-1$) line, whose critical density is well below the typical gas density in jets ($\sim 7.3 \times 10^3$ \cmc, see Table \ref{tab:lines}), while the SO and SiO lines may be sub-thermally excited if the gas density is lower than their critical density ($\sim 7.7 \times 10^5$ \cmc\, and $\sim 1.6 \times 10^6$ \cmc, see Table \ref{tab:lines}). To estimate the uncertainty affecting the estimates of the column density in the non-LTE case we use the statistical equilibrium, one-dimensional radiative transfer code RADEX adopting plane parallel slab geometry \citep{vandertak07}. We find that the SO and SiO column densities may be underestimated by a factor of up to 5 if $n_{\rm H_2}=10^5$ \cmc. 
Moreover, the column densities of CO, SO, and SiO are overestimated by a factor of $\sim 1.5$ if the gas temperature is lower than assumed ($20$ K), and are underestimated by a factor of $\sim1.6$, $\sim3.5$ if the temperature is higher (200 K and 500 K, respectively). These uncertainties are illustrated in Fig. \ref{fig:conv-fac} in the Appendix.

If lines are optically thick, the beam averaged column density derived in the optically thin limit is only a lower limit on the true column density. Previous studies have shown that jet emission may be optically thick even at high velocities \citep[e.g., ][]{cabrit12,podio15}. Therefore, in Appendix \ref{app:uncertainties}, we propose a criterion to flag lines that are (or may be) optically thick by using the ratio of the brightness temperature of the CO ($2-1$), SO ($5_6 - 4_5$), and SiO ($5-4$)  lines in the HV range ($T^{\rm CO}_{\rm B}$, $T^{\rm SO}_{\rm B}$, $T^{\rm SiO}_{\rm B}$, respectively). 
Based on the discussion in Appendix \ref{app:uncertainties}, we note that the opacity of the CO and SiO lines is constrained using the following criteria.
1) If $0.7 \times T^{\text{CO}}_{\text{B}} \le T^{\text{SiO}}_{\text{B}} \le T^{\text{CO}}_{\text{B}}$, both lines are very likely optically thick and the beam averaged $N_{\rm CO}$ and $N_{\rm SiO}$ derived in the optically thin limit are both considered as strict lower limits. No constraint on the SiO/CO abundance ratio can be obtained in this case.
2) If $T^{\text{SiO}}_{\text{B}} > T^{\text{CO}}_{\text{B}}$, CO ($2-1$) is optically thin and SiO ($5-4$) may be optically thick. We then consider the derived $N_{\rm SiO}$ as an inclusive lower limit and we can derive a lower limit on the SiO/CO abundance ratio.
3) If $T^{\text{SiO}}_{\text{B}} < 0.7 \times T^{\text{CO}}_{\text{B}}$, the SiO ($5-4$) line is considered to be optically thin whereas the CO ($2-1$) line may be optically thick. We then consider the derived $N_{\rm CO}$ as an inclusive lower limit and we obtain  an upper limit on the SiO/CO abundance ratio.
The brightness temperature of the SO ($5_6-4_5$) line is always much smaller than the brightness temperature of CO ($2-1$), suggesting that the SO emission is optically thin. 

Based on the above criteria, the brightness temperature ratios $T^{\text{SiO}}_{\text{B}}/ T^{\text{CO}}_{\text{B}}$ measured at the SiO emission peak in the B and R spectra in Fig. \ref{fig:spec1} indicate that for all but two jets in our sample (IRAS4A1 and IRAS4A2), that is, for $\sim 82\%$ of the SiO knots, the SiO ($5-4$) emission is (or may be) optically thick.
Therefore, the estimated SiO column densities are strict (for L1448-NB knot R, SerpM-SMM4b knot R, SerpM-S68N knot B, SerpS-MM18a, and IRAS2A1 knot B) or inclusive (for all the other jet lobes) lower limits. For CO, we also obtain strict lower limits for the five sources above, while we estimate an upper limit on the CO abundance for the jet lobes where no CO is detected in the B and R knots (the "CO-poor" jets: SerpM-S68Nb knot B, SVS13B knots B and R, L1448-NB knot B, and SerpM-S68N knot R). 
Based on the estimated column densities and on the assumption that the emission in the three tracers originates from the same jet component, the abundances of SO and SiO are derived as \Xso = \Xco $\times$ \Nso/\Nco\, and \Xsio = \Xco $\times$ \Nsio/\Nco,\, where \Xco $=10^{-4}$ is the assumed CO abundance in the jet with respect to H$_2$\footnote{\label{note:CO-abu} The adopted value of [CO]/[H] $= 5 \times 10^{-5}$ is based on the correlation of the total CO abundance (gas+ice) and the visual extinction $A_{\rm V}$ in the well shielded ISM \citep{whittet10}
using a standard $N_{\rm H} / A_{\rm V} = 2 \times 10^{21}$ cm$^{-2}$ mag$^{-1}$. Since high-velocity protostellar jets are launched well inside the CO ice-line, all CO is in the gas phase. If the jet is launched from inside the dust sublimation radius, 
the CO abundance could be larger or smaller by a  factor of $\ge 3$ \citep[see, e.g., ][]{glassgold91,tabone20}}.

\begin{figure*}
\centering
\includegraphics[width=0.7\textwidth]{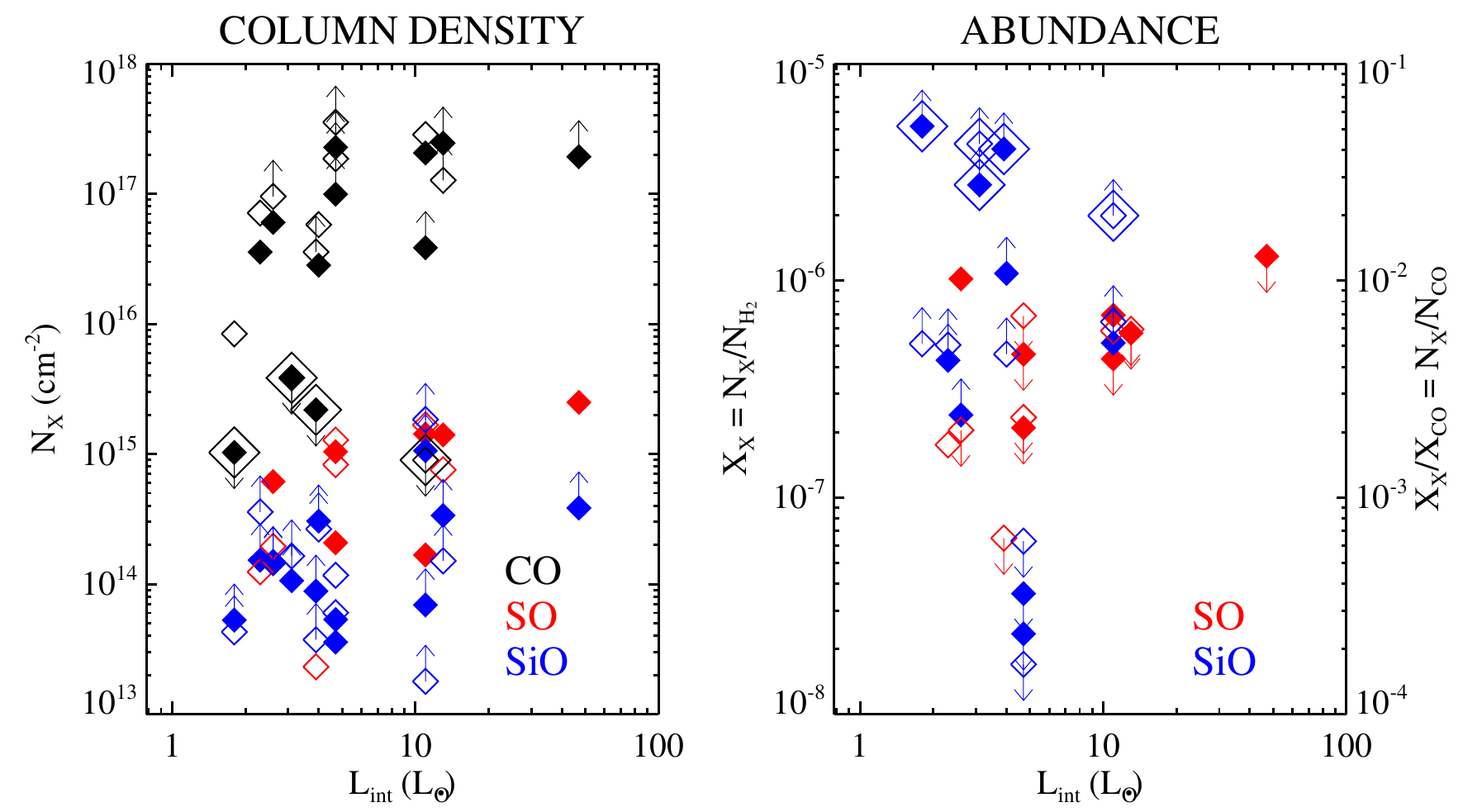}
\caption{Beam-averaged column densities ($N_{\rm X}$ in \cms, {\it left}), and SO and SiO abundances ({\it right}) versus the source internal luminosity ($L_{\rm int}$ in L$_{\odot}$).
Molecular abundances are with respect to H$_2$ on the left axis (assuming [CO]/[H$_2$] $= 10^{-4}$) and with respect to CO on the right axis. The values are inferred for the 12 Class 0 protostars driving an SiO jet detected at $> 10\sigma$ in the integrated maps (see Table \ref{tab:jet-occurrence}) at the position of the closest blue- and red-shifted SiO knots, B and R (filled and empty small diamonds, respectively; see Tables \ref{tab:fluxes} and \ref{tab:jets-energetics}). Black, red, and blue symbols are for CO, SO, and SiO, respectively. Lower and upper limits are indicated by upward and downward arrows. The jet knots where no CO is detected (the "CO-poor" jets) are indicated by larger empty diamonds. For these knots we derive upper limits on \Nco, and lower limits on \Xsio.}
\label{fig:jet-abundances}
\end{figure*}

The estimated molecular column densities and abundances averaged over the beam and the HV range for the blue- and red-shifted inner knots of the high-velocity jets are summarized in Table \ref{tab:jets-energetics} and in Figure \ref{fig:jet-abundances}, where the values are shown as a function of the source internal luminosity, \lint. As the CO abundance in jets is uncertain (see footnote \ref{note:CO-abu}), in Figure \ref{fig:jet-abundances}, we also plot the SO and SiO abundance with respect to CO (\Xso/\Xco\,=\,\Nso/\Nco, and \Xsio/\Xco\,=\,\Nsio/\Nco).  
We find that the beam-averaged column density of SO is \Nso$\sim 10^{13} - 3 \times 10^{15}$ \cms, and \Nsio\, goes from a minimum of $4 \times 10^{13}$ \cms\, to  $> 2 \times 10^{15}$ \cms\, for the jets where the SiO emission is (or could be) optically thick. 
The CO column density ranges from $\sim 10^{16}$ \cms\, up to $> 3 \times 10^{17}$ \cms, with the exception of the "CO-poor" jets for which we derive an upper limit of a few $10^{15}$ \cms. The abundance of SO with respect to H$_2$ goes from values $< 10^{-7}$ to  $10^{-6}$. For SiO, the inferred abundances, \Xsio, are lower limits for the jets where the SiO emission is (or could be) optically thick (\Xsio\, ranges from values larger than a few $10^{-7}$ to values larger than a few $10^{-6}$ for the "CO-poor" jets), whereas for the two jets where  CO and SiO are optically thick in both lobes (SerpS-MM18a, IRAS2A1), it was not possible to derive an estimate of \Xsio. Low SiO abundances are found only for IRAS4A1 and IRAS4A2 (\Xsio$\le2-6 \times 10^{-8}$).

\subsection{Jet energetics}
\label{sect:jet-energetics}

For the 12 Class 0 protostars driving an SiO jet detected at $>10\sigma$ (see Table \ref{tab:jet-occurrence}), we estimate the jet mass-loss and momentum rates, and the jet mechanical luminosity from the CO beam-averaged column density inferred from the high-velocity CO emission at the position of the innermost blue- and red-shifted SiO knots, B and R.
As discussed in Sects. \ref{sect:kinematics} and \ref{sect:jet-abundances}, while CO emission at low velocity probes the outflowing entrained material, the emission at high velocity is more collimated and has the same width and displacement as the SiO emission (see the width of the HV emission in Fig. \ref{fig:jet-width-hv} and the PV of SiO and CO emission in Fig. \ref{fig:PV-block1}). Hence, we assume that the bulk of the CO HV emission originates from the jet and that the contribution from entrained gas is negligible. 
To further minimize the possible contribution from entrained material, the CO column densities and mass-loss rates are inferred close to the driving source, that is, at the position of the B and R inner knots. In fact, at larger distances, the jet could be more affected by the interaction with the surrounding medium, hence, by gas entrainment, and part of the emission could be filtered out by the interferometer. 
This approach also guarantees that the jet properties are derived via the same methodology with regard to all the jets in our sample, including those where only a single red-shifted and blue-shifted knot is detected.

The mass-loss rate of the molecular jet is typically estimated assuming that the mass in the jet flows at constant density and speed along the jet axis over the beam length. However, the gas in the SiO knots where the gas column density is estimated is highly compressed by shocks, therefore the mass-loss rate is corrected for a factor of $1/\sqrt{C} \sim1/3$, where C is the compression factor \citep[e.g., ][]{hartigan94}. Taking shock compression into account, we infer the mass-loss rate as $\dot{M}_{\rm jet} = 1/\sqrt{C} \times m_{H_2} \times (N_{\rm CO}/X_{\rm CO}) \times b_t \times V_{\rm tan}$ \citep[e.g., ][]{lee07b,podio15}, where $m_{H_2}$ is the mass of molecular hydrogen, $N_{\rm CO}$ the CO beam-averaged column density over the HV range, $X_{\rm CO} = 10^{-4}$ the assumed CO abundance (see footnote \ref{note:CO-abu}), $b_t$ the linear size of the transverse beam (see Appendix \ref{app:jet-width}), and  $V_{\rm tan}$ the tangential component of the jet velocity, $V_{\rm jet}$, that is,$V_{\rm tan} = \sqrt{V_{\rm jet}^2 - V_{\rm rad}^2}$.

The true jet velocity, \Vjet, as well as its tangential component, $V_{\rm tan}$, can be recovered by correcting the observed radial velocity, $V_{\rm rad}$, for the jet inclination.  However, estimates of the inclination are available only for a few jets in our sample
and they are obtained using different methods:
for SerpM-SMM4, \citet{aso18} suggest that the two lobes have different inclination based on the assumption that they have the same intrinsic momentum ($i\sim36\degr$ and $i\sim70\degr$ from the line of sight for the blue and red lobe, respectively), which gives deprojected velocities of $\sim105$ \kms\, and $\sim 50$ \kms, respectively;
for SVS13B \citet{seguracox16} estimates a disk inclination of $\sim 71\degr$, which implies $V_{\rm jet} = 74$ \kms\, for the blue lobe, and $49$ \kms\, for the red lobe if we deproject the $V_{\rm rad}$ at the emission peak (see Table \ref{tab:fluxes}), and velocities up to $150$ \kms\, if we consider the maximum $V_{\rm rad}$ in our spectra (in agreement with \citealt{bachiller98b});
for L1157 the  model by \citet{podio16} indicates precession on a cone inclined by $73\degr$ to the line of sight, which implies $V_{\rm jet} \sim 87$ \kms\, and $\sim 137$ \kms\, in the blue and red lobe, respectively;
finally, for L1448-C the jet velocity derived from proper motion studies is $V_{\rm jet} \sim 98 \pm4$ \kms\, and $\sim 78\pm1$ \kms\, for the blue- and red-shifted lobes, respectively ($i=34\pm4\degr$ and $i=46\pm5\degr$ from the plane of sky, \citealt{yoshida20}).
In conclusion, even if the estimated jet inclinations are obtained with different methods and are affected by large uncertainties, the derived \Vjet\, values are always consistent with $V_{\rm jet} = 100 \pm 50$ \kms.
Moreover, this value is also consistent with observations of other prototypical Class 0 jets where SiO proper motions have been measured (e.g.,  
\Vjet\,$\sim 115\pm50$ \kms\, for HH 212, \citealt{lee-cf15}, and $\sim 114 \pm 50$ \kms\, for HH 211, \citealt{jhan16}).
Therefore, we assume that the jet velocity is $100$ \kms\, for all the jets in our sample. This ensures that the same method to derive $\dot{M}_{\rm jet}$  is applied to all the jets in the sample, without making assumptions on the jet inclination when estimates are not available or very uncertain\footnote{Rough estimates of the inclination are available for a few other jets, e.g. IRAS4A and IRAS4B1, but are affected by even larger uncertainties. For example, for IRAS4B1 early studies suggest that the jet is almost perpendicular to the plane of the sky based on its morphology ($i\sim0\degr$, \citealt{maret09}; $i\sim15-30\degr$,  \citealt{yildiz12}) or on VLBI H$_2$O water maser observations ($i\sim10-35\degr$, \citealt{desmurs09}), while \citet{marvel08} found the maser outflows to be nearly in the plane of the sky ($i \sim77\degr$). Similarly for IRAS4A \citet{yildiz12} suggest an inclination of $\sim45-60\degr$ to the line of sight, \citet{koumpia16} of $\sim70\degr$, and \citet{marvel08} of $\sim88\degr$.}.

The determination of the mass-loss rate allows us to estimate the momentum rate, $\dot{P}_{\rm jet} = \dot{M}_{\rm jet} \times V_{\rm jet}$, and the jet mechanical luminosity $L_{\rm jet} = 1/2 \times \dot{M}_{\rm jet} \times V_{\rm jet}^2$ at the position of the innermost blue- and red-shifted SiO knots, B and R.
As the inferred mass-loss and momentum rates, and mechanical luminosities depend linearly on the estimated beam-averaged CO column density, \Nco, they are affected by the uncertainties on \Nco\, due to the assumed temperature and LTE conditions. Moreover, as we assume optically thin emission to derive \Nco, the estimated \mjet, \pjet, and \ljet\, values should be regarded as lower limits for all jets where the comparison of CO and SiO spectra indicates that the CO emission is (or could be) optically thick (i.e., when $T_{\rm B}^{\rm SiO}$/$T_{\rm B}^{\rm CO} \le1$, see Sect. \ref{sect:jet-abundances}).
Finally, the \mjet, \pjet, and \ljet\, values are uncertain by a factor of 3-10 due to the assumption on the CO abundance, on \Vjet, and on the compression factor, which depends on the unknown shock parameters (magnetic field and shock speed, \citealt{hartigan94}). 
This, however, does not affect the general trends and the comparison with the mass accretion rates and the mass-loss rates estimated for Class II sources, which are discussed in Sect. \ref{sect:disc-ejec-accr}.

\begin{figure*}
\centering
\includegraphics[width=0.7\textwidth]{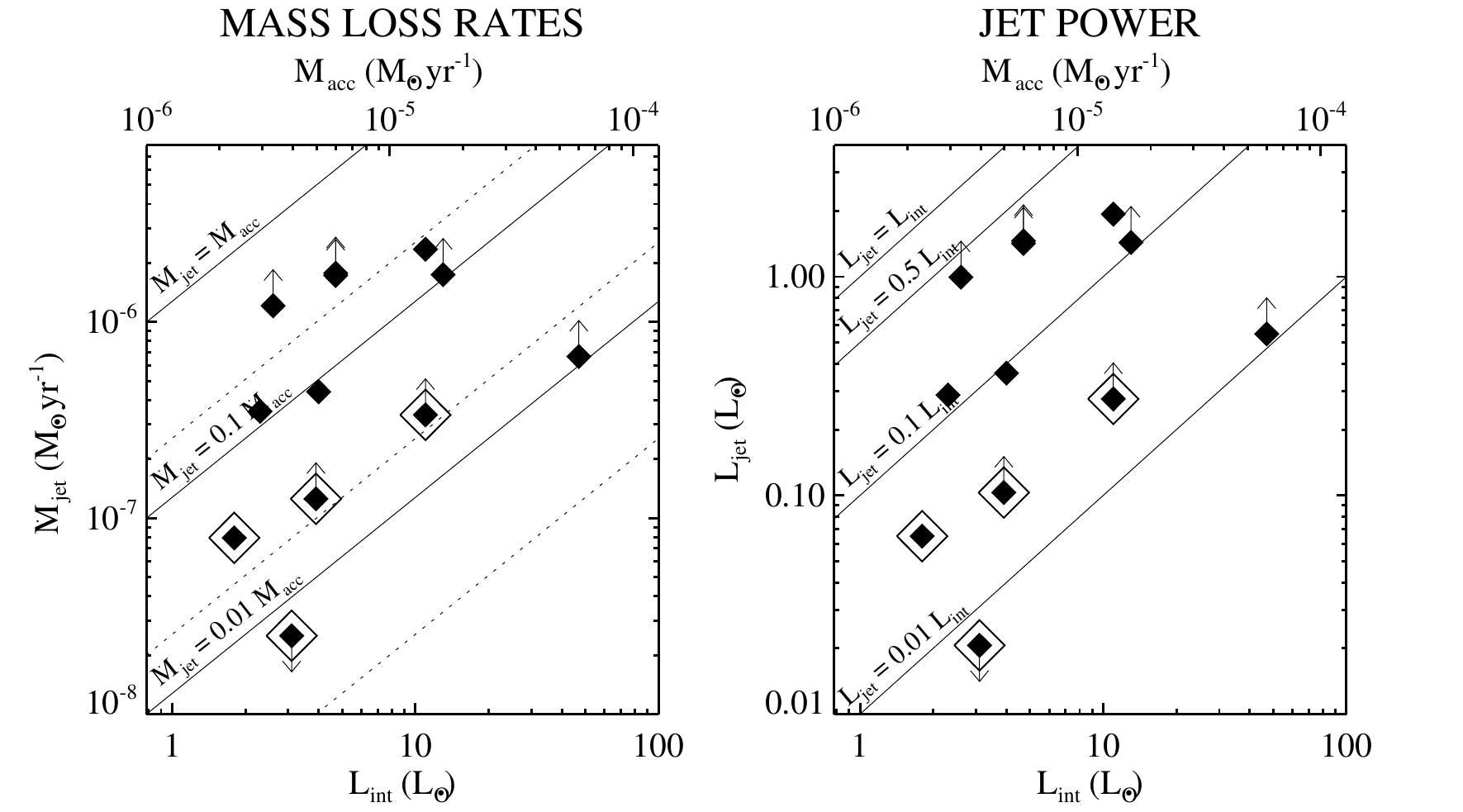}
\caption{Two-sided jet mass-loss rates (\mjet\, in \msolyr, {\it left}), and jet mechanical luminosities (\ljet\, in \lsol, {\it right}) versus the source internal luminosity ($L_{\rm int}$ in L$_{\odot}$). The values are inferred for the 12 Class 0 protostars driving an SiO jet detected at $> 10\sigma$ in the integrated maps (see Table \ref{tab:jet-occurrence}). $\dot{M}_{\rm jet}$ and \ljet\, are the sum over the closest blue- and red-shifted SiO knots, B and R (see Tables \ref{tab:fluxes} and \ref{tab:jets-energetics}). Lower and upper limits are indicated by upward and downward arrows. The jet knots where no CO is detected (the "CO-poor" jets) are indicated by larger empty diamonds. For the jets with only one "CO-poor" lobe, the total \mjet\, and \ljet\, is dominated by the estimate obtained for the other lobe, hence, they show up as a firm value or lower limit (depending if CO in the other lobe is optically thin or thick). The exception is SVS13B for which no CO is detected in both lobes, hence, only an upper limit is derived for the total \mjet\, and \ljet.
The black lines in the left panel indicate $\dot{M}_{\rm jet} = [0.01, 0.1, 1] \times \dot{M}_{\rm acc}$, with $\dot{M}_{\rm acc}$ estimated from \lint, assuming a protostellar mass of $0.05$ \msol\, or $0.25$ \msol\, (solid and dotted lines, respectively). The $\dot{M}_{\rm acc}$ values labeled on the upper x-axis correspond to $M_{*} = 0.05$ \msol. The black solid lines in the right panel indicate $L_{\rm jet} = [0.01, 0.1, 0.5, 1] \times L_{\rm int}$.}
\label{fig:jet-energetics}
\end{figure*}

The estimated mass-loss and momentum rates, and jet mechanical luminosities for the blue and red lobes of the jets are summarized in Table \ref{tab:jets-energetics}. Figure \ref{fig:jet-energetics} shows the sum of the jet mass-loss rate, \mjet, and of the jet mechanical luminosity, \ljet, on the innermost knots of the two jet lobes, B and R. For the jets with one "CO-poor" lobe, the other lobe shows CO emission in the HV range (sometimes possibly thick), hence the total \mjet\, and \ljet\, shows up as a firm value (SerpM-S68Nb) or lower limit (L1448-NB, SerpM-S68N) in Fig. \ref{fig:jet-energetics}. The exception is SVS13B for which no CO is detected in both lobes and we could only infer an upper limit on the total mass-loss rate and mechanical luminosity. We find that the two-sided jet mass-loss rates (sum over the blue and red inner knots) span from values of $\sim 7 \times 10^{-8}$ \msolyr\, up to $\sim 3 \times 10^{-6}$ \msolyr. Consequently, the total jet momentum rates vary between  $\sim 7 \times 10^{-6}$ \msolyr \kms\, up to $\sim 3 \times 10^{-4}$ \msolyr \kms, while the jet mechanical luminosity summed on the two lobes (i.e. the total jet power) is between 0.06 \lsol\, and $2$ \lsol. The exception is the "CO-poor" jet SVS13B for which we obtain \mjet\,$<2  \times 10^{-8}$ \msolyr, \pjet\,$<2  \times 10^{-6}$ \msolyr \kms, and \ljet\,$< 0.02$ \lsol. The obtained values are discussed in Sect. \ref{sect:disc-ejec-accr}.


\begin{table*}
\caption[]{\label{tab:jets-energetics} Beam-averaged molecular column densities and abundances, mass-loss and momentum rates, and jet mechanical luminosities for the 12 sources driving an SiO jet detected at $>10\sigma$ in the integrated maps (see Table \ref{tab:jet-occurrence}).}
\small{
\begin{tabular}[h]{ccc|ccc|cc|ccc}
\hline
\hline
Source & & \lint   & \Nco$^a$         & \Nso$^a$         & \Nsio$^a$        & \Xso$^b$     & \Xsio$^b$   & \mjet$^c$ & \pjet$^c$ & \ljet$^c$ \\
       & & (\lsol) & (10$^{17}$ \cms) & (10$^{14}$ \cms) & (10$^{14}$ \cms) & ($10^{-7}$) & ($10^{-7}$) & (10$^{-7}$ \msolyr) & (10$^{-5}$ \msolyr \kms) & (\lsol)\\
\hline
SerpM-S68Nb & B &  1.8 & $<0.01$ &    --  & $\ge0.5$ &      --  &  $>51.5$ &  $<0.08$ &  $<0.08$ & $<0.006$  \\  
            & R &      &    0.08 &    --  & $\ge0.4$ &      --  & $\ge5.1$ &      0.7 &      0.7 &  0.06 \\ 
    IRAS4B1 & B &  2.3 &     0.4 &    --  & $\ge1.5$ &      --  & $\ge4.3$ &      1.2 &      1.2 &  0.1 \\ 
            & R &      &     0.7 &    1.2 & $\ge3.6$ &      1.7 & $\ge5.0$ &      2.3 &      2.3 &  0.2 \\ 
SerpM-SMM4b & B &$<2.6$&     0.6 &    6.1 & $\ge1.4$ &     10.2 & $\ge2.4$ &      3.4 &      3.4 &  0.3 \\ 
            & R &      &  $>1.0$ &    1.9 &   $>1.5$ &   $<2.0$ &      --  &   $>8.7$ &   $>8.7$ & $>0.7$ \\ 
     SVS13B & B &  3.1 & $<0.04$ &    --  & $\ge1.1$ &      --  &  $>27.6$ &   $<0.1$ &   $<0.1$ & $<0.01$ \\ 
            & R &      & $<0.04$ &    --  & $\ge1.6$ &      --  &  $>42.7$ &   $<0.1$ &   $<0.1$ & $<0.01$ \\ 
    L1448NB & B &  3.9 & $<0.02$ &    --  & $\ge0.9$ &      --  &  $>40.4$ &  $<0.07$ &  $<0.07$ & $<0.006$  \\ 
            & R &      &  $>0.4$ &    0.2 &   $>0.4$ &   $<0.6$ &      --  &   $>1.2$ &   $>1.2$ & $>0.1$ \\ 
      L1157 & B &  4.0 &     0.3 &    --  & $\ge3.0$ &      --  & $\ge10.8$&      1.6 &      1.6 &  0.1 \\ 
            & R &      &     0.6 &    --  & $\ge2.7$ &      --  & $\ge4.6$ &      2.9 &      2.9 &  0.2 \\ 
    IRAS4A1 & B &$<4.7$& $\ge1.0$ &   2.1 &      0.4 & $\le2.1$ & $\le0.4$ & $\ge4.0$ & $\ge4.0$ & $\ge0.3$ \\ 
            & R &      & $\ge3.5$ &   8.3 &      0.6 & $\le2.3$ & $\le0.2$ &  $\ge14$ &  $\ge14$ & $\ge1.1$ \\ 
    IRAS4A2 & B &  4.7 &$\ge2.3$ &   10.4 &      0.5 & $\le4.6$ & $\le0.2$ & $\ge9.5$ & $\ge9.5$ & $\ge0.8$ \\ 
            & R &      &$\ge1.9$ &   12.8 &      1.2 & $\le6.9$ & $\le0.6$ & $\ge7.8$ & $\ge7.8$ & $\ge0.6$ \\ 
    L1448-C & B & 11   &     2.1 &   14.2 &$\ge10.6$ &      6.9 & $\ge5.1$ &      9.7 &      9.7 &  0.8 \\ 
            & R &      &     2.9 &   16.7 &$\ge18.4$ &      5.9 & $\ge6.4$ &     13.8 &     13.8 &  1.1 \\ 
 SerpM-S68N & B & 11   &  $>0.4$ &    1.7 &   $>0.7$ &   $<4.3$ &      --  &   $>3.3$ &   $>3.3$ & $>0.3$ \\ 
            & R &      & $<0.01$ &    --  & $\ge0.2$ &       -- &  $>19.9$ &  $<0.08$ &  $<0.08$ & $<0.006$\\ 
SerpS-MM18a & B & 13   &  $>2.5$ &   14.0 &   $>3.4$ &   $<5.7$ &      --  &  $>11.5$ &  $>11.5$ & $>0.9$ \\ 
            & R &      &  $>1.3$ &    7.6 &   $>1.5$ &   $<5.9$ &      --  &   $>5.9$ &   $>5.9$ & $>0.5$ \\ 
    IRAS2A1 & B & 47   &  $>1.9$ &   24.9 &   $>3.8$ &  $<12.9$ &      --  &   $>6.7$ &   $>6.7$ & $>0.5$ \\ 
\hline
\end{tabular}\\
$^a$ beam-averaged CO, SO, and SiO column densities in the jet (\Nco, \Nso, \Nsio) are derived from the CO, SO, and SiO line intensity integrated on the high-velocity range at the position of the blue-shifted and red-shifted SiO emission knots closest to the driving sources, B and R (see Table \ref{tab:fluxes} and the spectra in Fig. \ref{fig:spec1}). We assume LTE, optically thin emission at $T_{\rm K}$=100 K. The lower and upper limits reported in the table refer to the cases where the CO ($2-1$) and/or SiO ($5-4$) emission is (or could be) optically thick based on the $T_{\rm B}^{\rm SiO}$/$T_{\rm B}^{\rm CO}$ ratio (see Sect. \ref{sect:jet-abundances}).\\
$^b$ the abundances of SO and SiO are derived as \Xso = \Xco $\times$ \Nso/\Nco\, and \Xsio = \Xco $\times$ \Nsio/\Nco\, assuming $X_{\rm CO} = N_{\rm CO}/N_{\rm H_2} =  10^{-4}$.\\
$^c$ Mass-loss and momentum rates (\mjet, \pjet), and the jet mechanical luminosity (\ljet) are derived from \Nco\, assuming \Vjet $= 100$ \kms.\\
}
\end{table*}

\section{Discussion}
\label{sect:discussion}

\subsection{Considering whether outflows and jets could be ubiquitous at the protostellar stage}

As summarized in Sect. \ref{sect:detection-rate}, Table \ref{tab:jet-occurrence}, and Figure \ref{fig:jet-occurrence}, outflow emission in CO ($2-1$) is detected in all the Class 0 and Class I protostars in our sample (21 and 3 sources, respectively), indicating that the outflow phenomenon is ubiquitous in the CALYPSO sample of protostars.

High-velocity collimated jets emitting in the SiO ($5-4$) line are detected in 67\% of the sample, which is a higher detection rate than previously found. 
For example, \citet{gibb04} found that only 45\% of the Class 0 sources in their sample are associated with SiO jets, likely due to the lower angular resolution and sensitivity of their observations.

SO ($5_6-4_5$) jet and outflow emission is detected in 52\% of the Class 0 protostars and 67\% of the Class I sources, and for all of them, the spatio-kinematical distribution of the SO emission follows that of SiO (see the position-velocity diagrams in Fig. \ref{fig:PV-block1} and line spectra in Fig. \ref{fig:spec1}). 
In five more sources (L1527, GF9-2, L1157, L1448-2A, SerpS-MM18b), that is, 24\% of the Class 0 protostars, the detected SO ($5_6-4_5$) emission is compact and  elongated perpendicularly to the jet or outflow PA with a velocity gradient along the same direction (perpendicular to the jet or outflow).
This indicates that the SO emission does not probe ejection but, instead, could trace the inner flattened envelope, the disk, or the accretion shock at disk-envelope interface \citep[e.g., ][]{sakai14a,maret20}. Unfortunately, with the exception of L1527 whose SO emission has been analyzed in detail by \citet{maret20}, for the other sources, the SO emission is too weak to allow for a study of the gas kinematics to assess their origin. Follow-up observations at higher angular resolution and sensitivity would be crucial for achieving an understanding of the origin of SO ($5_6-4_5$) in these sources.
Therefore, based on our survey, we find that SiO emission unambiguously probes the jet, while SO may also be a probe of the inner envelope and disk.

Despite the small size of our sample, Fig. \ref{fig:jet-occurrence} indicates that the detection rate of jets increases with the source internal luminosity  up to $\sim 80\%$ for sources with $L_{\rm int} >1$ \lsol.
The internal luminosity, \lint, is a probe of the mass accretion rate onto the source, $\dot{M}_{acc}$, and the mass ejection rate is expected to be proportional to the mass accretion rate ($\dot{M}_{\rm jet} \sim 0.01-0.3 \dot{M}_{\rm acc}$, according to magnetohydro-dynamical models of the jet launch and observations of Class II sources, e.g., \citealt{ferreira06,hartigan95}). Hence, the fact that the jet detection rate increases with \lint\, confirms the correlation between accretion and ejection rates found for more evolved sources and suggests that the jets driven by the less-accreting and less-ejecting protostars in our sample may remain undetected at the sensitivity level of our observations.

Finally, it is interesting to note that high-velocity collimated emission in all three jet tracers (CO, SO, and SiO) is detected for the protostellar candidates, IRAS4B2 and SerpM-SMM4b, and only in CO for the candidate SerpS-MM18b. This strongly supports their identification as protostars. 

\subsection{Jet velocity}

The median radial velocity of the SiO jets detected in the CALYPSO sample, as derived at the position of the innermost SiO knots, is about 30 \kms\, (see Fig. \ref{fig:distri-vrad}).
In contrast, \citet{2018A&A...609A..87N} find a median radial velocity of 70 \kms\, for the high-velocity components (HVC) seen in the [OI] line at 6300~$\AA$ in a large sample of Class II sources. The difference in velocity between protostellar and Class II jets may depend on the different tracers (SiO and [OI], respectively). However, if we assume that both SiO and [OI] probe the collimated jet launched directly by the disk and not entrained gas, 
this difference indicates that the velocity of the jet increases from the protostellar stage to the Class II stage by a factor of about $2$. Because of the selection bias of the HVC component in Class II sources (an [OI] jet seen on the plane of the sky would not be considered as jet component), this number should be considered as an upper limit. In the following, we assume that the CALYPSO sample probes a Class 0 population that will eventually form Class II objects with a similar IMF as the sample analyzed in \citet{2018A&A...609A..87N}. Adopting the model of \citet{bontemps96} to describe the growth of a protostar, and assuming that the median age of the Class 0 protostars in our CALYPSO sample is half of the duration of the Class 0 phase ($\sim 3 \times 10^4$~yrs, according to the former reference), the median mass of Class 0 protostars is $0.3 M_{*}$, where $M_{*}$ is the mass of the final star. The Keplerian velocity at a given distance from the central object would then increase by a factor of $\sim 1.8$ between samples of Class 0 and Class II objects. According to disk winds models of the jet launch, the velocity of the jet scales with the Keplerian velocity at the launching radius \citep[see, e.g.,][]{ferreira06}. In this scenario, the increase of the jet velocity with age is consistent with a jet launched from the same region of the disk around a central object of increasing mass.

Moreover, for at least 33\% of the detected SiO bipolar jets the radial velocity of the two lobes differ by a factor of $1.4-2.1$ (and $7.8$ for SerpM-S68Nb). The occurrence and degree of velocity asymmetries inferred for the protostellar jets in the CALYPSO sample are in agreement with what was found for optical jets from T Tauri stars. For example, \citet{hirth94} find that about 53\% of their sample of 15 bipolar optical jets show a difference in velocity between the two jet lobes of a factor of $1.4-2.6$  \citep[see also, ][]{hirth97,melnikov09,podio11}.

\subsection{Jet collimation}

The CALYPSO maps and position-velocity diagrams of the detected jets, and the estimates of their widths (see Figs. \ref{fig:jets1}, \ref{fig:PV-block1}, \ref{fig:jet-width-main}, and \ref{fig:jet-width-hv}) show, on a statistical basis, that protostellar molecular flows exhibit an onion-like structure with nested components of different velocities that are highlighted by different lines: CO ($2-1$) probes slow and wide angle outflows ($\sim 8-35\degr$ opening angle), while the high-velocity jet is collimated and best traced by SiO ($\sim 10\degr$ opening angle) but is also seen in CO and SO when  high-velocity components are selected. Interestingly, higher angular resolution observations toward the HH 212 Class 0 system show a similar onion-like structure on a smaller scale (<200~au) with SiO tracing a collimated jet launched from $\sim 0.3~$au, and SO tracing a broader and slower outflow, possibly associated with an MHD disk wind launched from a larger radius in the disk \citep{lee17b,2017A&A...607L...6T,lee18b}. In the latter studies, rotation signatures detected in both the jet and the outflow were used to derive the launching zone. We note, however, that our CALYPSO resolution is too low to uncover rotation signatures and derive the launching radius.

The jet width, 2\Rjet, derived from the spatial profile of the SiO ($5-4$) line can be compared with the results obtained for atomic jets driven by Class I and Class II sources. The lowest dashed green curve in Figure \ref{fig:jet-width-main} shows the typical collimation inferred for Class I and Class II high-velocity atomic jets on the same scale probed by our observations, that is, $300-1500$ au, with an opening angle of $\sim 3\degr$ and a width of $\sim80$ au at $800$ au distance \citep[e.g., ][]{agraamboage11,dougados00,reipurth00,woitas02}. Figures \ref{fig:jet-width-main} and \ref{fig:jet-width_all} show that except for IRAS4B1, SVS13B, and L1448-NB, most of the high-velocity SiO jets in our sample appear systematically broader than Class I and Class II atomic jets. This could be due to the lower resolution of the CALYPSO observations ($\sim 0\farcs4-1.1\arcsec$, i.e., $\sim120-480$ au at $d=293-436$ pc) compared to studies of Class I and II sources ($\sim 0\farcs1$, i.e. $14-46$ au $d=140-460$), as well as to projection effects, and to the lower temperature regime probed by SiO emission with respect to the atomic lines, which makes SiO sensitive to emission by bow-shocks wings further out than optical lines in atomic jets. Higher angular resolutions observations are required to probe the collimation properties of SiO jets down to $<100$ au scale as for atomic jets from Class II sources. To date jet width measurements on $<100$ au scales were obtained only for two protostellar jets observed at a few au resolution with ALMA, that is, HH 212 \citep{lee17b} and B355 \citep{bjerkeli19}, who find that the jet width is comparable or even smaller than what was found for Class II atomic jets \citep[see, e.g., the recent review by ][]{lee20}.

\subsection{Origins of gas phase SiO}

The SiO abundances inferred for the high-velocity jets in the CALYPSO sample are summarized in Table \ref{tab:jets-energetics} and Figure \ref{fig:jet-abundances}. 
For eight out of the twelve SiO jets detected at $>10\sigma$ in the integrated maps, we infer a lower limit on the SiO abundance which ranges from $>2.4 \times 10^{-7}$ up to $> 5 \times 10^{-6}$. These values are larger than what was estimated by previous low resolution observations \citep[e.g., ][]{gibb04} and they require for  $>1\%$ up to $>10\%$ of elemental Silicon is released in the gas phase, assuming the silicon solar abundance, [Si/H]$_{\odot} \sim 3.5 \times 10^{-5}$ \citep{holweger01}.

Two types of scenarios have been invoked to account for the release of silicon into the gas phase, and the subsequent formation of SiO, in jets \citep[see, e.g., ][]{cabrit12}:
(i) shock processing of silicate grains in a dusty wind launched from the disk \citep[e.g., ][]{panoglou12} and
(ii) silicon release at the base of dust-free jets launched from within the dust-sublimation radius of silicates \citep[e.g., ][]{1991ApJ...373..254G,tabone20}.
In the following, we discuss these two scenarios by comparing the SiO abundances inferred for the jets in the CALYPSO sample with what is predicted by models of shocks and dust-free jets.

In the first scenario, SiO jets trace dusty material, either in a magneto-hydrodynamical disk wind launched from beyond the dust sublimation radius of silicates ($R_{\rm sub} \simeq 0.15\,{\rm au} \sqrt{L_{\rm bol}/L_{\odot}}$, i.e. $\sim 0.5$~au for a bolometric luminosity $L_{\rm bol} \sim 10$ $L_{\odot}$, \citealt{2016A&A...585A..74Y}) or envelope material entrained by the jet. Most of the silicon is initially locked in the grain cores but may be released in the gas phase in the shocks occurring along the jet. Models of C-type magnetized shocks show that due to ion-neutral decoupling, grain cores are sputtered and silicon is released into the gas phase \citep{1997A&A...321..293S}. Subsequent reactions with O$_2$ or OH produce SiO in the warm post-shock region. Stationary C-shock models  predict an SiO abundance ($X_{\rm SiO} = N_{\rm SiO}/N_{\rm H_2}$) in the range of $8 \times 10^{-8} \le X_{\rm SiO} < 6 \times 10^{-7}$ for shock velocities of $20-50$ \kms\,  and pre-shock densities of $n_{\rm H} = 10^4 - 10^6$ cm$^{-3}$ \citep{2008A&A...482..809G}. The upper edge of the predicted values is at the lower edge of the lower limits on \Xsio\, inferred for the jets in the CALYPSO sample (\Xsio\, from $>2.4 \times 10^{-7}$ up to $> 5 \times 10^{-6}$). Moreover, the timescale to sputter Si from grains and produce large abundance of SiO ($> 10^{-7}$) in C-type shocks is $>100$ years, unless the pre-shock density is high ($n_{\rm H} = 10^6$ cm$^{-3}$). This timescale may be too long to account for the observed SiO abundances in the knots close to the source, which have short dynamical timescales ($< 100$ years).
For example, the SiO abundance in the L1157 jet is $> 10^{-6}$, which implies that at least $2\%$ of elemental silicon is released in the gas phase in the knot B located at a distance of $\sim 80$~au from the driving source (see Tables \ref{tab:fluxes} and \ref{tab:jets-energetics}), indicating that this large abundance is reached in $\sim 5$~yr if the jet travels at $\sim 100$ \kms\, \citep{podio16}. On the other hand, C-type shock models assuming that $5\%-10\%$ of the silicon is initially locked in the dust mantles in the form of SiO \citep{gusdorf08b}, or models including dust shattering and grain vaporization at high density ($n_{\rm H} \ge 10^{5}$ cm$^{-3}$, \citealt{guillet11}) enhance the release of silicon into the gas phase (up to $\sim 8\%$) and the formation of SiO (up to an abundance of $4 \times 10^{-7} - 10^{-6}$) on a timescale of $\le 10$ years.
Dust processing could efficiently produce SiO also in J-type shocks, though requiring higher shock velocities and high magnetization \citep{guillet09}. In both C-type and J-type shock models, the fraction of silicon released into the gas phase and converted into SiO depends crucially on the shock velocity and pre-shock density. However, whatever the considered processes and shock parameters (pre-shock density and shock velocity), the lower limits on the SiO abundance inferred for the jets in our sample (from $>2.4 \times 10^{-7}$ up to $> 5 \times 10^{-6}$) are at the upper edge or larger than the values predicted by shocks models ($10^{-6}$ at most).

Alternatively, if SiO jets are launched within the dust-sublimation radius of silicates, the majority of silicon is released into the gas phase at the base of the jet.  However, pioneering models of dust-free stellar winds have shown that the abundance of molecules, such as SiO and CO, can be severely reduced when a far ultra-violet (FUV) field is present for two reasons: (i) in the absence of dust, the FUV field can more easily penetrate the unscreened flow and photodissociate molecules; (ii) H$_2$ formation on grains, the starting point of molecule synthesis, is severely reduced \citep{1991ApJ...373..254G}. On the other hand, recent models of laminar dust-free disk winds show that if the jet is launched from the disk within the dust-sublimation radius, CO and SiO may be abundant if the mass-loss rate is $\gtrsim 10^{-6}$ \msolyr\, and the temperature is $T \gtrsim 800$~K \citep{tabone20}. In the optimal case, SiO is the main carrier of elemental Si with SiO/CO ratio as high as $0.1$. For lower mass-loss rates, the abundance of SiO and CO is predicted to drop by several orders of magnitudes. Our observations show that two of the jets that are SiO rich also have high observed mass-loss rates (as derived from CO) (SerpM-SMM4b, L1448-C), in agreement with the predictions by dust-free disk winds. However, even jets with observed mass-loss rates $< 10^{-6}$  \msolyr\, (namely, SerpM-S68Nb, IRAS4B1, SVS13B, L1448-NB, L1157, and SerpM-S68N) show large SiO abundances (\Xsio\, from $> 5 \times 10^{-7}$ to $> 5 \times 10^{-6}$). This could be due to the presence of a non-vanishing fraction of surviving dust or to the impact of shocks which are expected to increase the SiO abundance by compressing and heating the gas \citep{tabone20}.
In conclusion, our finding that protostellar jets are SiO rich on a large sample of sources support the pioneering result of \citet{cabrit12} who found a SiO abundance which accounts for up to 40\% of elemental silicon in the jet from HH 212. In this context, dust-free jets can be a viable scenario to explain SiO-rich jets.
\subsection{Ejection and accretion properties}
\label{sect:disc-ejec-accr}

The estimated mass-loss rates, \mjet, and jet mechanical luminosities, \ljet, are key to the understanding of the role of jets in the energy budget of the star-formation process. To date, these parameters have been estimated only for a handful of Class 0 protostellar jets \citep[e.g., ][and references therein]{lee20}. 
In the following, we compare the mass-loss rates estimated for the Class 0 sources in the CALYPSO sample with the \mjet\, derived for Class II sources and with estimates of the mass accretion rates.

There is increasing observational evidence that accretion and ejection in young stars are episodic \citep[e.g., ][]{audard14,plunkett15b}. However, an empirical correlation of the time-averaged CO outflow force with the quiescent (current) accretion luminosity is observed in Class 0 sources \citep[e.g., ][]{bontemps96}. Similarly, a correlation between the time-averaged mass-loss rates of atomic jets and the accretion rates has been observed for T Tauri stars \citep[e.g., ][]{hartigan95,ellerbroek13}. Therefore, even if outbursts dominate the integrated momentum injected in outflows, it appears that the time-averaged results correlates with the quiescent level of accretion. The reason for such a behavior is unclear. Simulations proposed to explain the knotty structure of jets show that even if the mass-loss rate of the jet is constant, the knots can be produced by a small periodical variation of the ejection velocity \citep[e.g., ][]{raga90,lee04}.

Our goal is to use the empirical correlations between ejection and accretion rates as a proxy for investigating potential similarities or differences in the ejection mechanism with source age by making a comparison with similar correlations found for other samples. 
Figure \ref{fig:jet-energetics} shows that the two-sided jet mass-loss rates, \mjet, inferred from high-velocity CO emission, range from $7 \times 10^{-8}$ \msolyr\, up to $3 \times 10^{-6}$ \msolyr\, for \lint\,$\sim 1-50$ \lsol. These values are in agreement with those found for a few other molecular jets from Class 0 protostars \citep[e.g., ][ and references therein]{lee20} and are larger by up to five orders of magnitude than the \mjet\, values estimated for atomic jets driven by Class II sources (from $\sim 10^{-11}$ to a few $10^{-8}$ \msolyr, e.g., \citealt{hartigan95,coffey08,podio12}). This indicates that the ejection rate decreases from a few $10^{-6}$ to $10^{-11}$ \msolyr\, as the source evolves from the Class 0 to the Class II stages and accretion proceeds at a slower pace.
Observational studies of Class II sources compare the jet mass-loss rate, inferred from the luminosity of atomic emission lines, with the mass accretion rate, derived from veiling or Hydrogen emission lines \citep[e.g., ][]{hartigan95,muzerolle98c} and show that the ejection and accretion rates are correlated; in particular, \mjet/\macc\, varies between 0.01 and 0.3 \citep{hartigan95,coffey08,cabrit07a,ellerbroek13}.
To compare the ejection properties of the Class 0 protostellar jets detected by CALYPSO with those of the Class II atomic jets, we derive the mass accretion rates of the protostars in our sample from their internal luminosity. Assuming that the source internal luminosity is due to the gravitational energy released by the accretion onto the protostar, that is, $L_{\rm int} = L_{\rm acc}$, the source mass accretion rate (\macc) can be estimated as $\dot{M}_{\rm acc} = L_{\rm int} \, R_* / (G \, M_*)$.
We assume that the stellar radius is $R_* = 2$ \rsol\, \citep{stahler88} and for the protostellar mass we assume a range of values $M_* = 0.05-0.25$ \msol\, \citep[e.g., ][]{yen17}. This range of masses encompasses the kinematic estimates obtained for a few Class 0 protostars, including three sources in our sample (L1157, IRAS4A2, L1448-C), from the fit of the rotation curves of their disk \citep[e.g., ][]{choi10,kwon15,lee20}. The estimated mass accretion rates are highly uncertain because they depend strongly on the protostellar properties which are unknown for most of the sources in our sample, and because accretion may be episodic and  characterized by accretion bursts. However, Fig. \ref{fig:jet-energetics} shows that assuming low protostellar mass, $M_* = 0.05$ \msol, the estimated  mass-loss rates are 10\%-50\% of the mass accretion rates for two-sided jets, while for the monopolar jet IRAS2A1 and the jets with one "CO-poor" lobe the mass-loss rate is 1\%-10\% of \macc. The exception is SVS13B, where both jets lobes are "CO-poor", for which we find  $\dot{M}_{\rm jet} <0.01 \dot{M}_{\rm acc}$. If we assume that the protostellar mass is  $M_* = 0.25$ \msol, the inferred \macc\, values are lower by a factor of 5. Therefore, for three jets (SerpM-SMM4b, IRAS4A1, and IRAS4A2), we find $\dot{M}_{\rm jet} \ge \dot{M}_{\rm acc}$, while for the rest of our sample we find $\dot{M}_{\rm jet} \sim 0.01-0.5 \dot{M}_{\rm acc}$. 
Despite the large spread of \mjet\, values (by about 1 order of magnitude at a given \lint) our estimates indicate that the jet mass-loss rates increases with the protostellar internal luminosity between $1$ and $50$ \lsol, hence, with the mass accretion rate (between $\sim 1.3 \times 10^{-6}$ \msolyr\, and $\sim 6 \times 10^{-5}$ \msolyr, respectively, for $M_* = 0.05$ \msol). Moreover, the correlation between the mass accretion rate and the mass ejection rate holds from the early protostellar stage probed by our CALYPSO survey ($10^4$ years) to evolved T Tauri stars of 1 Myr \citep[e.g., ][]{hartigan95}.

The total jet power, that is, the sum of the jet mechanical luminosity over the innermost knots along the blue and red lobes (B and R), \ljet, is 10\%-50\% of the source internal luminosity, \lint, for most of the jets in our sample (7 out of 12 jets), confirming the results found by \citet{lee20} collecting all the recent studies of six Class 0 protostellar jets from the literature. 
Despite the uncertainties affecting our estimates of \ljet\, mainly due to the uncertainty on the estimated CO column density, and on the assumed jet velocity, compression factor, and CO abundance, the correlation between \ljet\, and \lint\, suggests that the gravitational energy released by accretion onto the protostar could be efficiently extracted and converted into mechanical energy transported by the jet.
The exceptions are the jets where one lobe is "CO-poor", as in the case of SerpM-S68N and Nb, L1448-NB, and in the case of the monopolar jet IRAS2A1, which exhibit lower luminosity jets (i.e., $L_{\rm jet} \ge 0.01-0.1 \, L_{\rm int}$).
Finally, CO is not detected in both lobes of the SVS13B jet, therefore we infer an upper limit on its mechanical luminosity, $L_{\rm jet} <0.02$ \lsol, i.e. $< 0.01$ \lint. This value is much lower than what is derived for all the other jets in the sample, which could be explained by, for instance, an abnormally low CO abundance in this jet.

\section{Conclusions}
\label{sect:conclusions}

In this paper, we present a statistical survey of the occurrence and properties of jets and outflows in a sample of 21 Class 0 protostars mainly located in Perseus, Taurus, and Serpens. Our analysis is based on IRAM-PdBI observations of CO ($2-1$), SO ($5_6-4_5$), and SiO ($5-4$) taken in the context of the CALYPSO Large Program.
The main results of our analysis may be summarized as follows:
\begin{itemize}

\item[-] The observed tracers show the following differentiation: SiO ($5-4$) probes the collimated jet; CO ($2-1$) traces wide angle outflows at low velocity and the collimated jet at high velocities, where it is co-spatial with SiO; SO ($5_6-4_5$) is associated to the jet similarly to SiO in 52\% of the sources (e.g., IRAS4A1, IRAS4A2, L1448-C, SerpS-MM18a, IRAS2A1).
In 25\% of the sources (L1527, GF9-2, L1157, L1448-2A, SerpS-MM18b), the SO emission probes a compact circumstellar region with a small velocity gradient perpendicular to the jet and outflow axis ($|V - V_{\rm sys}| < 3$ \kms). This suggests that in these sources either SO traces the inner envelope or the disk, or the accretion shock at the interface between them, as in the case of L1527 \citep{sakai14a,maret20}.\\

\item[-] Collimated high-velocity jets traced by SiO ($5-4$) are detected in 67\% of the sources and 79\% of these also show jet emission in SO ($5_6-4_5$). The detection rate of jets increases with \lint, which is a probe of the mass accretion rate onto the protostar. 
This confirms the expected correlation between the mass accretion and the mass ejection rate (which, in turn, is proportional to the brightness of the emission lines). Hence, the non-detection of jet emission associated with the less-accreting sources could be an observational bias and deeper observations could demonstrate that jets are ubiquitous in our sample.
We detect for the first time high-velocity collimated SiO and SO jets in IRAS4B1, L1448-NB, and SerpS-MM18a, and in two protostellar candidates (IRAS4B2, and SerpS-MM18b, with the latter only in CO), supporting their identification as Class 0 protostars.\\

\item[-] 
Slow outflow emission in CO ($2-1$)  is detected in 100\% of the 21 Class 0 protostars suggesting that ejection phenomena are ubiquitous at the protostellar stage. We report for the first time the detection of a CO outflow for SerpS-MM22 and SerpS-MM18b.\\


\item[-] 
The median radial velocity of the detected SiO protostellar jets is $30$ \kms, that is, about two times smaller than the median radial velocity of atomic jets driven by Class II sources \citep{2018A&A...609A..87N}. Assuming that the velocity of the jet scales with the Keplerian velocity at the launching point, the increase of the jet velocity with age is consistent with a jet launched from the same region of the disk around a central object of increasing mass.
Moreover, at least 33\% of the detected SiO bipolar jets show a velocity asymmetry between the two lobes by a factor of $1.3-2.1,$ with the exception of SerpM-S68Nb, which shows a larger difference (by a factor of $7.8$). The occurrence and degree of velocity asymmetries inferred for the protostellar jets in the CALYPSO sample are in agreement with what was found for optical atomic jets from T Tauri stars \citep{hirth94}. The similarity in knot spacings and velocity asymmetries between Class 0 and Class II jets suggests that the jet launching mechanism in protostars of $10^4$ yr might be similar to that in Class II sources ($10^6$ yr).\\

\item[-] 
We find that 50\% of the 12 SiO jets detected in our sample show non-straight jets and this might indicate precession or wiggling.\\

\item[-] The observed protostellar flows have an onion-like structure: SiO ($5-4$) emission (the "jet probe") is more collimated than SO ($5_6-4_5$) emission, which in turn is narrower than CO ($2-1$) (the "outflow probe"), with median opening angles of 10$^{\circ}$, 15$^{\circ}$, and 25$^{\circ}$, respectively. However, high-velocity CO emission is as collimated as SiO. This indicates that low-velocity CO probes entrained material in the outflow, while high-velocity CO traces the collimated jet.
At scales larger than $300$ au, most of the high-velocity SiO jets are broader ($\sim 4\degr-12\degr$ collimation) than Class I and Class II atomic jets ($\sim 3\degr$ collimation). This could be due to projection effects as well as to contamination by the bow-shock wings at the low temperature probed  by  SiO.\\


\item[-]
We find that SiO ($5-4$) is optically thick in 26\% of the inner jet knots, and possibly thick in another 56\%. At least 67\% (8/12) of the jets are SiO rich (\Xsio\, goes from  $> 2.4 \times 10^{-7}$ to $> 5 \times 10^{-6}$), which requires that $>1\%-10\%$ of silicon is released in the gas phase, confirming the pioneering result by \citet{cabrit12} for the HH 212 jet.
This is difficult to explain in a scenario where dusty material launched from outside the dust sublimation radius is processed in shocks, especially in the inner knots which have short dynamical timescales ($\le 10$ years) \citep[see, e.g., shock models by][]{gusdorf08a,gusdorf08b}. On the other hand, formation of SiO in dust-free jets can be a viable scenario to explain SiO-rich jets for mass-loss rates $\ge 10^{-6}$ \msolyr.\\

\item[-] The mass-loss rates of the detected Class 0 molecular jets, \mjet, as derived from high-velocity CO emission, range from $7 \times 10^{-8}$ \msolyr\, up to values of $\sim 3 \times 10^{-6}$ \msolyr\, for internal luminosities of the driving protostars of $\sim 1-50$ \lsol. These \mjet\, values are larger by up to five orders of magnitude than those measured for atomic jets driven by Class II sources (from $\sim 10^{-11}$ to a few $10^{-8}$ \msolyr, \citealt{hartigan95,coffey08,podio12}). Moreover, despite the uncertainties affecting the estimates of the mass accretion rates for Class 0 protostars, due to the unknown protostellar mass and radius, we find that $\dot{M}_{\rm jet} \sim 0.1 - 0.5 \dot{M}_{\rm acc}$ for most of the jets, with the exception of the "CO-poor" and monopolar jets for which $\dot{M}_{\rm jet} \sim 0.01 - 0.1 \dot{M}_{\rm acc}$. These \mjet/\macc\, ratios are similar to those found for atomic Class II jets ($\dot{M}_{\rm jet} \sim 0.01 - 0.3 \dot{M}_{\rm acc}$, \citealt{hartigan95,cabrit07a,coffey08,podio12}) and indicate that the correlation between ejection and accretion holds over the whole star-formation process, from protostellar objects of 10$^4$ years to pre-main sequence stars of 1 Myr.\\

\item[-] The total jet power ($L_{\rm jet} = 1/2 \times \dot{M}_{\rm jet} \times V_{\rm jet}^2$) is 10\%-50\% of the source internal luminosity for $\sim 60\%$ of the jets in the sample. This indicates that the gravitational energy released by accretion onto the protostar could be efficiently extracted and converted into mechanical energy in the jet. For "CO-poor" and monopolar jets the jet power is lower ($\ge 1\%-10\%$ of the internal luminosity, with the exception of SVS13B, for which $L_{\rm jet } < 1\% L_{\rm int}$).\\

\end{itemize}

\begin{acknowledgements}
  We thank the IRAM staff for their support in carrying out the CALYPSO observations and the INSU “Action Spécifique ALMA” for their financial support to the CALYPSO collaboration. We are grateful to the annonymous referee for their instructive comments and suggestions.
  This work was also supported by the PRIN-INAF 2016 "The Cradle of Life - GENESIS-SKA (General Conditions in Early Planetary Systems for the rise of life with SKA)", and the European MARIE SKŁODOWSKA-CURIE ACTIONS under the European Union's Horizon 2020 research and innovation program, for the Project “Astro-Chemistry Origins” (ACO), Grant No 811312. L.P. acknowledges the European Union FP7, GA No. 267251. B.T. acknowledges support from the research program Dutch Astrochemistry Network II with project number 614.001.751, which is (partly) financed by the Dutch Research Council (NWO). 
\end{acknowledgements}

\bibliographystyle{aa} 
\bibliography{mybibtex.bib} 

\begin{appendix}

\section{Synthesized beam sizes and angular resolution}
\label{app:beam}

In Tables \ref{tab:beam_CO}, \ref{tab:beam_SO}, and \ref{tab:beam_SiO}, we report  the synthesized beam size ($b_{\rm maj} \times b_{\rm min}$, in $\arcsec$) and position angle (PA in $\degr$), and the corresponding angular scale (in au), for each of the sixteen targeted fields in Table \ref{tab:sample}, and for the three molecular tracers (CO ($2-1$), SO ($5_6-4_5$), and SiO ($5-4$)). We also report the velocity resolution and rms noise per channel of the line cubes.

\begin{table*}
\caption{Synthesized beam size and PA, corresponding angular scale, velocity resolution, and rms noise per channel for the CO ($2-1$) line cubes.}
\begin{center}
\begin{tabular}{lccccc}
\hline \hline
Source  & $b_{\rm maj} \times b_{\rm min}$ & (PA) & Res & $\Delta V$ & rms \\
& ($\arcsec \times \arcsec$) & ($\degr$) & (au) & (\kms) & (mJy beam$^{-1}$) \\
\hline
L1448-2A   & $0.62\times0.46$ & ($-145$) & 156 & 3.25 & 3 \\
L1448-NB1  & $0.55\times0.39$ & ($-143$) & 136 & 3.25 & 3 \\
L1448-C    & $0.80\times0.70$ & ($+187$) & 219 & 3.25 & 6 \\
IRAS2A1    & $0.62\times0.44$ & ($-130$) & 153 & 3.25 & 5 \\
SVS13B     & $0.57\times0.33$ &  ($+28$) & 127 & 3.25 & 3 \\
IRAS4A1    & $0.68\times0.44$ &  ($+47$) & 160 & 3.25 & 2 \\
IRAS4B1    & $0.59\times0.35$ & ($-150$) & 133 & 3.25 & 3 \\
IRAM04191  & $1.07\times0.83$ &  ($+31$) & 132 & 1.63 & 4 \\
L1521-F    & $1.14\times0.92$ &  ($+23$) & 143 & 3.25 & 2 \\
L1527      & $0.57\times0.36$ &  ($+41$) &  63 & 3.4  & 2 \\
SerpM-S68N & $1.21\times0.45$ &  ($+14$) & 322 & 1.0  & 6 \\
SerpM-SMM4a& $0.92\times0.51$ &  ($+30$) & 299 & 3.25 & 2 \\
SerpS-MM18a& $0.84\times0.49$ &  ($+27$) & 225 & 3.25 & 2 \\
SerpS-MM22 & $0.91\times0.43$ &  ($+22$) & 219 & 3.25 & 2 \\
L1157      & $0.88\times0.60$ &  ($+12$) & 256 & 3.25 & 4 \\
GF9-2      & $0.88\times0.60$ &  ($+18$) & 344 & 3.25 & 2 \\
\hline
\end{tabular}
\end{center}
\label{tab:beam_CO}
\end{table*}

\begin{table*}
\caption{Synthesized beam size and PA, corresponding angular scale, velocity resolution, and rms noise per channel for the SO ($5_6-4_5$) line cubes.}
\begin{center}
\begin{tabular}{lccccc}
\hline \hline
Source  & $b_{\rm maj} \times b_{\rm min}$ & (PA) & Res & $\Delta V$ & rms \\
& ($\arcsec \times \arcsec$) & ($\degr$) & (au) & (\kms) & (mJy beam$^{-1}$) \\
\hline
L1448-2A   & $0.60\times0.39$ & ($+36$) & 142 & 3.4 & 3 \\
L1448-NB1  & $0.77\times0.69$ & ($+43$) & 214 & 3.4 & 3 \\
L1448-C    & $0.60\times0.39$ & ($+36$) & 142 & 3.4 & 2 \\
IRAS2A1    & $0.78\times0.69$ & ($+44$) & 214 & 3.4 & 3 \\
SVS13B     & $0.67\times0.50$ & ($+37$) & 170 & 3.4 & 2 \\
IRAS4A1    & $0.69\times0.55$ & ($+34$) & 180 & 3.4 & 2 \\
IRAS4B1    & $0.69\times0.56$ & ($+32$) & 182 & 3.4 & 2 \\
IRAM04191  & $0.87\times0.72$ & ($+13$) & 111 & 1.0 & 7 \\
L1521-F    & $0.99\times0.72$ & ($+26$) & 118 & 1.0 & 5 \\
L1527      & $1.10\times0.81$ & ($+16$) & 132 & 1.0 & 8 \\
SerpM-S68N & $1.10\times0.57$ & ($+23$) & 345 & 1.0 &10 \\
SerpM-SMM4a& $1.13\times0.68$ & ($+34$) & 382 & 3.4 & 4 \\
SerpS-MM18a& $1.30\times0.61$ & ($+197$)& 311 & 3.4 & 4 \\
SerpS-MM22 & $1.24\times0.58$ & ($+199$)& 297 & 1.0 & 9 \\
L1157      & $0.59\times0.46$ &  ($-1$) & 183 & 3.4 & 3 \\
GF9-2      & $0.84\times0.73$ & ($+13$) & 371 & 1.0 & 6 \\
\hline
\end{tabular}
\end{center}
\label{tab:beam_SO}
\end{table*}

\begin{table*}
\caption{Synthesized beam size and PA, corresponding angular scale, velocity resolution, and rms noise per channel for the SiO ($5-4$) line cubes.}
\begin{center}
\begin{tabular}{lccccc}
\hline \hline
Source  & $b_{\rm maj} \times b_{\rm min}$ & (PA) & Res & $\Delta V$ & rms \\
& ($\arcsec \times \arcsec$) & ($\degr$) & (au) & (\kms) & (mJy beam$^{-1}$) \\
\hline \hline
L1448-2A   & $0.60\times0.39$ & ($-144$)& 142 & 3.4 & 3 \\
L1448-NB1  & $0.77\times0.69$ & ($+43$) & 214 & 3.4 & 3 \\
L1448-C    & $0.60\times0.39$ & ($+36$) & 142 & 3.4 & 3 \\
IRAS2A1    & $0.78\times0.69$ & ($+44$) & 214 & 3.4 & 4 \\
SVS13B     & $0.67\times0.50$ & ($+37$) & 170 & 3.4 & 2 \\
IRAS4A1    & $0.69\times0.55$ & ($+34$) & 180 & 3.4 & 3 \\
IRAS4B1    & $0.69\times0.56$ & ($+32$) & 182 & 3.4 & 2 \\
IRAM04191  & $0.93\times0.73$ & ($+19$) & 115 & 1.0 & 6 \\
L1521-F    & $1.01\times0.73$ & ($+28$) & 120 & 1.0 & 4 \\
L1527      & $0.64\times0.51$ & ($+51$) &  80 & 3.4 & 2 \\
SerpM-S68N & $1.13\times0.62$ & ($+25$) & 365 & 3.4 & 3 \\
SerpM-SMM4a& $1.05\times0.64$ & ($+30$) & 357 & 3.4 & 3 \\
SerpS-MM18a& $1.05\times0.60$ & ($+22$) & 278 & 3.4 & 3 \\
SerpS-MM22 & $1.25\times0.60$ & ($+20$) & 303 & 3.4 & 5 \\
L1157      & $0.59\times0.46$ & ($-181$)& 183 & 3.4 & 3 \\
GF9-2      & $0.86\times0.74$ & ($+13$) & 378 & 1.0 & 5 \\
\hline
\end{tabular}
\end{center}
\label{tab:beam_SiO}
\end{table*}

\section{Position-velocity diagrams}
\label{app:jet-pv}

In this section, we present the position-velocity diagrams extracted over a 1 pixel (i.e., $0\farcs14$) slice along the direction of the SiO jet, or of the CO outflow  when no SiO jet was detected. The adopted position angles are given in Table \ref{tab:jet-occurrence}.
The extracted PV diagrams extend over the full field-of-view of our observations but in Figure \ref{fig:PV-block1}, we only show the inner $5\arcsec-20\arcsec$ where the jet and outflow emission is detected.

\begin{figure*}
\centering
\includegraphics[width=.45\textwidth]{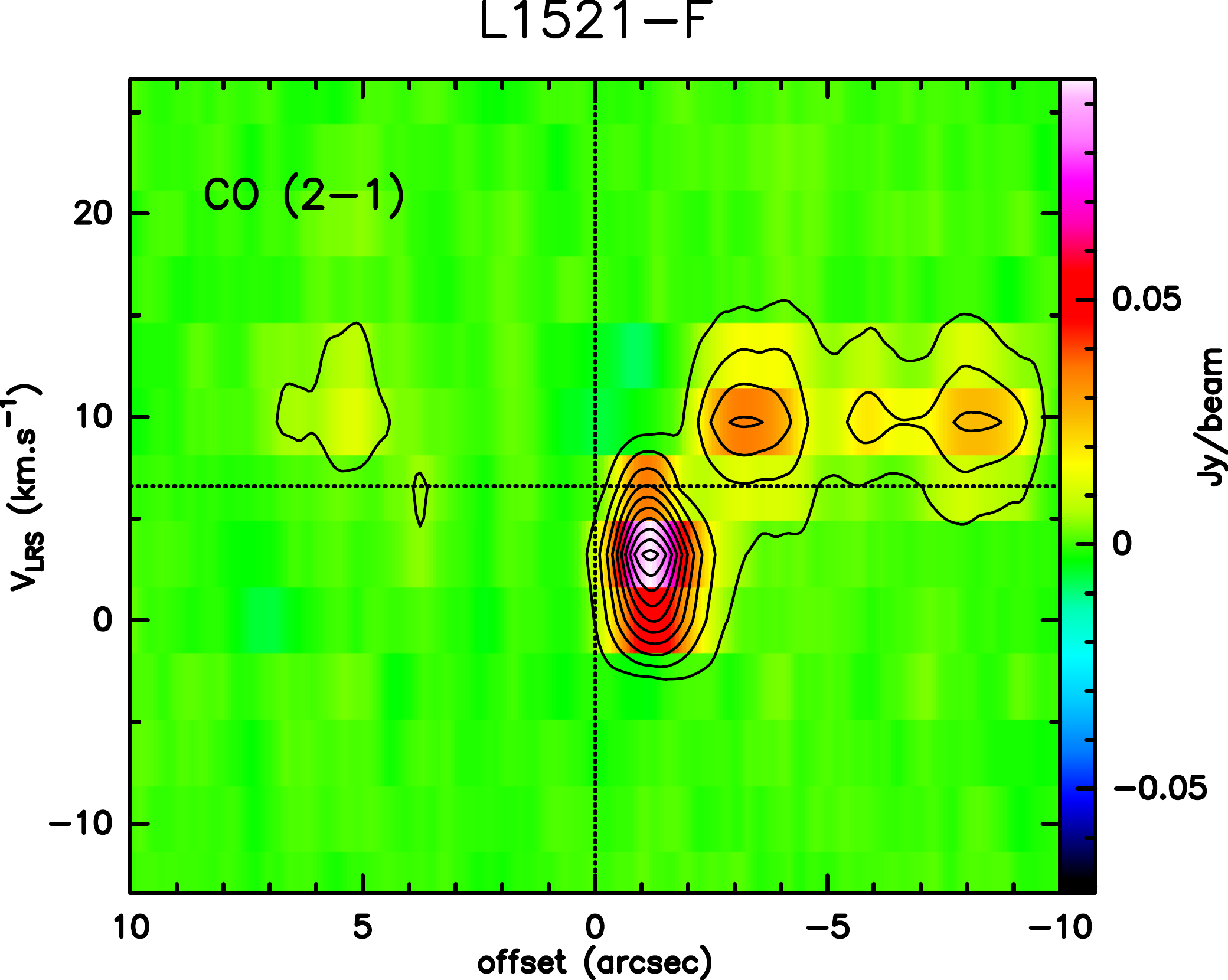}
\includegraphics[width=.45\textwidth]{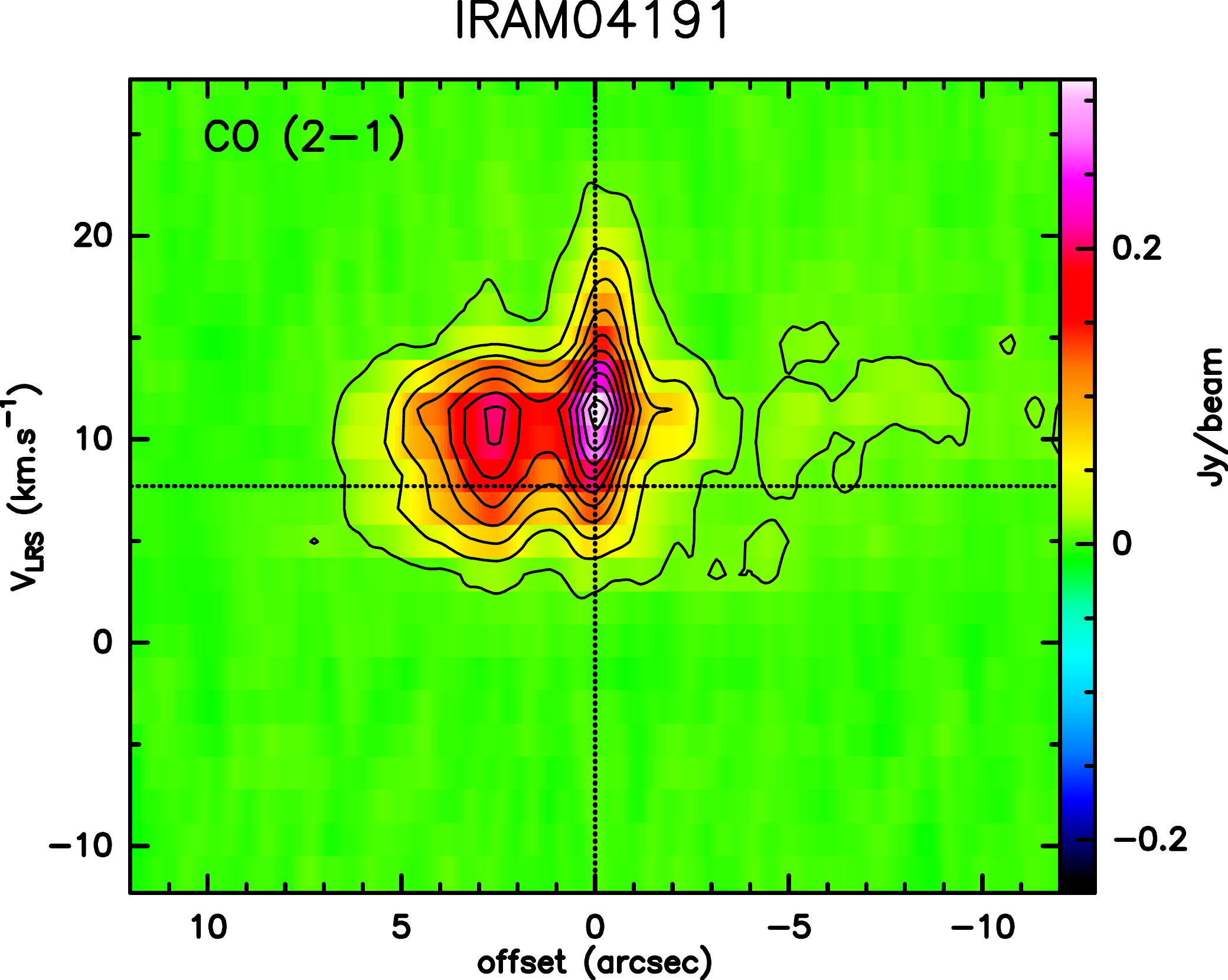} \\
\includegraphics[width=.45\textwidth]{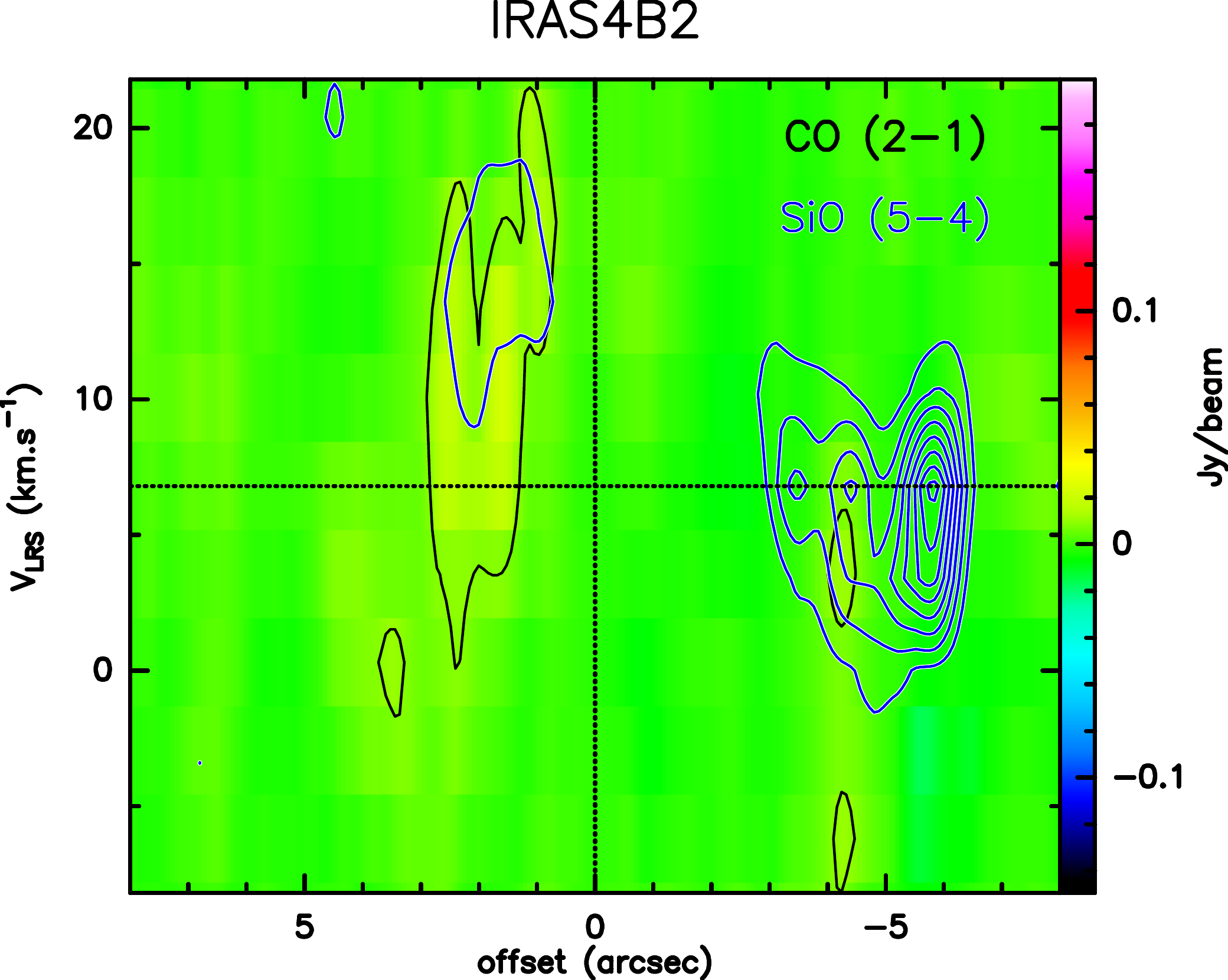}
\includegraphics[width=.45\textwidth]{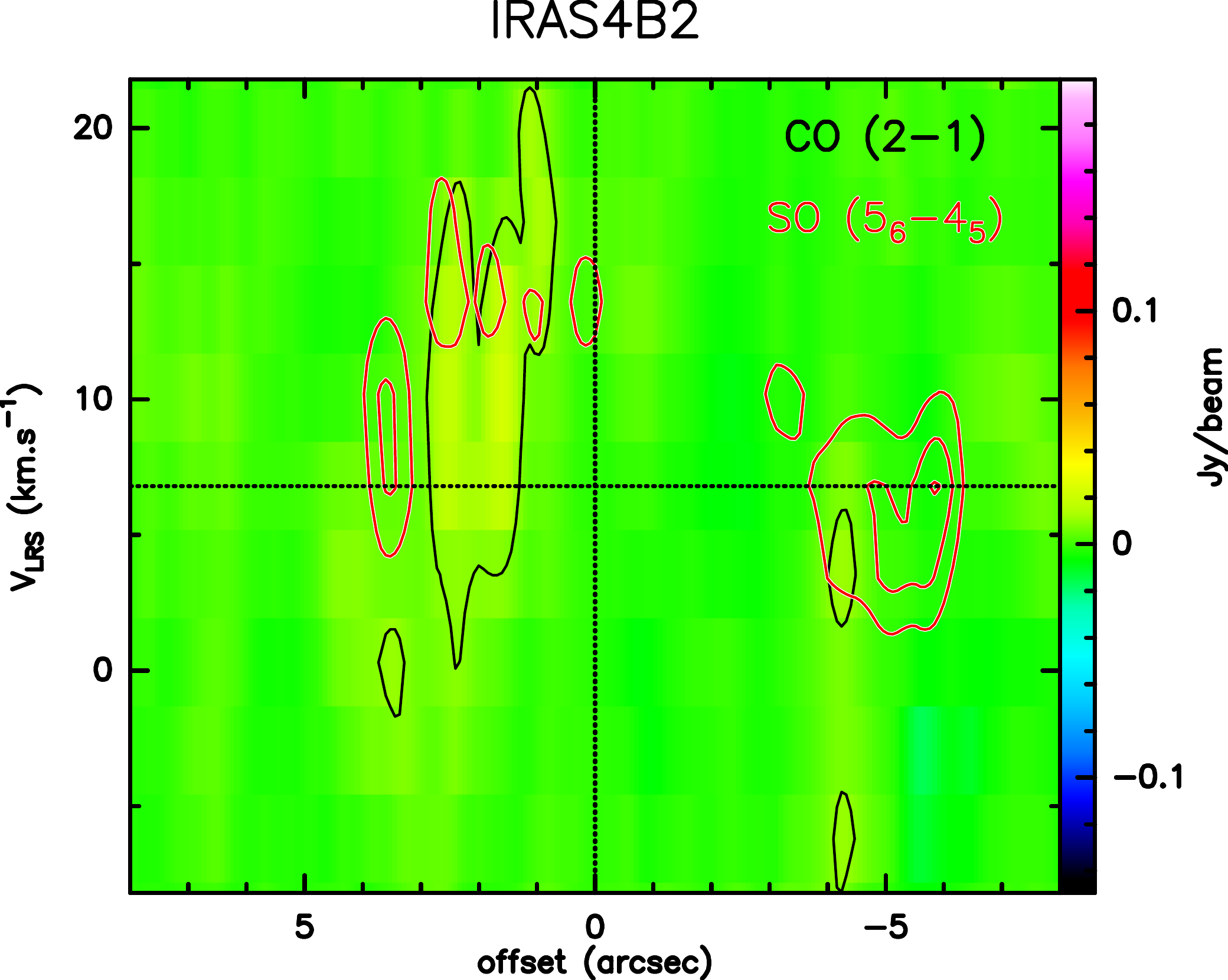} \\ \includegraphics[width=.45\textwidth]{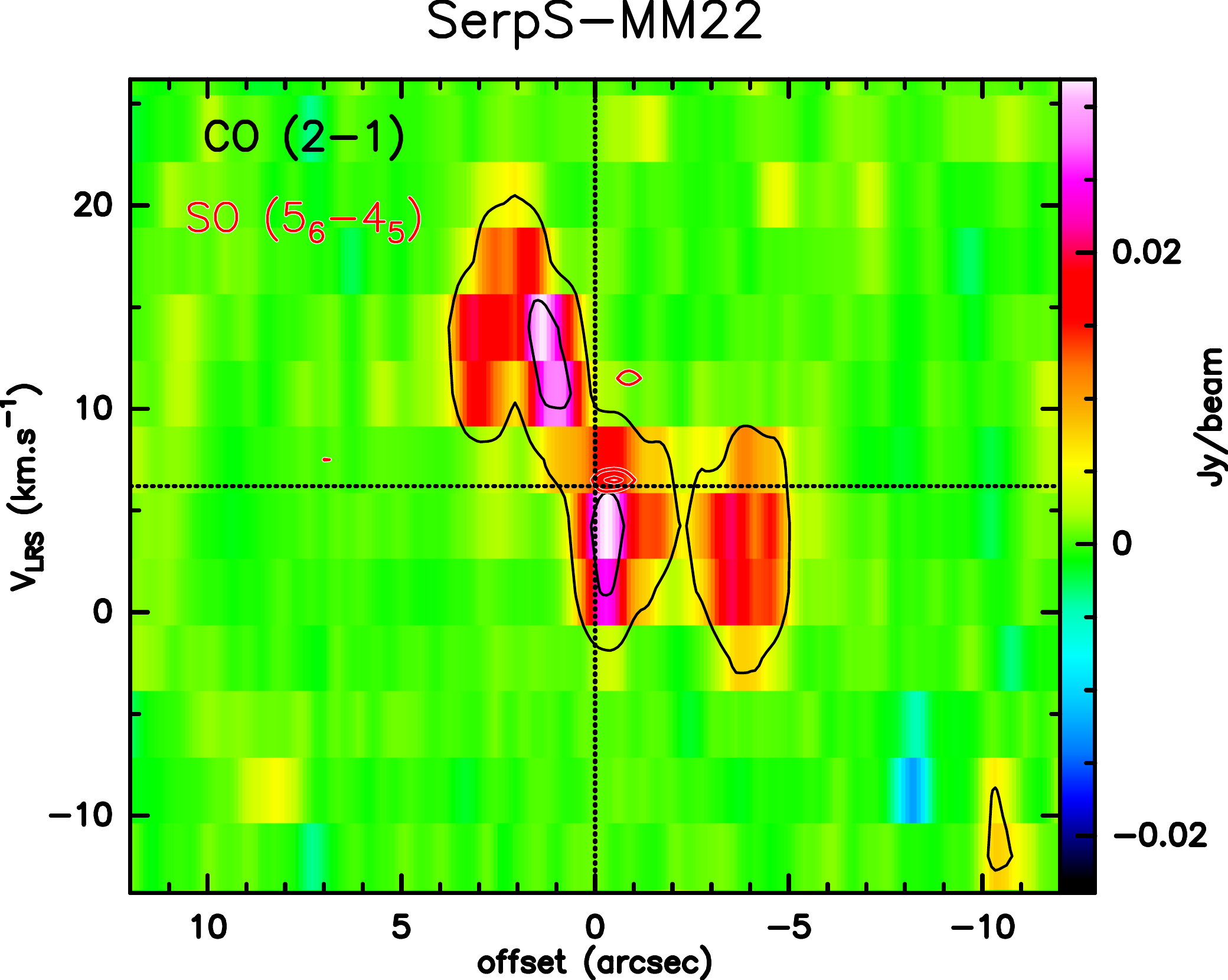}
\includegraphics[width=.45\textwidth]{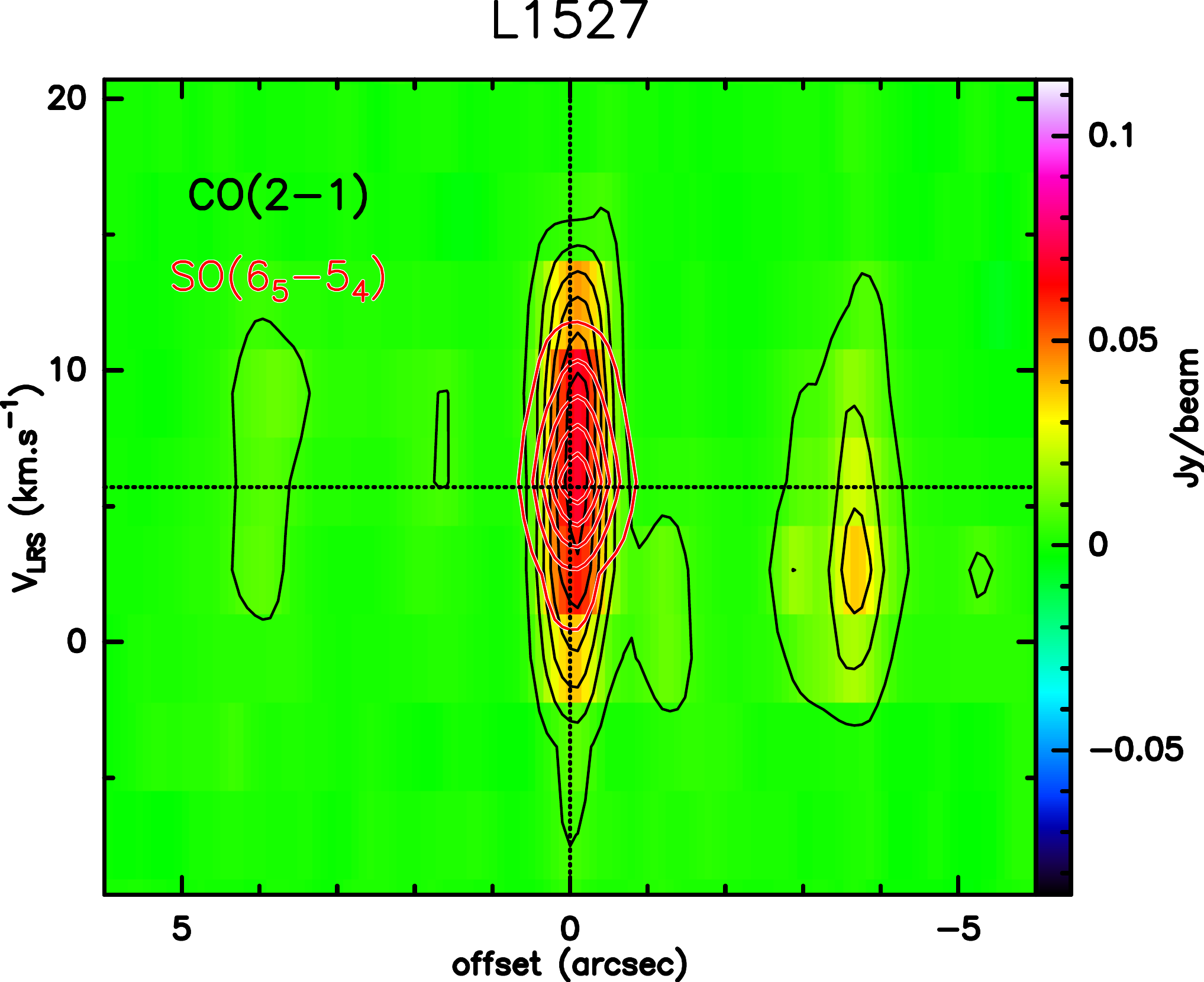}
\caption{Position-velocity diagrams of CO ($2-1$) (color map and black contours), SiO ($5-4$) (blue contours), and SO ($5_6-4_5$) (red contours) along the PA of the jet and outflow (PA reported in Table \ref{tab:jet-occurrence}). The name of the source is indicated in the upper part of each panel. Contours start at $3\sigma$. Only detected lines are shown.}
\label{fig:PV-block1}
\end{figure*}

\begin{figure*}
\setcounter{figure}{0}
\centering
\includegraphics[width=.45\textwidth]{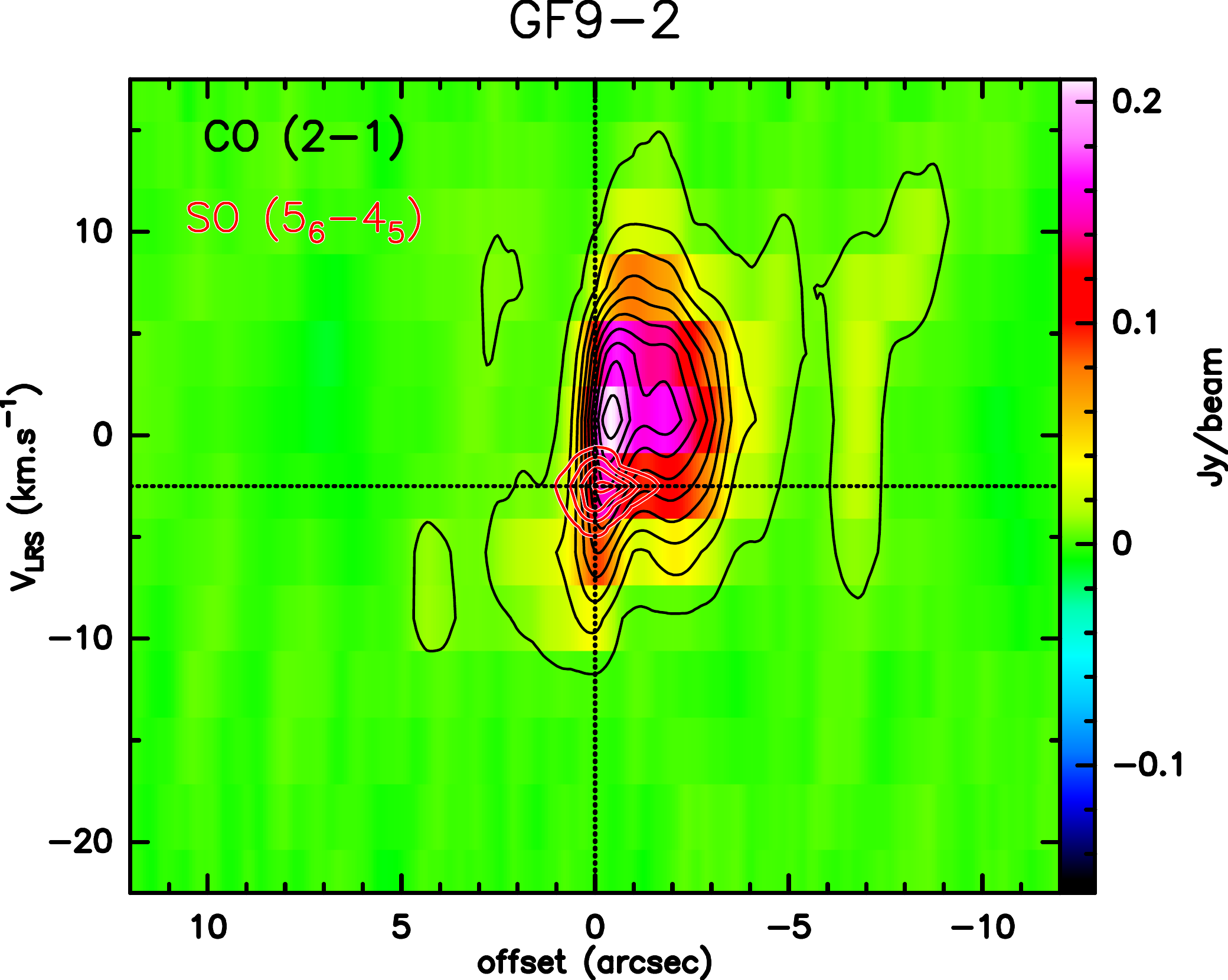}
\includegraphics[width=.45\textwidth]{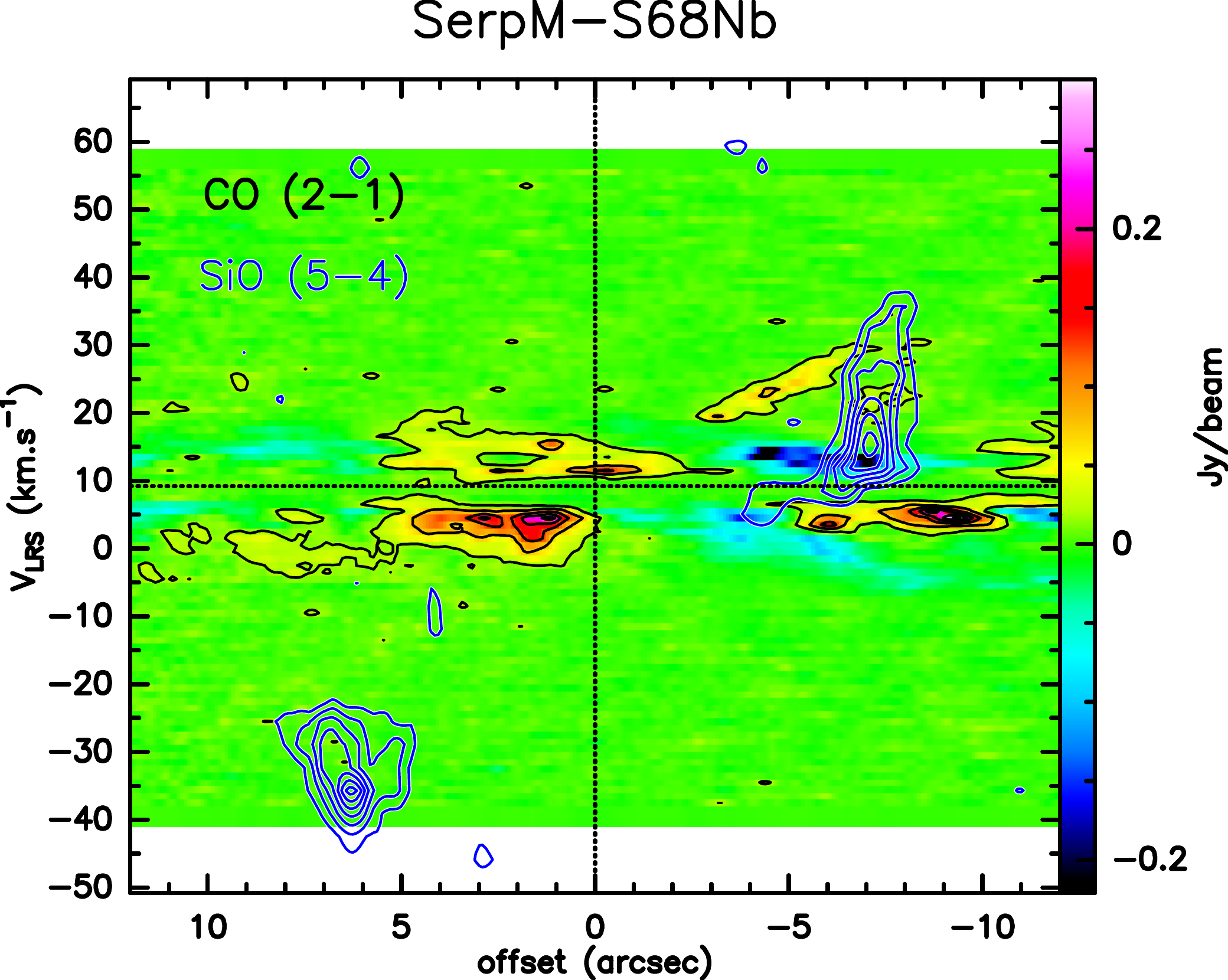}\\
\includegraphics[width=.45\textwidth]{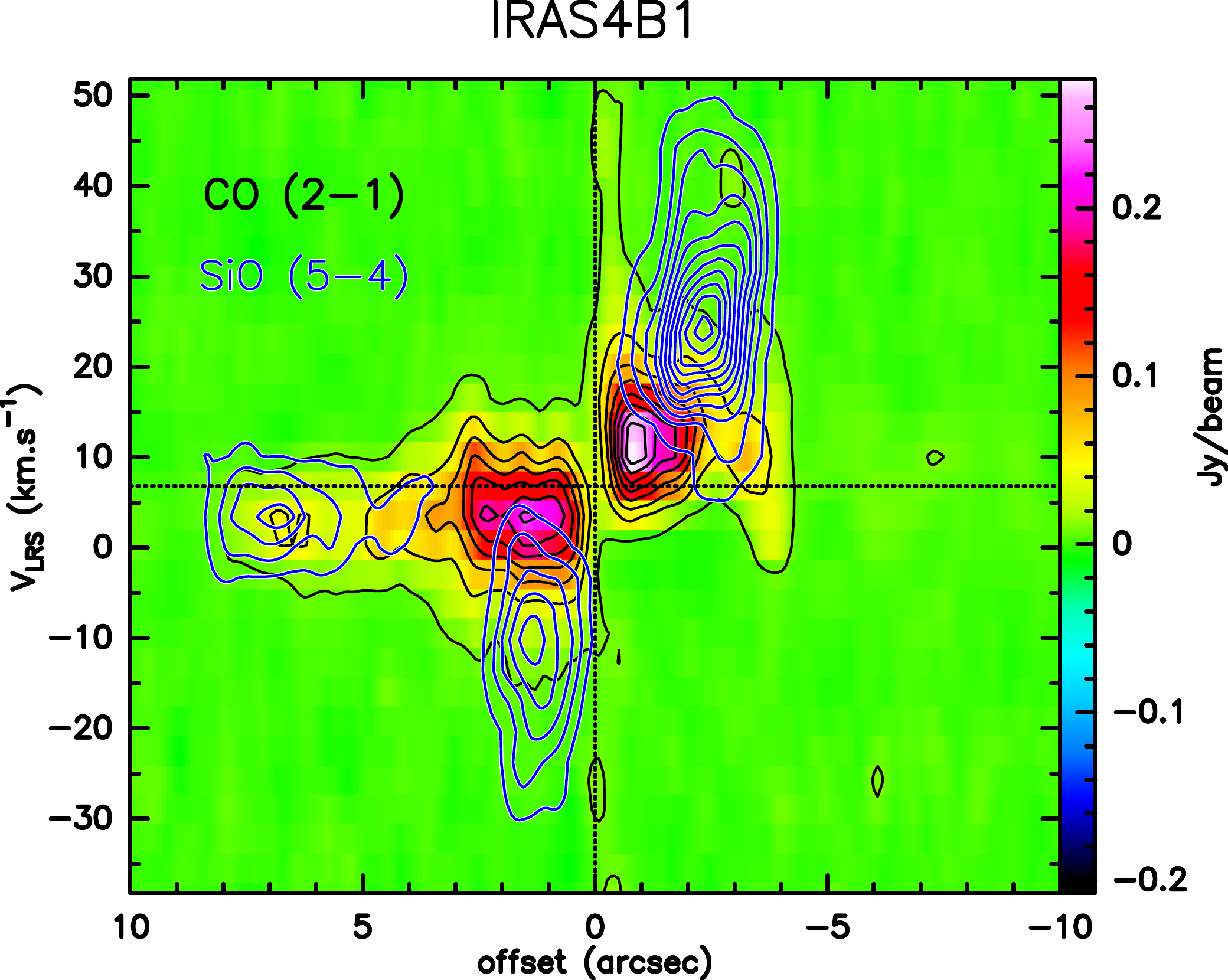}
\includegraphics[width=.45\textwidth]{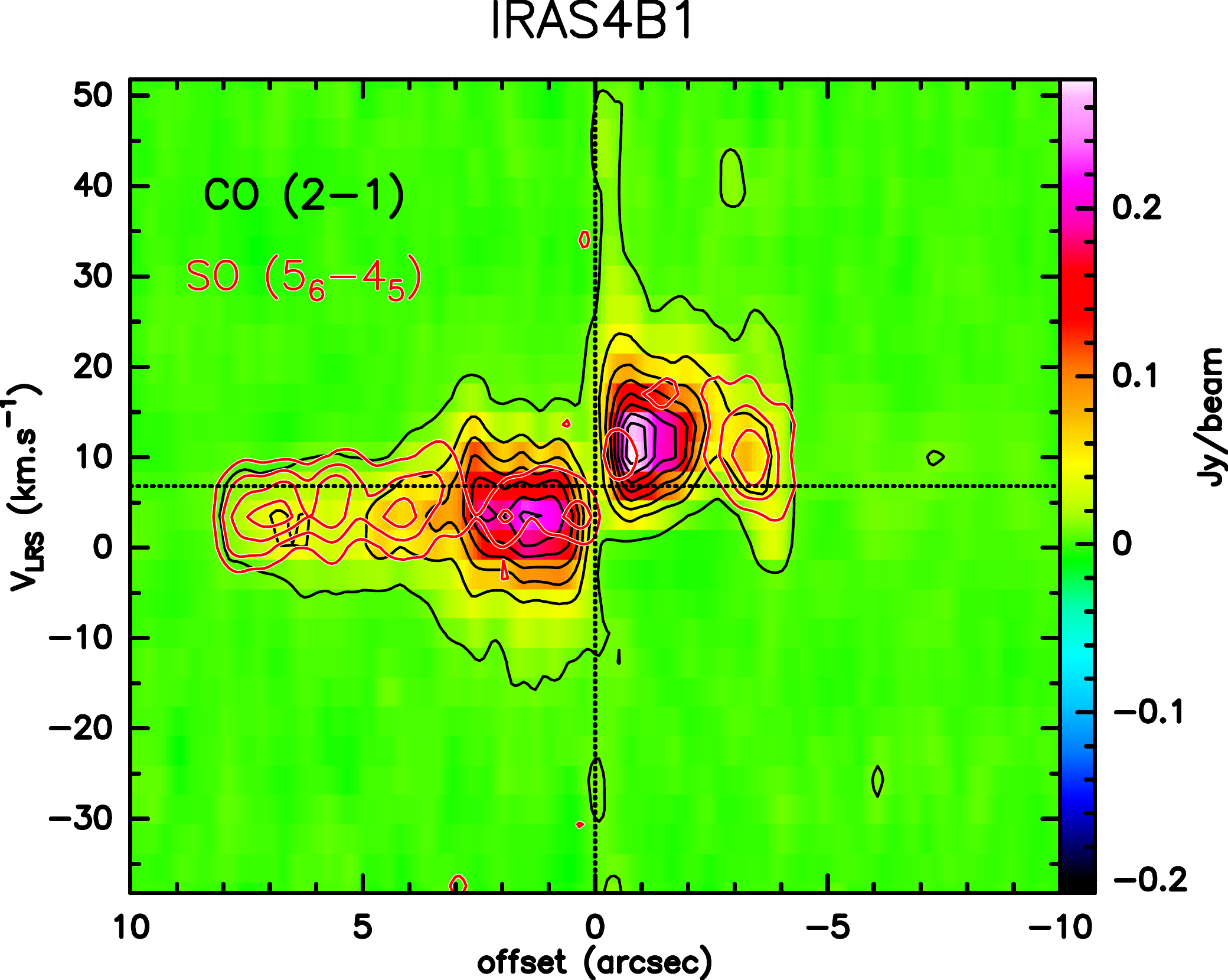} \\
\includegraphics[width=.45\textwidth]{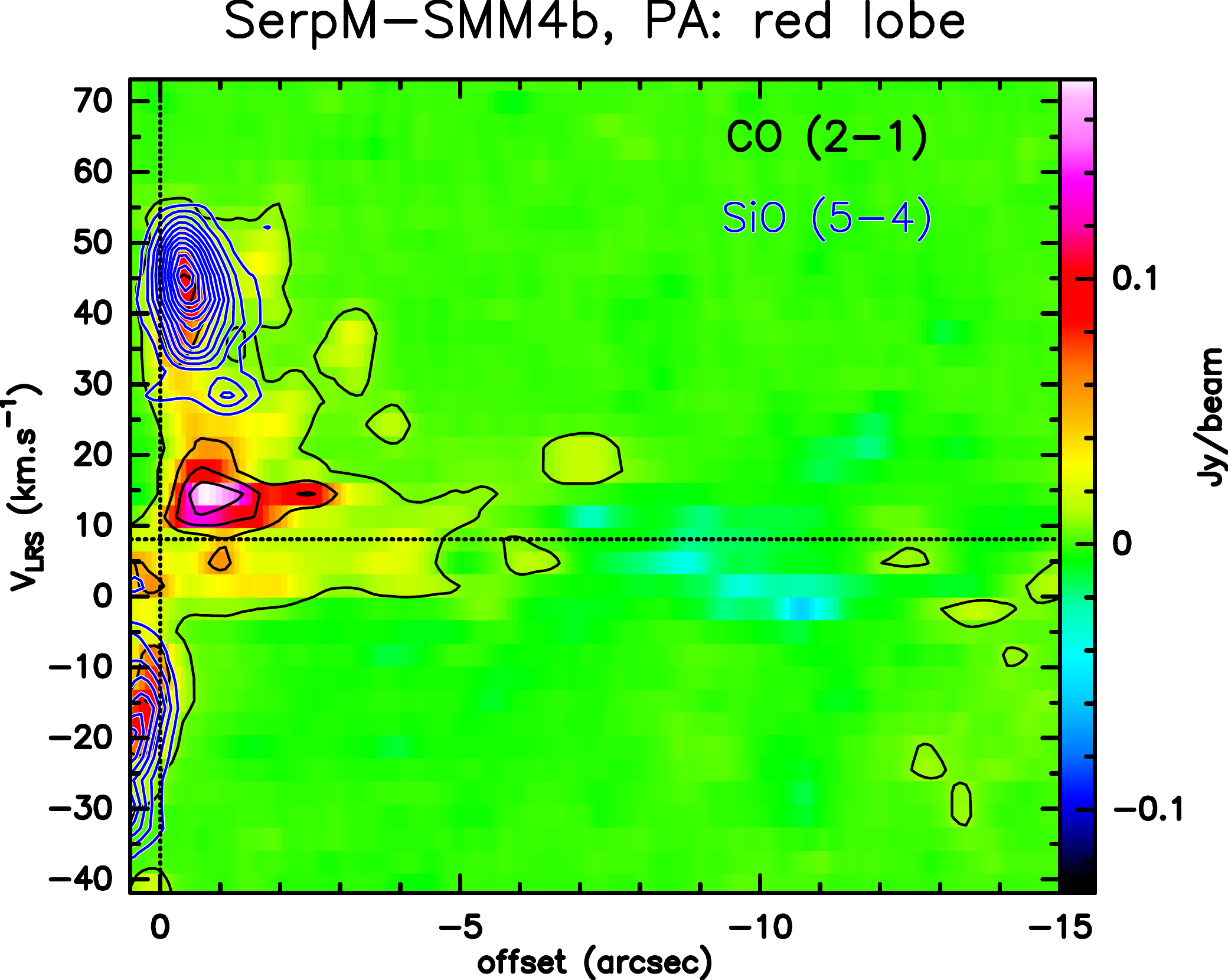}
\includegraphics[width=.45\textwidth]{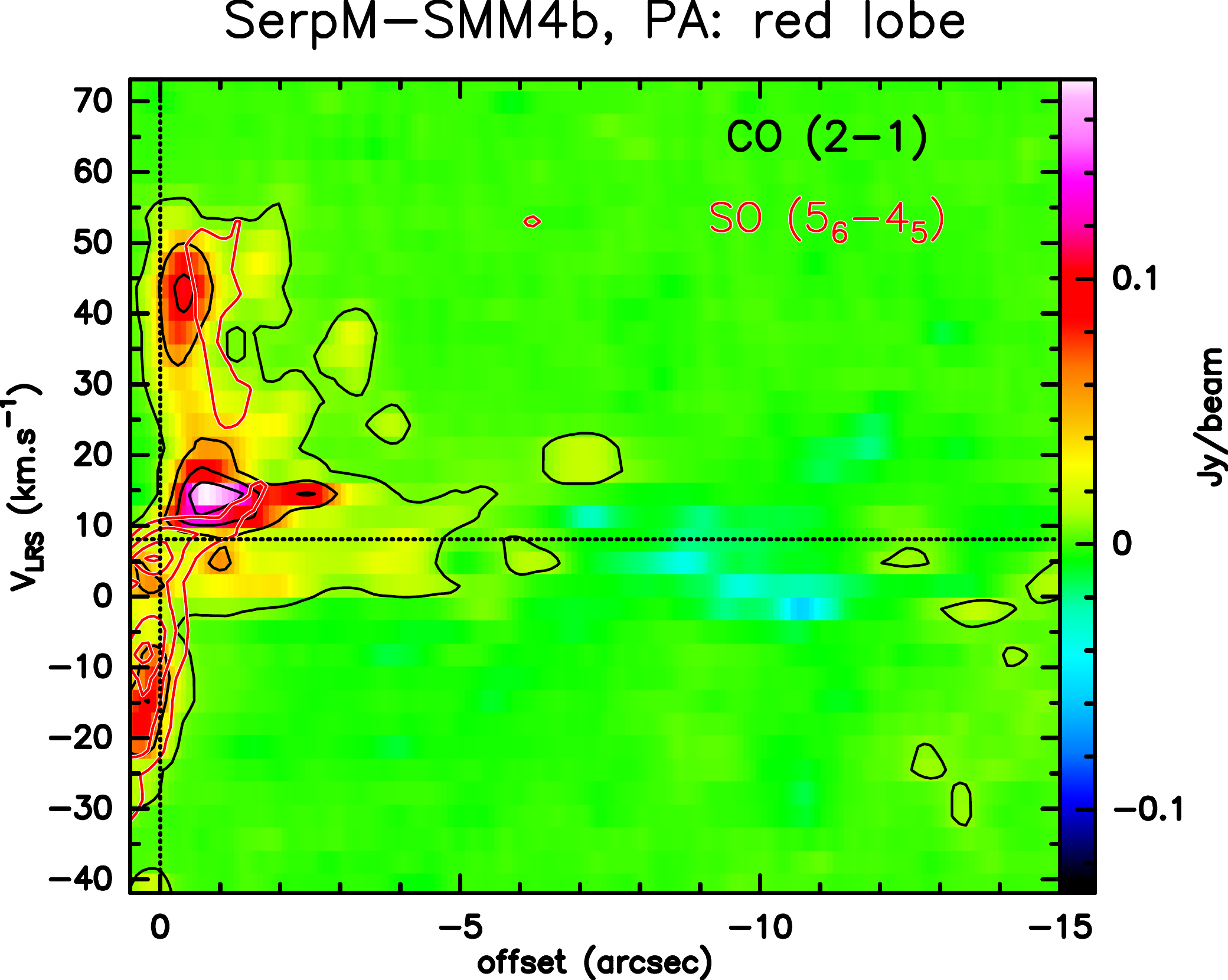} \\
\caption{{\it Continued}}
\label{fig:PV-block2}
\end{figure*}

\begin{figure*}
\setcounter{figure}{0}
\centering
\includegraphics[width=.45\textwidth]{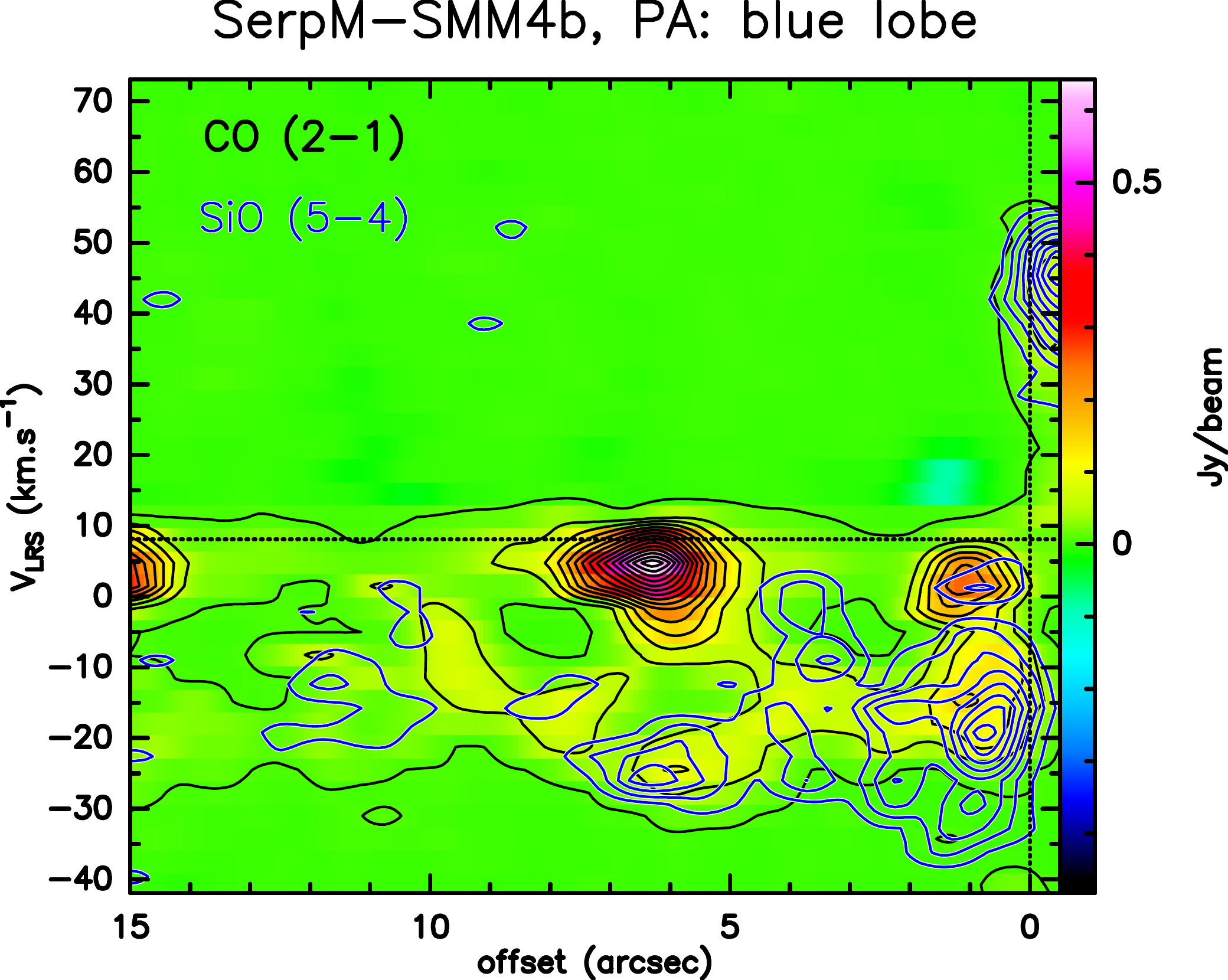}
\includegraphics[width=.45\textwidth]{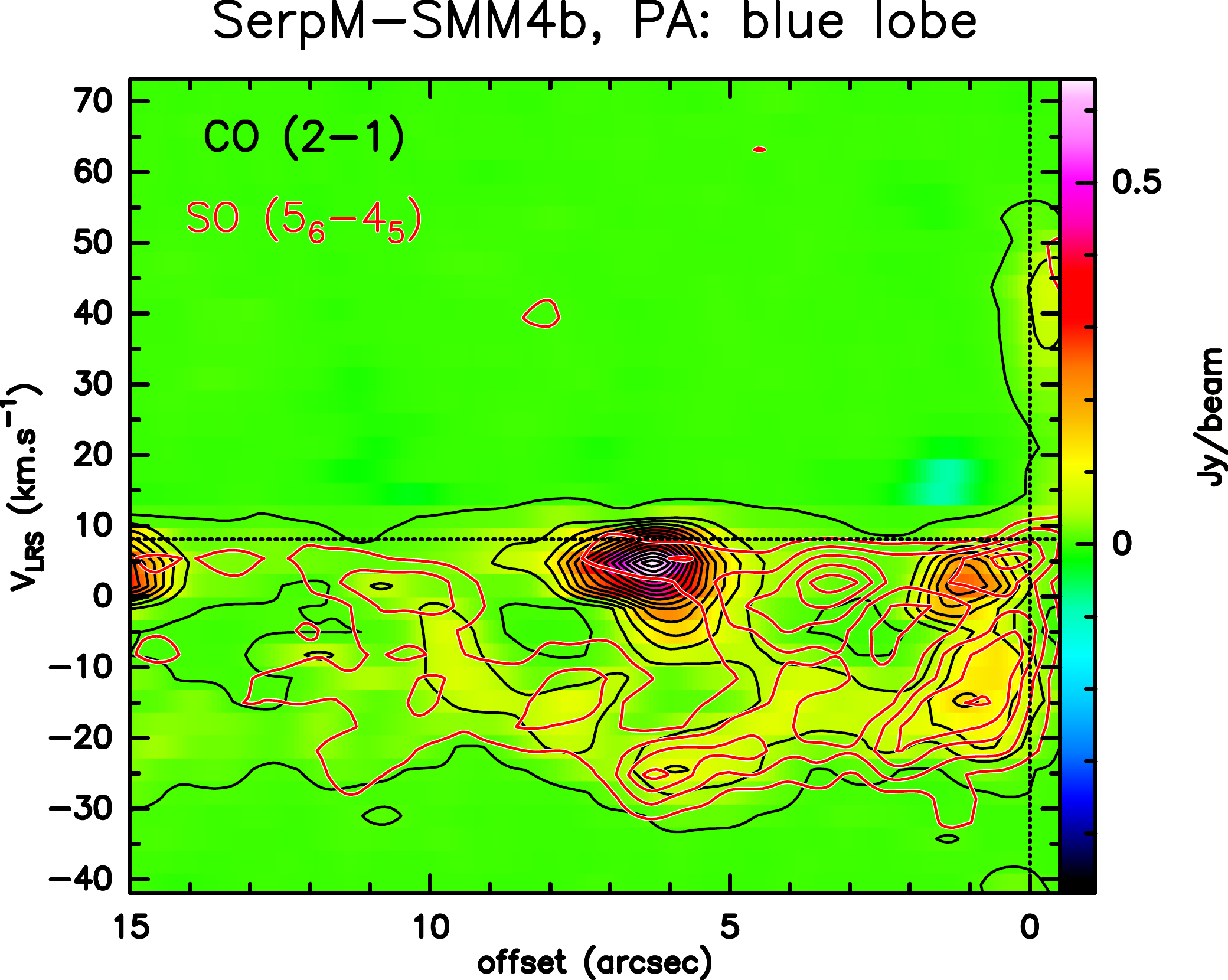} \\
\includegraphics[width=.45\textwidth]{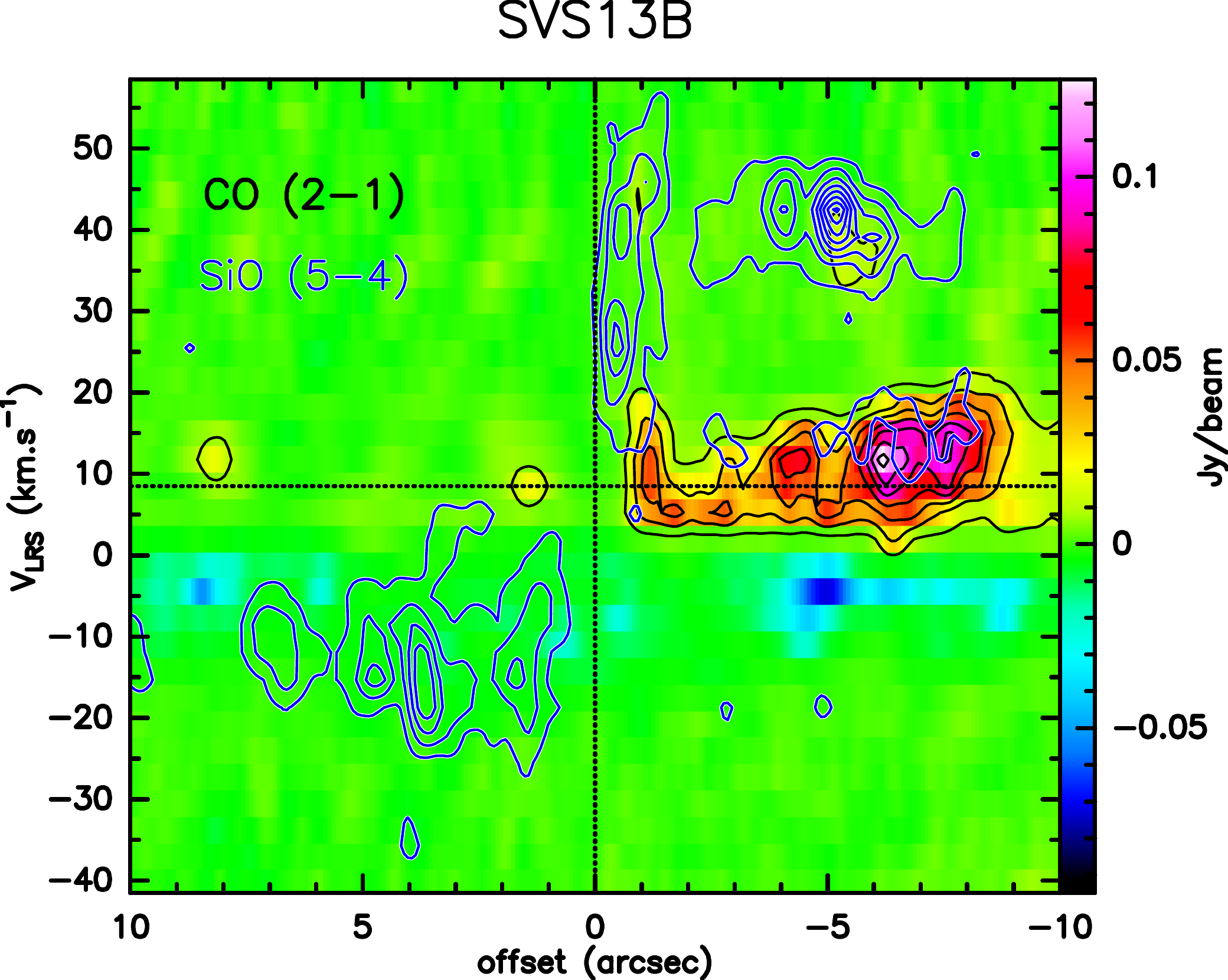}
\includegraphics[width=.45\textwidth]{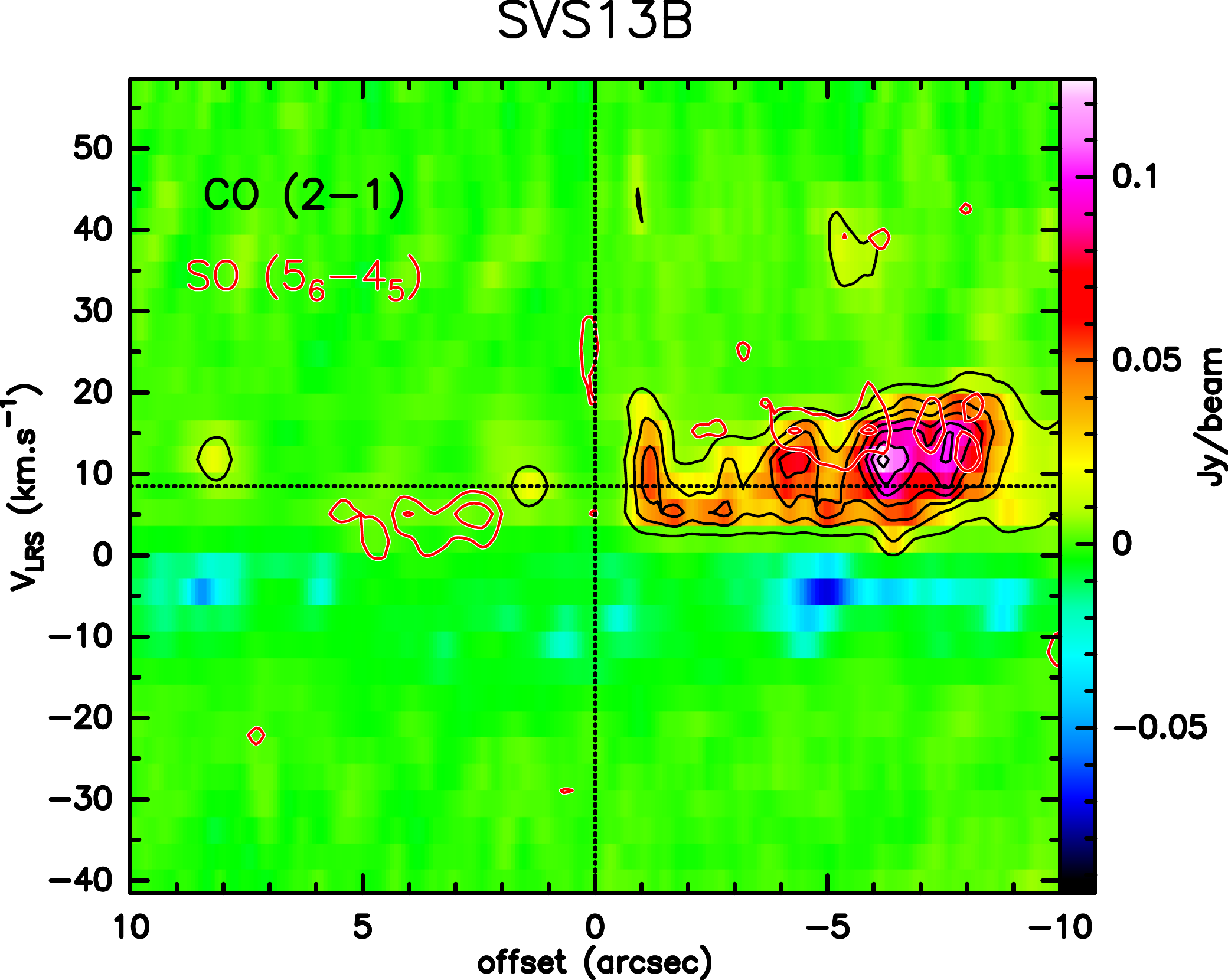}\\
\includegraphics[width=.45\textwidth]{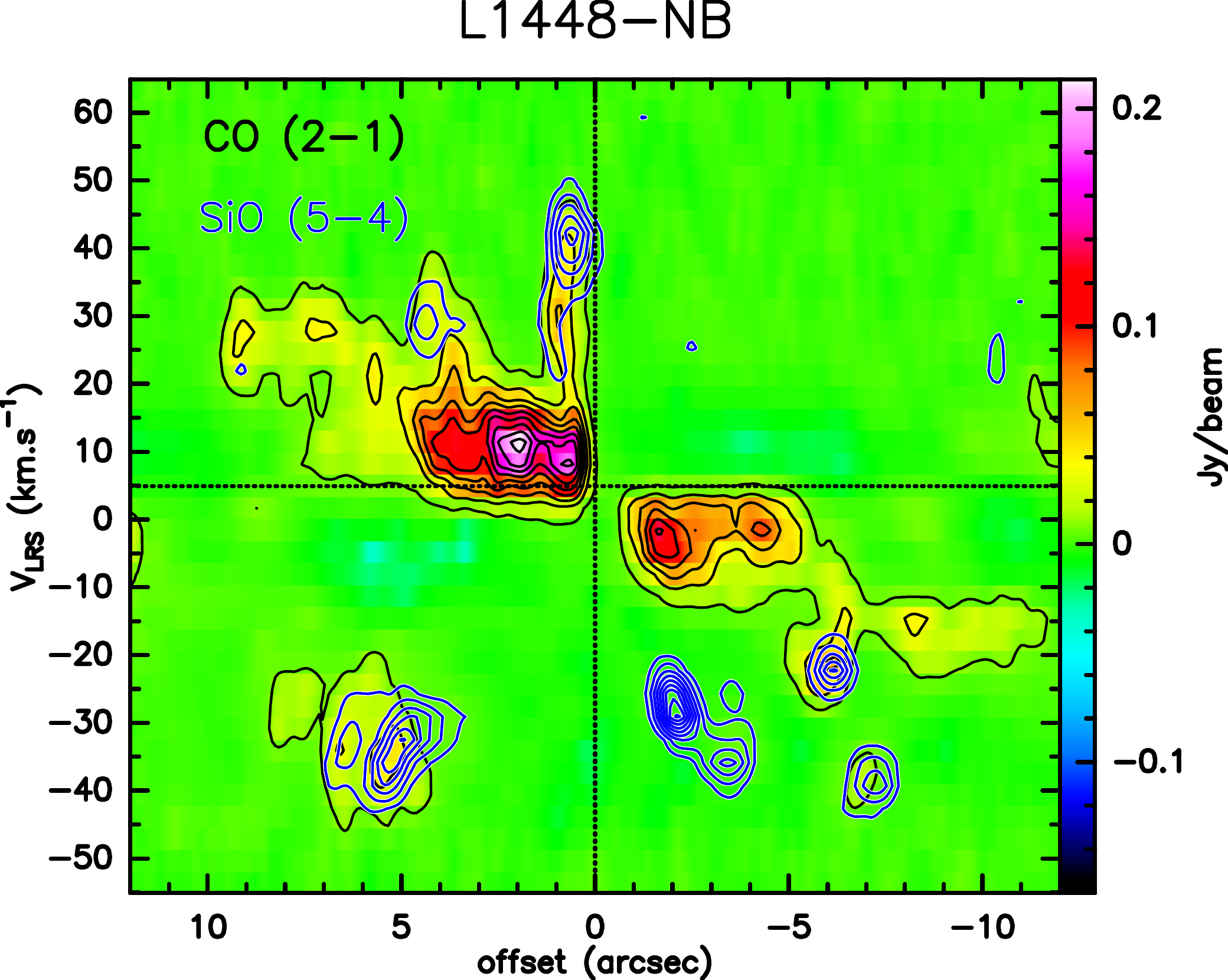}
\includegraphics[width=.45\textwidth]{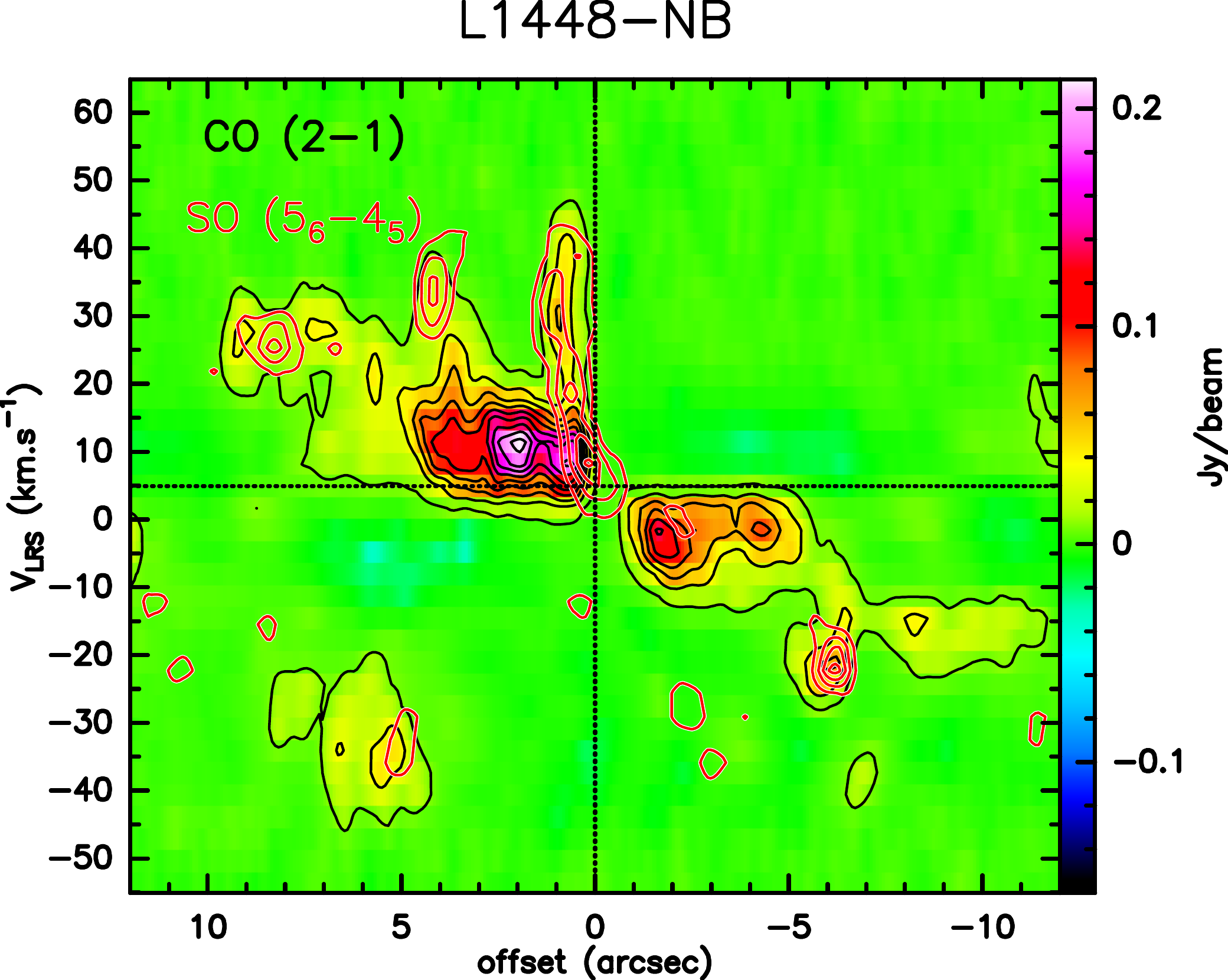}\\
\caption{{\it Continued}}
\label{fig:PV-block3}
\end{figure*}

\begin{figure*}
\setcounter{figure}{0}
\centering
\includegraphics[width=.45\textwidth]{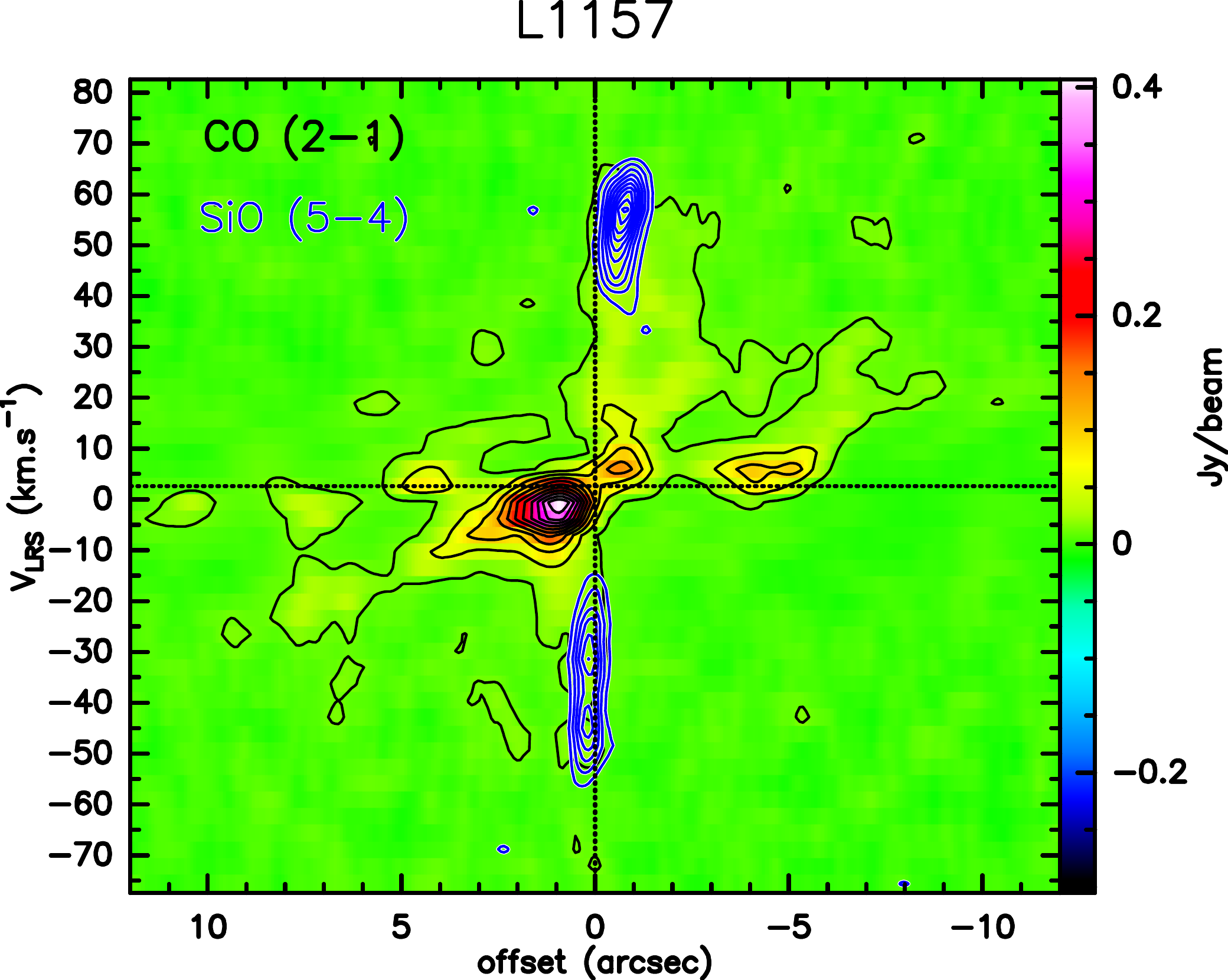}
\includegraphics[width=.45\textwidth]{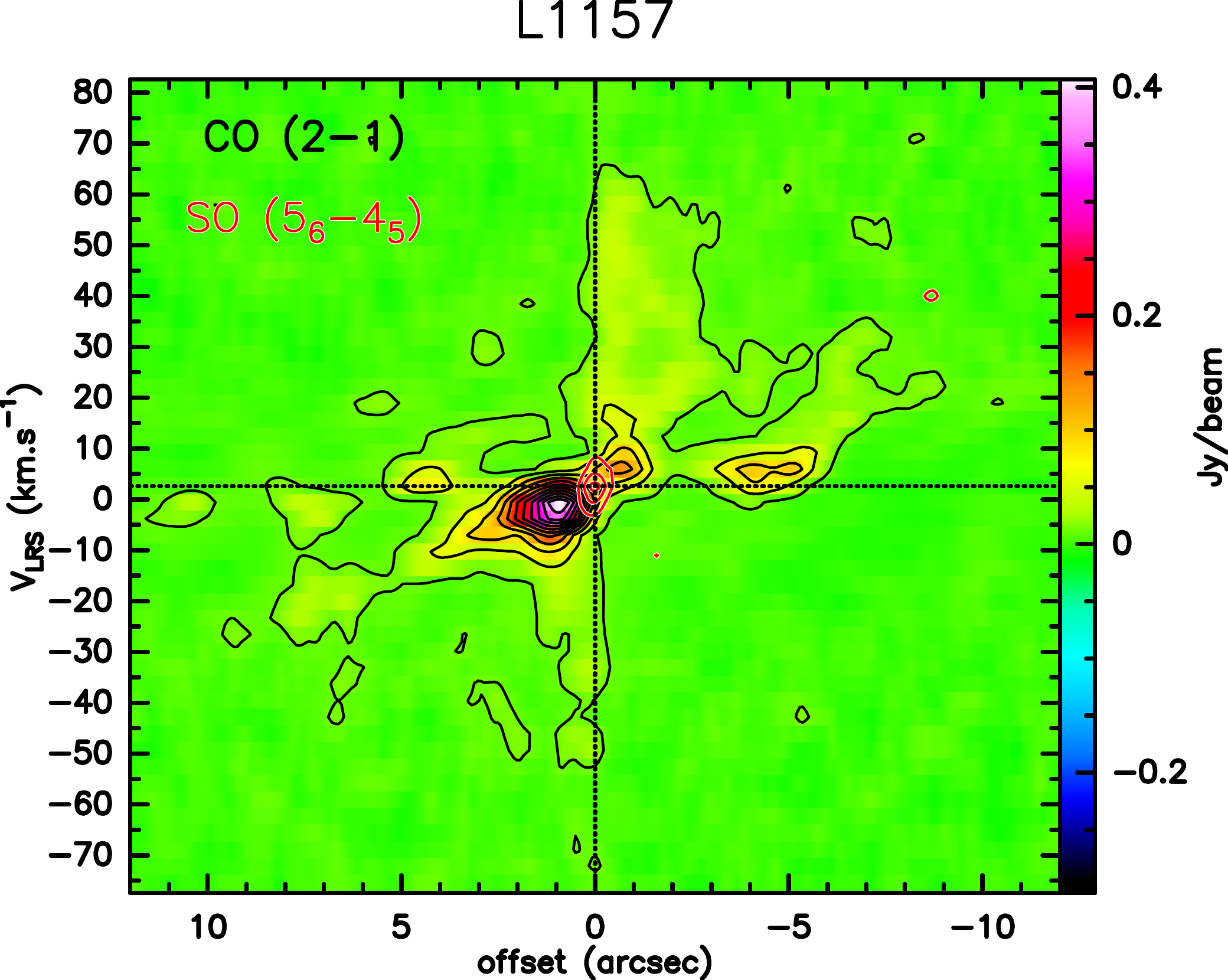}  \\
\includegraphics[width=.45\textwidth]{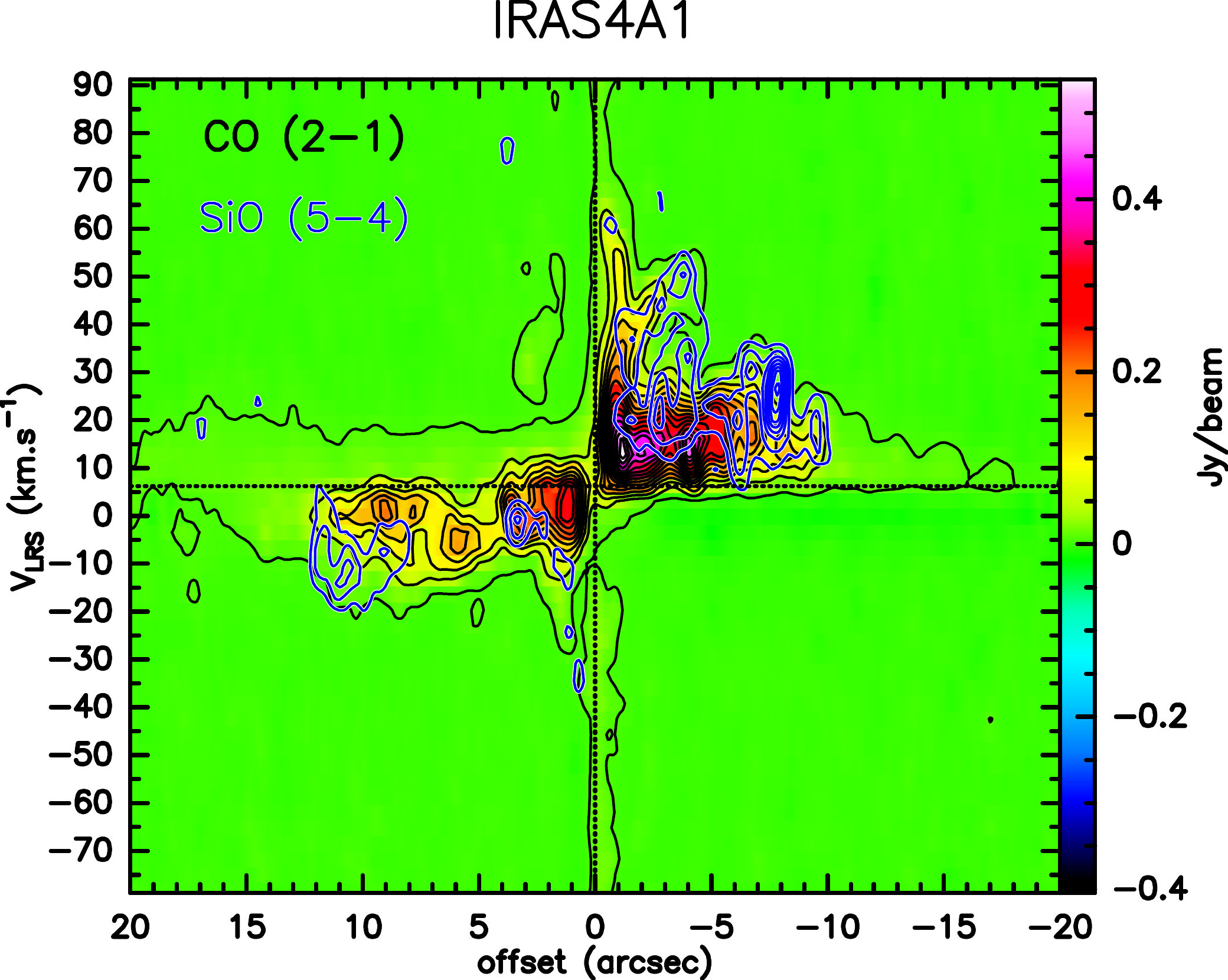}
\includegraphics[width=.45\textwidth]{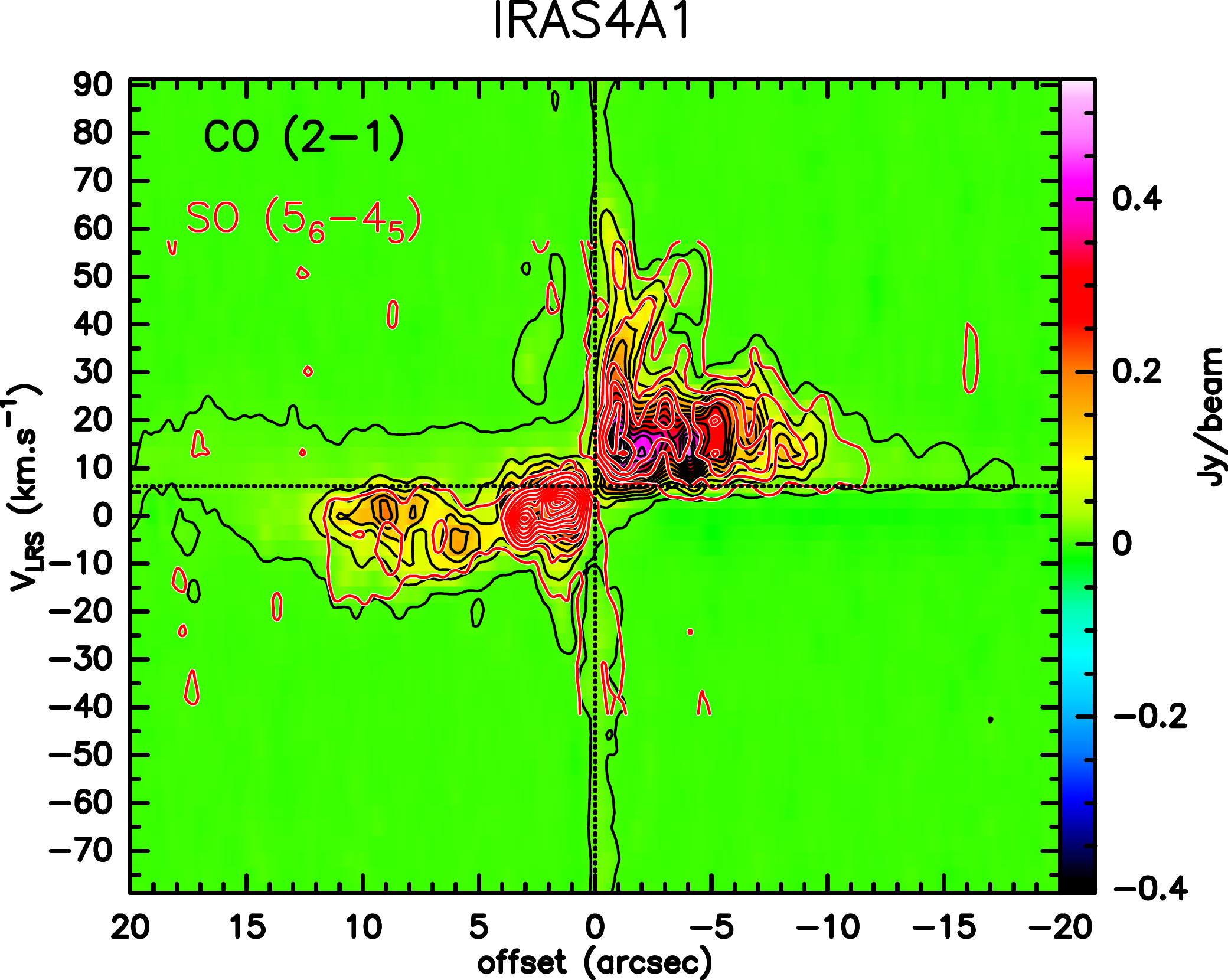}\\
\includegraphics[width=.45\textwidth]{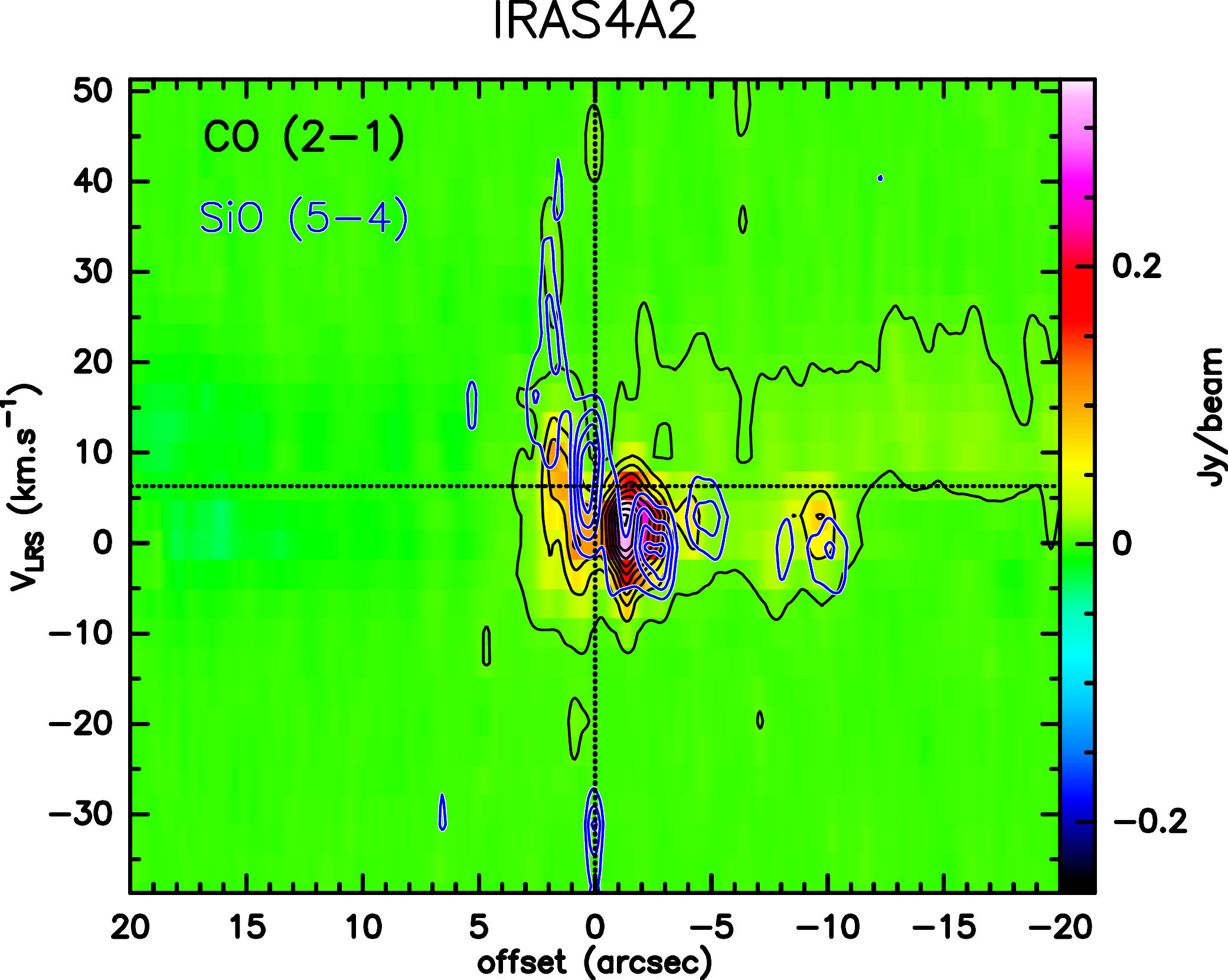}
\includegraphics[width=.45\textwidth]{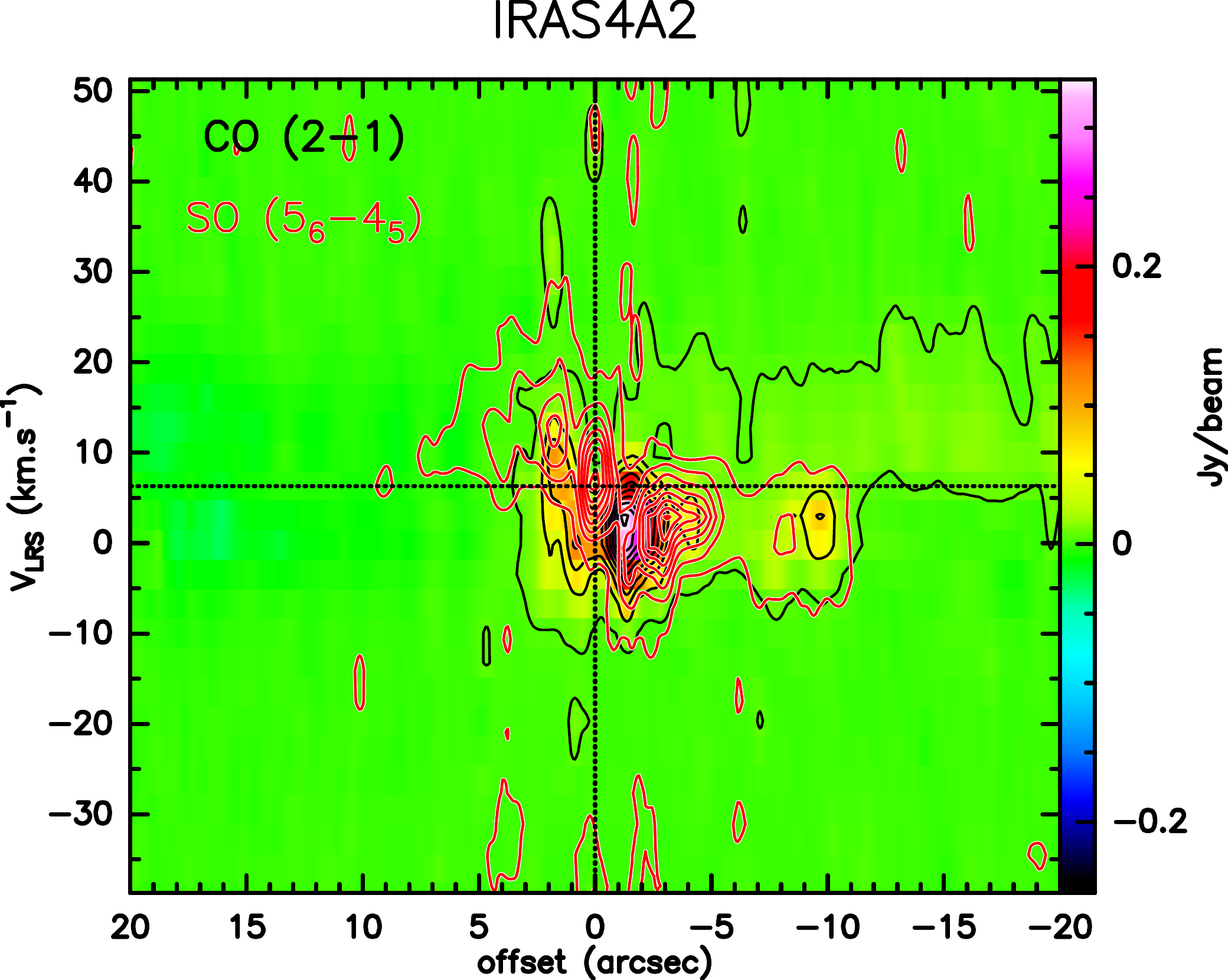} \\
\caption{{\it Continued}}
\label{fig:PV-block4}
\end{figure*}

\begin{figure*}
\setcounter{figure}{0}
\centering
\includegraphics[width=.45\textwidth]{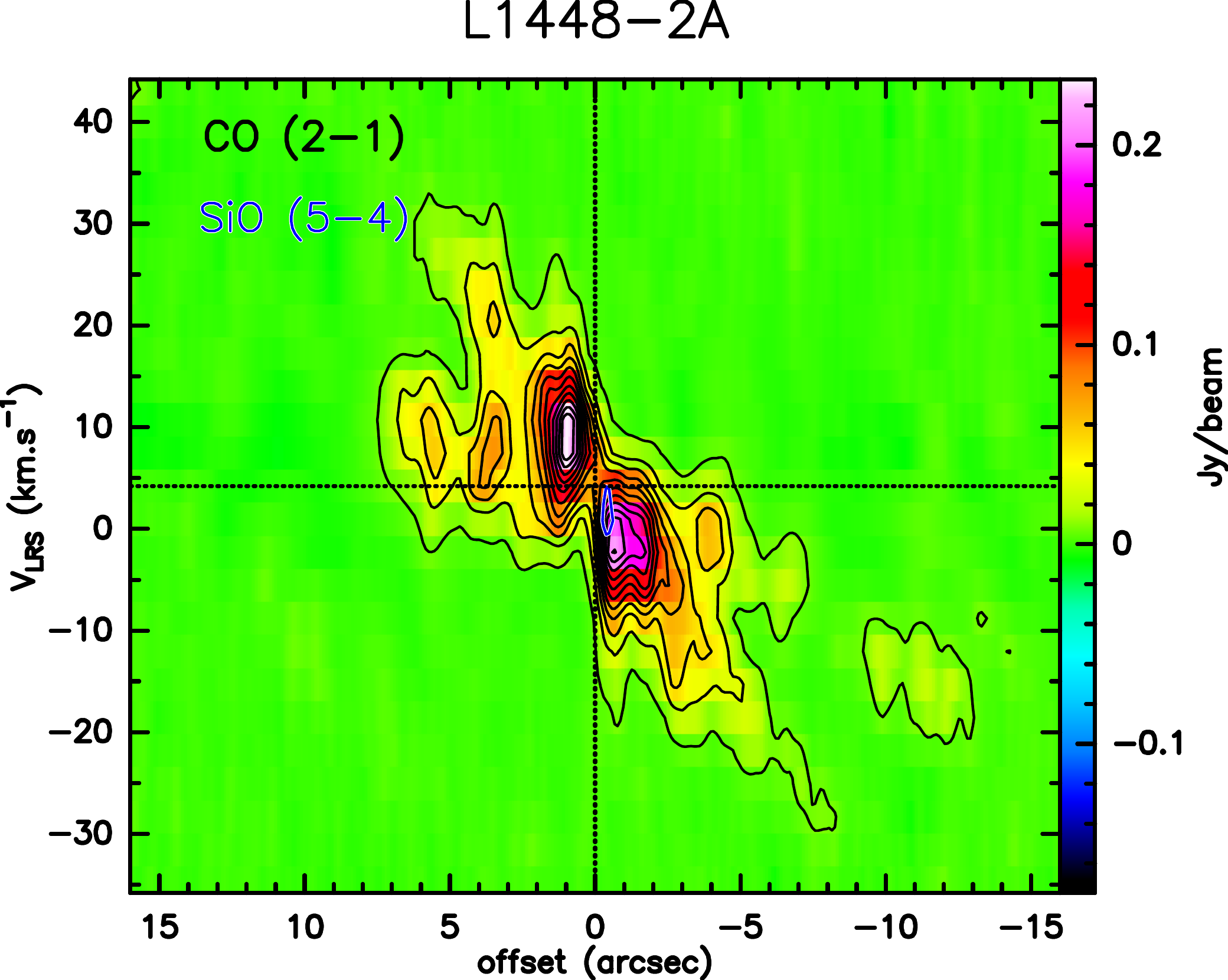}
\includegraphics[width=.45\textwidth]{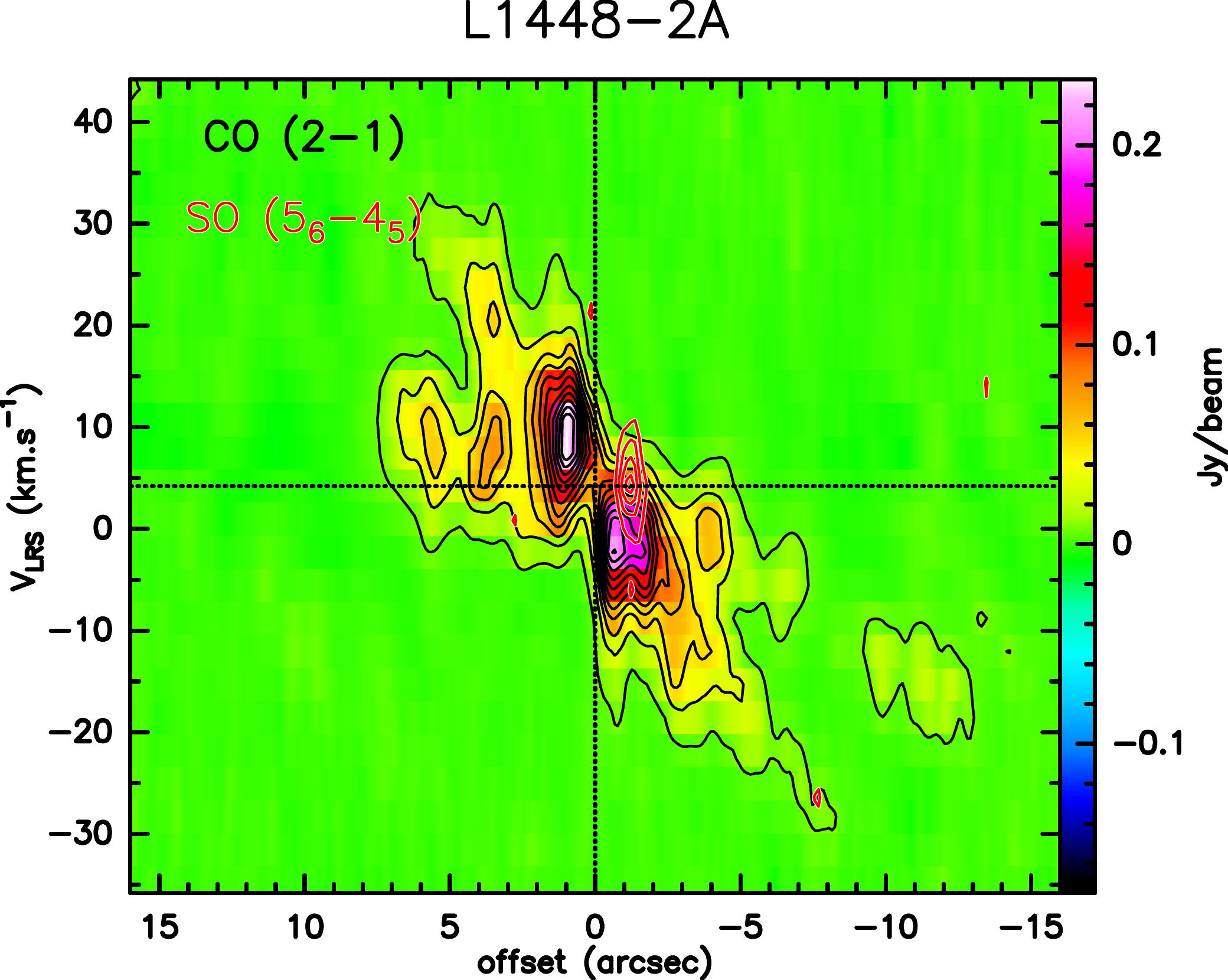}\\
\includegraphics[width=.45\textwidth]{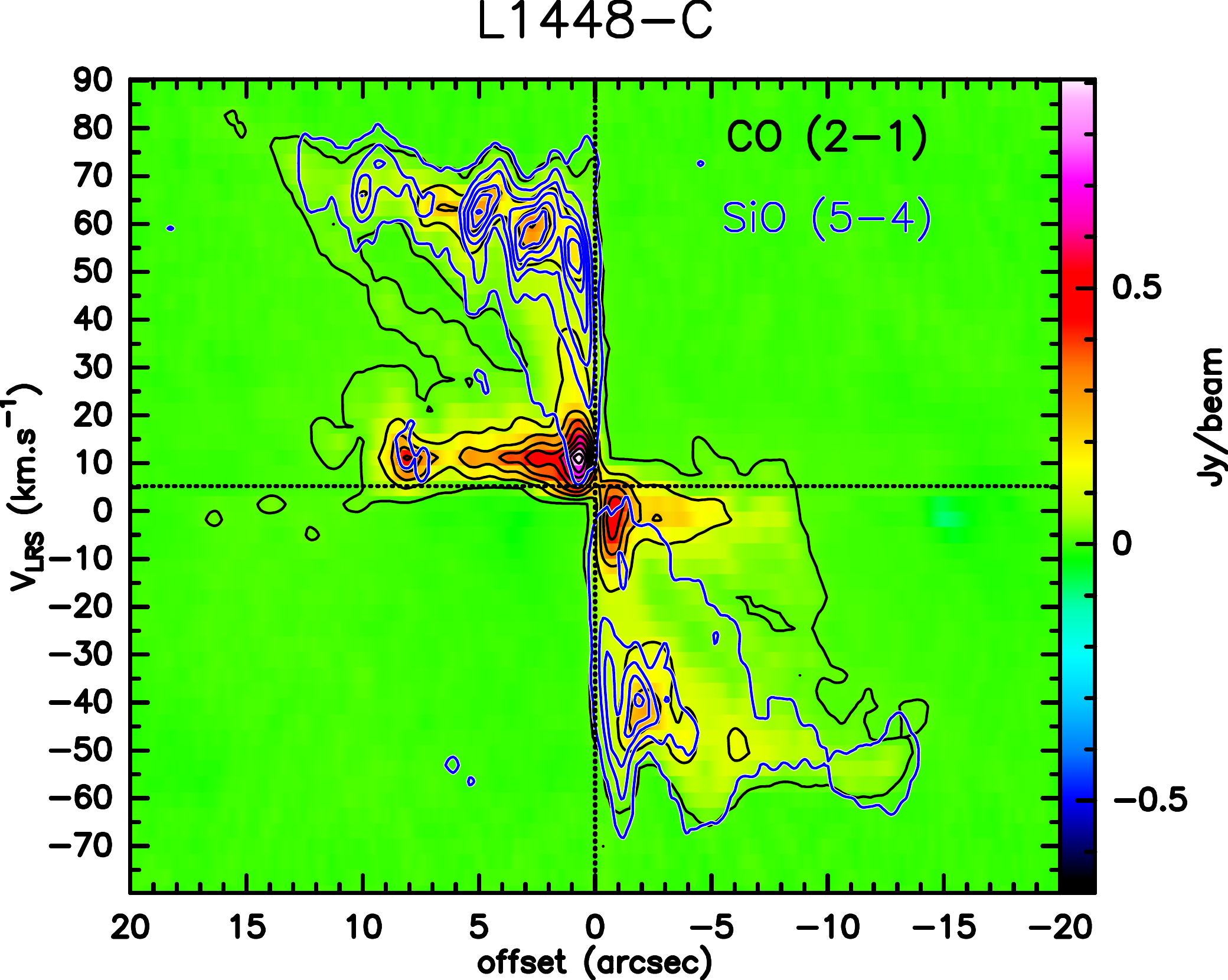}
\includegraphics[width=.45\textwidth]{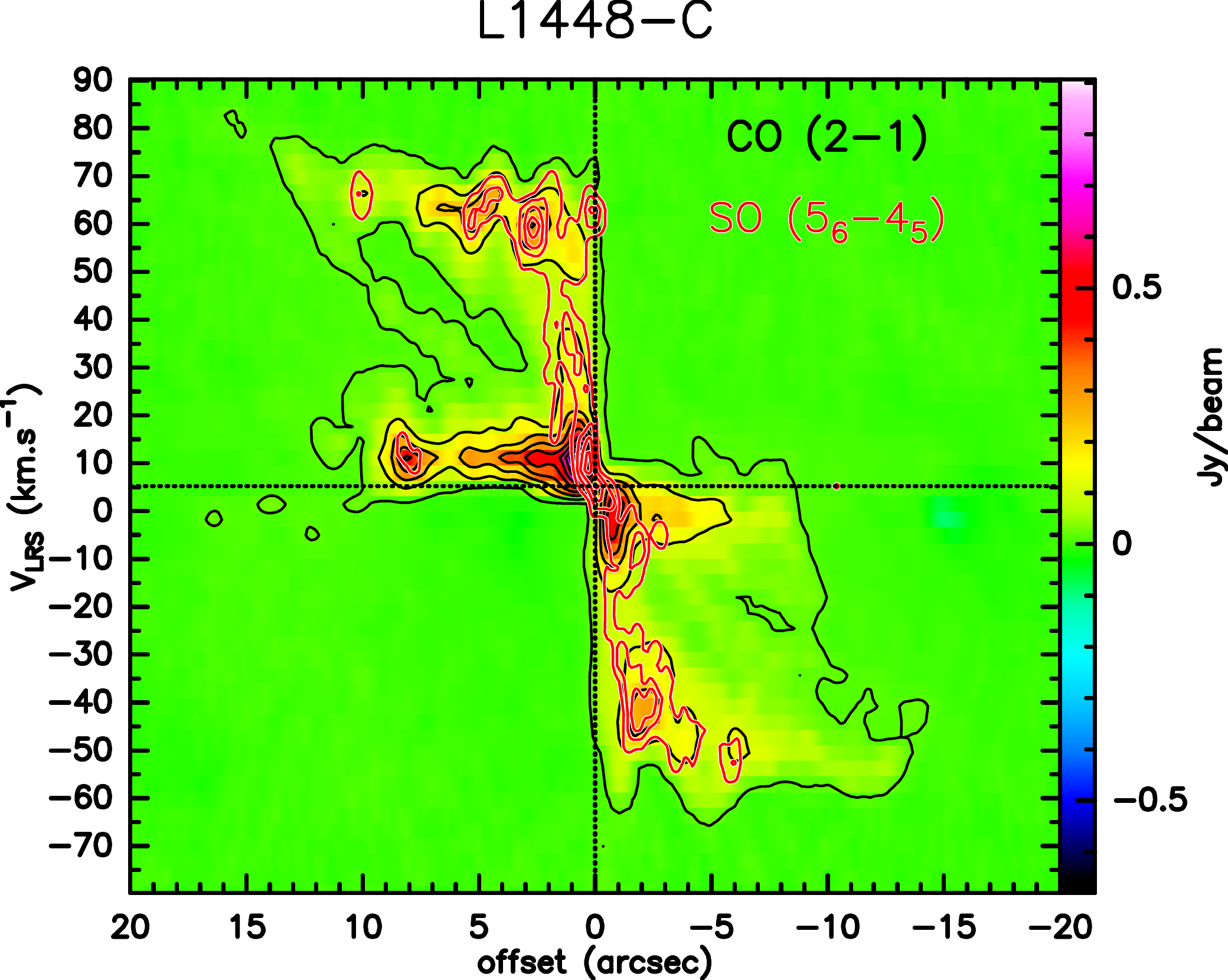} \\
\includegraphics[width=.45\textwidth]{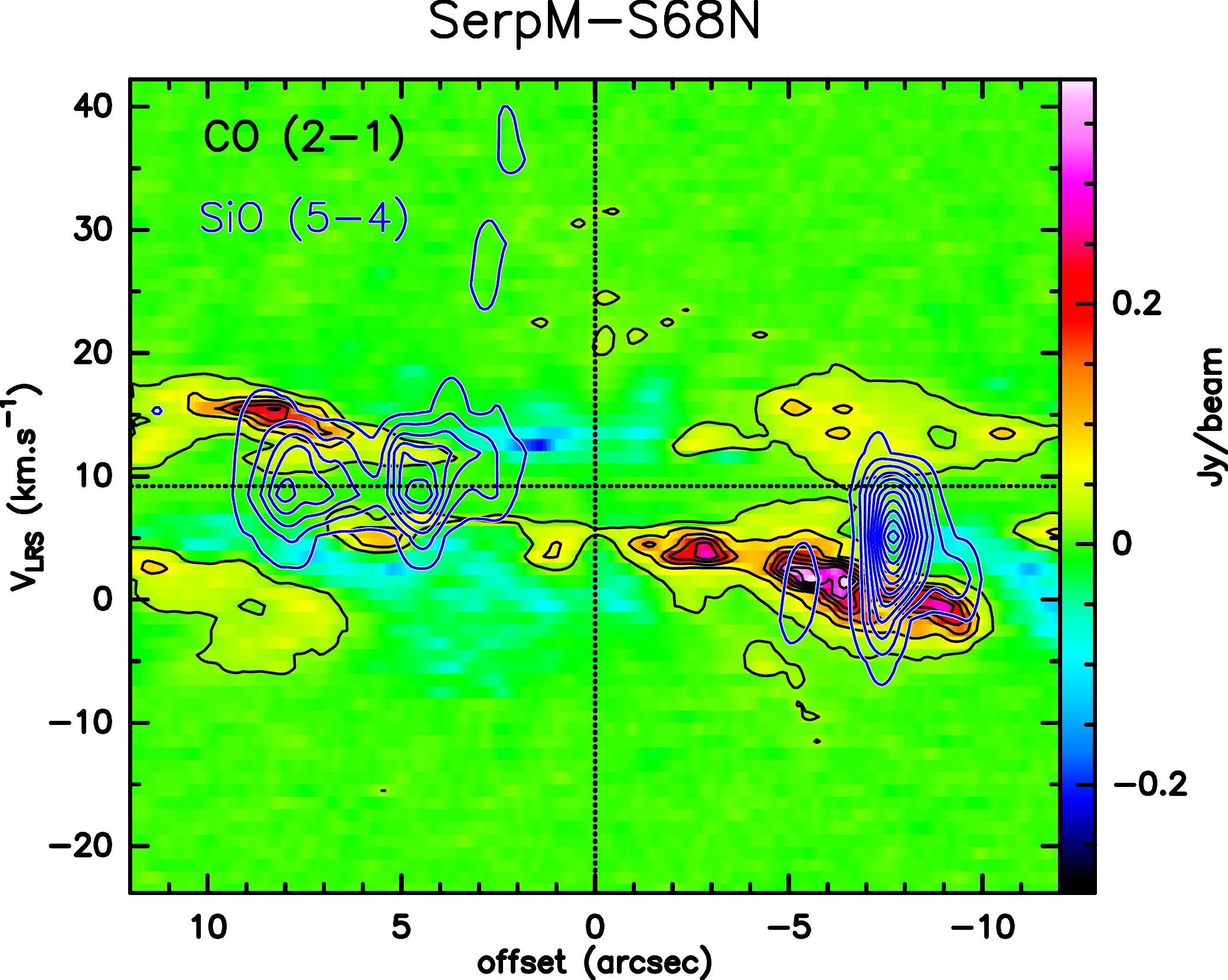}
\includegraphics[width=.45\textwidth]{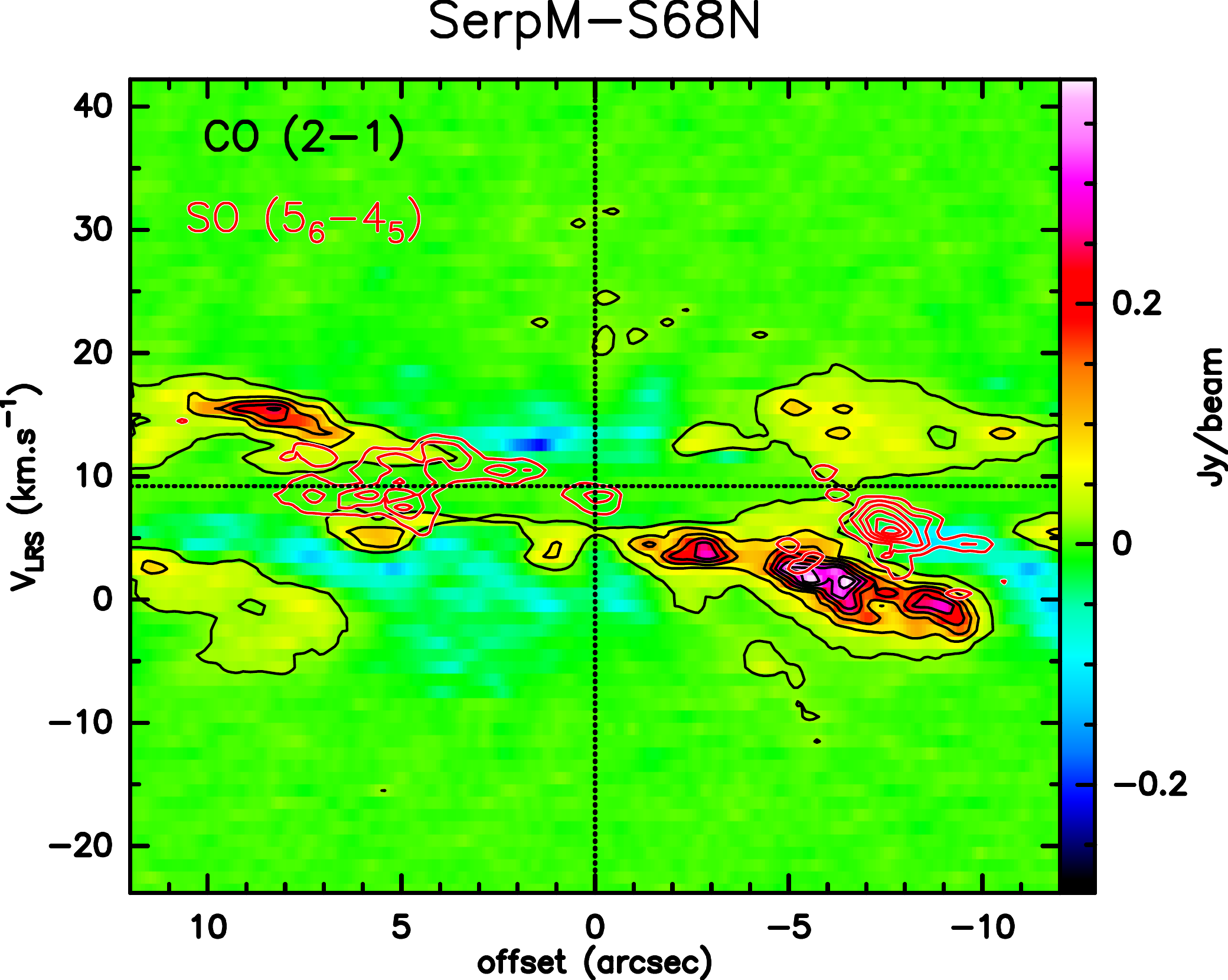}
\caption{{\it Continued}}
\label{fig:PV-block5}
\end{figure*}

\begin{figure*}
\setcounter{figure}{0}
\centering
\includegraphics[width=.45\textwidth]{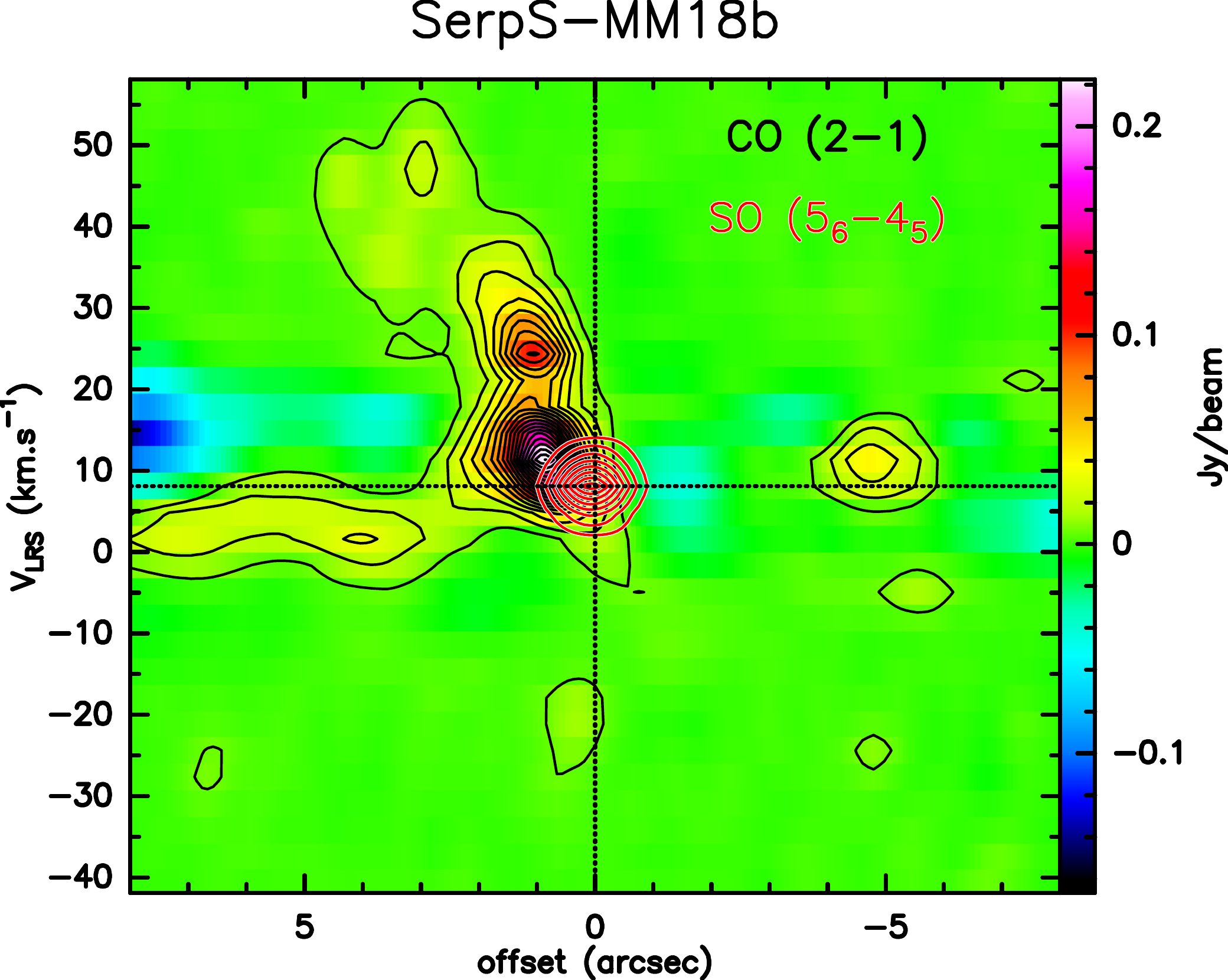} \\
\includegraphics[width=.45\textwidth]{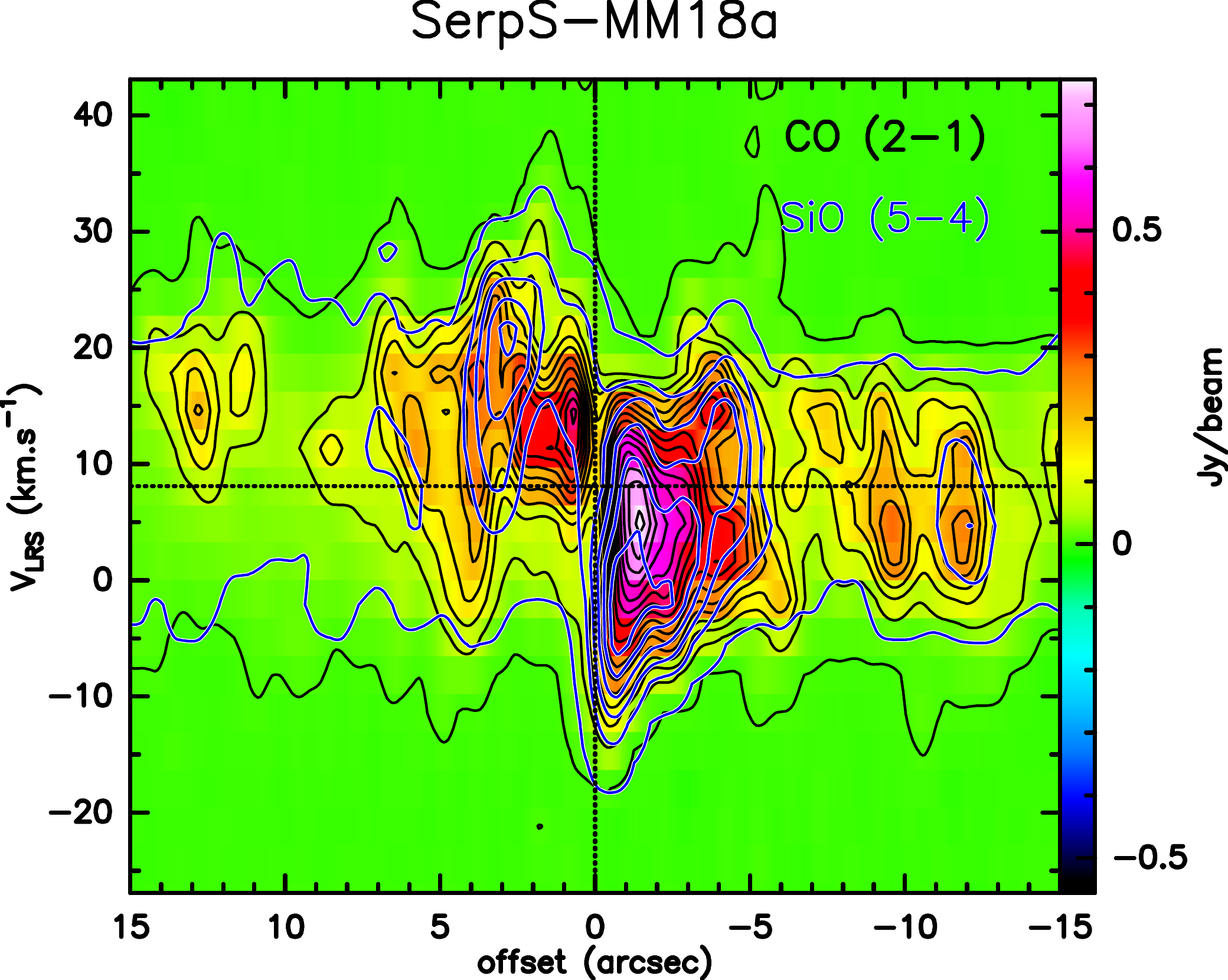}
\includegraphics[width=.45\textwidth]{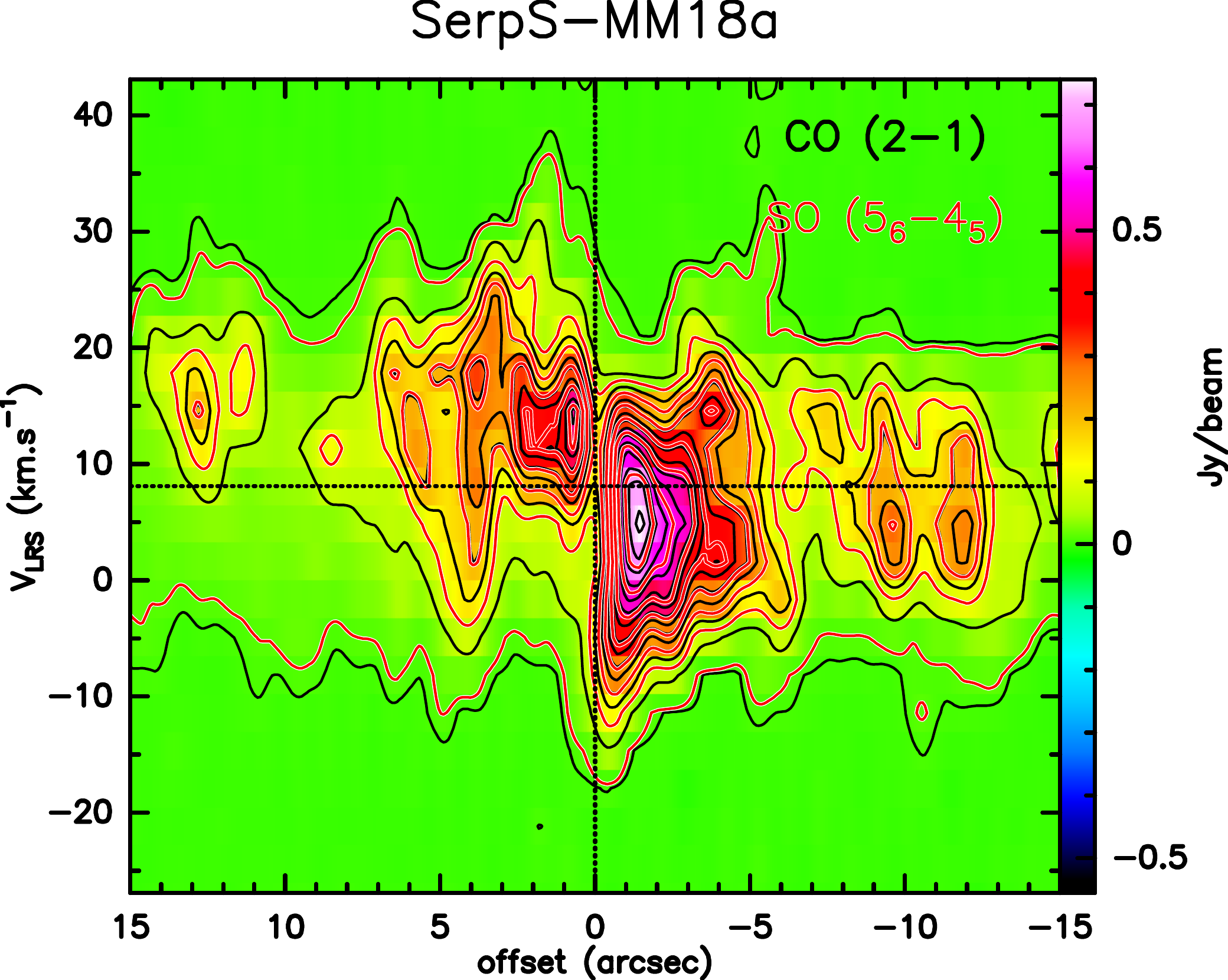} \\
\includegraphics[width=.45\textwidth]{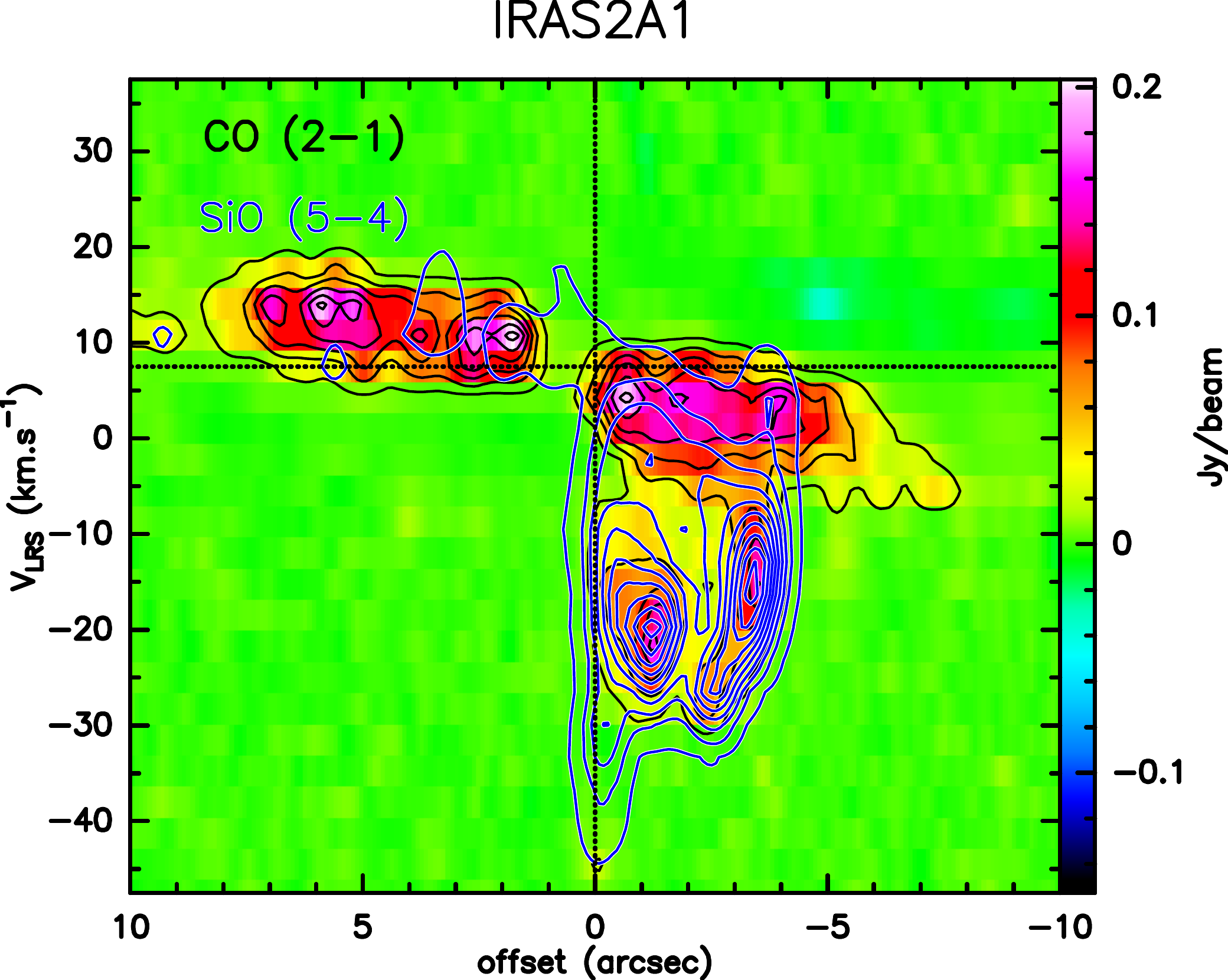}
\includegraphics[width=.45\textwidth]{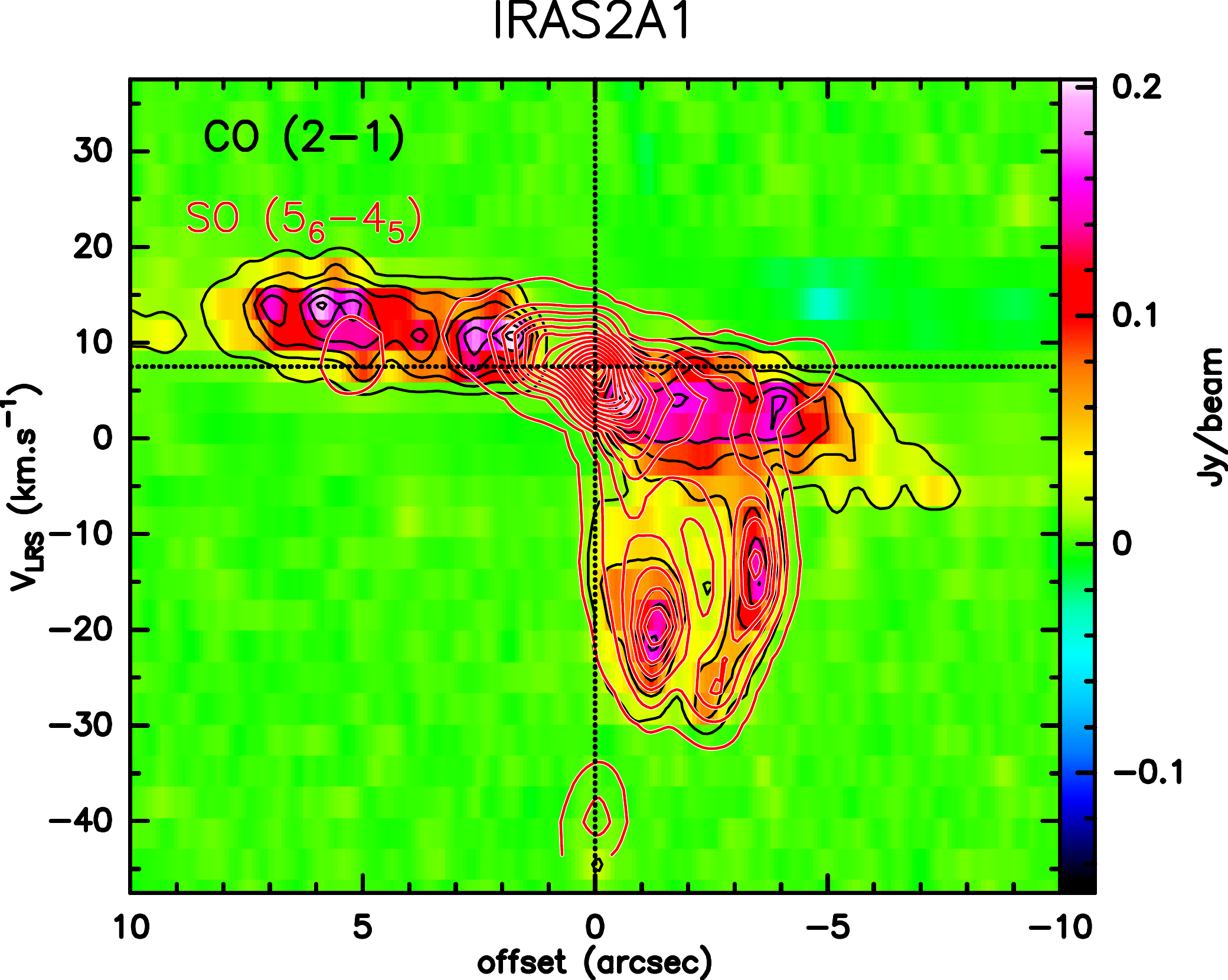} \\
\caption{{\it Continued}}
\label{fig:PV-block6}
\end{figure*}


\section{CO, SO, and SiO spectra}
\label{app:spectra}

We extracted CO ($2-1$), SO ($5_6-4_5$), and SiO ($5-4$) spectra for the 12 sources associated with an SiO jet detected at $>10\sigma$ in the integrated maps which are listed in Table \ref{tab:fluxes} and \ref{tab:jets-energetics}. For each source the spectra are extracted from the pixel at the positions of the blue-shifted and red-shifted SiO emission peaks located closest to the driving source, B and R.
The spectra are shown in  Fig. \ref{fig:spec1} for sources with increasing internal luminosity ($L_{\rm int}$). The figure shows for each source and lobe the high-velocity range over which the line intensity is integrated. The RA and Dec offsets of the blue-shifted and red-shifted SiO knots where spectra are extracted (B and R positions), the high-velocity ranges, and the integrated line fluxes are listed in Table \ref{tab:fluxes}.

\begin{figure*}
\centering
\includegraphics[width=.45\textwidth]{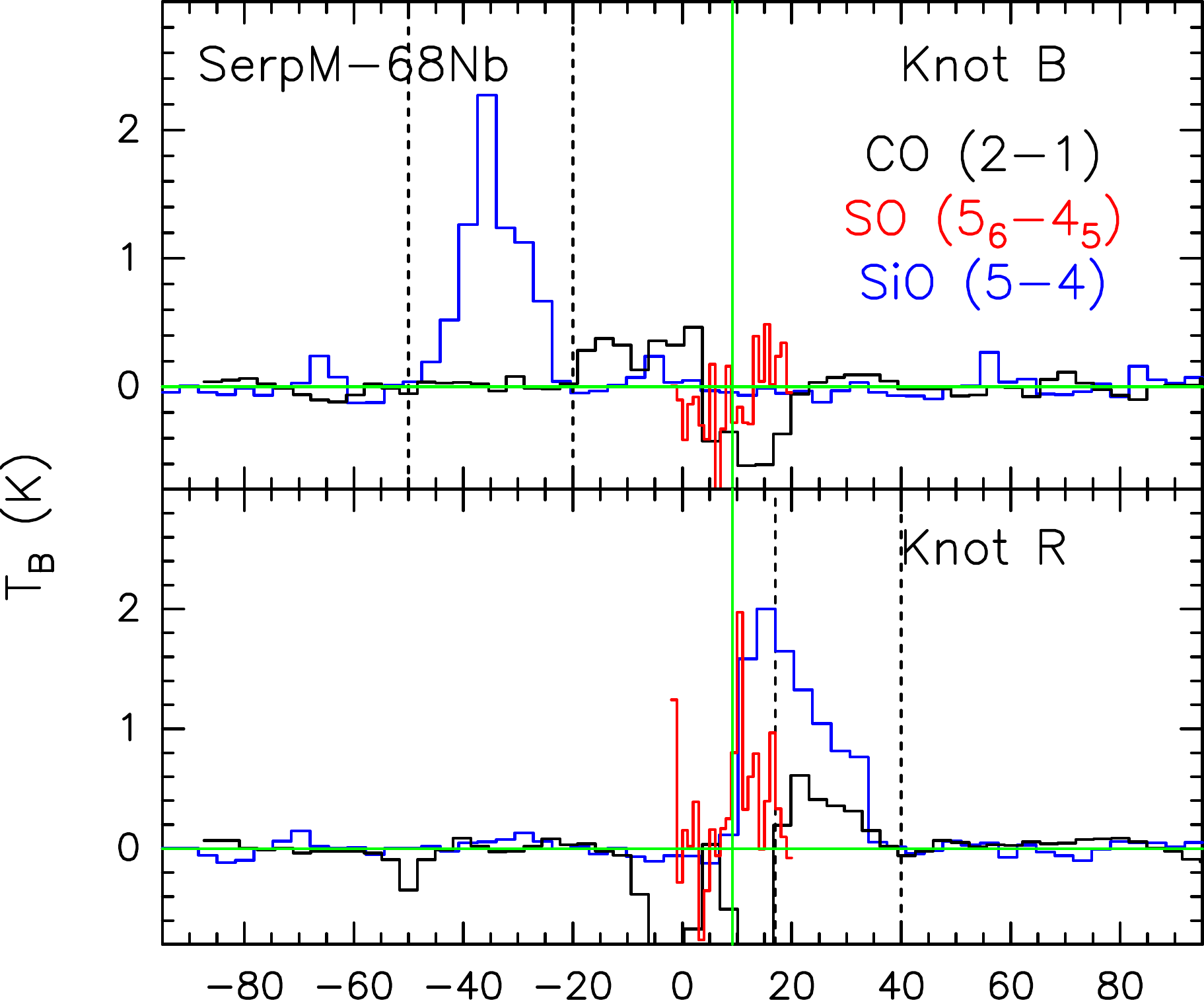} 
\vspace{.5cm}
\includegraphics[width=.45\textwidth]{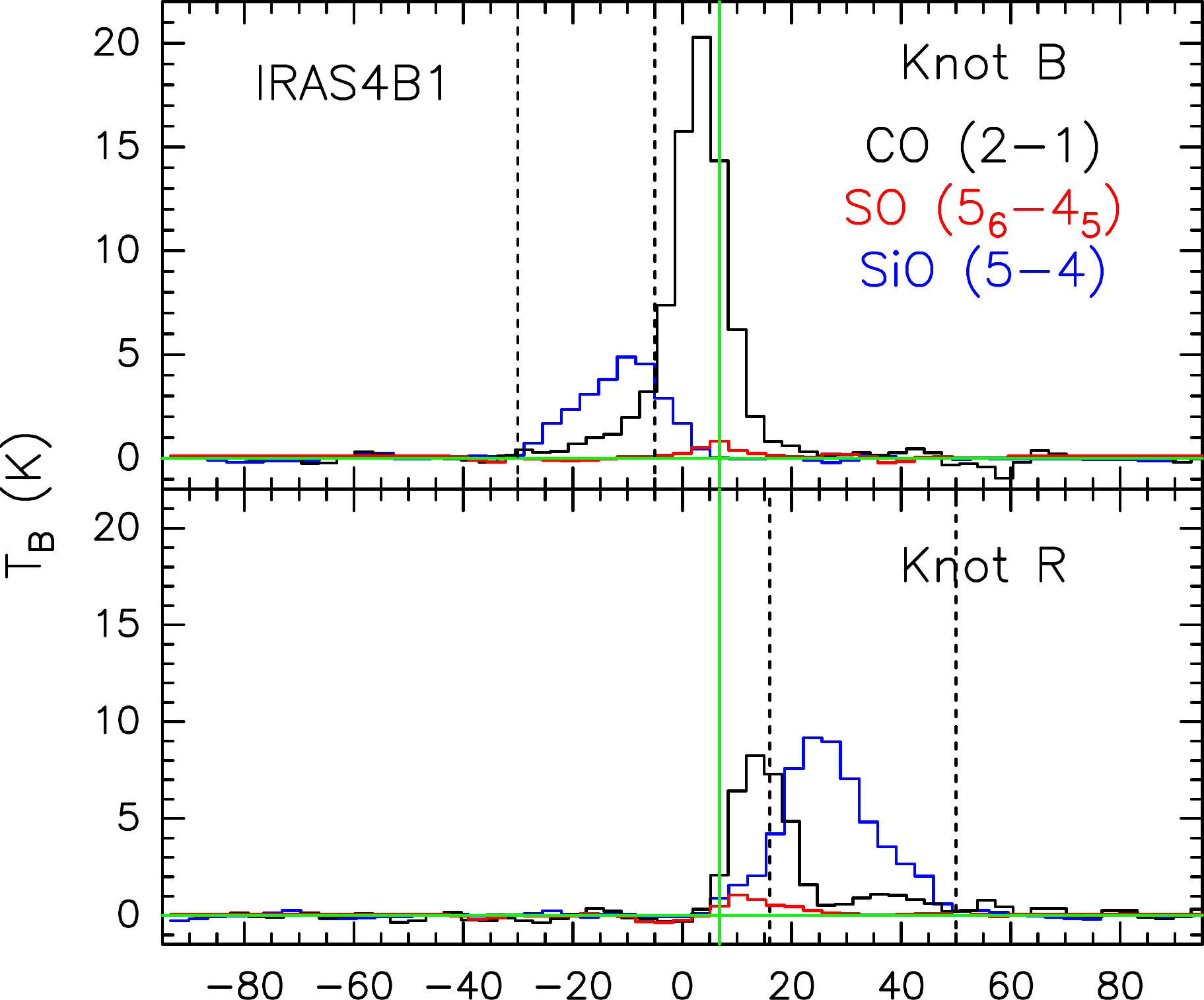}
\includegraphics[width=.45\textwidth]{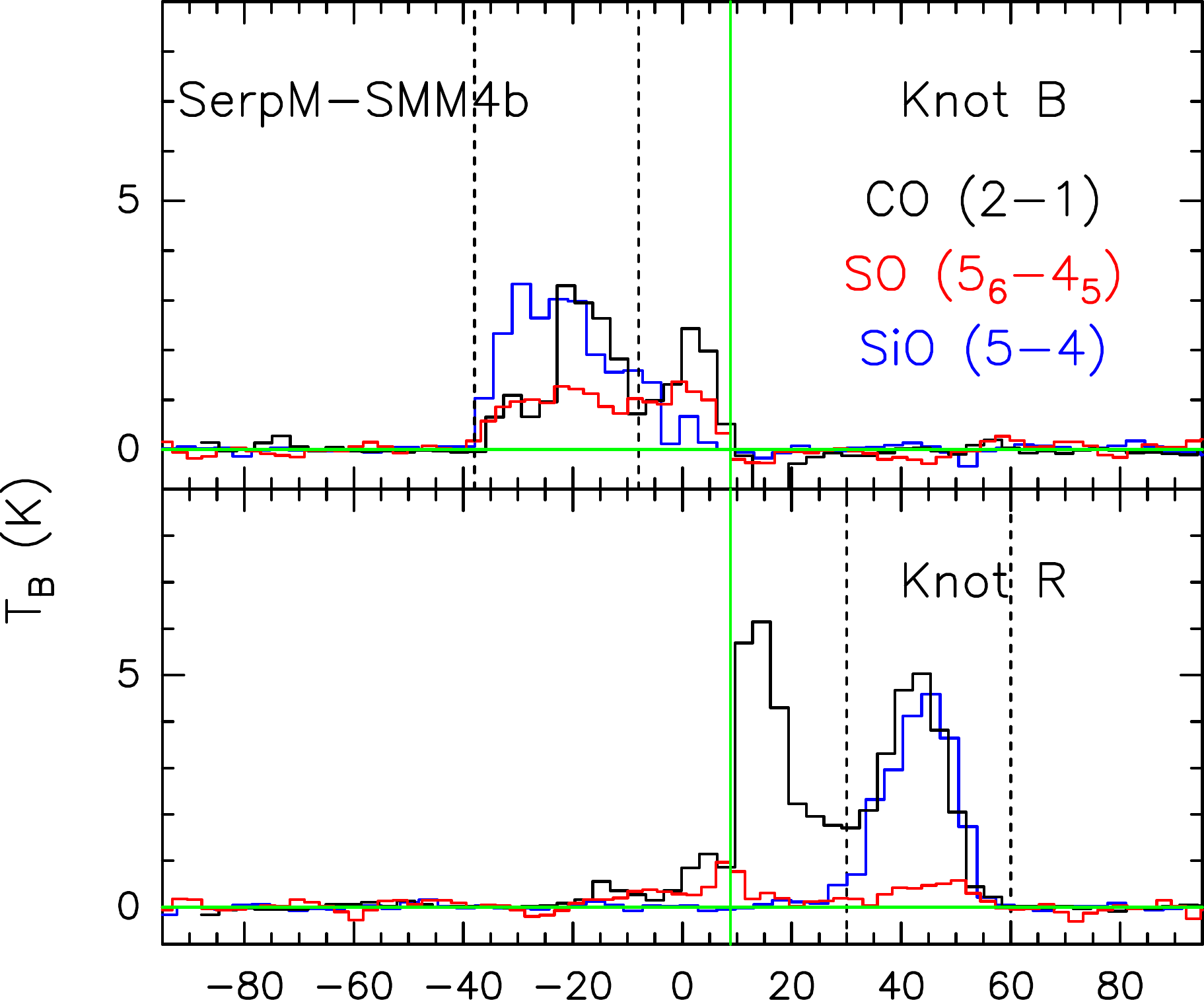} 
\vspace{.5cm}
\includegraphics[width=.45\textwidth]{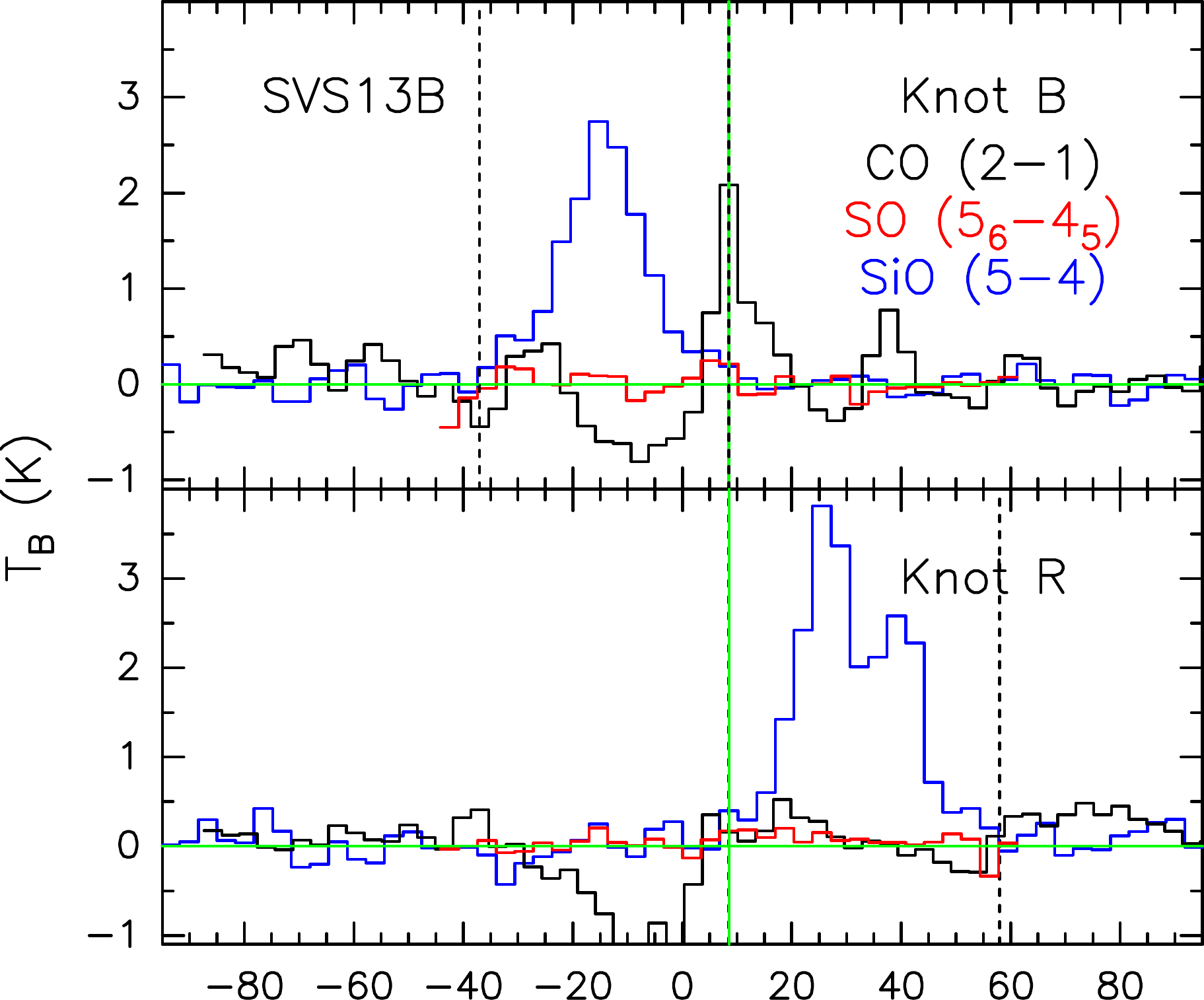} 
\includegraphics[width=.45\textwidth]{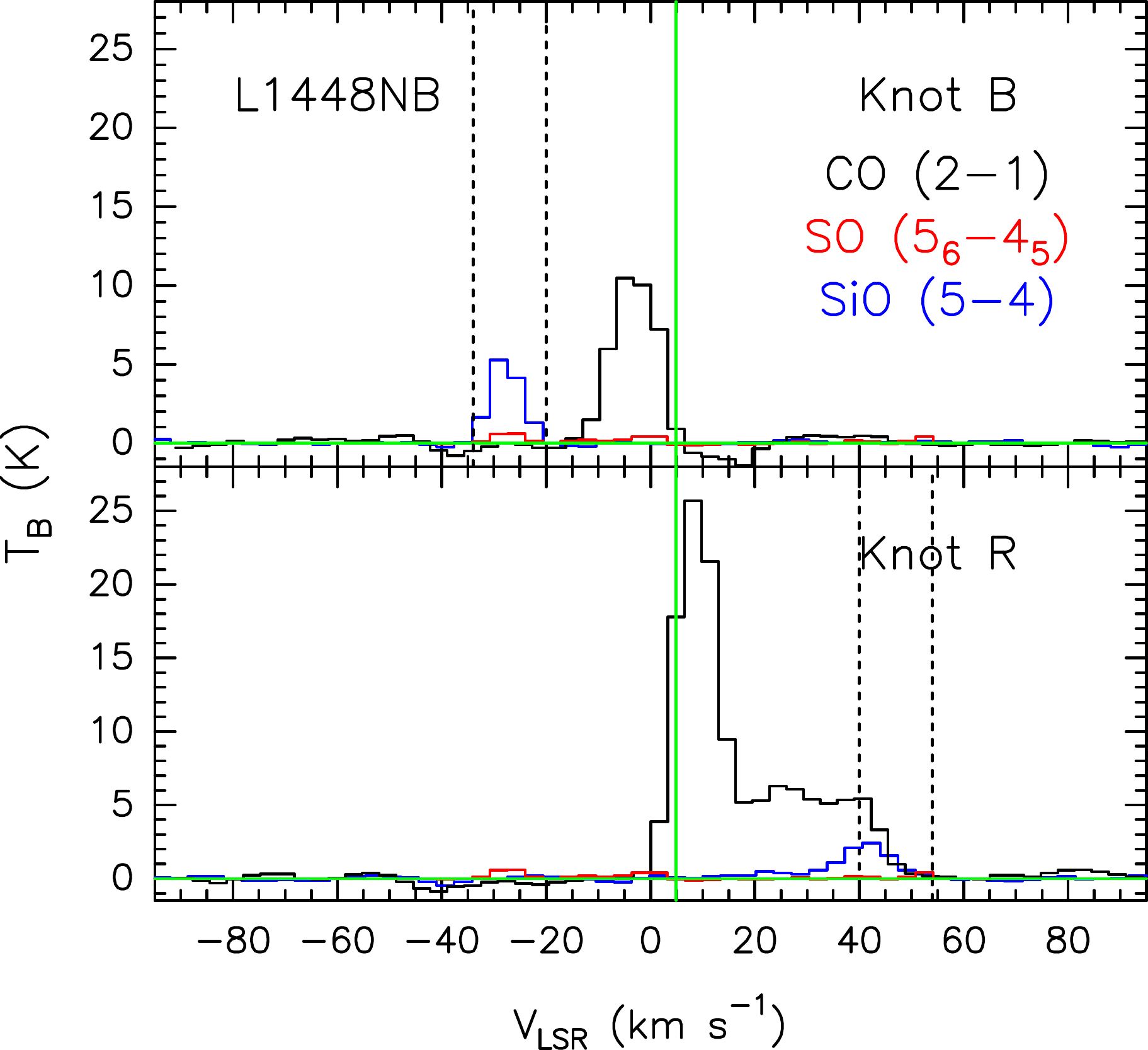} 
\includegraphics[width=.45\textwidth]{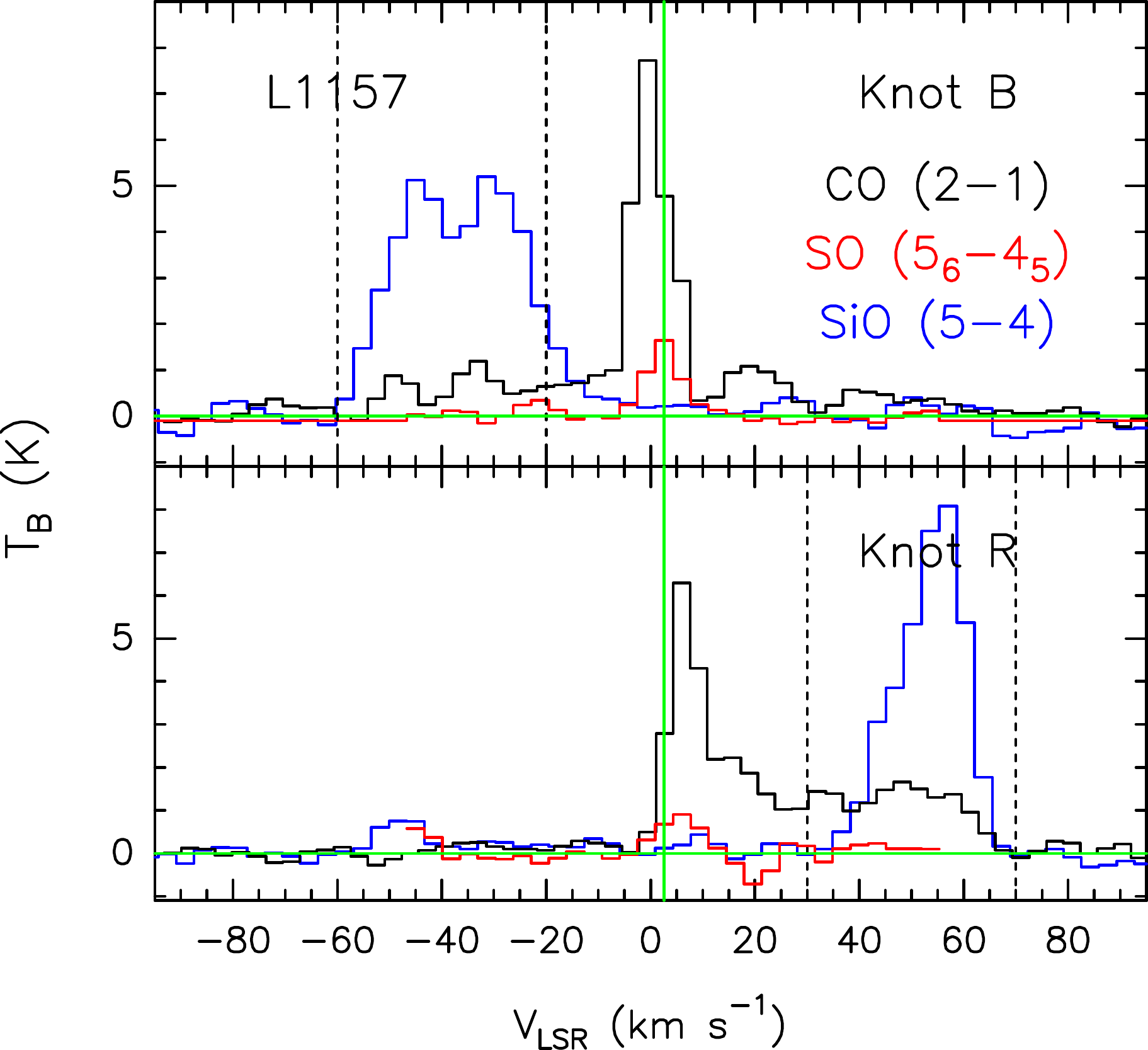} 
\caption{Spectra of CO (black), SO (red), and SiO (blue) for the 12 sources driving an SiO jet which are listed in Table \ref{tab:jet-occurrence}. For each source, the upper and lower panels show the spectra extracted at the position of the blue-shifted and red-shifted SiO knots located closest to the driving source, B and R. The source name is labeled in the top-left corner, and the sources are ordered by increasing internal luminosity ($L_{\rm int}$). The horizontal and vertical green lines indicate the baseline and systemic velocity, $V_{\rm sys}$, as listed in Table \ref{tab:sample}. The vertical black dotted lines indicate the high-velocity range over which the line intensity is integrated, as reported in Table \ref{tab:fluxes}.}
\label{fig:spec1}
\end{figure*}

\begin{figure*}
\setcounter{figure}{0}
\centering
\includegraphics[width=.45\textwidth]{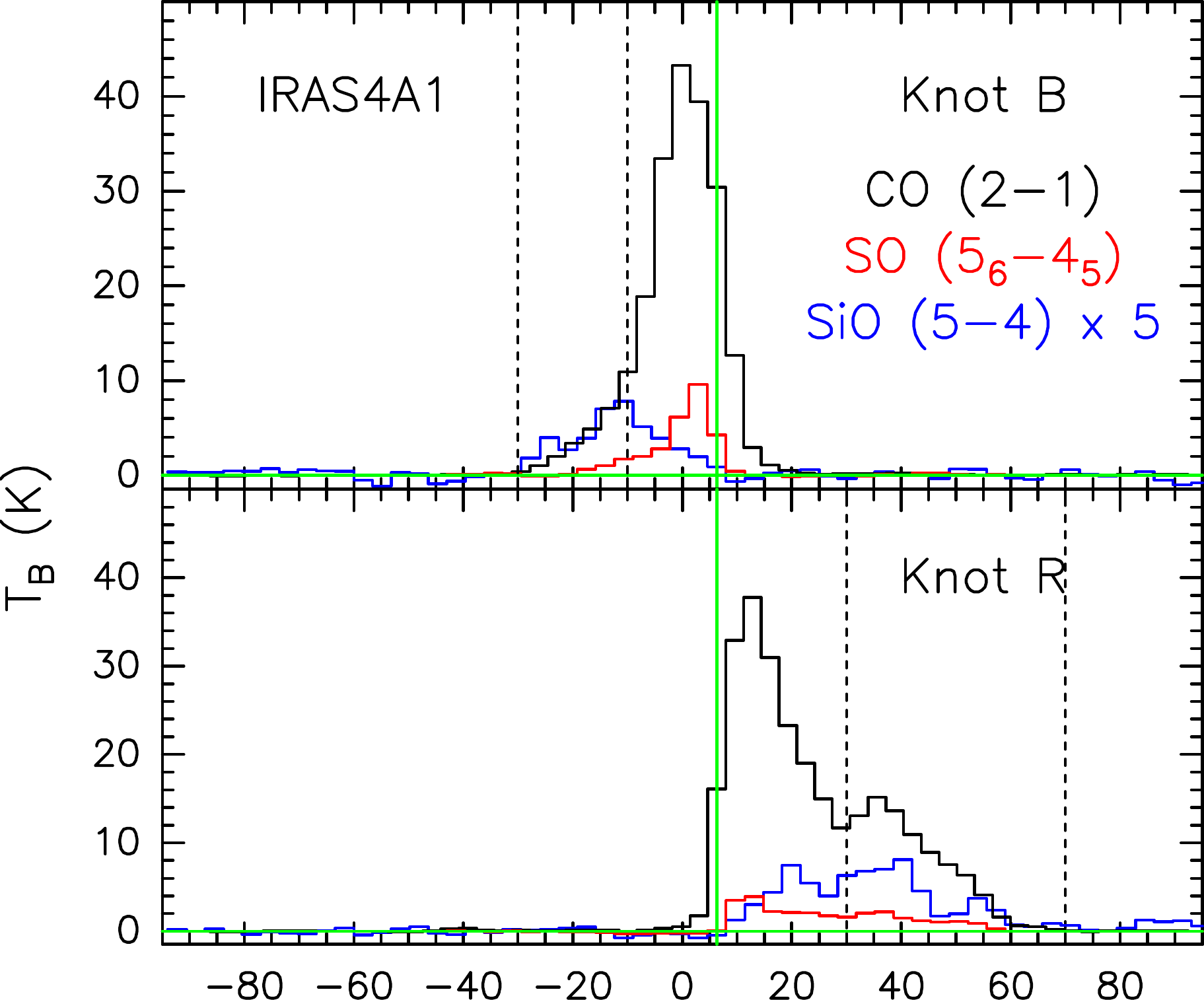} 
\vspace{.5cm}
\includegraphics[width=.45\textwidth]{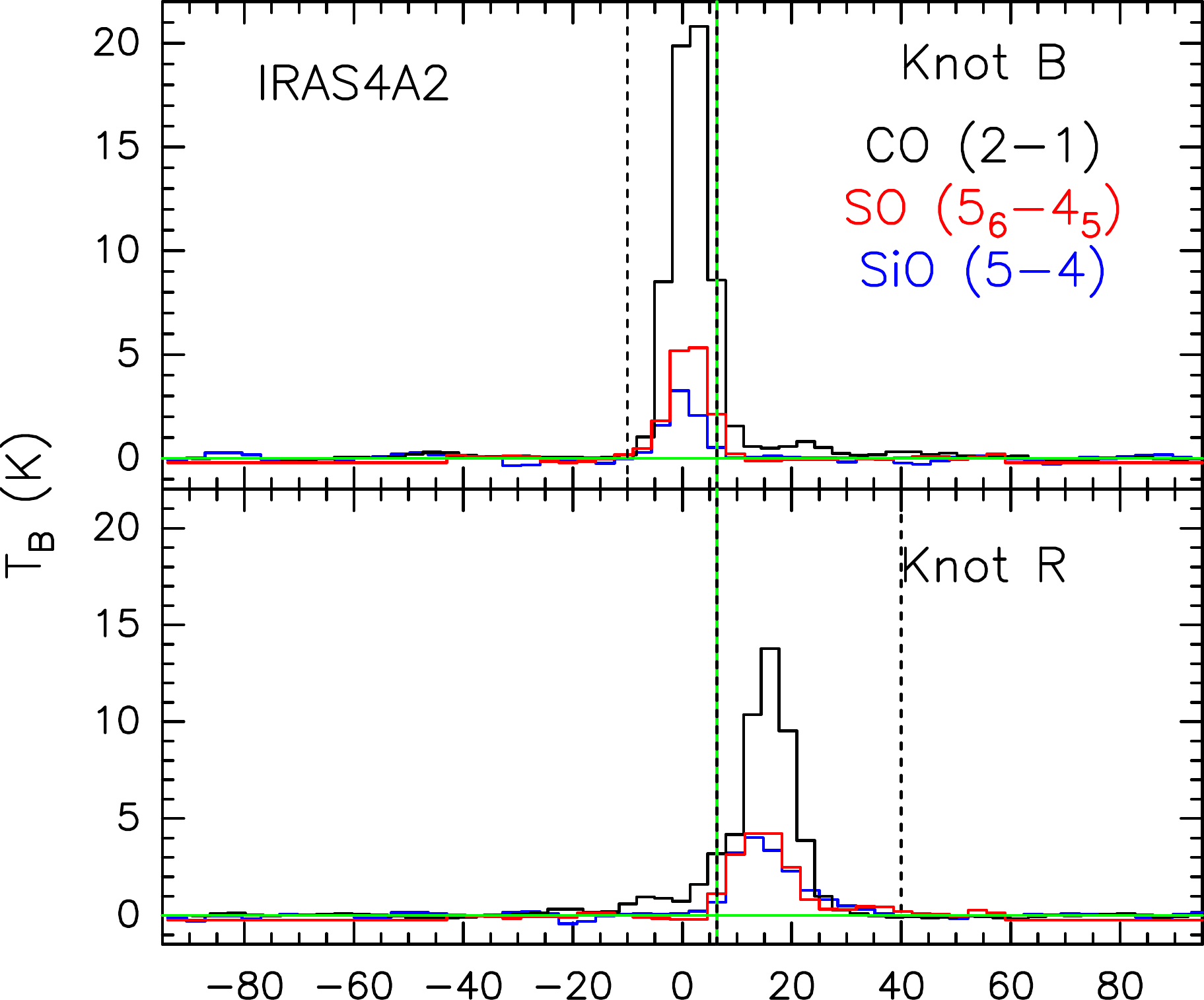} 
\includegraphics[width=.45\textwidth]{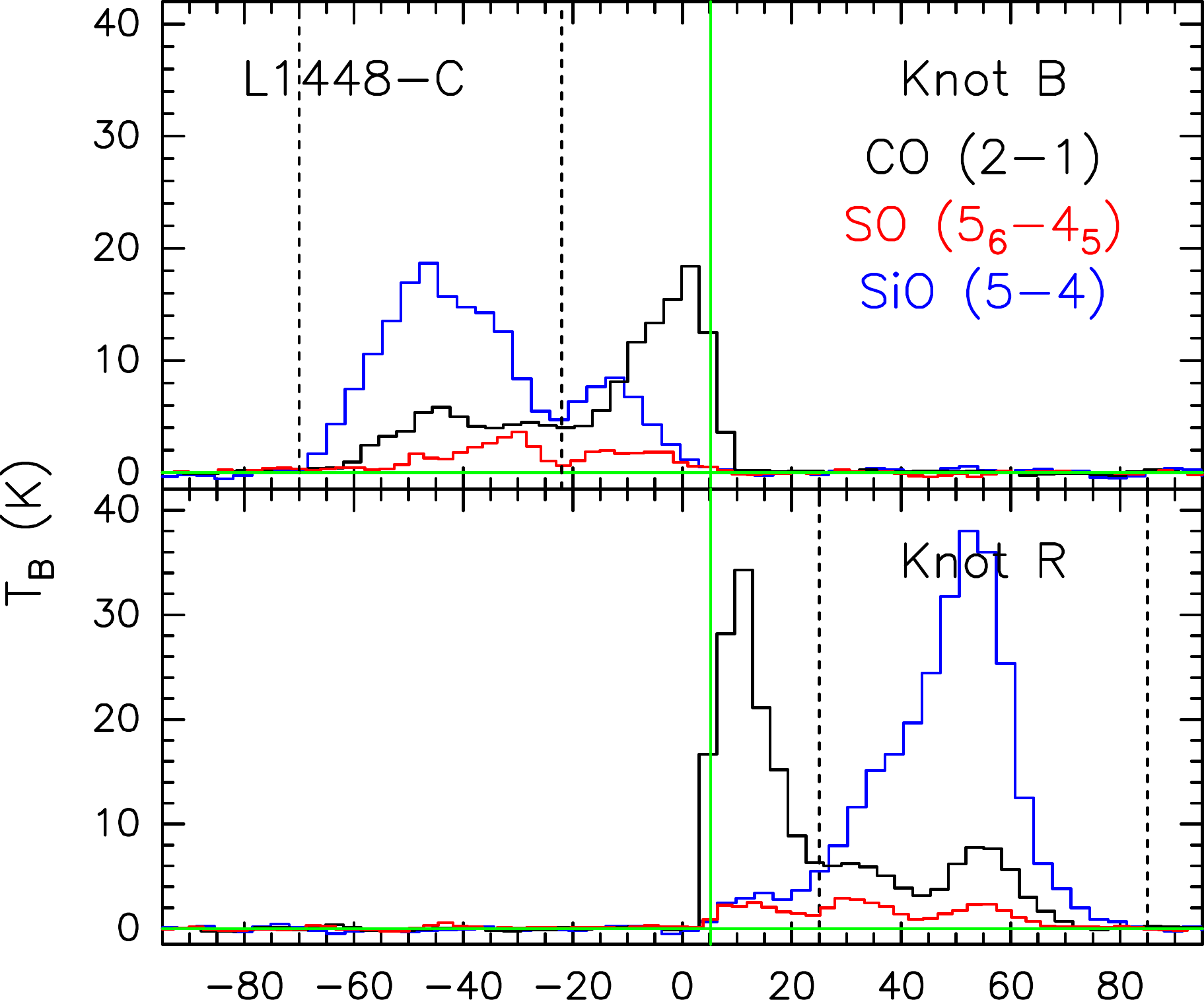} 
\vspace{.5cm}
\includegraphics[width=.45\textwidth]{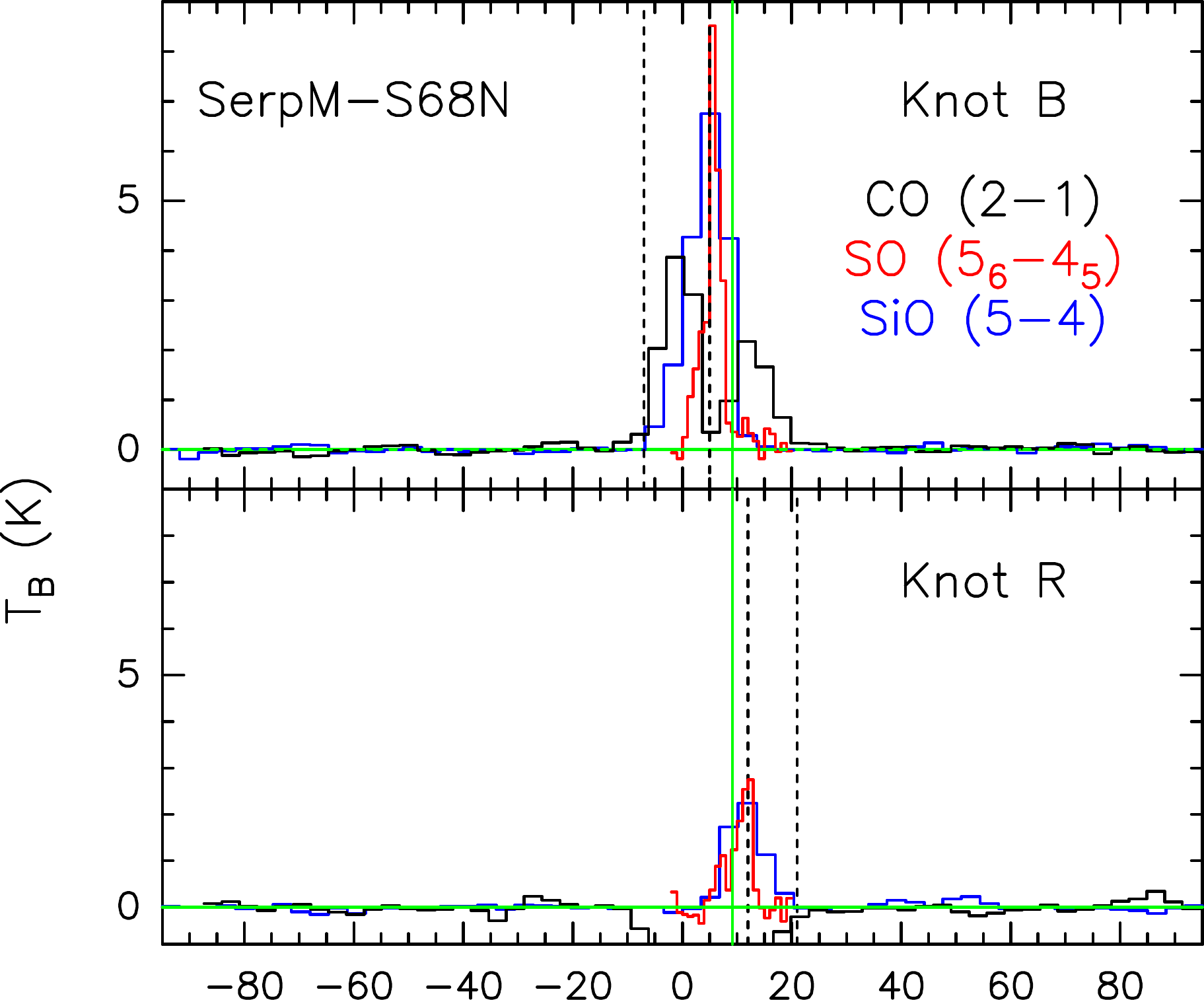} 
\includegraphics[width=.45\textwidth]{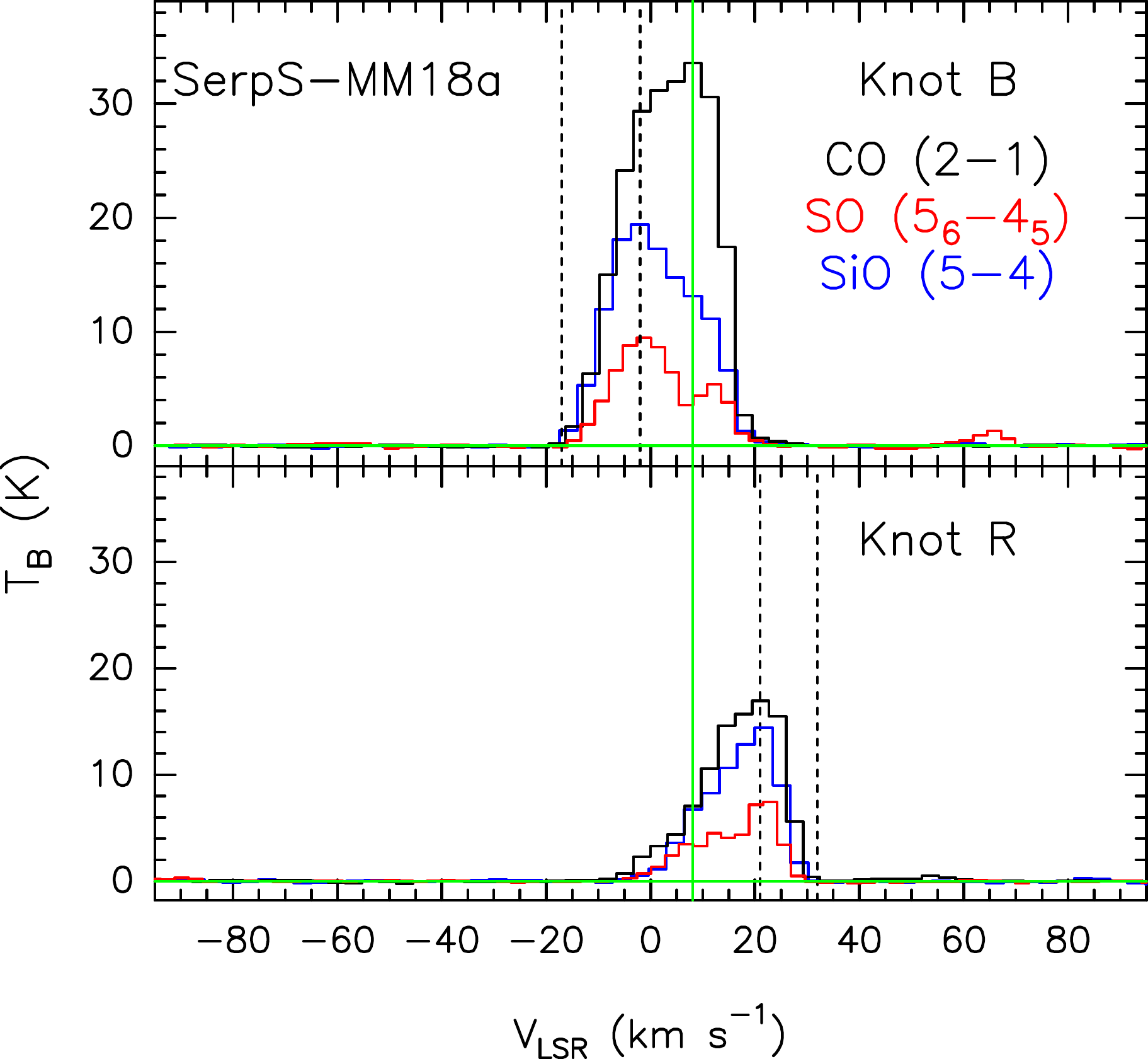}
\includegraphics[width=.45\textwidth]{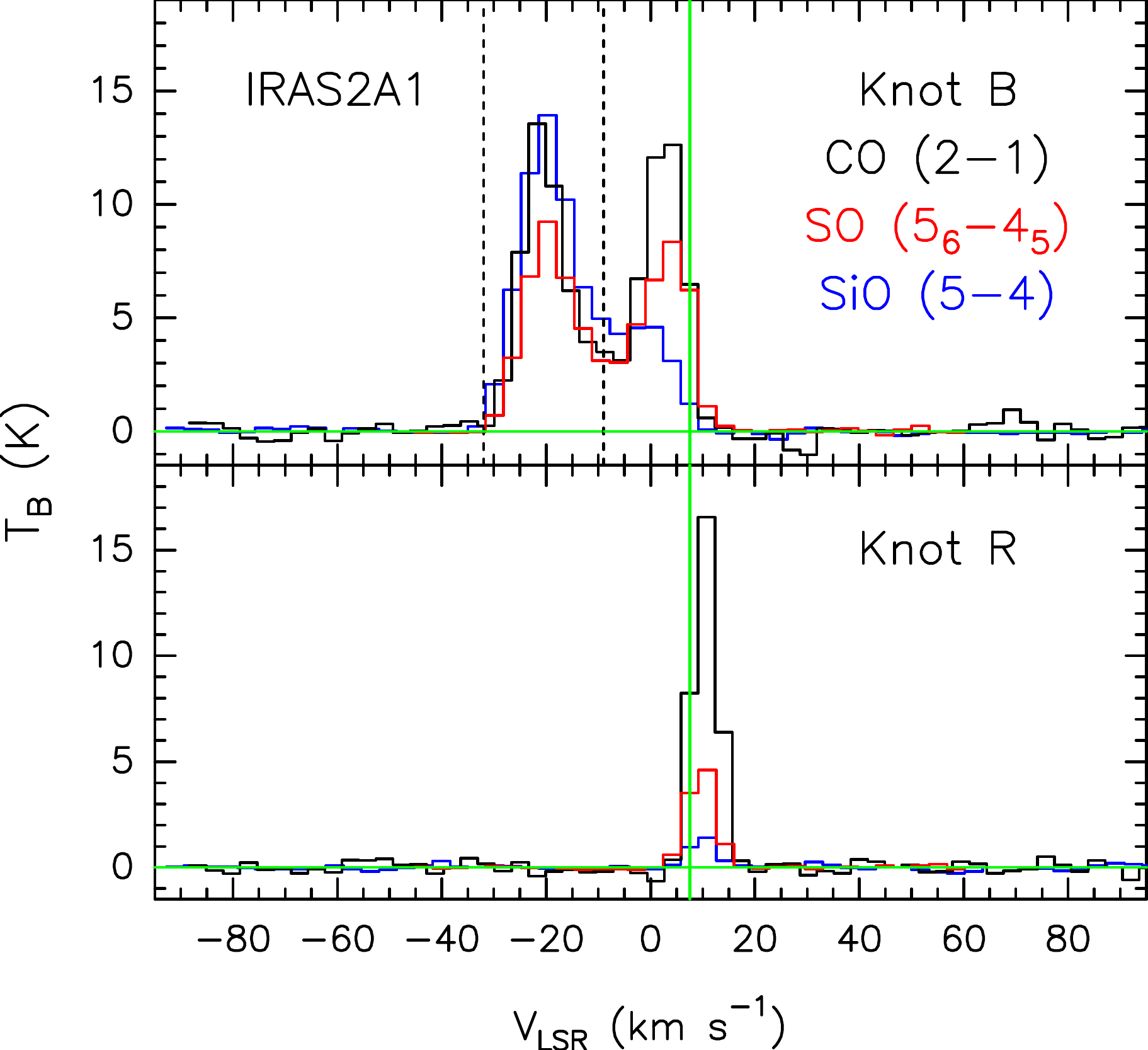} 
\caption{{\it Continued}.}
\label{fig:spec2}
\end{figure*}



\section{Measurement of the jet radius}
\label{app:jet-width}

In this appendix, the methodology used to estimate the jet radius, \Rjet, is explained.
We assume that the jet is spatially resolved along its axis and has a Gaussian emission profile across its axis with a FWHM, which is equal to twice the jet radius, 2\Rjet. The FWHM of the observed emission, FWHM$_{\rm obs}$, is the convolution of the intrinsic jet width, 2\Rjet, and the transverse size of the beam, $b_{\rm t}$, that is:
\begin{equation}
\label{eq:fwhm-obs}
    FWHM_{\rm obs}^2 = (2 R_{\rm jet})^2 + b_{\rm t}^2,
\end{equation}
where $b_{\rm t}$ is given by 
\begin{equation}
    b_{\rm t}^2 = \frac{1}{\frac{cos^2{\theta}}{b_{\rm maj}^2} + \frac{sin^2{\theta}}{b_{\rm min}^2}},
    \label{eq:transverse-beam}
\end{equation}
where $\theta$ is the position angle of the beam with respect to the jet axis, and $b_{\rm maj}$, and $b_{\rm min}$ are the major and minor axis of the elliptical beam, respectively. 

The FWHM of the observed line emission, FWHM$_{\rm obs}$, is obtained by fitting the velocity integrated jet and outflow emission shown in Fig. \ref{fig:jets1} perpendicular to the jet PA (see Table \ref{tab:jet-occurrence}) with a Gaussian profile. This procedure is applied to the line emission in the three tracers (CO ($2-1$), SO ($5_6-4_5$), and SiO ($5-4$)), and at each position along the jet axis, that is, for increasing distance from source. Then the jet width, 2\Rjet, is obtained from FWHM$_{\rm obs}$ by using  Eq. \ref{eq:fwhm-obs} and Eq. \ref{eq:transverse-beam}, that is:\ 
\begin{equation}
\label{eq:Rjet}
    2 R_{\rm jet} = \sqrt{FWHM_{\rm obs}^2 - b_{\rm t}^2}
.\end{equation}

\begin{figure*}
\centering
\includegraphics[width=\textwidth]{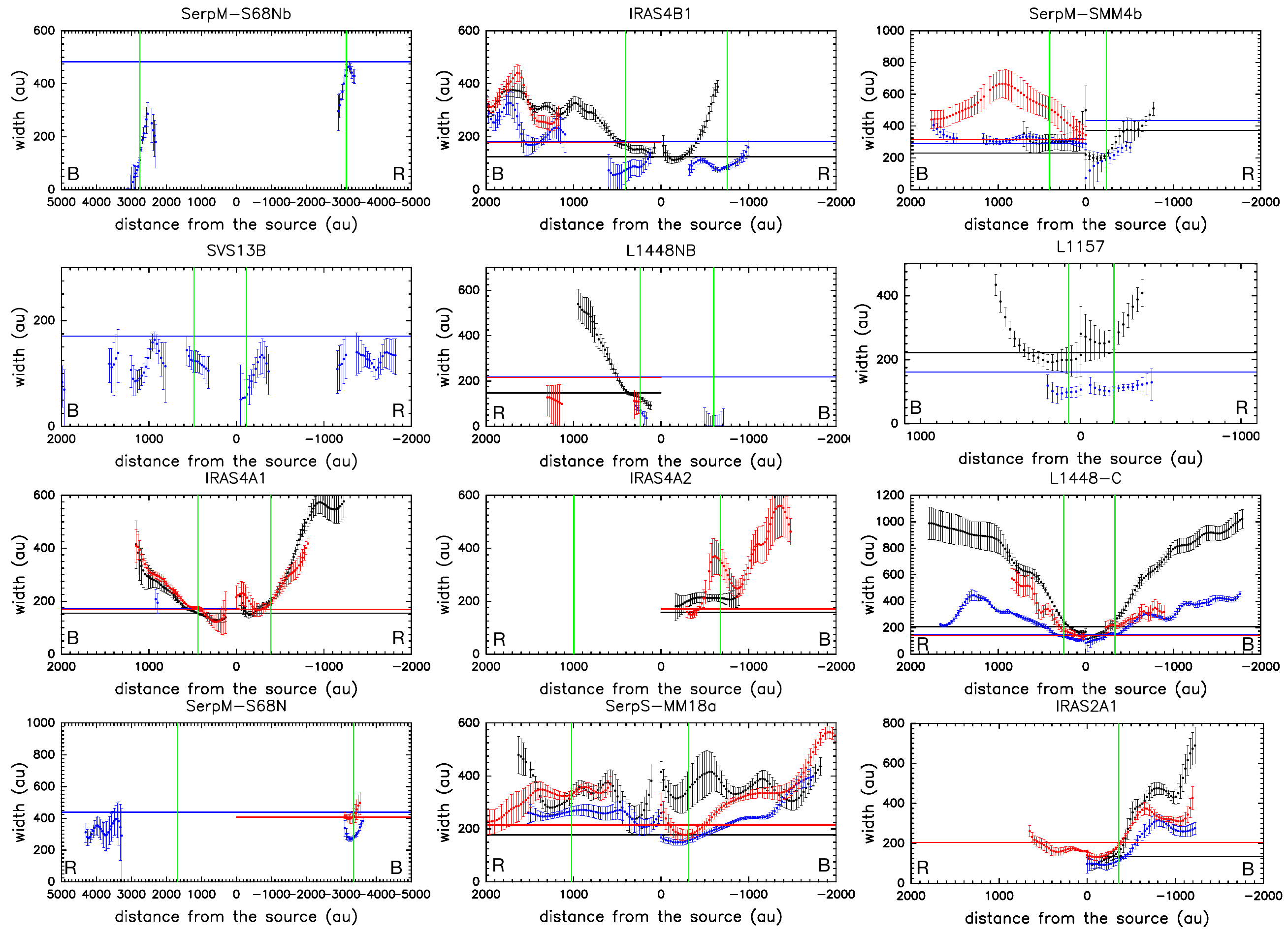} 
\caption{Deconvolved widths of the 12 jets detected in the CALYPSO sample within 5000 au from protostars. Widths of CO ($2-1$), SO ($5_6-4_5$), and SiO ($5-4$) emission are plotted in black, red, and blue, respectively. The flow width has been measured only at the positions where the transverse profile shows a peak emission larger that $10\sigma$. 
Flows which show spatially resolved structures strongly deviating from a Gaussian profile have been discarded. The positions of the blue- and red-shifted SiO knots located closest to the driving protostar, B and R, where the line spectra have been extracted (see Fig. \ref{fig:spec1}), are indicated by vertical green lines.  
No measurements of the width of the SiO emission at the position of the inner B and R knots is derived for the jets of IRAS4A1 and IRAS4A2, whose SiO emission is detected at $>10\sigma$ only at larger distances from the protostar in the terminal bow shocks which are not located along the jet axis.
The size of the beam transverse to the jet axis for CO ($2-1$), SO ($5_6-4_5$), and SiO ($5-4$) are indicated by black, red, and blue horizontal lines, respectively.}
\label{fig:jet-width_all}
\end{figure*}

If the signal-to-noise ratio (S/N) is high enough, a jet width smaller than the beam size can be inferred by measuring the (small) difference betwen FWHM$_{\rm obs}$ and $b_{\rm t}$. For example, a jet diameter of $\sim45$\% the beam size would result in a FWHM$_{\rm obs} \sim 1.1 b_{\rm t}$.
The estimated jet widths, 2 \Rjet, as a function of the distance from the driving source for the 12 sources associated with an SiO jet detected at $>10\sigma$ in the integrated maps are shown in Fig. \ref{fig:jet-width_all}.


\section{Uncertainties on derived column densities}
\label{app:uncertainties}

\begin{figure}
\includegraphics[width=.49\textwidth]{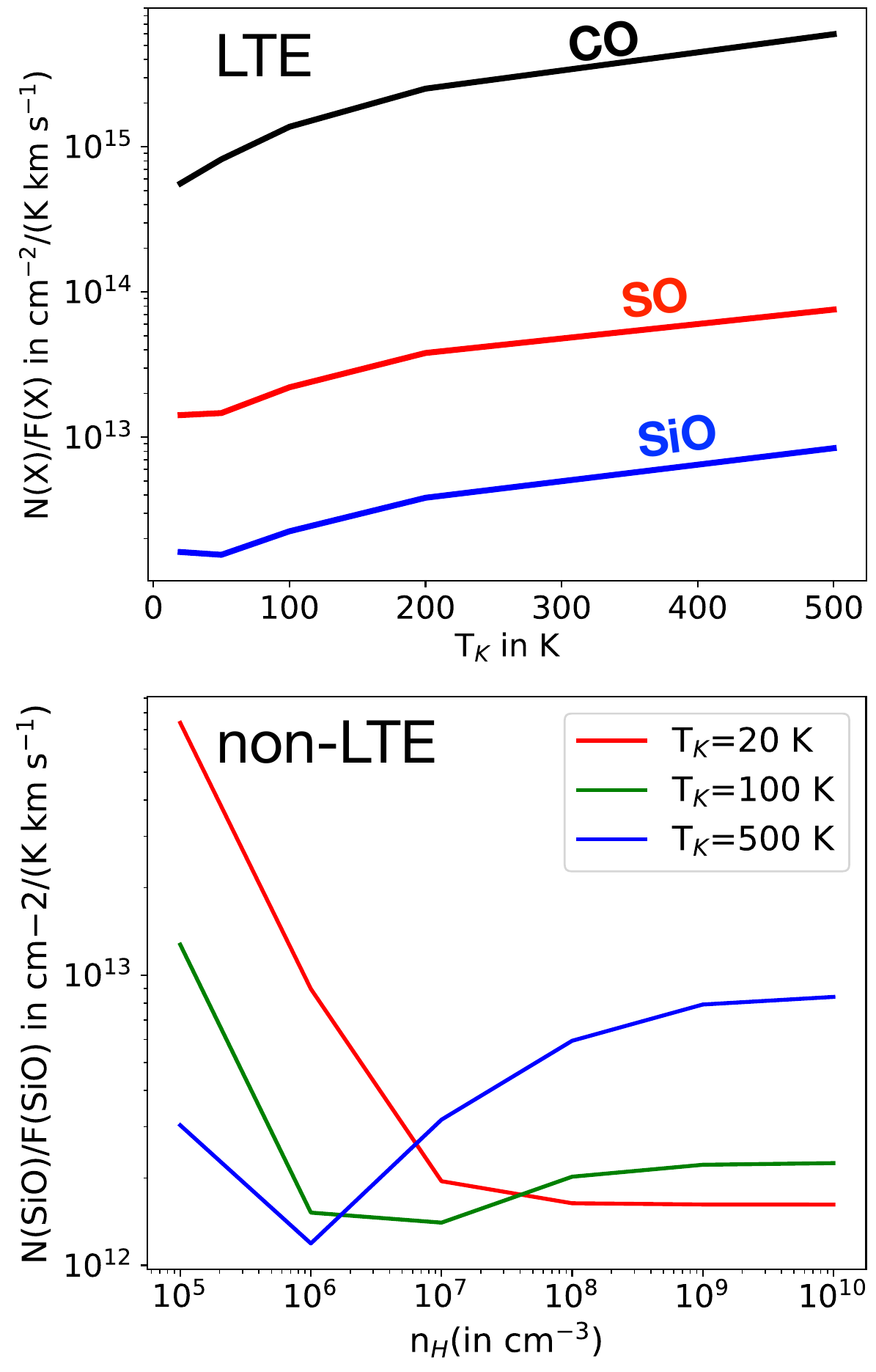} 
\caption{Conversion factors between line intensity and derived column density assuming optically thin emission. {\it Top panel:} CO ($2-1$) (black), SO ($5_6-4_5$) (red), and SiO ($5-4$) (blue) conversion factor assuming LTE. {\it Bottom panel:} SiO ($5-4$) conversion factor for various kinetic temperatures as a function of the gas density computed by RADEX (non-LTE).}
\label{fig:conv-fac}
\end{figure}

Through this work, column densities of CO and SiO have been derived assuming that rotational levels are populated according to a Boltzman distribution (LTE) with a single temperature $T_{\rm K} = 100$~K. Given the low critical density of CO ($2-1$), CO is expected to be at LTE and the uncertainties on $N_{\rm CO}$ are mostly due to the uncertainties on the temperature. Figure \ref{fig:conv-fac}-a shows that the error on the true $N_{\rm CO}$ remains within a factor of 3 for a kinetic temperature between 50 and 100 K. On the contrary, SiO is expected to be out of LTE and the derived column density depends not only on the temperature but also on the density. Figure \ref{fig:conv-fac}-b shows that for a temperature of 100 K and densities above $10^6$cm$^{-3}$, errors due to the LTE assumption remain within a factor of 2.

\begin{figure}
\centering
\includegraphics[width=.49\textwidth]{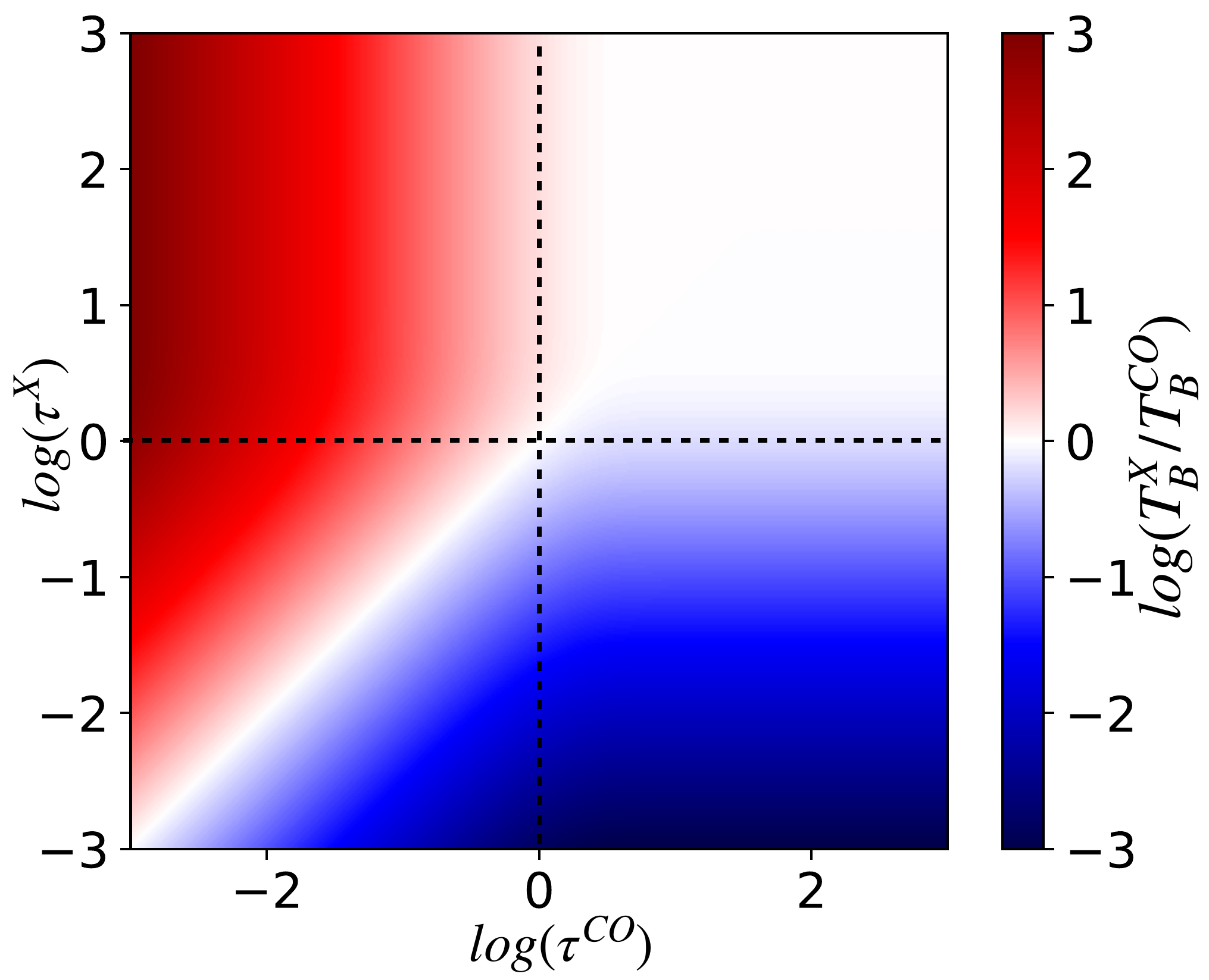} 
\caption{Line ratio as a function of the opacity of each line, $\tau^{\rm CO}$ and $\tau^{\rm X}$, assuming that levels are in LTE at the same temperature. The dashed vertical line and the dashed horizontal line delimit the region where the CO line and the $X$ line are optically thick, respectively. $T_{\rm B}^{\rm X}/T_{\rm B}^{\rm CO} \simeq 1$ only if both lines are optically thick or if $\tau^{\rm X} \simeq \tau^{\rm CO}$.}
\label{fig:Tratio-vs-tau}
\end{figure}

If lines are optically thick, the beam averaged column density derived in the optically thin limit is only a lower limit on the true column density. We propose a criterion to flag lines that are, or may be, optically thick using the brightness temperature of CO ($2-1$), SiO ($5-4$), and SO ($5_6 - 4_5$) lines in the HV range. In the large velocity gradient (LVG) approximation, and assuming that the rotational levels are in LTE at a temperature $T_{\text{K}}$, the brightness temperature is given by:
\begin{equation}
    T_{\text{B}} = f (1-\exp{(-\tau)}) T_{\text{K}},
    \label{eq:Tmb}
\end{equation}
where $\tau$ is the opacity of the line and $f$ is the filling factor. Equation (\ref{eq:Tmb}) shows that for an optically thin emission, the beam corrected brightness temperature $T_{\text{B}}/f$ is strictly smaller than $T_{\text{K}}$ whereas $T_{\text{B}}/f = T_{\text{K}}$ only if the line is optically thick ($\tau >> 1$). $T_{\text{B}} /f = T_{\text{K}}$ would thus be the direct evidence of optically thick emission. At the position of the extracted spectra, jets are however marginally resolved in the transverse direction and the beam may comprise multiple smaller scale structures along the jet. The filling factor, $f,$ is thus uncertain. Moreover, the kinetic temperature is poorly known. Alternatively, we propose a method that is independent of $f$ and $T_{\text{K}}$. Assuming that the emission of all lines originates from the same gas component in the jet and that levels are in LTE at the same temperature, the lines ratio:
\begin{equation}
  \frac{T^{\text{X}}_{\text{B}}}{T^{\text{CO}}_{\text{B}}} = \frac{1-\exp{(-\tau^{\text{X}})}}{1-\exp{(-\tau^{\text{CO}})}},
    \label{eq:Tmb-ratio}
\end{equation}
where X denotes either the SiO ($5-4$) or the SO ($5_6-4_5$) line, depends only on the relative opacity of the $X$ and CO line.
Figure \ref{fig:Tratio-vs-tau} shows that ${T^{\text{X}}_{\text{B}}} = {T^{\text{CO}}_{\text{B}}}$ if either of the following is true: both lines are optically thick (upper right quadrant), or the opacity of both lines are exactly equal (thin region along the bottom right diagonal), a case that is unlikely.
If ${T^{\text{X}}_{\text{B}}} > {T^{\text{CO}}_{\text{B}}}$ [resp. ${T^{\text{X}}_{\text{B}}} < {T^{\text{CO}}_{\text{B}}}$], CO [resp. $X$] is optically thin whereas $X$ [resp. CO] can be either optically thin or thick. 

 This analysis suggests that we can use the peak temperature ratio in the high-velocity range to identify lines that may be optically thick. However, the assumption of LTE does not always apply for the SiO and SO lines. Due to the high critical densities of the SiO and SO lines, the excitation temperature may be smaller than the kinetic temperature for densities smaller or about the critical density. Complementary non-LTE models using the RADEX code indicate that when both $X$ and CO lines are optically thick, the ratio $T^{\text{X}}_{\text{B}}/T^{\text{CO}}_{\text{B}}$ lays between $\sim 0.7$ and $1$ for densities of $n_{\text{H}} \ge 10^6$~cm$^{-3}$. The effects of non-LTE are then included in our analysis by considering that:
 \begin{itemize}
     \item  if $0.7 \le \frac{T^{\text{X}}_{\text{B}}}{T^{\text{CO}}_{\text{B}}} \le 1$, both line are optically thick
      \item if $\frac{T^{\text{X}}_{\text{B}}}{T^{\text{CO}}_{\text{B}}} > 1$, the CO line is optically thin and the $X$ line may be optically thick
     \item if $\frac{T^{\text{X}}_{\text{B}}}{T^{\text{CO}}_{\text{B}}} < 0.7$, the $X$ line is optically thin and the CO line may be optically thick
 \end{itemize}


\section{Notes on individual sources}
\label{app:notes-on-sources}

\subsection{Sources in Perseus}

The CALYPSO sample comprehends many sources located in Perseus ($d=293$ pc, \citealt{ortiz-leon18a}) in the active star-formation sites L1448 (2A, N, and C objects) and NGC1333 (IRAS2A, SVS13, IRAS4A, and IRAS4B). 

\subsubsection{L1448-2A}

L1448-2A was first identified as a Class 0 protostar based on far-infrared and (sub-)millimeter observations by \citet{olinger99} who also reported a parsec scale bipolar outflow detected in CO ($1-0$) emission.
The continuum emission at 94 and 231 GHz observed in the CALYPSO dataset show a secondary peak west of L1448-2A and along its equatorial plane, where a secondary source named L1448-2Ab was detected with the VLA \citep{tobin16a}. Hence, \citet{maury19} indicates L1448-2Ab as a candidate protostellar companion.
A velocity gradient orthogonal to the outflow has been detected in the molecular line emission by \citet{tobin18}.

The large-scale outflow cavities associated to L1448-2A have been mapped in CO ($2-1$) at moderate ($\sim 4\arcsec$) and high ($\sim 0\farcs5$) angular resolution by \citet{stephens17a} and \citet{kwon19}, respectively, and in scattered light by \citet{tobin07}. The modeling of the scattered light suggests moderate inclination: $i_{\rm jet} = 57\degr$ (with respect to the line of sight).

Our CALYPSO maps show the outflow cavities in CO ($2-1$) along PA$_{\rm outflow} = 118\degr$, in agreement with what was found by \citet{stephens17a} and \citet{kwon19}. SiO ($5-4$) emission is barely detected at $\sim 6\sigma$ and possibly reveals a jet driven by the protostellar companion, L1448-2Ab. Finally, compact and low-velocity ($\pm 3.5$ \kms\, with respect to systemic velocity) SO ($5_6-4_5$) emission is detected towards both the components L1448-2A and 2Ab.

\subsubsection{L1448-N}  
 L1448-N is a multiple low-mass protostellar system, composed by three dust continuum sources detected at 345, 231, and 94 GHz, i.e. L1448-NB, L1448-NW, and L1448-NA \citep{lee15,maury19}.
 L1448-NB is resolved into two components, NB1 and NB2, detected at 1.3~mm with the PdBI and ALMA, while the VLA at 8 mm suggests that NB2 itself is a binary, as well as L1448-NW \citep{maury19,tobin16a}.
 \citet{lee15} show that the three dust continuum sources, L1448-NB,  L1448-NW, and L1448-NA are associated with slow outflowing gas traced by CO ($2-1$) up to velocities of $\pm 10$ \kms\, with respect to systemic ($V_{\rm sys} \sim +4.9$ \kms).
 The CALYPSO observations do not show blue- and red-shifted emission associated to NA and NW, possibly due to the fact that they lie outside of the primary beam at 231 GHz and/or to the higher angular resolution of the CALYPSO observations.
 For L1448-NB, instead, CO ($2-1$) probes the large scale outflow previously detected by \citet{lee15}  -- for velocities $< 20$ \kms\, with respect to systemic velocity -- whereas SiO ($5-4$) is detected only at high-velocities (up to 40-50 \kms\, with respect to $V_{\rm sys}$) and probes, for the first time, the collimated jet (see the integrated line maps, PV diagrams, and line spectra in Figs. \ref{fig:jets1}, \ref{fig:PV-block1}, and \ref{fig:spec1}).
The red lobe of the jet is detected also in SO ($5_6-4_5$) and in the high-velocity CO ($2-1$) emission ($|V_{\rm LSR}-V_{\rm sys}|> 20$ \kms), while the jet blue lobe is detected only in SiO ($5-4$).

\subsubsection{L1448-C}

L1448-C is a well-known low-mass Class 0 protostar, first detected as a continum source in the radio \citep{curiel90}, then in the millimeter by \citet{bachiller91a}, \citet{barsony98}, \citet{maury10}.
The analysis of the CALYPSO data revealed a candidate disk in L1448-C, marginally resolved at $\sim 40-50$ au in the continuum \citep{maury19}, and associated with Keplerian rotation out to $\sim200$ au seen with the $^{13}$CO and C$^{18}$O ($2-1$) lines \citep{maret20}.

The protostar drives a powerful outflow first reported by \citet{bachiller90b}, then mapped at higher angular resolution in CO and SiO with PdBI by \citet{guilloteau92} and \citet{bachiller95}, showing terminal velocities up to $\pm 70$ \kms. 
\citet{girart01} first reported
proper motions of up to $\sim (0\farcs12 \pm 0\farcs6)$ yr$^{-1}$, which imply an outflow inclination with respect to the plane of the sky of $21\degr$$^{+18\degr}_{-6\degr}$, and a large deprojected velocity of $\sim 180\pm70$ \kms. Recent multi-epoch observations with the SMA at higher angular resolution ($\sim 0\farcs5$) report smaller proper motions of $\sim0\farcs06$ yr$^{-1}$ and $\sim0\farcs04$ yr$^{-1}$ for the blue- and red-shifted jet, respectively, hence larger inclination angles ($34\degr\pm4\degr$ and $46\degr\pm5\degr$, respectively) and smaller deprojected velocities of $98\pm4$ and $78\pm1$ \kms\, \citep{yoshida20}.
These values are consistent within the uncertainties with the value of $100$ \kms\, assumed to derive $\dot{M}_{\rm jet}$ for all the jets in our sample (see Sect. \ref{sect:jet-energetics} for a discussion). 

Our CALYPSO maps and position-velocity diagrams (see Fig. \ref{fig:jets1} and Fig. \ref{fig:PV-block1}) beautifully show the bipolar flow out to $\sim 15\arcsec$ distance with a nested structure, where CO ($2-1$) at low velocities (up to $15$ \kms\, with respect to $V_{\rm sys}$) probes the wide-angle outflow cavities, while SiO as well as high-velocity CO and SO emission (from 15 to 75 \kms\, with respect to $V_{\rm sys}$) trace the collimated jet. 
The onion-like structure of the flow is well illustrated in Fig. \ref{fig:jet-width_all} which show the width in the three tracers, while Fig. \ref{fig:jet-width-hv} shows that at high velocity CO is as collimated as SiO.
The presence of two well separated velocity components in the CO and SO emission is clearly visible also in the line spectra presented in Fig. \ref{fig:spec1}.
The estimated jet mass-loss rate and mechanical luminosity are $\sim 2.7 \times 10^{-6}$ \msolyr, and 2.2 \lsol, and are in agreement within a factor of 2 with the values estimated by \citet{yoshida20} from their SMA data (see the discussion on the uncertainty affecting the estimated jet parameters due to the assumption on the CO  abundance,  on \Vjet,  and  on  the  compression  factor in Sect. \ref{sect:jet-energetics}).

In the same field, 8$\arcsec$ (i.e., $2000$ au) southeast to L1448-C, \citet{maury19} detected a second continuum source at 94 and 231 GHz, L1448-CS, which is also detected in the mid-infrared (MIR) with {\it Spitzer} \citep{jorgensen06}. L1448-CS is brighter than L1448-C in the MIR, while it is weaker than it in the millimeter and submillimeter bands \citep[see, also, ][]{jorgensen07,maury10,hirano10}. Hence, \citet{maury19} classified it as a Class I source. 
Our CALYPSO maps in Fig. \ref{fig:jets1} clearly show that there is SO ($5_6-4_5$) and SiO ($5-4$) red-shifted emission associated with L1448-CS, along the direction almost perpendicular to that of the jet from L1448-C (PA$_{\rm jet} = +60\degr$). However, L1448-CS lies almost along the red-shifted lobe of the jet driven by L1448-C, and the faint and compact ($<2\arcsec$) emission from its red-shifted jet overlaps on it. Therefore, we were unable to derive the properties of the jet driven by L1448-CS.

\subsubsection{IRAS2A}

The IRAS2A1 protostar is located in the Perseus cluster known as NGC1333. A secondary source $\sim 0\farcs6$ south of IRAS2A1 is detected at 8~mm, 1~cm, and 4~cm with VLA \citep[VLA2, ][]{tobin15} and at ALMA waveleghts (Maury et al. in prep) but is not detected with our CALYPSO data \citep{maury19}.
A detailed analysis of the outflow, hot-corino, and rotation signature in the envelope based on the CALYPSO data are reported by \citet{codella14a}, \citet{maret14}, and \citet{maury14}.
In the past, two perpendicular outflows, directed NE-SW (PA$_{\rm outflow} \sim + 205\degr$) and SE-NW (PA$_{\rm outflow} \sim -75\degr$) have been detected in several tracers (e.g., CO, CS, SO, CH$_3$OH) using both
single-dishes and interferometers, and both apparently originating from IRAS2A1 \citep[e.g., ][]{bachiller98a,knee00,jorgensen04a,jorgensen04b,jorgensen07,wakelam05,persson12,plunkett13}. The CALYPSO observations in the three line tracers (CO, SO, and SiO) have been analyzed by \citet{codella14a}, who confirmed the occurrence of two jets, the brighter one along the NE-SW direction associated with IRAS2A1, and the fainter one along the SE-NW direction. The latter is possibly driven by the protostellar candidate VLA2, as suggested by \citet{tobin15}.
The maps in Fig. \ref{fig:jets1} show that the wider-angle CO outflow driven by IRAS2A1 extends out to $\sim 2500$ au distance from the driving protostar, while the collimated jet detected in SiO and SO extends only out to $1200$ au. The extracted spectra in Fig. \ref{fig:spec1} show that the emission in the two lobes is strongly asymmetric in all the tracers. The SiO emission in the blue-shifted lobe is $\sim 15$ times brighter than in the red-shifted one, and shows a low-velocity (LV: $V_{\rm LSR} - V_{\rm sys}$ up to $\sim -16$ \kms)) and a high-velocity component (HV: $V_{\rm LSR}  - V_{\rm sys} = -40, -16$ \kms), while only the LV component is detected in the red-shifted one (LV: $V_{\rm LSR} - V_{\rm sys}$ up to $\sim +13$ \kms).
As discussed by \citet{codella14a}, this indicates that on small scales ($<3\arcsec$) the jet from IRAS2A1 is intrinsically monopolar and intermittent in time, on dynamical scales less than 100 yr.

\subsubsection{SVS13}

SVS13A (Class I) and SVS13B (Class 0) are two protostars located in the NGC1333-SVS13 sub-cluster, which is composed of at least five objects, imaged, for example, by the CALYPSO PdBI continuum survey (\citealt{maury19}; see also, e.g., \citealt{looney00,tobin18}).
More specifically, SVS13A and SVS13B are the brightest ones in the mm-spectral window.
SVS13A is in turn a 0$\farcs$3 binary source (VLA4A, VLA4B; \citealt{anglada00}), not disentangled with the continuum CALYPSO images \citep{maury19}, while no multiplicity has been so far revealed for SVS13B.

In the jet context, SVS13A has been extensively analyzed in the CALYPSO paper by \citet{lefevre17}, where
it has been shown that the binary drives at least two extended jets pointing in slightly different directions (their position angles differ by $\sim 20\degr$, while their orientations differ only by $9\degr$ in 3D):
(i) a molecular jet with CO bullets and SiO and SO counterparts on small-scale (few arcsec); along with (ii) an atomic jet associated with the well known Herbig-Haro 7—11 chain. 

The SVS13B SiO jet has been first discovered by \citet{lefloch98} with the IRAM-30m, then mapped at higher angular resolution with the IRAM-PdBI by \citet{bachiller98b}, who reported bipolar emission in SiO ($2-1$) along PA$\sim 160\degr$ reaching velocities up to $-20$ \kms\, and $>+40$ \kms\, (in $V_{\rm LSR}$). In their maps, the blue lobe extends out $\sim 25\arcsec$, while the red lobe consists of a bright knot at $\sim 5\arcsec$ distance.
H$_2$ emission associated with the SVS13B jet was first reported at 2.12~$\mu$m by \citet{hodapp95}, then with {\it Spitzer} by \citet{maret09}, extending for more than $0.3$~pc in the south without appreciable loss of collimation \citep{bachiller98b}. 
The CALYPSO map of SiO ($5-4$) reveals the knotty structure of the collimated bipolar jet in the inner $8\arcsec$ along PA$_{\rm jet} = 167\degr$ (see Fig. \ref{fig:jets1}). While on large scales, the blue lobe of the SiO jet is much more extended than the red one \citep{bachiller98b}, when mapped on small scales the SiO jet is extremely symmetric, with three knots symmetrically displaced at $\sim 1\arcsec$, $5\arcsec$, and $7\arcsec$ distance along the blue and the red lobe. 
\citet{seguracox16} estimate a disk inclination of $\sim71\degr$. By deprojecting   
the peak radial velocities of the innermost SiO knots B and R reported in Table \ref{tab:fluxes} we obtain velocities of $74$ \kms\, and 49 \kms, respectively. However,  the maximum $V_{\rm rad}$ in our spectra may imply velocities up to 150 \kms.
Assuming an average deprojected velocity of $100$ \kms\, as in \citet{bachiller98b}, the inner three knots have dynamical timescales of 15, 70, and 100 years. The jet emission is also detected at 5 to $10\sigma$ in the CO and SO emission lines (see Fig. \ref{fig:jets1}).

\subsubsection{IRAS4A}

The IRAS4A protostellar system, located in the NGC1333-IRAS4 sub-cluster, is composed by two Class 0 objects IRAS4A1 and IRAS4A2  separated by about 1$\farcs$8 \citep[see, e.g., ][and references therein]{looney00,maury19, tobin18}.
The IRAS4A system is associated with an extended  arcminute-scale bipolar outflow observed in several tracers, such as CO and SiO \citep[e.g., ][and references therein]{blake95,lefloch98,choi01,choi06,yildiz12}. 

\citet{choi05} used SiO emission as observed with the VLA at 2$\arcsec$ spatial resolution to reveal two different jets driven by A1 and A2. Two southern blue-shifted lobes are well disentangled, while the northern red-shifted lobes are not easily distinguishable, being located on the same line-of-sight, and showing a peculiar bending toward NE at 20$''$ from the protostars. 
Only very recently, using S-bearing species emission imaged with IRAM-NOEMA, \citet{taquet20} were able to distinguish the A1 jet from the A2 one also in the northern lobe.
Thanks to the CALYPSO dataset, \citet{santangelo15} imaged the region at 1$\arcsec$ scale and characterised the jets: A1 drives a fast collimated jet associated with bright H$_2$ emission \citep[e.g., ][and references therein]{maret09}, while A2 is feeding a slower jet, with a large spatial extent and a S--shape pattern, probably due to precession.
Estimates of the inclination of the two jets are very uncertain. Due to the large extent of the collimated jets, and, at the same time, the high line-of-sight velocities \citet{yildiz12} suggest that the jets have an inclination of $\sim45\degr-60\degr$ to the line of sight. \citet{koumpia16} and \citet{marvel08} indicate larger inclination of $\sim70\degr$ and $\sim88\degr$, respectively.
Finally, \citet{santangelo15} suggest that the difference in velocity and extent of the two jets, with the jet driven by A1 being much faster ($V_{\rm rad} \sim -16$ \kms, and $+34$ \kms\, in the B and R knots) and much more extended (up to $4\arcmin$) than the jet driven by A2 ($V_{\rm rad} \sim 6-7$ \kms, extent of $\sim 20\arcsec$)  may be due to different inclinations.
 
\subsubsection{IRAS4B1}

IRAS4B is part of the multiple system NGC1333-IRAS4 \citep[see, e.g., ][]{looney00,tobin18,maury19}. 
The primary targeted by the CALYPSO observations is IRAS4B1, a Class 0 protostar, also named Per-emb-13. Our CALYPSO data revealed hot-corino emission (\citealt{maret20}; see also \citealt{sakai06,bottinelli07}) and rotation perpendicular to the jet direction on a few hundred au scale \citep{maret20}.

IRAS4B1 is associated with an outflow roughly located along the N-S direction, which has been first revealed using CO and CS by \citet{blake95}. Later on, \citet{choi01} imaged a symmetric bipolar outflow in HCN located along PA$_{\rm outflow} \sim 180\degr$, also detected in H$_2$CO, CH$_3$OH, SO, and SiO by \citet{jorgensen07}. Each lobe is $\sim 10\arcsec$ long, but the resolution of these previous observations (3$\arcsec$--4$\arcsec$) did not allow the authors to image the jet and measure its width and collimation.
The CALYPSO maps in Fig. \ref{fig:jets1} reveal for the first time the SiO collimated bipolar jet extending up to $\sim 4\arcsec$, i.e. 1200 au, along PA$_{\rm jet} \sim +165\degr$ with terminal projected velocities up to $-35$ \kms\, and $+45$ \kms\, with respect to systemic in the blue lobe and red lobe, respectively. At larger distances of $\sim 4\arcsec-8\arcsec$ (i.e., $1200-2400$ au), SiO traces two shock structures which have lower terminal velocities ($\pm 15$ \kms\, with respect to systemic). These are coincident with the structures detected in several molecular tracers by  \citet{choi01} and \citet{jorgensen07} along the N-S direction. These terminal shocks are symmetrically offset with respect to the PA of the inner SiO jet, indicating that the jet is precessing. 
As shown by the position-velocity diagrams in Fig. \ref{fig:PV-block1} and the jet width in Fig. \ref{fig:jet-width_all} the bulk of CO emission is at low velocity and less collimated than SiO, hence it probes the outflow. At high velocities, however, it is only slightly wider than SiO, therefore we assume that the bulk of the HV emission originates from the jet with little or no contamination from the outflow (see Fig. \ref{fig:jet-width-hv}). At difference with SiO and CO, SO emission is detected only in the terminal shock structures with terminal velocities of $\pm 15$ \kms\, with respect to systemic. The inner SiO knots are located at about $\sim 2\arcsec$ distance from source, indicating a very short dynamical timescale of $\sim 30$ years for an assumed deprojected velocity of $100$ \kms\, (which also implies that the jet is close to the plane of the sky, in agreement with the observations of maser outflows  by \citealt{marvel08}). However, if the jet lies close to the line of sight \citep[e.g., ][]{yildiz12,desmurs09}, the dynamical timescale may be a factor of $\sim 10$ larger.

\subsubsection{IRAS4B2}

The candidate protostellar object IRAS4B2, located 
$10\arcsec$ east to IRAS4B1, is detected at 94 and 231 GHz by \citet{maury19} (also called IRAS 4BE, IRAS 4B', IRAS4C, and IRAS 4BII; see \citealt{looney00}). At difference with IRAS4B1, this secondary source is not associated with COM emission \citep{desimone17,belloche20}, and is not detected at 70~$\mu$m in the Herschel HGBS maps \citep{sadavoy14}. Therefore, \citet{maury19} suggest that IRAS4B2 is either a very low luminosity source or a candidate first hydrostatic core.

The ejection activity of IRAS4B2 has been poorly characterized up to date. Recently, faint CO ($2-1$) and ($3-2$) bipolar emission possibly tracing an outflow has been reported in the context of the SMA MASSES survey \citep{lee-k16} and of the CARMA TADPOL survey \citep{hull14}. In both cases the outflow is poorly resolved because observations are at $2\farcs5-3\farcs5$ resolution.
Our CALYPSO maps show faint but well-collimated, bipolar emission with a knotty structure in all three tracers (CO ($2-1$), SO ($5_6-4_5$), and SiO ($5-4$)) along PA$_{\rm jet/outflow} \sim -99\degr$ with terminal velocities of up to $12$ \kms\, with respect to systemic. The detection of the outflow and jet is convincing and strongly supports the identification of IRAS4B2 as a low-luminosity protostar.

\subsection{Sources in Taurus}

The CALYPSO sample comprehend three sources located in Taurus ($d=140$ pc), the two very low luminosity objects (VeLLOs) IRAM04191 and L1521-F, and the prototype of the so-called warm carbon chain chemistry (WCCC) L1527.

\subsubsection{IRAM04191}

The IRAM 04191+1522 (here called IRAM04191) Class 0 source located in the southern part of the Taurus molecular cloud was first discovered by \citet{andre99}. After being observed with {\it Spitzer}, it has been classified as a very low luminosity object by \citet{dunham06} (\lint $=0.05$ \lsol).
The source is associated with a rotating infalling envelope detected in molecular lines by \citet{belloche02} and \citet{lee05}, and a bipolar outflow first imaged in CO ($2-1$) by \citet{andre99}.
\citet{lee02,lee05} and \citet{andre99} observations in CO ($1-0$), ($2-1$) show that the outflow extends up to $\sim 2\arcmin$ distance with velocities of up to $\pm 7$ \kms\, with respect to systemic, while strong shocks along it are probed by CS and HCO$^+$. 
Interestingly, when observed at higher angular resolution with CALYPSO only the red-shifted outflow lobe is detected in the CO ($2-1$) line up to $\sim 4\arcsec$ distance (i.e., $\sim 560$ au) with terminal velocities of $+13$ \kms\, with respect to systemic. Therefore, the outflow is monopolar on small scales.
No jet is detected in the SiO ($5-4$) and SO ($5_6-4_5$) lines.

\subsubsection{L1521-F}

The L1521-F core was mapped by \citet{mizuno94} and \citet{onishi99} in H$^{13}$CO$^+$, and by \citet{codella97} in NH$_3$ emission at 23.7 GHz. Then {\it Spitzer} observations in the mid-infrared revealed the presence of a very low luminosity object ($L_{\rm int} = 0.035$ \lsol) embedded in the core  \citep{bourke06}, also observed at mm wavelengths by \citet{maury10}. 
The {\it Spitzer} observations also show a bipolar nebula in scattered light roughly oriented east-west which may trace an outflow cavity.
The region is also associated with blue- and red-shifted emission seen in CO and HCO$^+$ with SMA and ALMA \citep{takahashi13,tokuda14,tokuda16}. However, the morphology of the molecular emission is very complex and far to be clearly understood. \citet{takahashi13} first claimed a compact and poorly collimated outflow detected in CO ($2-1$) along PA$=75\degr$, but \citet{tokuda14,tokuda16} suggest the opposite PA based on CO ($3-2$) and HCO$^+$ ($3-2$) emission.
Our CALYPSO map of CO ($2-1$) confirms complex structures with no clear displacement of the blue-shifted and red-shifted emission, thus following \citet{tokuda16}, we tentatively report PA$\sim +260\degr$ based on the displacement of the brightest blue-shifted structure detected south-west to the source.
No SiO ($5-4$) and SO ($5_6-4_5$) emission is detected in our maps.



\subsubsection{L1527}

L1527 is a Class 0/I protostar and is considered the prototype of warm carbon chain chemistry \citep{sakai14a}.
This is one of the two sources in the sample which is associated with a Keplerian rotating disk \citep{maret20}, first mapped by \citet{tobin12}. 
The disk, the rotating infalling envelope, and the transition between them have been mapped in several molecular tracers by \citet{sakai14a}, \citet{ohashi14}, \citet{sakai14b}.
A wide-angle and slow ($V_{\rm max} \sim 7$ \kms\, with respect to systemic) outflow extending out to $\sim 150\arcsec$ (i.e. $2 \times 10^4$ au) was first imaged in CO ($3-2$) by \citet{hogerheijde98}, then confirmed by an X-shaped structure seen at $\sim 1.5\arcsec$ scale by \citet{tobin12}.
Our observations at much higher angular resolution are clearly affected by filtering and no well-defined structure is observed in CO ($2-1$).
No SiO ($5-4$) emission is detected, while SO ($5_6-4_5$) probes the Keplerian disk as shown by \citet{maret20}.

\subsection{Sources in Serpens}
Our CALYPSO sample includes five sources located in the Serpens Main region ($d=436$ pc, \citealt{ortiz-leon17,ortiz-leon18b}), and three sources located in the Serpens South region ($d=350$ pc, Palmeirim et al., in prep).

\subsubsection{SerpM-S68N}
S68N \citep{mcmullin94}, also called Serp emb 8 or SMM9 is a Class 0 protostar located in the Northern part of the Serpens Main region. At about $10\arcsec$ North-East from S68N is located another Class 0 protostar, named S68Nb by \citet{maury19}, while previously classified as S68Nc by \citet{williams00} (same naming also in \citealt{dionatos10b}). This source is also called Serp emb 8 (N) \citep[eg.][]{hull14,tychoniec19}. Observations in the (sub)millimeter have revealed that both sources drive slow CO outflows \citep{hull14} and high-velocity molecular jets \citep{tychoniec19,aso19}. Regarding S68N, our CALYPSO maps show two symmetric SiO knots also detected in SO at a distance of $\sim 7''$ from the source in each lobe, indicative of a past ejection burst. 
The low velocity of the SiO/SO jet, together with the overlap between red-shifted and blue-shifted emission in each lobe suggest that the jet lies close to the plane of the sky. This probably explains why the CO counter parts of the SiO and SO knots are not clearly visible. Regarding S68Nb  a pair of symmetric knots in SiO, with no counterpart in SO, is detected at 7$''$ from the source. Surprisingly, the velocity of these knots is very different with the red-shifted knot emitting between 3 and 27 \kms, and the blue-shifted knot between -35 and -48 \kms\, with respect to the systemic velocity. This may point toward a jet misalignment at the emission of the bullet.


\subsubsection{SerpM-SMM4}
SerpM-SMM4 is a continuum condensation located in the central part of Serpens Main region \citep{casali93}. Subarcsecond observations in the submillimeter have spatially resolved this clump into two Class 0 protostars: SerpM-SMM4a, the brightest continuum source, and SerpM-SMM4b located at $4\arcsec$ South-West from the main source \citep{aso19, maury19}. Our CALYPSO maps probe the outflows emanating from both SMM4a and SMM4b, and the jet powered by SerpM-SMM4b, recently imaged by ALMA \citep{aso18}.

Regarding SerpM-SMM4a, a broad and faint CO outflow is detected in the Northern part whereas the Southern outflow remains undetected, in line with the observations by \citet{aso18}.
On the contrary, the fainter continuum source SerpM-SMM4b drives a bright molecular jet with the red-shifted part seen in CO ($2-1$) and SiO ($5-4$), while the blue-shifted one is seen also in SO ($5_6-4_5$), with projected velocities up to $\pm 40$ \kms\, with respect to $V_{\rm sys}$ in both lobes. Interestingly, the red-shifted and the blue-shifted lobes are not aligned and their PAs deviate by 35$^{\circ}$, a feature that is also observed in other Class 0 objects (IRAS4A). The presence of a pair of molecular bullets, symmetric with respect to the position of the source in the longitudinal PV diagram (Fig. \ref{fig:PV-block1}), suggests that both lobes are actually launched from the vicinity of the same source.
\citet{aso18} suggest that the two jet lobes have different inclination ($i\sim36\degr$ and $i\sim70\degr$ from  the line of sight for the blue and red lobe, respectively). This translates the estimated $V_{\rm rad}$ of $\sim -39$ \kms\, in knot B, and $+36$ \kms\, in knot R, in a jet velocity of $\sim105$ \kms\, and $\sim50$ \kms, respectively. These values imply that also the jet driven by SerpM-SMM4b is asymmetric in velocity ($V_{\rm rad, f}/V_{\rm rad, s} = 2.1$), as found for four other jets in our sample (SVS13B, L1157, IRAS4A1, and SerpM-S68Nb).

\subsubsection{SerpS-MM18}

SerpS-MM18 is a continuum condensation classified as a Class 0 source and located in the Serpens-South region of the Aquila rift complex \citep{maury11,gutermuth08}. Our CALYPSO continuum data have resolved this source as a binary system separated by $10\arcsec$: SerpS-MM18a, the brightest source in the millimeter continuum, and SerpS-MM18b \citep{maury19}.

\citet{plunkett15b} report low- and high-velocity CO ($2-1$) emission towards SerpS-MM18a, which probe the wide-angle outflow and collimated jet, respectively. In the CALYPSO survey, the spectacular jet is detected for the first time in the SO ($5_6-4_5$) and SiO ($5-4$) lines with both lobes extending from the source up to $\sim 20''$ in the three tracers. The relatively low velocity of the emission (up to $20$ \kms\, with respect to systemic, in agreement with \citealt{plunkett15b}), together with overlap between red-shifed and blue-shifted emission suggest that the jet lies close to the plane of the sky.

Moreover, we report the first detection of the jet emanating from SerpS-MM18b. We unveiled red-shifted emission in the CO ($2-1$) line extending to the North-East with a maximum velocity of $\sim50$ \kms\, with respect to systemic. Interestingly, no SiO emission is detected. SO is detected at low velocity ($<1.5$ \kms) in the inner 2$''$ around the continuum peak of SerpS-MM18b. The emission is elongated along the axis perpendicular to the jet and a transverse velocity gradient is detected. We thus conclude that SO traces the inner envelope or the disk associated to SerpS-MM18b.

\subsubsection{SerpS-MM22}
SerpS-MM22 is a Class 0 source located in the Serpens-South region of the Aquila rift complex \citep{maury11}. We report here the first detection of a CO bipolar outflow with velocities up to 10 \kms\, with respect to the systemic velocity in the inner 4$''$ around the driving source. Faint SO emission at systemic velocity is tentatively detected on the source and along the outflow direction. Here, SiO remains undetected.

\subsection{L1157}

The L1157-mm Class 0 protostar ($d = 352$ pc, \citealt{zucker19}) drives an episodic and precessing jet \citep{gueth96,podio16}, which created the archetypal chemically rich outflow \citep[e.g., ][]{bachiller01}.
The bright shocks along the bipolar outflow have been observed in several molecular tracers since \citet{gueth98}, but the jet has been elusive until the IRAM-30m observations of \citet{tafalla15}, who reported the detection of SiO high-velocity  emission in the inner $\sim 10\arcsec$ around L1157-mm supporting the occurrence of a jet. Finally, based on the CALYPSO maps, \citet{podio16} imaged the first $200$ au of the bipolar high-velocity protostellar jet. This allowed us to reveal for the first time the jet triggering the rich chemistry observed in the last ten years with a dramatic increase of the abundances of a large number of molecular species (such as CH$_3$OH, H$_2$O, H$_2$CO; e.g., \citealt{codella10,lefloch18}). 
The detected SO ($5_6-4_5$) emission is compact ($<1\arcsec$) around the protostar and shows a velocity gradient perpendicular to the jet direction, suggesting that the emission originates from the inner envelope or disk rather than the jet and outflow.
A detailed analysis of the high-velocity collimated jet detected in CO and SiO in the context of CALYPSO is presented by \citet{podio16}. In particular, the  S-shaped morphology and the radial velocities measured at the base and in the shocks along the L1157 outflow 
indicate that the jet precesses counter-clockwise on a cone inclined by $73\degr$ to the line of sight with an opening angle of $8\degr$, implying deprojected velocities of $\sim 87$ \kms\, and $\sim 137$ \kms\, in the blue and red lobe, respectively. Therefore, the jet of L1157 is asymmetric both in the morphology and in velocity, as discussed further in Sect. \ref{sect:kinematics}.

\subsection{GF9-2}

This source is located in the filamentary dark cloud GF9 (LDN 1082) and it was first cataloged as a transitional object between the pre-stellar and protostellar Class 0 phase by \citet{wiesemeyer99}.
Outflow signatures were first reported by \citet{furuya03}, who detected weak red-shifted CO emission and a weak H$_2$O maser. Recently, \citet{furuya19} reported the first resolved map of the outflow driven by GF9-2 in CO ($3-2$). The outflow PA, extent, and velocity inferred from the CO ($2-1$) map obtained with CALYPSO are in good agreement with what was estimated by \citet{furuya19}. The outflow is strongly asymmetric with the blue-shifted lobe which is less extended and slower ($<1\arcsec$, i.e., $<500$ au, $|V_{\rm LSR} - V_{\rm sys}$| up to $6$ \kms) than the red-shifted one ($\sim 8\arcsec$, i.e., $\sim 4000$ au, $|V_{\rm LSR} - V_{\rm sys}$| up to $16$ \kms).
No jet was detected in the SiO ($5-4$) line, while the SO ($5_6-4_5$) shows a velocity gradient roughly perpendicular to the jet direction, which suggests emission from the flattened envelope rotating about the jet axis. 


\end{appendix}

\end{document}